\documentclass[12pt,a4paper,twoside]{report}

\usepackage{a4,isolatin1,fancyhdr,german}
\usepackage{epsfig,psfrag}
\usepackage{amssymb,latexsym,mathrsfs}
\newfont{\hellenic}{cmmi10 at 12pt}
\newcommand{\lenis}{{\raisebox{0.7ex}{\symbol{45}}$\!\!$}}

\renewcommand{\thechapter}{\Roman{chapter}}
\renewcommand{\thesection}{\arabic{section}}

\newcommand{\chap}[3]{\chapter[$\,\,$ #1]{#2}\label{#3}{}\fancyhead[CE]{#1}}
\newcommand{\sekt}[3]{\section[#1]{#2}\label{#3}{}\fancyhead[CO]{#1}}

\newenvironment{pf}{\noindent{\bf Proof:}}{\newline\kbn}

\newcounter{zaehler} %dreckiger zeahler fuer zwischendurch

\newcommand{\theoname}{Theorem}
\newcommand{\corname}{Corollary}
\newcommand{\lemname}{Lemma}
\newcommand{\propname}{Proposition}
\newcommand{\defname}{Definition}

\newtheorem{theo}{\theoname}[chapter]
\newtheorem{cor}[theo]{\corname}
\newtheorem{lemma}[theo]{\lemname}
\newtheorem{prop}[theo]{\propname}
\newtheorem{defi}[theo]{\defname}

\newcommand{\Diff}{{\mbox{\it Diff}}}
\newcommand{\diff}{{\mbox{\it Vect}}}
\newcommand{\PSL}{{\mbox{\it PSL}}}
\newcommand{\PSU}{{\mbox{\it PSU}}}
\newcommand{\SL}{{\mbox{\it SL}}}
\newcommand{\SU}{{\mbox{\it SU}}}
\newcommand{\PSO}{{\mbox{\it PSO}}}
\newcommand{\SO}{{\mbox{\it SO}}}
\newcommand{\III}{{\mbox{\it III}}}
\newcommand{\opo}{{1\!+\!1}}
\newcommand{\opd}{{3\!+\!1}}

\newfont{\lokcal}{cmsy10}
\newcommand{\Vir}{{\mbox{{$\mathcal{V}$}{\it ir}}}}

\newcommand{\Einsop}{\leavevmode{\rm 1\mkern  -4.4mu l}}

\newcommand{\Seins}{\mathsf{S}^1}

\newcommand{\coset}{{\copyright{}}}
\newcommand{\smcoset}{\mbox{\scriptsize \copyright{}}}

\newcommand{\closure}[1]{{#1}^-}

\newcommand{\lok}[1]{{\mathcal #1}}
\newcommand{\Hilb}[1]{{\mathscr #1}}

\newcommand{\Geb}[1]{{\mathcal #1}}

\newcommand{\Name}[1]{{\sc #1}} 
\newcommand{\symb}[1]{{\mathsf #1}} 

\newcommand{\lie}[1]{{\mathfrak #1}}
\newcommand{\Loop}[1]{{\mathcal{L}#1}}
\newcommand{\luup}[1]{{l\lie{#1}}}

\newcommand{\dopp}[1]{{\mathbb #1}}

\newcommand{\komm}[2]{{\left[ #1 , #2 \right]}} 
\newcommand{\antkomm}[2]{{\left\{ #1 , #2 \right\}}} 
\newcommand{\ket}[1]{{\left| #1 \right\rangle}}
 
\newcommand{\norm}[1]{{\left\| #1 \right\|}} 
\newcommand{\betrag}[1]{{\left| #1 \right|}} 
\newcommand{\klammer}[1]{{\left( #1 \right)}}
\newcommand{\menge}[1]{{\left\{ #1 \right\}}}

\newcommand{\skalar}[2]{{\langle #1 , #2 \rangle }}

\newcommand{\dolch}[1]{{#1 \!\!\! /}}

\newcommand{\zendot}{{ \,\, .}}
\newcommand{\zencom}{{ \,\, ,}}

\newcommand{\kbn}{$\square$}

\title{Structure of \Name{Coset} Models}
\author{S\"oren K\"oster}
\begin{document}
%>>>>>>>>>>>>>>>>>>>>>>>>>>>>>>>>>>>>>>>>>>>>>>>>>>>>
%<<<<<<<<<<<<<<<<<<<<<<<<<<<<<<<<<<<<<<<<<<<<<<<<<<<<

%%%%%%%%%%%%%%%%%%%%%%%%%%%%%%%%%%%%%%%%
%%%% Titelseiten und dergleichen %%%%%%%
%%%%%%%%%%%%%%%%%%%%%%%%%%%%%%%%%%%%%%%%

%\input{starterkit.tex}

\begin{titlepage}
  \begin{center}
    \vspace*{\fill}
    {\Huge\bfseries Structure of Coset Models}
    \vfill
    \vfill
    \vfill
    Dissertation\\
    zur Erlangung des Doktorgrades\\
    der Mathematisch-Naturwissenschaftlichen Fakult\"{a}ten\\
    der Georg-August-Universit\"at zu G\"{o}ttingen\\
    \vspace{1cm}
    vorgelegt von\\
    \vspace{0.5cm}
    \textsc{S\"oren K\"oster}\\
    \vspace{0.5cm}
    aus\\
    Aachen\\
    \vspace{1cm}
    G\"{o}ttingen 2003\\
  \end{center}
\end{titlepage}

\thispagestyle{empty}
\vspace*{\fill}
\vfill
\noindent D 7\\
Referent: \textsc{Prof.~Dr.~Karl-Henning Rehren}\footnote{Institut
  f\"{u}r Theoretische Physik der Universit\"{a}t G\"{o}ttingen} \\
Korreferent:  \textsc{Prof.~Dr.~Detlev Buchholz}\footnote{Institut
  f\"{u}r Theoretische Physik der Universit\"{a}t G\"{o}ttingen}\\
Tag der m\"{u}ndlichen Pr\"{u}fung: 3. Juni 2003 
\vfill

\cleardoublepage

\thispagestyle{empty}

\vspace*{\fill}
\vfill

\begin{quote}
{\hellenic
  \symbol{6}\symbol{29}\symbol{23}\symbol{17}\symbol{18}\symbol{34}}$\acute{\mbox{\hellenic \symbol{19}}}${\hellenic \symbol{11}\symbol{38}}
 {\hellenic  \symbol{14}}$\acute{\mbox{\hellenic \symbol{17}}}$, 
 {\hellenic  \symbol{111}\hspace{0.3em} {$\!\!\!$\raisebox{0.7ex}{\symbol{127}}\raisebox{0.7ex}{\symbol{45}}$\!\!$}\symbol{19}\symbol{22}\symbol{11}\symbol{19}},
 {\hellenic
   \symbol{14}$\acute{\mbox{\hellenic \symbol{34}}}$\symbol{111}\symbol{19}\symbol{28}\lenis{\hspace{0.3em}}}
 {\hellenic  \hspace{0.3em} {\raisebox{0.7ex}{\symbol{45}}$\!\!$}$\acute{\mbox{\hellenic \symbol{11}}}$\symbol{23}},
 {\hellenic  \symbol{34}\lenis\symbol{19}}
 {\hellenic
   \symbol{22}$\acute{\mbox{\hellenic \symbol{34}}}$\symbol{21}\symbol{21}\symbol{111}\symbol{19}} 
 {\hellenic  \symbol{28}$\grave{\mbox{\hellenic \symbol{11}}}$}
 {\hellenic  \hspace{0.3em} {\raisebox{0.7ex}{\symbol{45}}$\!\!$}$\acute{\mbox{\hellenic \symbol{11}}}$\symbol{23}\symbol{33}}
 {\hellenic  \hspace{0.3em} {\raisebox{0.7ex}{\symbol{45}}$\!\!\!\;$}$\acute{\mbox{\hellenic \symbol{111}}}$\symbol{32}\symbol{34}\symbol{27}\symbol{18}\symbol{11}\symbol{19}}.
% {\hellenic  \symbol{}\symbol{}\symbol{}\symbol{}\symbol{}\symbol{}\symbol{}}

 {\hellenic
   \symbol{5}\symbol{111}\symbol{21}\symbol{19}\symbol{28}\symbol{34}\symbol{19}\symbol{11}\symbol{38}}
{\hellenic
  \symbol{90}} \cite[(516a4)]{Ppol}.
%\vspace{1ex}
%
% Gew\"ohnung also, meine ich, wird er n\"otig haben um das obere zu
% sehen.
%
% Politeia VII (516a4) \cite{Ppol}.
\end{quote}

\vspace*{\fill}
\vfill

\clearpage

%\chapter*{}
%  \label{chap-abstract}
\thispagestyle{empty}
%\addcontentsline{toc}{chapter}{{\glqq{}Zusammenfassung\grqq{} and Abstract}}

\vspace*{\fill}
\vfill
\noindent {\bf Zusammenfassung:} 
Wir untersuchen Einbettungen 
lokaler, chiraler, konformer  Quantentheorien $\lok{C}\subset\lok{B}$,
die mit einer gegebenen Untertheorie $\lok{A}\subset\lok{B}$ vertauschen;
die Untertheorien $\lok{C}\subset\lok{B}$ werden als
\Name{Coset}-Modelle bezeichnet. %2 
Die meisten Ergebnisse dieser Arbeit sind modellunabh\"angig, jedoch
wird diese Untersuchung motiviert durch  Einbettungen von
Stromalgebren und deren \Name{Coset}-Modelle. %3 

Wir zeigen, dass es zu jeder gegebenen Untertheorie %4
$\lok{A}\subset\lok{B}$ eine eindeutige, innere Darstellung
$U^\lok{A}$ gibt, die die konforme Symmetrie auf der
Untertheorie verwirklicht. Die lokalen %5
beobachtbaren Gr\"o{\ss}en von $\lok{B}$, die mit $U^\lok{A}$
vertauschen, bilden das maximale \Name{Coset}-Modell $\lok{C}_{max}$.

Unter der Annahme, dass $U^\lok{A}$ durch Integrale eines %6
$\lok{A}$ innewohnenden Quantenfeldes erzeugt wird, zeigen wir: Die %7
Einbettung der Untertheorie und ihrer 
\Name{Coset}-Modelle steht in unmittelbarer Analogie zur Inklusion
chiraler Theorien in einer $\opo$-dimensionalen lokalen,
konformen Quantentheorie. Die lokalen beobachtbaren Gr\"o{\ss}en %8
 von $\lok{C}_{max}$ zu einem bestimmten Gebiet sind
gerade diejenigen Gr\"o{\ss}en von $\lok{B}$, die mit den lokalen
beobachtbaren Gr\"o{\ss}en der Untertheorie $\lok{A}$ zu demselben
Gebiet vertauschen.

Wir geben einige Anwendungen unserer Ergebnisse und diskutieren m\"ogliche %9
Verallgemeinerungen unserer Vorgehensweise.
 
\vspace{3ex}

{\bf Abstract:} 
We study inclusions of local, chiral, conformal %1
quantum theories $\lok{C}\subset\lok{B}$ which commute with a
given subtheory $\lok{A}\subset\lok{B}$. These subtheories %2
$\lok{C}\subset\lok{B}$ are called 
{\em Coset models}. 
Most of our results are model-independent, although our analysis is %3
motivated by the inclusions of current
algebras and their \Name{Coset} models.

We prove that to every given $\lok{A}\subset\lok{B}$ there is a unique, %4
inner representation $U^\lok{A}$ which
implements conformal symmetry on the subnet. The local observables of %5
$\lok{B}$ which commute with $U^\lok{A}$ form the maximal \Name{Coset}
model $\lok{C}_{max}$. 

Assuming $U^\lok{A}$ to be generated by integrals of a quantum field %6
affiliated with the subnet $\lok{A}\subset\lok{B}$, we show: The %7
inclusion of the subnet and of its \Name{Coset} models 
is directly analogous to the inclusion of chiral observables in a
local, conformal theory in $\opo$ dimensions. The local observables of %8
the maximal \Name{Coset} model associated with a given region are
found to be characterised by their commuting with the local
observables of $\lok{A}$ associated with the very same region.

We give applications and  discuss possible generalisations of our %9
methods.
\vspace*{\fill}
\vfill

%\addcontentsline{toc}{chapter}{{\glqq{}Zusammenfassung\grqq{} and  Abstract}}

\cleardoublepage

\pagenumbering{roman}

\tableofcontents
\fancyhead[RE]{}
\fancyhead[LO]{}

%%%%%%%%%%%%%%%%%%%%%%%%%%%%%%%%%%%%%%%%%%%%%%%%%%%%%%%%%
%%%% ENDE Starterkit %%%%%%%%%%%%%%%%%%%%%%%%%%%%%%%%%%%%
%%%%%%%%%%%%%%%%%%%%%%%%%%%%%%%%%%%%%%%%%%%%%%%%%%%%%%%%%

%\input{Intro.tex}
%%%%%%%%%%%%%%%%%%%%%%%%%%%%%%%%%%%%%%%%%%%%%%%%%%%%%%%%%%
%% EINLEITUNG %%%%%%%%%%%%%%%%%%%%%%%%%%%%%%%%%%%%%%%%%%%
%%%%%%%%%%%%%%%%%%%%%%%%%%%%%%%%%%%%%%%%%%%%%%%%%%%%%%%%%
\chap{Introduction}{Introduction}{cha:intro}
\fancyhead[CO]{Introduction} %header centre, odd pages
\fancyhead[RE]{\thechapter}  %header right, even pages
\fancyhead[LO]{\thechapter}  %header left, odd pages
\pagenumbering{arabic}

%\chapter[Introduction]{Introduction}
%\label{cha:intro}

The first quarter of the twentieth century saw two great revolutions
is physics: the uncovering of quantum physics and the discovery of
relativity. The quest for unifications of both in systems with
infinitely many degrees of freedom, the development of {\em
  relativistic quantum field theories}, has led to remarkable
successes, especially in elementary particle
physics. The description of the physical laws valid in this context
and our present understanding of the fundamental interactions, mainly
summarised in the standard model for the strong, weak and
electro-magnetic forces, has not yet reached the stage of a
mathematically consistent theory. On the other hand, the underlying
physical principles are clear:

 The primary objects are expectation values of observable quantities. Not
all observables are {\em compatible}, ie some sets of observables can
not be measured simultaneously with arbitrary accuracy, 
like eg position and momentum due to \Name{Heisenberg}'s
uncertainty relation. The laws of physics are independent of the
choice of frame of reference, ie they may be formulated covariantly
with respect to the relativistic spacetime symmetry group. The
principle of {\em locality} states that it is meaningful to talk of
observables which can be measured in bounded spacetime regions and
that observables are always compatible, if they are spacelike
separated.  

There are two concise mathematical frameworks which capture possible
ways to formulate relativistic quantum field theories. In the first, one
assigns to points $x^\mu$ in spacetime operator-valued distributions
$\phi(x^\mu)$, the {\em quantum fields}, which may be viewed as quantised
versions of the field strengths known from classical field
theories. The principles of relativistic quantum field theory are
reflected in certain requirements on the quantised fields,
mostly known as \Name{Wightman}'s axioms \cite{GW65}. Successes of
this framework are summarised eg in \cite{rJ65,SW64,iT65,BLT75}.

Another approach was proposed by \Name{Haag and Kastler}
\cite{HK64}. Here, one selects regions $\Geb{O}$ in spacetime and
assigns to them  (topological $*$-) algebras $\lok{A}(\Geb{O})$ of
bounded operators. The {\em local algebras} $\lok{A}(\Geb{O})$ are
regarded as the algebras of  observations possible within the region
$\Geb{O}$ and their elements are called the {\em local
  observables}. The principles of relativistic quantum field theory
can be expressed in this setting in a natural manner, leading to the
framework of {\em local quantum physics}. This way of describing
relativistic quantum field theory can be regarded as an extension of
the picture of quantum fields and it has the advantage that one does
not need to deal with particular sets of ``coordinates'' of the theory
(the quantum fields). Especially structural and conceptual problems can be
discussed within the setting of local quantum physics successfully (cf
\cite{rH92,dB00}).      

The physics of a local quantum theory $\lok{A}$ is mainly encoded in
the {\em isotony} inclusions of local algebras: if a region
$\Geb{O}_1$ is contained in another region $\Geb{O}_2$ the property of
isotony means that the local algebra $\lok{A}(\Geb{O}_1)$ is
contained in the algebra $\lok{A}(\Geb{O}_2)$.  Put differently:
isotony says 
that the amount of possible observations increases with the
localisation region they can be made in. This appears to be an almost
trivial statement, but the isotony inclusions give structure of
physical relevance to the local algebras which, ignoring this
substructure, are (essentially) all isomorphic on quite general
grounds \cite{BDF87}. By isotony, a local quantum theory is a {\em net of
  local algebras}.

Within the framework of \Name{Haag and Kastler} it is natural to
consider subtheories, ie inclusions of one local quantum theory,
$\lok{A}$, in another one, $\lok{B}$, given by the inclusion of their
local algebras: $\lok{A}(\Geb{O}) \subset\lok{B}(\Geb{O})$. The origin
of these inclusions is encoded in two properties: they are covariant
with respect to the action 
of the spacetime symmetry group and they are consistent with the
net-structure of the subtheory, ie for $\Geb{O}_1\subset\Geb{O}_2$ we
have: $\lok{A}(\Geb{O}_1)\subset\lok{A}(\Geb{O}_2)$.

Obviously, there are non-trivial subtheories like, for example,
observables $\lok{A}$ included as gauge invariants in a local
quantum theory $\lok{F}$ with a compact gauge symmetry. This example
gives one motivation for 
studying subtheories:  By  the
\Name{Doplicher-Roberts} reconstruction \cite{DR90} all the relevant
information is in fact encoded in $\lok{A}$ already. This perspective
is taken eg in \cite{hA92}.

One might expect
that the interaction between any subset of observables with the
ambient system does not admit the subset to give rise to a local
quantum theory essentially different from the whole theory. One could
argue, for 
example, that local observables associated with energy and momentum
already generate the whole net $\lok{A}$ and that the remaining
freedom is taking local extensions like the local quantum theory
$\lok{F}$ above (cf \cite{sD92,rC95}).

Actually, it seems that in most studies on subtheories
$\lok{A}\subset\lok{B}$ the ``energy content'' of $\lok{B}$ is already
contained in $\lok{A}$. In this work we are interested in cases
where this is {\em not} the case. Rather, we want to look at the
situation where the inclusion of $\lok{A}$ in $\lok{B}$ leaves enough
space for other subtheories $\lok{C}\subset\lok{B}$ which commute with
all of $\lok{A}$, ie the local subalgebras $\lok{C}(\Geb{O})$ fulfill:
\begin{displaymath}
  \lok{C}(\Geb{O}) \subset \lok{A}(\Geb{O})' \cap\lok{B}(\Geb{O}) =:
  \lok{C}_{\Geb{O}} \zendot 
\end{displaymath}
Such subtheories we call \Name{Coset} {\em  models associated
  with $\lok{A}\subset\lok{B}$}, and the algebras $\lok{C}_{\Geb{O}}$
are called the {\em local relative commutants}. Admittedly quite
trivial examples of this
structure are tensor products of local quantum theories defined by
$\lok{B}(\Geb{O}):= \lok{A}(\Geb{O})\otimes \lok{C}(\Geb{O})$, but
there are examples which are less simple.

We are interested in typical \Name{Coset} models $\lok{C}$, in a
description of the relative position of $\lok{A}$, $\lok{C}$ in the
ambient theory $\lok{B}$, and in objects naturally associated with
these inclusions. For example, one may ask immediately: Is there a
maximal \Name{Coset} model and how is it characterised? The local
relative commutants $\lok{C}_{\Geb{O}}$ are upper bounds for the local
algebras of all \Name{Coset} models, and if they define a \Name{Coset}
model themselves, then this is maximal, obviously.

As covariance is automatic, one has to show that
the $\lok{C}_{\Geb{O}}$ increase with $\Geb{O}$. Since for
$\Geb{O}_1\subset\Geb{O}_2$ we have
$\lok{A}(\Geb{O}_1)'\supset\lok{A}(\Geb{O}_2)'$, we need arguments
which ensure that this inclusion is inverted upon intersection with
$\lok{B}(\Geb{O}_1)$ and $\lok{B}(\Geb{O}_2)$, respectively. For a
general subtheory $\lok{A}\subset\lok{B}$ it is not obvious how to
obtain such arguments, and this task will be
referred to as the {\em isotony problem}.

Results of \Name{Carpi and Conti} \cite{CC01} indicate that in $\opd$
dimensions the structure of subnets essentially reduces to the tensor 
product scenario $\lok{A}\subset\lok{B} = \lok{A}\otimes \lok{C}$
(under some additional assumptions). In these cases the isotony
problem is absent and, in fact, it had not to be discussed explicitly
in \cite{CC01}.   

In two dimensions, however, we know of many subtheories
$\lok{A}\subset\lok{B}$ which have interesting \Name{Coset} models and
a less simple relative position for the pair $\lok{A}$, $\lok{C}$ in
$\lok{B}$. Characteristic for these examples is their high spacetime
symmetry: they are not only covariant with respect to scale
transformations $x^\mu\mapsto \lambda x^\mu$, but in fact covariant
with respect to the whole stabiliser group of light-like directions,
the {\em conformal group}. Conformal symmetry facilitates analysing
such models a great deal. A physicist
interested in relativistic 
quantum field theory has to ask for reasons justifying investigations
in this setting. 

The classical equations of motion of a typical massless field theory
(containing no dimensional parameters),
like \Name{Maxwell}'s 
equations, are not only \Name{Poincar\'e} and scale invariant, but
actually invariant with respect to the whole conformal
group. Initially, the hope was that in the high energy limit of a
quantum field theory, where masses of particles  do not make much of a
difference, there holds an asymptotic conformal symmetry. The
appearance of renormalisation scale dependent anomalous dimensions  
showed that scale symmetry (and {\em a posteriori} conformal symmetry)
could not hold at the quantum level generally, but it could be valid
at values of the coupling constant stable under the action of the
renormalisation group (cf eg \cite{FST89,iT82} and references
therein). 

Renewed interest arose when
\Name{Belavin, Polyakov and Zamolodchikov} \cite{BPZ84} took
seriously the connection between conformally covariant quantum field
theories in $\opo$ dimensions and two-dimensional statistical systems
undergoing a second order phase transition, like eg the critical
\Name{Ising} model, and exhibited how conformal symmetry in low
dimensions allows one to obtain soluble models. And then, much of the
 work undertaken in string theory  was made possible by
 the essential part that conformal symmetry in $\opo$ dimensions was
 able to play in this context as well (cf \cite{GSW87}). 

From the perspective of local quantum physics there are some points of
special interest: many of the soluble models can be shown to fit into
this framework and thus they admit valuable case studies, which may teach us
something about situations in four dimensions. Moreover, some new
structures arise in this context, like non-trivial braid group
statistics which is realised eg by quasi-particles in the fractional
\Name{Hall} effect. Both aspects led and lead to further insights into
the structures of relativistic quantum field theory (cf eg \cite{iT00,iT94}). 
  
The models which inspire this work are generated by conserved
currents in $\opo$ dimensions, $j^{a\mu}$.  The currents transform in
the adjoint representation of a compact, global gauge group, which is
indicated by the label $a$. The \Name{Lie} algebra $\lie{g}$ of the
gauge group determines, up to the normalisation of the current
two-point functions, the commutation relations between the currents
and hence the whole model, called the {\em current algebra}
$\lok{A}_{\lie{g}}$. By an inclusion of compact \Name{Lie} algebras, 
$\lie{h}\subset\lie{g}$, one gets an inclusion of quantum field
theories, the {\em current subalgebra}
$\lok{A}_{\lie{h}}\subset\lok{A}_{\lie{g}}$. As many conformal fields
in $\opo$  dimensions, the currents decompose into two commuting parts, each
depending on one light cone 
coordinate, $t\pm x$, only; these generate the {\em chiral current
  algebras}. Hence, it suffices to look at the chiral inclusions only.

Infinitesimal coordinate transformations are
implemented by a quadratic function of the currents, the stress-energy tensor
$\Theta^\lie{g}$. Since the current subalgebra possesses its own
stress-energy tensor, $\Theta^\lie{h}$, the field
$\Theta^\lie{g}-\Theta^\lie{h}$ commutes with all of
$\lok{A}_\lie{h}$. $\Theta^\lie{g}-\Theta^\lie{h}$ is a stress-energy
tensor itself, generically it does
not vanish and hence generates a non-trivial \Name{Coset} model
associated with the current subalgebra
$\lok{A}_{\lie{h}}\subset\lok{A}_{\lie{g}}$.
$\Theta^\lie{g}-\Theta^\lie{h}$  is called the \Name{Coset} {\em
  stress-energy tensor} and subtracting $\Theta^\lie{h}$ from
$\Theta^\lie{g}$ is usually referred to as the \Name{Coset} {\em
  construction}. 

\Name{Goddard, Kent and Olive} \cite{GKO86} used the \Name{Coset}
construction to obtain the {\em discrete series} of chiral
stress-energy tensors, which are of special interest due to a
remarkable classification result \cite{FQS86}. In fact, it proved
possible to determine the features of the local quantum theories
generated by these stress-energy tensors  by
studying the respective current algebras
$\lok{A}_\lie{h}$, $\lok{A}_\lie{g}$ and the relative position of
$\lok{A}_\lie{h}$ in $\lok{A}_\lie{g}$ \cite{tL94, KL02}. This was made
possible, to a large extend, by works of \Name{Xu} (\cite{fX00,fX99}
in particular), which are complementary to this work as they deal with
particular inclusions of current algebras and their \Name{Coset}
models in the setting of  local quantum physics.

More generally, one of the driving forces in the investigations on
current subalgebras and \Name{Coset} models was the interest in
obtaining new models. A natural thing to do is to search for
additional fields in the current algebra $\lok{A}_\lie{g}$ which
commute with all 
of $\lok{A}_\lie{h}$, and to determine the algebra generated by
these. This approach leads to {\em $\mathcal{W}$-algebras}. We will 
discuss \Name{Coset} models from the perspective of local quantum
physics and therefore we do not want to refer to particular field
coordinates in \Name{Coset} models. We will not comment further on the
interesting achievements obtained on $\mathcal{W}$-algebras or on the
related contributions to the understanding of current subalgebras and
their \Name{Coset} models from this side; we rather refer to
\cite{BS93,EHH93,jF97,gW97} and references therein.

In the following, we want to broaden the perspective, to achieve clarity
 through a search for intrinsic structures and to shed new
light on old problems. We intend to work in the spirit of the
{\em general theory of quantised fields}, which ``... analyzes the
notions which are at the 
basis of all previously analyzed models'',  as \Name{Res Jost}
captured it \cite{rJ65}.

The next chapter provides a summary on chiral
conformal quantum field theory in the sense of \Name{Haag and
  Kastler}. We will give a technical formulation of the questions
raised in this introduction, state and discuss our main assumptions and obtain
first, completely model independent results. In particular, we will
see that for any chiral subtheory $\lok{A}\subset\lok{B}$ there is a
unique implementation $U^\lok{A}$ of conformal
transformations on $\lok{A}$ by unitary operators which are affiliated with
$\lok{A}$ in a global sense; this {\em inner-implementing
  representation} $U^\lok{A}$ is the natural abstract counterpart of
the stress-energy tensor of a current subalgebra. We prove that the
local operators in $\lok{B}$ which commute with $U^\lok{A}$ form the
maximal \Name{Coset} model $\lok{C}_{max}$ associated with
$\lok{A}\subset\lok{B}$. 

Thus, we are led to study subnets and their \Name{Coset} models by
analysing the action of the inner-implementing representation
$U^\lok{A}$ of the subtheory $\lok{A}$ on the local observables of the
ambient theory $\lok{B}$. This ansatz is new and the essence of this
work. In particular, the isotony problem may be dealt with in a
natural manner.

The chiral current subalgebras $\lok{A}_\lie{h}\subset\lok{A}_\lie{g}$
are discussed as examples satisfying our
assumptions in chapter \ref{cha:currsub}. We give the relation between
the inner implementation $U^{\lok{A}_\lie{h}}$ of
conformal transformations and the stress-energy tensor
$\Theta^\lie{h}$. This relation allows to establish some special
properties of $U^{\lok{A}_\lie{h}}$ (lemmas \ref{lem:autUA},
\ref{lem:isoUA} in chapter \ref{cha:netend}), which we believe to hold
in general.  
We take the presence of a stress-energy tensor as the
{\em Additional Assumption} for the subsequent analysis.         

Chapter \ref{cha:netend} constitutes the main part of this work: It
provides a solution of the isotony problem and thus establishes the
identity of the local algebras of the maximal \Name{Coset} model
$\lok{C}_{max}$ and 
the local relative commutants of a chiral subtheory
$\lok{A}\subset\lok{B}$. This way, we prove that the local 
operators in the ambient theory $\lok{B}$ which are contained in some
\Name{Coset} model may be characterised entirely in terms of local
data, namely their commutativity with the respective local operators
of the subtheory $\lok{A}$. Hence, all \Name{Coset} models are of a
{\em local nature}.

It is shown that through the action of $U^\lok{A}$ on the
chiral theory $\lok{B}$ one may construct
a $\opo$-dimensional {\em quasi-theory} $\lok{B}^\opo$ containing the
original net $\lok{B}$ as ``time zero subtheory''; this construction
may be regarded as {\em chiral holography}. Taking
$\lok{A}_{max}$ as the subtheory of $\lok{B}$ 
consisting of all local observables on which $U^\lok{A}$ implements
the conformal symmetry, $\lok{B}^\opo$ contains $\lok{C}_{max}$ and
$\lok{A}_{max}$ as subtheories of chiral observables, each  depending on one
light cone coordinate in the {\em holographic} $\opo$-dimensional
spacetime only. This establishes $\lok{A}_{max}$ as a natural object
connected with a subtheory $\lok{A}\subset\lok{B}$ and gives a
straightforward interpretation for chiral  
subtheories and their \Name{Coset} models as fixed-points with respect
to a classical spacetime symmetry in a suitably enlarged ambient
theory.

Chapter \ref{cha:finind} is devoted to making contact with related
works in the field. We spell out how the inclusion of a subtheory
$\lok{A}\subset\lok{B}$ and its \Name{Coset} models $\lok{C}$ may be
seen as  {\em localised representations} of the
tensor product of 
the subtheories in the sense of \Name{Doplicher, Haag and
  Roberts}. This formulation is used to get some more insights,
mainly under the assumption that the localised representation has
finite statistics. Moreover, we give a \Name{Coset} construction for some
normal {\em canonical tensor product subfactors} as introduced by
\Name{Rehren} \cite{khR00}, when we revisit current subalgebras.

The limitations of the additional assumption on the presence of a
stress-energy tensor are the issue of chapter \ref{cha:noset}: we prove
that a class of models do not contain a stress-energy tensor. 
The concluding chapter briefly summarises the main results and gives
an outlook on possible generalisations of the discussions
given here. The appendix contains a few technical lemmas and
miscellaneous results.   
%%%%%%%%%%%%%%%%%%%%%%%%%%%%%%%%%%%%%%%%%%%%%%%%%%%%%%%%
%%%%% ENDE EINLEITUNG %%%%%%%%%%%%%%%%%%%%%%%%%%%%%%%%%%
%%%%%%%%%%%%%%%%%%%%%%%%%%%%%%%%%%%%%%%%%%%%%%%%%%%%%%%%

%\input{CosPa.tex}
%%%%%%%%%%%%%%%%%%%%%%%%%%%%%%%%%%%%%%%%%%%%%%%%%%%%%%%%%%
%% KAPITEL COSETPAIRS %%%%%%%%%%%%%%%%%%%%%%%%%%%%%%%%%%%%
%%%%%%%%%%%%%%%%%%%%%%%%%%%%%%%%%%%%%%%%%%%%%%%%%%%%%%%%%%
\chap{{Coset} pairs of chiral subtheories}{{Coset} pairs of
chiral subtheories}{cha:cospa} 

%%%%%%%%%%%%%%%%%%%%%%%%%%%%%%%%%%%%%%%%%%%%%
%%%%%% ADJUSTING HEADER %%%%%%%%%%%%%%%%%%%%%
%%%%%%%%%%%%%%%%%%%%%%%%%%%%%%%%%%%%%%%%%%%%%

%\fancyhead{} % clears all
\fancyhead[LE]{\thepage} %header left, even pages
%\fancyhead[CE]{\chaptermark} %header middle, even pages
\fancyhead[RE]{\thechapter.\thesection} %header right, even pages
\fancyhead[LO]{\thechapter.\thesection} %header left, odd pages
%\fancyhead[CO]{} %header middle, odd pages 
\fancyhead[RO]{\thepage} %header right, odd pages

%\chapter[\Name{Coset} pairs]{\Name{Coset} pairs of chiral conformal subtheories }
%\label{cha:cospa}

%%%%%%%%%%%%%%%%%%%%%%%%%%%%%%%%%%%%%%%%%%%%%%%%%%%%%%%%%%%
%%%%%%%%

We introduce conformal quantum field theories in one
chiral dimension, their subnets and the associated \Name{Coset}
models, which are the objects of this work. We discuss our main
assumptions and obtain model
independent results. In particular, section \ref{sec:bosug} is devoted
to the construction of a representation $U^\lok{A}$ which 
implements chiral conformal symmetry on any given chiral  subtheory
$\lok{A}\subset\lok{B}$, is affiliated with
$\lok{A}$ in a global sense and uniquely determined by these two
properties. This representation forms the foundation 
of the subsequent analysis.

\sekt{Assumptions, conventions, and notions}{Assumptions, conventions,
  and notions}{sec:basic} 
% \section{Assumptions, conventions, and notions}
% \label{sec:basic}

%%%%%%%%%%%%%%%%%%%%%%%%%%%%%%%%%%%%%%%%%%
%%%%%% INTRO %%%%%%%%%%%%%%%%%%%%%%%%%%%%%
%%%%%%%%%%%%%%%%%%%%%%%%%%%%%%%%%%%%%%%%%%

\subsection{Chiral conformal symmetry}
\label{sec:introcongeo}

Conformal transformations of spacetime are required to leave the
\Name{Lorentz} metric invariant up to a \Name{Weyl} scaling. In $\opo$
dimensions, this group is infinite-dimensional and contains localised
diffeomorphisms. In positive-energy representations, only a
finite-dimensional subgroup of this symmetry group remains unbroken,
the 
group of {\em global  conformal transformations}. If one restricts
attention to this subgroup, the structures are similar to the situation 
  of conformal symmetry in higher dimensions \cite{BGL93}. When we
  speak of conformal symmetry in the following, we mean symmetry with
  respect to global conformal transformations.

Conformally covariant  models are given as local nets on \Name{Minkowski} 
space. In this context, conformal transformations have an action  through
mapping the  observables associated with a given localisation region to the
 observables of the transformed region, as long as this is 
 contained in \Name{Minkowski} space. Following general arguments
 these nets may be extended to theories over  the  {\em conformal
covering} of \Name{Minkowski} space, on which the
conformal transformations  act as a proper spacetime symmetry
group. The extensions then form {\em conformal quantum field
  theories}. This
scheme has been discussed for 
the general case by \Name{Brunetti, Guido and Longo} \cite{BGL93}
(cf\ \cite{iS71}, \cite{LM75}) and
for the chiral case by \Name{Fredenhagen and J\"or{\ss}} \cite{FJ96}.

In $\opo$ dimensions, the light-cone (at the origin) consists of the
two (chiral) light-rays, which  provide the right and 
the left light-cone coordinates on \Name{Minkowski} space. 
Each conformal transformation in $\opo$ dimensions may be
represented as a product of two commuting chiral conformal
transformations, which for themselves act on one light-cone coordinate
only. 

Chiral conformal quantum field theories arise in $\opo$-dimensional
models as subsystems of observables which are invariant 
with respect to chiral coordinate transformations on the other
light-cone coordinate. Examples of this structure are stress-energy
tensors, the $U(1)$ current and its conformally covariant derivatives 
 (cf chapters \ref{cha:currsub} and \ref{cha:noset}). 
In the following we restrict our attention to one chiral sector.

The conformal
transformations of the chiral light-ray, 
$\dopp{R}$, form a group isomorphic to $\PSL(2,\dopp{R})$, which is the
factor group of $\SL(2,\dopp{R})$ modulo the equivalence relation
$\symb{A}\sim -\symb{A}$. Such a matrix $\symb{A}=\left( {a \atop c}
  {b \atop d}\right)$ acts on the light-cone coordinate $x\in\dopp{R}$
as follows:
\begin{equation}
  \label{eq:pslaction}
  g_{\symb{A}} x = \frac{ax+b}{cx+d}\zendot
\end{equation}

Some one-parameter subgroups of $\PSL(2,\dopp{R})$ of particular
interest are the {\em translations}, $T(a) x = x+a$, {\em scale
  transformations}, 
$D(t) x = e^t x$, often called {\em dilatations}, and {\em special
conformal transformations}, $S(n)x = x/(1+nx)$. 

A chiral conformal transformation maps exactly one point to
$\infty$ and assumes a definite asymptotic value at $\pm\infty$, 
and $\PSL(2,\dopp{R})$ has its natural geometric action on the compactified
light-ray $\Seins \cong \dopp{R} \cup \{\infty\}$. This
compactification is achieved by means of a \Name{Cayley}
transformation:
\begin{equation}
  \label{eq:cayley}
  z = \frac{ix+1}{-ix+1} \zendot
\end{equation}
When we speak of the light-ray, $\dopp{R}$, as a subset of the
conformal covering space, $\Seins$, then we mean the inclusion of the
corresponding image under the \Name{Cayley} transformation. The positive
light-ray, $\dopp{R}_+$, corresponds to the upper half-circle,
$\Seins_+$, and $\dopp{R}_-$ to the lower
half-circle, $\Seins_-$. The point
$-1\in\Seins$ corresponds to the point at infinity of $\dopp{R}$, the
point $1$ is the image of the origin on the light-ray. 

The  \Name{Cayley} transformation induces an
isomorphism\footnote{$\PSL(2,\dopp{R})$ is isomorphic to the proper,
orthochronous \Name{Lorentz} group in $2+1$ dimensions,
$\SO(2,1)_+^\uparrow$, too.} $\PSL(2,\dopp{R})\cong
\PSU(1,1)$. The latter is the factor group of the group
$\SU(1,1)$ consisting of complex matrices which have determinant $1$ and leave
invariant $diag(+1,-1)$, divided by the relation $\symb{A}\sim
-\symb{A}$. Elements of $\SU(1,1)$ have the form $\symb{A} =
\left({\alpha \atop 
    \bar{\beta}}{\beta \atop 
    \bar{\alpha}}\right)$ with $|\alpha|^2-|\beta|^2 =1$; their action
on $z\in\Seins$ reads:
\begin{equation}
  \label{eq:psuaction}
  g_{\symb{A}} z = \frac{\alpha z + \beta}{\bar{\beta} z +
  \bar{\alpha}} \zendot
\end{equation}
The one-parameter group of {\em rigid conformal 
rotations} %Ind: ``rigid conformal rotations'' 
acts by multiplication with a phase: $R(\varphi)z = e^{i\varphi}z$,
$\varphi\in[-\pi,\pi[$.

Implementations of chiral conformal symmetry are given by unitary,
strongly continuous representations of the universal covering
group of $\PSL(2,\dopp{R})$, which we denote by
$\PSL(2,\dopp{R})^\sim$. There is
a local identification between one-parameter subgroups of
$\PSL(2,\dopp{R})$ and of $\PSL(2,\dopp{R})^\sim$ by the identity of their
\Name{Lie} algebras. We denote the one-parameter subgroups in
$\PSL(2,\dopp{R})^\sim$ corresponding to the subgroups of
$\PSL(2,\dopp{R})$ introduced above by $\tilde{T}$, $\tilde{D}$,
$\tilde{S}$, and $\tilde{R}$, respectively. The covering projection from
$\PSL(2,\dopp{R})^\sim$ onto $\PSL(2,\dopp{R})$ will be written $\symb{p}$. 

We adopt the physicists' convention on elements of the \Name{Lie} 
algebra which allows us to use the same symbols for the generators of
one-parameter subgroups in $\PSL(2,\dopp{R})$, for the corresponding
generators in
$\PSL(2,\dopp{R})^\sim$ and their self-adjoint representatives as
generators of unitary one-parameter groups in a unitary, strongly
continuous representation of $\PSL(2,\dopp{R})^\sim$. The generator of
translations, the 
{\em momentum operator}, will be denoted by $P$, the generator of special
conformal transformations by $K$, and the generator of rigid conformal
rotations, the {\em conformal {Hamilton}ian}, by $L_0$. By the
parametrisations above we have the identity
\begin{equation}
  \label{eq:L0PK}
  2L_0 = P - K \zendot
\end{equation}
$L_0$, $P$, and $-K$ all are positive operators in a representation,
if one of them is positive (proposition \ref{prop:poengcond}).

The localisation regions are open, non-dense intervals
contained in the circle, called the {\em proper intervals}. A
connected, open subset $I$ of $\Seins$ is a proper interval, denoted
by $I\Subset\Seins$, if its {\em causal complement} $I':=
\Seins\setminus\overline{I}$ is not the empty set.  The action of the
global conformal symmetry group on points in the circle as in
(\ref{eq:psuaction}) induces an action of $\PSL(2,\dopp{R})$ on 
the set of proper intervals. For $g\in \PSL(2,\dopp{R})$ the image of
an interval $I\Subset \Seins$ under this action will be denoted $gI$. 

Occasionally, we will encounter operators which are phases, for
example cocycles of ray-representations. In order to distinguish the
set of these phases from the chiral conformal covering space we will
use the notation $\dopp{C}_1\equiv \{z\in\dopp{C}, \,\, |z|^2=1\}$.

%%%%%%%%%%%%%%%%%%%%%%%%%%%%%%%%%%%%%%%%%%%%%
%%%%%% NETS, SUBNETS, COSETS %%%%%%%%%%%%%%%
%%%%%%%%%%%%%%%%%%%%%%%%%%%%%%%%%%%%%%%%%%%%%
\subsection{Chiral nets, chiral subnets and their \Name{Coset} models} 
\label{sec:genass}

After having clarified the geometric situation, we now state the
\Name{Haag-Kastler} axioms of local quantum physics \cite{HK64} in a
form adequate for chiral conformal quantum field theory
\cite{GL96,FJ96,FG93}. 

\clearpage
\begin{defi}\label{def:chcotheo}
  A {\bf chiral conformal theory} (in short: {\bf chiral net }) 
  $\lok{B}$ is given by a map assigning to each proper interval,
$I\Subset\Seins$, a \Name{v.Neumann} algebra, $\lok{B}(I)$, of bounded
operators on a separable \Name{Hilbert} space, $\Hilb{H}$, fulfilling
the following properties:
\begin{enumerate}
\item {\em Isotony\label{ax:iso}:}
If $I_1\subset I_2$,  we have
$\lok{B}(I_1)\subset\lok{B}(I_2)$. 
\item {\em Locality\label{ax:loc}:} For $I_1\subset I_2'$,
  $\lok{B}(I_1)$ commutes with 
  $\lok{B}(I_2)$, ie $\lok{B}(I_1)\subset \lok{B}(I_2)'$.
\item {\em Covariance\label{ax:cov}:} There is a strongly continuous,
  unitary representation, $U$, of $\PSL(2,\dopp{R})$ on $\Hilb{H}$
  such that the adjoint action $Ad_{U(g)}$, $g\in \PSL(2,\dopp{R})$,
  defines for each $I\Subset\Seins$ an isomorphism
  $\alpha_g\restriction{}{\lok{B}(I)}$ from $\lok{B}(I)$ onto
  $\lok{B}(gI)$; $\alpha_g$ stands for the action of  $Ad_{U(g)}$ on
  the net of  local algebras: 
  \begin{equation}
    \label{eq:autocov}
    Ad_{U(g)} \lok{B}(I) = \alpha_g(\lok{B}(I)) = \lok{B}(gI)\, , \,\,
    I\Subset\Seins  \zendot
  \end{equation}
\item {\em Positivity of energy\label{ax:poseng}:} The spectrum of the
  momentum operator $P$ is positive in the representation $U$.
\item {\em Vacuum\label{ax:vac}:} The space of $U$-invariant vectors
  is one-dimensional. We choose a unit vector, the vacuum $\Omega$,
  which is assumed to be cyclic for the \Name{v.Neumann} algebra
  $\bigvee_{I\Subset\Seins}\lok{B}(I)$. 
\end{enumerate}
\end{defi}

Remark: The set of proper
    intervals in $\Seins$ is not directed with respect to the partial
    order defined by inclusion and thus is not a net in the proper
    sense of the word. Same holds true for the  set of local algebras. It
    would be rigorous to call them {\em precosheaves} (cf eg
    \cite{GL96}), but for three reasons we will use the term {\em net}
    for chiral conformal theories: They are completely
    determined by their restrictions to the light-ray, which are
    genuine nets of local algebras. We
    regard chiral conformal theories as models for general structures
    connected with local quantum theories in ``realistic spacetimes'',
    which are given by nets of local algebras. Finally, speaking
    of nets of local algebras is common practise in the literature on
    chiral conformal 
    theories.

The fundamental object of this study is an inclusion of a chiral
conformal theory $\lok{A}$ in a theory $\lok{B}$ as just
introduced. We adopt the following definition \cite{rL01}:

\begin{defi}\label{def:chsubnet}
 A {\bf chiral conformal subtheory } (short: {\bf chiral subnet})
 $\lok{A}$ embedded in $\lok{B}$, written 
as $\lok{A}\subset \lok{B}$, is given by a map from the set of proper
intervals to \Name{v.Neumann} algebras, $I \mapsto
\lok{A}(I)$, with the following properties:
\begin{enumerate}
\item {\em Inclusion\label{ass:inc}:} $\lok{A}(I)\subset \lok{B}(I)$ for $I\Subset \Seins$.
\item {\em Isotony\label{ass:iso}:} If $I_1\subset I_2$, then
  $\lok{A}(I_1)\subset\lok{A}(I_2)$.
\item {\em Covariance\label{ass:cov}:} For all $g\in \PSL(2,\dopp{R})$ and
  $I\Subset\Seins$ we have:  $\lok{A}(gI) = \alpha_g
  (\lok{A}(I))$. 
\end{enumerate} 
\end{defi}
We say that $\lok{A}$ is {\em non-trivial}, if its local algebras are
different from $\dopp{C}\Einsop$ and do not coincide with the local
algebras of $\lok{B}$.

The goal of this work is to find and to establish typical features
related with the following objects:

\begin{defi}\label{def:cospa}
Any chiral subnet $\lok{C}\subset\lok{B}$ is called a
{\bf {Coset} model } 
associated with a given chiral subnet $\lok{A}\subset\lok{B}$, if we
have $\lok{C}(I)\subset \lok{A}(I)'$ for one and hence for all
$I\Subset\Seins$.  

The  chiral subnet defined by $\lok{A}\coset\lok{C}(I):=
  \lok{A}(I)\vee\lok{C}(I)$ is called a {\bf Coset pair}.
The {\bf local relative commutants} $\lok{C}_I$, $I\Subset\Seins$,  of
 $\lok{A}\subset\lok{B}$ are given by  $\lok{C}_I:=\lok{A}(I)'\cap\lok{B}(I)$.
\end{defi}

Obviously, the  local relative commutants contain the local algebras
of any \Name{Coset} model:
  \begin{equation}
    \label{eq:defcospa}
    \lok{C}_I= \lok{A}(I)'\cap\lok{B}(I) \supset \lok{C}(I) \, , \,\,
     I\Subset\Seins \zendot
  \end{equation}
Note that, a priori, the  local relative commutants $\lok{C}_I$ do not
define a  \Name{Coset} model, because isotony is not known to hold; we
refer to this as the {\em isotony problem}.
 
We have chosen the term ``\Name{Coset} model'' rather than
``\Name{Coset} theory'' since typically one constructs the
\Name{Coset} subnet from a specific chiral subnet and analyses its
features in this setting rather than defining it abstractly. 

A general, yet not obvious consequence of the assumptions is that
there always is a representation $U^\lok{A}$ of
$\PSL(2,\dopp{R})^\sim$ which implements conformal covariance on the
subtheory $\lok{A}$ and whose unitaries are {\em globally inner} in
$\lok{A}$, ie the operators 
$U^\lok{A}(\tilde{g})$, $\tilde{g}\in \PSL(2,\dopp{R})^\sim$, are
contained in the \Name{v.Neumann} algebra
$\bigvee_{I\Subset\Seins}\lok{A}(I)$. With a slight abuse of notation,
we denote the {\em global algebra} of the subnet
$\lok{A}\subset\lok{B}$ by $\lok{A}:=\bigvee_{I\Subset\Seins}\lok{A}(I)$.
 It contains all local observables of the subtheory $\lok{A}$ as well
 as all observables which are weak 
limits of local observables of $\lok{A}\subset\lok{B}$ but not local
themselves; the latter we call {\em genuine global observables}. \label{ind:globalg} 

$U^\lok{A}$ is constructed and some desirable properties of it are derived
in section \ref{sec:bosug}. There is shown, for example, that
there is exactly on inner-implementing representation $U^\lok{A}$ for
$\lok{A}$, put differently: $U^\lok{A}$ is {\em the} inner
implementation of conformal symmetry for $\lok{A}\subset\lok{B}$.
Furthermore, the operators
$U^\lok{A}(\tilde{g})\neq\Einsop$  are genuine global observables in
$\lok{A}$. We have the following important consequence:

\begin{lemma}\label{lem:cosmax}
$\lok{C}_{max}(I):=\{U^\lok{A}(\tilde{g}),\,
\tilde{g}\in\PSL(2,\dopp{R})^\sim\}' \,\cap\,\lok{B}(I)$ defines a
\Name{Coset} model 
associated with $\lok{A}\subset \lok{B}$. Every \Name{Coset} model
$\lok{C}$ associated with $\lok{A}\subset\lok{B}$ satisfies
$\lok{C}(I)\subset \lok{C}_{max}(I)$. 
The {\bf maximal Coset model} $\lok{C}_{max}$ associated to a subnet
$\lok{A}\subset\lok{B}$  satisfies
$\lok{C}_{max}(I)=\lok{A}'\cap\lok{B}(I)$.
\end{lemma}

\begin{pf} Obviously this definition yields a subtheory
$\lok{C}_{max}\subset\lok{B}$. Since the operators of a local algebra
of $\lok{C}_{max}$ commute with the inner implementation of $\lok{A}$,
we deduce from locality of $\lok{B}$ that $\lok{C}_{max}$  is in fact a
\Name{Coset} model.

Let $\lok{C}$ be any \Name{Coset} model, $I, J$ proper intervals
satisfying $I\subset J$ and $I'\cup J =\Seins$. By isotony of
$\lok{C}$, locality and weak additivity (see below) for chiral
subtheories we have:  
$
  \lok{C}(I)\subset (\lok{A}(I')\vee\lok{A}(J))' =
  \lok{A}'\subset \{U^\lok{A}(\tilde{g}),
  \tilde{g}\in \PSL(2,\dopp{R})^\sim\}'
$.
\end{pf}

The characterisation of
$\lok{C}_{max}$ as a subtheory which commutes with a representation of
$\PSL(2,\dopp{R})^\sim$ is analogous to that of {\em maximal chiral
  observables} 
in a $\opo$-dimensional conformal theory \cite{khR00}. 
It turns out
that this analogy is quite complete; see section
\ref{cha:netend}.\ref{sec:chihol}.  

%%%%%%%%%%%%%%%%%%%%%%%%%%%%%%%%%%%%%%%%%
%%%%%%% CONSEQUENCES OF ASSUMPTIONS %%%%%%%
%%%%%%%%%%%%%%%%%%%%%%%%%%%%%%%%%%%%%%%%%%

\subsection{Discussing the assumptions}
\label{sec:disass}

% In definition \ref{def:chcotheo} one could assume $U$ to be a
% representation of $\PSL(2,\dopp{R})^\sim$ instead of
% $\PSL(2,\dopp{R})$, but it was shown by \Name{Guido and Longo}
% \cite{GL95} that the kernel of the covering projection $\symb{p}$
% coincides with the kernel of $U$ in this setting, ie $U$ is a
% representation of $\PSL(2,\dopp{R})$ even if we drop this as an
% explicit assumption.

\subsubsection{Geometric modular action on chiral nets}

The vacuum, $\Omega$, is cyclic and separating for each local algebra
$\lok{B}(I)$, $I\Subset\Seins$, by  locality and the \Name{Reeh-Schlieder} theorem
\cite{RS61,hjB68}  (cf \cite{FG93,FJ96}), which means
that the closure of 
$\lok{B}(I)\Omega$ coincides for every $I\Subset\Seins$ with the whole
\Name{Hilbert} space, $\Hilb{H}$, and for a local observable
$B\Omega=0$ implies $B=0$.

Hence, it is possible to
apply  {\em Tomita-Takesaki theory}\footnote{See \cite{BR87}, \cite{KR86},
  \cite{SZ79}, \cite{sS81}, \cite{mT70}.}, also called {\em modular
  theory}, which is of particular use for local quantum physics (cf
\cite{hjB00}). 
 The fundamental structures  of modular theory are contained in the
 {\em Tomita-Takesaki  theorem}: Given a
\Name{v.Neumann} algebra $\lok{M}$ with a cyclic and separating vector
$\Omega$, there is a positive, invertible operator, $\Delta$, called {\em
  modular operator}, and
an anti-unitary involution $J$, called {\em modular conjugation},
such that $J\lok{M}J = \lok{M}'$ and $\sigma_t(\lok{M}):=
\Delta^{it}\lok{M}\Delta^{-it}= \lok{M}$, $t\in\dopp{R}$. The automorphism
group $\sigma$ is 
called the {\em modular group}. The operators $J, \Delta$ form the
{\em modular data} of the pair $(\lok{M},\Omega)$ and they satisfy:
$J\Omega = \Omega = \Delta \Omega$, $J=J^*=J^{-1}$,
$\Delta^{it}=J\Delta^{it}J$. $J$ and $\Delta$ are given by the
polar decomposition of {\em Tomita's operator}
$S=J\Delta^{\frac{1}{2}}$, which is defined  
densely by: $S M \Omega = M^*\Omega$, $M\in \lok{M}$. 

% Closely related is the {\em KMS condition}, which consists of two
% parts: For every pair 
% $M_1, M_2\in\lok{M}$ the functions 
% $f(t):=\langle\Omega, M_1\sigma_t M_2\Omega\rangle$ possess analytic
% continuations for $-1<Im(t)<0$; on the boundary we have: $f(t-i) =
% \langle\Omega, 
% M_2\sigma_{-t}M_1\Omega\rangle$. If a one-parameter group of
% automorphisms of $\lok{M}$ fulfills these conditions, it coincides
% with the modular group $\sigma$.

Every local algebra of $\lok{B}$ has its modular data because of the
\Name{Reeh-Schlieder} theorem, but modular theory becomes really
useful for studies on quantum field theories, if one can make contact
with the geometric net structure underlying the theory. In case one
has such a geometrical interpretation of modular data, the theory is
said to have the {\em Bisognano-Wichmann
property}\footnote{\Name{Bisognano and Wichmann} were the first to
  establish such a connection \cite{BW75, BW76}.}. In general, such
links are hard to establish, but conformally covariant theories are a
remarkable exception \cite{BGL93}. Taking 
$\lok{M}=\lok{B}(\Seins_+)$ yields in our setting: the modular group 
is directly related to the scale transformations according to  
$\Delta^{it}= U(D(-2\pi t))$, and the modular
conjugation $J$ implements the
reflection $x\mapsto -x$, $x\in\dopp{R}$ \cite{FG93,FJ96}. 

\Name{Guido, Longo and Wiesbrock} have shown
the following: A local net $\lok{B}$ on the chiral light-ray, which is
covariant with respect to a representation $U$ of the translation-dilatation
group\footnote{This group is given as semi-direct product of
  the translations $T$ and the dilatations $D$  with the relation
  $D(t)T(a)D(-t)= T(e^{t}a)$.}, extends to a conformal
net, if and only if we have $U(D(-2\pi t))= \Delta^{it}$ for the
modular operator of the algebra $\lok{B}(\Seins_+)$ \cite[theorem
1.4]{GLW98}.  Thus 
one can not have dilatation-translation covariance and the
\Name{Bisognano-Wichmann} property in the indicated form without
having conformal covariance.

The local algebras $\lok{B}(I)$, $I\Subset\Seins$, are {\em continuous from
the inside} as well as {\em from the outside}, that is: $\lok{B}(I)$
coincides with the intersection of all local algebras assigned to
proper intervals $J$ containing $\overline{I}$ and is generated by all
its local 
subalgebras assigned to proper intervals $J$ with
$\overline{J}\subset I$, respectively. Continuity from the inside
implies {\em weak
additivity}, ie  $\lok{B}(I)$ is generated by the subalgebras
$\lok{B}(J_i)$ for each covering $\bigcup_i J_i = I$ \cite{FJ96}. The
crucial continuity argument leading to these properties 
depends on scale invariance and stems from \cite[lemma II.2.2]{LRT78}.    

Uniqueness of the vacuum (up to scalar multiples) implies that the
local algebras are factors, to be precise type $\III_1$
factors \cite{wD75}. Type $\III$ factors $\lok{M}$ have the specific
property that 
any of their non-trivial projections $P<\Einsop$ has infinite dimension and is
equivalent in $\lok{M}$ to $\Einsop$, ie there is an isometry
$W\in \lok{M}$ satisfying $P=WW^*$, $\Einsop=W^*W$. Using these
properties, it is
straightforward to show  that type $\III$
factors possess cyclic and separating vectors and that an algebraic
isomorphism between type $\III$ factors may always be implemented by a
unitary operator (eg \cite{jS67}).

According to \Name{Connes}' classification of type 
$\III$ factors \cite{aC73}, the $\III_1$ factors are characterised by
the following properties: the action of the modular group is outer, ie
there is no one-parameter group of unitaries in the algebra itself
which could implement the modular group, and the action of the modular
group is ergodic, ie every operator left invariant by the modular
group is a multiple of $\Einsop$. 

Ergodicity of the modular group may be deduced from the
\Name{Bisognano-Wichmann} property: If $B\in\lok{B}(\Seins_+)$
fulfills $\Delta^{it}B\Delta^{-it} =B$,  
then $B\Omega$ is in fact invariant with respect to the whole
representation $U$ (lemma \ref{lem:invvec}). By the separating
property of the vacuum, we get $B = U(g)BU(g)^*$ for all $g\in
\PSL(2,\dopp{R})$. Covariance, locality and irreducibility then force
$B$ to be a scalar multiple of $\Einsop$.  Factoriality is established
quite  easily and the outerness of the
modular group follows from results contained in \cite{aC73}, see
discussion in \cite{wD75}.

In fact, uniqueness of the vacuum is
equivalent to irreducibility of the net $\lok{B}$
($\bigvee_{I\Subset\Seins}\lok{B}(I)$ coincides with the algebra of
all bounded operators on $\Hilb{H}$),
factoriality of local algebras, and triviality of local algebras
associated with points ($\bigcap_{I\Subset\Seins,I\ni\zeta}\lok{B(I)}=
\dopp{C}\Einsop$) \cite{GL96}. 

By
covariance, the \Name{Bisognano-Wichmann} property of $\lok{B}$ means
in particular: the vacuum  
representation of $\lok{B}$ satisfies {\em Haag duality (on the
circle)}, namely we have  $\lok{B}(I)'=\lok{B}(I')$,
$I\Subset\Seins$. If we simply talk of
``\Name{Haag} duality'' in the following, we will always mean
``\Name{Haag} duality on $\Seins$''. In physical terms, \Name{Haag}
duality says that the local algebras can not be extended without
violating locality. It is a very useful feature of a theory, if one
wants to study its representations.

\subsubsection{Representations of chiral nets}

In $\Seins$ the causal complement of each
localisation region is again a localisation region. This is different
from the situation for  \Name{Poincar\'e} covariant theories in  $3+1$
dimensions or, indeed, for chiral nets on 
the light-ray: here the localisation regions are taken to be bounded
which results in unbounded causal complements. On the chiral
light-ray the causal complement of a bounded localisation region even
consists of two disconnected components; the same is true in
$\opo$-dimensional spacetime. It is a general feature of conformal
covering 
spaces that causal complements are themselves localisation regions
\cite{BGL93}. As a consequence of this, all locally normal 
representations turn
out to be {\em localisable} in the sense of \Name{Doplicher, Haag and
  Roberts} (\Name{DHR}) \cite{DHR69a}. 

A {\em representation $\pi$} of a chiral conformal theory is a set of
homomorphisms $\pi_I: \lok{B}(I) \rightarrow \pi_I(\lok{B}(I))$,
$I\Subset\Seins$, into 
algebras of bounded operators on some \Name{Hilbert} space
$\Hilb{H}_\pi$, where the $\pi_I$ are required to fulfill the
{\em consistency relation} $\pi_I\restriction{}{\lok{B}(J)}= \pi_J$ for $J\subset
I$. This family lifts
uniquely to a representation $\pi$ of $\lok{B}_{uni}$, the {\em universal
$C^*$-algebra} generated by all $\lok{B}(I)$, $I\Subset\Seins$,
and the $\pi_I$ are given in terms of
the embeddings $\iota_I: \lok{B}(I)\hookrightarrow \lok{B}_{uni}$
by $\pi_I = \pi\circ \iota_I$ \cite{kF90, GL92} (cf eg \cite{aS97}). 

Local normality of $\pi$ says that the local representations $\pi_I$
are normal (weak${}^*$ continuous). The physical relevance
of this property is discussed, for example, in \cite{rH92}. 
A representation $\rho$ is said to be {\em localised} in some
$I_0\Subset\Seins$, if we have $\Hilb{H}_\rho=\Hilb{H}$ and
$\rho\restriction{}{\lok{B}(I_0')} = id$. 
Any locally normal representation of a
chiral conformal net is unitarily equivalent to a localised
representation, ie localisable. This is due to the fact
that the local algebras are type $\III$ factors.

Local normality is automatic if the representation lives on a
separable \Name{Hilbert} space 
(cf \cite{GL96}). It follows as well, if a representation is covariant
with positive energy and the global algebra
$\pi(\lok{B}):=\bigvee_{I\Subset\Seins}\pi_I(\lok{B}(I))''$ possesses
a cyclic vector. This is
content of a theorem of \Name{Buchholz, Mack and Todorov}
\cite[theorem 1]{BMT88}; the original proof is very
brief and we consider it worth while to make 
available a detailed proof in appendix \ref{cha:app}.\ref{sec:BMT}. 

Locally normal representations of a local quantum theory are gathered in
unitary equivalence 
classes, the {\em superselection sectors}, and the
localisable representations form the class of \Name{DHR}
sectors. Conformal symmetry ensures that all sectors of a conformal
theory are of \Name{DHR} type. The fact that the set of localisation
regions is not directed, ie that 
there are pairs of proper intervals $I_1,I_2\Subset\Seins$ to which
there is no proper interval containing both of them, necessitates a
generalisation of the standard treatment of superselection theory of
\Name{DHR} sectors \cite{DHR69a,DHR69b,DHR71,DHR74}. Such a
generalisation has been established by \Name{Fredenhagen, Rehren and
  Schroer} (\Name{FRS}) \cite{FRS89,FRS92} (reviews eg \cite{KMR90,
  aS97}). Although the general structures of \Name{DHR} theory carry
over, there occur some striking differences. In particular, presence of
non-trivial braid group statistics prohibits the application of the
reconstruction method of \Name{Doplicher and Roberts} \cite{DR90},
which identifies in higher dimensions the theory of \Name{DHR} sectors
as a result of the action of a global gauge group on a field algebra
and leaving fixed the observables.

\subsubsection{{Haag} duality on the light-ray}

In both the \Name{DHR} theory and its generalisation by \Name{FRS},
\Name{Haag} duality is fundamental. But while \Name{Haag} duality on
$\Seins$ follows from the general assumptions, \Name{Haag} duality for
the restricted net on the light-ray may be violated. 
A bounded interval $I$ on the light-ray
$\dopp{R}$, denoted $I\Subset\dopp{R}$, has a causal complement
$I^\perp\subset \dopp{R}$ which, as a subset of the covering space
$\Seins$,  consists of two disjoint, proper intervals $I_\pm\Subset\Seins$:
$I^\perp := \Seins\setminus\overline{I}\cup \{\infty\} = I_+\dot{\cup}
I_-$. \Name{Haag} duality  
on the light-ray is valid, by definition, if we have: $\lok{B}(I)'=
\lok{B}(I_+)\vee\lok{B}(I_-)$. This identity does not hold in large
classes of models \cite{BS90,jY94}.

It is not difficult to show that  \Name{Haag} duality
on the light-ray for chiral conformal theories is equivalent
to {\em strong additivity} of local algebras (eg
\cite[lemma 1.3]{GLW98}):
 a theory $\lok{B}$  is said to be strongly
additive, if for every disjoint decomposition
$I=I_1\dot{\cup}\{\zeta\}\dot{\cup}I_2$, $\zeta\in I\Subset\Seins$, we have
$\lok{B}(I)=\lok{B}(I_1)\vee\lok{B}(I_2)$, ie the local algebra
$\lok{B}(I)$ is, as a \Name{v.Neumann} algebra, generated by its
subalgebras $\lok{B}(I_1)$ and $\lok{B}(I_2)$. If this property holds
for one such decomposition, then for all by covariance.

If a subnet $\lok{A}\subset\lok{B}$ is strongly additive, then the
isotony problem is absent, ie the local relative commutants are
automatically isotonuous. For a
pair $I_{1,2}$ of proper intervals, $\overline{I_1}\subset I_2$, the
latter possesses a disjoint decomposition
$I_2=I_l\dot{\cup}\{\zeta_l\}I_1\dot{\cup}\{\zeta_r\}\dot{\cup} I_r$,
where $\zeta_{l,r}$ denote the boundary points of $I_1$ and $I_{l,r}$ are
proper intervals. Locality implies under the assumption of strong
additivity of $\lok{A}$:
\begin{eqnarray}
  \label{eq:strongiso}
  \lok{C}_{I_1 }\subset \lok{A}(I_l)'\cap \lok{A}(I_1)' \cap
  \lok{A}(I_r)' = \lok{A}(I_2)'\zendot
\end{eqnarray}
Examples of strongly additive theories are the stress-energy tensors
with $c\leq 1$ \cite{KL02, fX03} and the chiral current algebras
\cite{BS90,vL97}. 

It is always possible to construct from a non-strongly additive chiral
conformal theory $\lok{B}$ a strongly additive one by taking the {\em
  dual net} $\lok{B}^d$ (cf \cite{GLW98}). One first defines the local
algebras of its restriction to the light-ray to be:
\begin{equation}
  \label{eq:dualnet}
\lok{B}^d(I) := \klammer{\lok{B}(I_+)\vee\lok{B}(I_-)}'\, , \,\,
I\Subset\dopp{R} \zendot
\end{equation}
As for its analogue in higher dimensions, the duality property for
half-lines\footnote{Corresponding to duality for wedges in higher
dimensions; this {\em wedge duality} usually is 
called {\em essential duality}. In our setting essential duality is
identical to \Name{Haag} duality on $\Seins$.}, ie
$\lok{B}(\dopp{R}_+)' = \lok{B}(\dopp{R}_-)$, ensures locality for the
dual net. The \Name{Bisognano-Wichmann} property of $\lok{B}$ yields
conformal covariance of the dual net \cite[theorem 1.4]{GLW98}. By
construction, both theories live on the same vacuum \Name{Hilbert} space.

The dual net $\lok{B}^d$ is an extension of $\lok{B}$ only upon
restriction to the light-ray and not as a net on $\Seins$ and it has,
in general, 
different physical properties. For example the superselection theory
might be different, in particular due to the possible occurrence of
soliton sectors \cite{GLW98}. So, absence of strong additivity in a chiral
conformal theory has to be taken seriously.

For the following another aspect is illuminating:  conformal
covariance is implemented on $\lok{B}^d$ by a representation $U^d$ of
$\PSL(2,\dopp{R})$, which agrees 
with the implementation on  $\lok{B}$, namely
$U$,  when restricted to the dilatation-translation
subgroup; $U^d$ and $U$ are different as representations of
$\PSL(2,\dopp{R})$,  if $\lok{B}$ is not strongly additive.
The 
following proposition is a simple variant of \cite[lemma 1.3]{GLW98}:
\begin{prop}\label{prop:dualcov}
Let $\lok{B}$ be a chiral conformal net, covariant with respect to the
representation $U$ of $\PSL(2,\dopp{R})$, and $\lok{B}^d$ its dual net,
covariant with respect to the 
representation $U^d$. Then the following are equivalent:
\begin{enumerate}
\item \label{propdualcovi} $\lok{B}$ satisfies \Name{Haag} duality on the light-ray.
\item \label{propdualcovii}$\lok{B}$ is strongly additive.
\item \label{propdualcoviii} $\lok{B}$ and $\lok{B}^d$ coincide.
\item \label{propdualcoviv}$U$ and $U^d$ coincide. 
\end{enumerate}
\end{prop}

\begin{pf}
The equivalence \ref{propdualcovi} $\Leftrightarrow$
\ref{propdualcovii} is established in \cite[lemma
1.3]{GLW98}. \ref{propdualcoviii} is equivalent with
\ref{propdualcovii} by the very definition of the dual net,
(\ref{eq:dualnet}), and \Name{Haag} duality.

The \Name{Bisognano-Wichmann} property of chiral conformal theories
establishes \ref{propdualcoviii} $\Rightarrow$
\ref{propdualcoviv}, because the one-parameter subgroups of
$\PSL(2,\dopp{R})$ which leave the boundary points of some proper
interval fixed coincide, when represented through $U$ and $U^d$, with
the modular groups of the respective local algebra, and together these
generate all of $\PSL(2,\dopp{R})$.  

The opposite direction is implied by modular theory, too. Covariance
allows us to deal with the situation for a single proper interval. For
$I\Subset\dopp{R}$ we have the inclusion
$\lok{B}(I)\subset\lok{B}^d(I)$. By assumption \ref{propdualcoviv}
and the \Name{Bisognano-Wichmann} property of $\lok{B}^d$ the
subalgebra $\lok{B}(I)$ is globally invariant with respect to the
action of the modular group of $\lok{B}^d(I)$. The remainder follows
by a standard argument: first \Name{Takesaki}'s
theorem \cite{mT72} on modular covariant subalgebras  ensures 
existence of a normal, faithful conditional expectation from
$\lok{B}^d(I)$ onto $\lok{B}(I)$, which leaves invariant the vacuum
state. From here an argument of \Name{Jones}
\cite{vJ83} establishes $\lok{B}(I)$
as the subalgebra of operators in $\lok{B}^d(I)$ which commute with
the projection onto the closure of $\lok{B}(I)\Omega$, which is
$\Einsop$ by the \Name{Reeh-Schlieder} theorem for $\lok{B}$. 
\end{pf}

\subsubsection{Modular covariance of chiral subnets}

The assumption of conformal covariance, definition 
\ref{def:chsubnet}.\ref{ass:cov}, introduces rich and powerful
structures into the investigations of a chiral  subnet
$\lok{A}\subset\lok{B}$. Basically, this assumption requires the restricted
nets $\lok{A}$ and $\lok{B}$ to have {\em simultaneous} conformal
extensions preserving the inclusions for the extended nets on
$\Seins$. This excludes interesting inclusions on the light-ray. One
example is given by the 
relation between the strongly additive dual net $\lok{B}^d$ and its
basis $\lok{B}$ (proposition \ref{prop:dualcov}): the restriction of
$\lok{B}$ to the light-ray is contained in the corresponding
restriction of $\lok{B}^d$, both possess conformal extensions, but
this extension can not be simultaneous, if
$\lok{B}\neq\lok{B}^d$. The conformally covariant
derivatives of the $U(1)$ current form a particular class of examples
of this structure \cite[corollary 2.11]{GLW98}.  

Conformal covariance of the subnet $\lok{A}\subset\lok{B}$ implies two
important facts: first, the chiral subnet possesses the
\Name{Reeh-Schlieder} property, ie the cyclic subspace generated by
the global algebra $\lok{A}$ from the vacuum coincides with the cyclic
subspace of any local algebra $\lok{A}(I)$, $I\Subset\Seins$. This is
a direct consequence of the \Name{Reeh-Schlieder} theorem as proved by
\Name{Borchers} \cite{hjB68}. We write  $e_\lok{A}$ for the cyclic
projection of $\lok{A}$, ie we have: $e_\lok{A}\Hilb{H} =
\overline{\lok{A}(I)\Omega}$. \label{ind:eA}

While the \Name{Reeh-Schlieder} property holds for
dilatation-translation covariant subnets on the light-ray as well, in
conformally covariant subnets each local inclusion
$\lok{A}(I)\subset\lok{B}(I)$, $I\Subset \Seins$, is globally
left invariant by the modular group of $\lok{B}(I)$. Such
{\em modular covariant subalgebras} have features which
will be crucial in the following (cf section
\ref{cha:finind}.\ref{sec:loccospa}): 
The cyclic projection $e_\lok{A}$ characterises completely the
subalgebra $\lok{A}(I)$ in $\lok{B}(I)$, namely we have by \Name{Takesaki}'s
theorem \cite{mT72} and an argument of \Name{Jones} \cite{vJ83}: 
\begin{equation}
  \label{eq:modinvsub}
  \lok{A}(I)=
\{e_\lok{A}\}'\cap\lok{B}(I) \, , \,\, I\Subset \Seins \zendot
\end{equation}
In particular there are no modular covariant subalgebras of $\lok{B}(I)$
which generate a dense subspace from the vacuum other than
$\lok{B}(I)$ itself.

The mapping $\lok{A}(I)\rightarrow\lok{A}(I)e_\lok{A}$ given by
$A\mapsto A e_\lok{A}$ is known to define an isomorphism of
\Name{v.Neumann} algebras (because of the separating property
of the vacuum, eg \cite{hjB97}). It is easy to see that
the net $\lok{A}e_\lok{A}$ defines a chiral conformal theory in its
vacuum representation, and it is readily checked that the inverse
isomorphisms 
$\pi_I: \lok{A}(I)e_\lok{A} \rightarrow \lok{A}(I)\subset\lok{B}(I)$
define a locally normal\footnote{Algebraic isomorphisms of
  \Name{v.Neumann} algebras are automatically ultra-weakly and
  ultra-strongly continuous \cite[I.4.3. corollary 1]{jD81}. Since the
  local algebras  are type $\III$ factors, the isomorphism may
  be implemented by a unitary operator from $\Hilb{H}$ onto
  $e_\lok{A}\Hilb{H}$ \cite[II.4.6. theorem]{jS67}.} representation of
$\lok{A}e_\lok{A}$ due to equation \ref{eq:modinvsub}. Because of
these local isomorphisms we will speak occasionally of ``representations
of $\lok{A}$'', although these actually are representations of
$\lok{A}e_\lok{A}$. We will \underline{never} speak of representations
of the global algebra $\lok{A}$ and so no confusion should arise.

\Name{Takesaki}'s theorem \cite{mT72} says the following as well: the
vacuum state, $\omega(.)= \langle\Omega, . \Omega\rangle$, is a
{\em product state} on every \Name{v.Neumann} algebra
$\lok{A}(I)\vee\lok{C}_I$:
\begin{equation}
  \label{eq:prodstate}
  \omega(AC) = \omega(A)\omega(C) \, ,\,\, A\in \lok{A}(I),\, C \in
  \lok{C}_I,\, I\Subset\Seins \zendot
\end{equation}
Because of this product state,  the mapping
$\lok{A}(I)\vee\lok{C}_I\ni\sum_i a_ic_i \mapsto \sum_i a_i\otimes c_i\in
\lok{A}(I)\otimes\lok{C}_I$ extends to an isomorphism from
$\lok{A}(I)\vee\lok{C}_I$ onto $\lok{A}(I)\otimes\lok{C}_I$. 
% $\lok{A}(I)\vee\lok{C}_I$ is a factor, because $\lok{A}(I)$ is. 

% As a subalgebra of $\lok{B}(I)$, the factor  $\lok{A}(I)\vee\lok{C}_I$
% is modular covariant. For subalgebras
% $\lok{N}\subset\lok{M}$, $\Omega$ cyclic and separating, $\lok{N}$
% modular covariant with respect to the modular data of
% $(\lok{M},\Omega)$, holds: the modular data $J, \Delta$ of
% $(\lok{M},\Omega)$ commute with the projection $e_\lok{N}$ onto the closure of
% $\lok{N}\Omega$ and coincide with the modular data of
% $(\lok{N}e_\lok{N},\Omega)$ (eg \cite{hjB97}). Now, if the action of
% the modular group of $\lok{M}$ is ergodic, then so is the action of
% the modular group of $\lok{N}e_\lok{N}$. This proves that
% $\lok{A}(I)\vee\lok{C}_I$ is, indeed, a type $\III_1$ factor and the
% isomorphy to $\lok{A}(I)\otimes\lok{C}_I$ may be implemented by a
% unitary operator from $\Hilb{H}$ onto
% $\overline{\lok{A}(I)\Omega\otimes\lok{C}_I\Omega}$. This proves that the local
% algebras $\lok{A}\coset\lok{C}(I)$ of any \Name{Coset} pair are in a
% ``spatial'' tensor product position. 

For \Name{Coset} pairs $\lok{A}\coset\lok{C}\subset\lok{B}$ the
product state property of the vacuum has the following consequence:
\begin{prop}\label{prop:cospatenpos}
  Let $\lok{A}\coset\lok{C}\subset \lok{B}$ be a \Name{Coset} pair. The
  chiral conformal theories
  $\lok{A}\coset\lok{C}e_{\lok{A}\smcoset\lok{C}}$ and
  $\lok{A}e_\lok{A}\otimes\lok{C}e_\lok{C}$ are unitarily
  equivalent. 
\end{prop}

\begin{pf}
  Straightforward verification shows
  $\lok{A}e_\lok{A}\otimes\lok{C}e_\lok{C}$ to be a chiral conformal
  theory with the obvious definitions: its vacuum is given by
  $\Omega\otimes\Omega$, the representation implementing covariance is
  $(Ue_\lok{A}\otimes Ue_\lok{C})(.)$, its representation space is
  $e_\lok{A}\Hilb{H}\otimes e_\lok{C}\Hilb{H}$.  The factoriality of
  the local algebras proves that $\Omega\otimes\Omega$ is (up to
  scalar multiples) the unique vacuum \cite[proposition 1.2]{GL96},
  \cite[IV.5., corollary 5.11]{mT79}. 
  
  $\Omega$ is separating for $\bigcup_{I\Subset\dopp{R}}
  \lok{A}\coset\lok{C}(I)e_{\lok{A}\smcoset\lok{C}}$, the union of all
  local algebras assigned to bounded intervals in $\dopp{R}$. Thus, we
  are allowed to define a linear operator $W$ densely by: 
  \begin{equation}
    \label{eq:defcospauni}
    W A C \Omega := A\Omega\otimes C\Omega\, , \,\, A\in \lok{A}(I),
    C\in \lok{C}(I), I\Subset\dopp{R} \zendot
  \end{equation}

  On the algebra   $\bigcup_{I\Subset\dopp{R}}
  \lok{A}\coset\lok{C}(I) e_{\lok{A}\smcoset\lok{C}}$ the vacuum is a
  product state (a corollary to \Name{Takesaki}'s theorem \cite{mT72}). Hence,
  $W$ is bounded and extends by continuity to an isometry, as one may
  readily verify.  

  We may check from the definition, that $W$ is an intertwiner: 
  \begin{equation}
    \label{eq:Wintertw}
    W A_1C_1 A_2C_2\Omega = (A_1\otimes C_1)\,\, W A_2C_2\Omega\, , \,\,
    A_{1,2}\in \lok{A}(I),\, C_{1,2}\in\lok{C}(I),\, I\Subset \dopp{R}
    .
  \end{equation}
  Thus, $WW^*$ and $W^*W$ commute with the respective  restricted nets
  on $\dopp{R}$, but these are irreducible, as follows from
  irreducibility of the respective conformal nets on 
    $\Seins$ using weak additivity, \Name{Haag} duality and the factor
    property of the algebras assigned to $\Seins_+$. Thereby, $W$ is
  a unitary operator. 

  $Ad_W$ induces a unitary equivalence of the respective local
  algebras associated with every $I\Subset\dopp{R}$ by its definition
  (\ref{eq:defcospauni}) and the separating property of the
  vacuum. Furthermore, $W$ is readily shown to be covariant. If we
  denote the covariance automorphisms of
  $\lok{A}e_\lok{A}\otimes \lok{C}e_\lok{C}$ by $\alpha^\otimes$,
  we have for $gI\Subset\dopp{R}$, $I\Subset\dopp{R}$:
  $\alpha^\otimes_g Ad_W\restriction  {\lok{A}\coset\lok{C}(I)}  = Ad_W 
  \alpha_g\restriction {\lok{A}\coset\lok{C}(I)}$. 

Using the \Name{Reeh-Schlieder}
  property of the local algebras, one may reconstruct the
  representations $(Ue_\lok{A}\otimes Ue_\lok{C})(.)$ and
  $U(.)e_{\lok{A}\smcoset\lok{C}}$ from the action of the
  automorphisms.
  This, in turn, proves that $W$ intertwines the representations
  $U(.)e_{\lok{A}\smcoset\lok{C}}$ and $(Ue_\lok{A}\otimes
  Ue_\lok{C})(.)$. Finally, we reconstruct the conformal models from
  their restrictions by applying conformal covariance.
\end{pf}

Because of this proposition we will denote the vacuum
representation of a \Name{Coset} pair
$\lok{A}\coset\lok{C}\subset\lok{B}$ by $\lok{A}\otimes\lok{C}$;
hence, we regard the inclusion $\lok{A}\coset\lok{C}\subset\lok{B}$ as
a locally normal representation of $\lok{A}\otimes\lok{C}$. \label{ind:ACtimes}

% The argument in the
% proof of proposition \ref{prop:cospatenpos} directly proves the
% respective {\em quasi-local algebras} 
%  of $\lok{A}\coset\lok{C}e_{\lok{A}\coset\lok{C}}$ and
%   $\lok{A}e_\lok{A}\otimes\lok{C}e_\lok{C}$ to be spatially
%   isomorphic. These $C^*$-algebras are simple \cite{hjB67}
%   \cite[appendix 
% II.D]{KMR90}, and thus the vacuum has to be separating for the
% quasilocal algebras, as the 
% elements which annihilate the vacuum form a closed two-sided
% ideal.   

% The argument covers the inclusion of the $\opo$ dimensional subnet
% given by left and right ``chiral observables'' \cite{khR00} of a $\opo$
% dimensional conformal theory as well. We will derive a connection
% between \Name{Coset} pairs and chiral observables in chapter{}
% \ref{cha:netend}.

Even in the case of {\em finite index} (see below) it is not clear,
whether the representation induced 
by the inclusion $\lok{A}\coset\lok{C}\subset\lok{B}$ has a spatial
decomposition into tensor products of representations of
$\lok{A}e_\lok{A}$ and $\lok{C}e_\lok{C}$, respectively. The examples
we study in chapter{} \ref{cha:finind} have such a spatial
decomposition and there are conditions which ensure
this structure \cite[lemma 27]{KLM01}.

\subsubsection{Split property for chiral subnets}

One of the properties which are expected to hold true in every
physically decent local quantum theory is the {\em split
property}. In physical terms, the split property ensures that
observables which are sufficiently spacelike separated do not only
commute, but actually become {\em statistically
independent}\footnote{For a discussion of different notions of
  statistical independence see eg \cite{sS90}.}. There are equivalent
formulations of this property, but the usual definition (adapted to
our context; cf \cite[definition 2.11]{FG93}) is: a
chiral net $\lok{B}$ has the split property, if for any pair
$I_{1,2}$ of proper intervals satisfying $\overline{I_1}\subset I_2$
there is a type $I$ factor $\lok{M}$ 
interpolating between $\lok{B}(I_1)$ and $\lok{B}(I_2)$, ie:
\begin{equation}
  \label{eq:splitprop}
  \lok{B}(I_1) \subset \lok{M} \subset \lok{B}(I_2)\, , \,\,
  \overline{I_1}\subset I_2\Subset \Seins \zendot
\end{equation}

The split property is implied by the condition of
{\em nuclearity} which, loosely speaking, says that finite volumes in
classical phase space should correspond to almost finite-dimensional parts in
state space of quantum physics. Nuclearity ensures, in particular,
decent thermodynamic properties of a theory. A technical formulation
was given and established for the free scalar Hermitian field by
\Name{Buchholz and Wichmann} \cite{BW86}. For a general summary on the
notions of split property and nuclearity see eg \cite[V.5]{rH92}.

The split property has far reaching consequences. First to
name
is the (almost) complete determination of the type 
of local algebras. \Name{Buchholz, D'Antoni and Fredenhagen}
\cite{BDF87} proved the local algebras to be the tensor product of
their centre and {\em the hyperfinite type $\III_1$ factor}; the latter is
unique up isomorphism \cite{uH87}. Another
  important consequence is a general quantum version of \Name{Noether}'s
  theorem \cite{BDL86, ADF87}. 

In \cite{BDF87} the formulation is given for nets in $\opd$-dimensional
\Name{Minkowski} spacetime, but it was translated for chiral
nets by \Name{Gabbiani and Fr\"ohlich} \cite[lemma 2.12, theorem
2.13]{FG93}. In typical chiral models $e^{-\beta L_0}$, $\beta>0$, is
trace-class and nuclearity follows from the asymptotic properties of
its trace in the limit $\beta\searrow 0$ (see section 
\ref{cha:noset}.\ref{sec:nucPn}); a very general discussion on this
aspect of nuclearity for chiral conformal theories is contained in
\cite[theorem 3.2., lemma 3.3]{ALR01}.

From the very formulation of the nuclearity condition
(see equations (\ref{eq:defnuccond}), (\ref{eq:nucbound})), it is
obvious, that any  
chiral subnet $\lok{A}\subset\lok{B}$ inherits nuclearity and hence
the split property. It is remarkable that the split property can be
shown directly to be passed on from $\lok{B}$ to the subnet $\lok{A}$
by standard methods \cite[proof of proposition 2.3]{CC01}. As every
\Name{Coset} model $\lok{C}$ is a chiral subnet, the following
proposition applies
to them in particular:
\begin{prop} \label{prop:subsplit}
 Let 
$\lok{A}\subset\lok{B}$ be  a chiral subnet. If  $\lok{B}$ has the
split property (is 
nuclear), then $\lok{A}$ has the split property (is nuclear).
\end{prop}

% In particular, heredity of the split property means: for each
% $I\Subset\Seins$ the algebras 
% $\lok{A}(I)$ and $\lok{B}(I)$ are algebraically (and spatially)
% isomorphic, if $\lok{B}$ is split- because both are isomorphic to the
% hyperfinite $\III_1$ factor. 

\begin{pf}
  There is nothing to prove with respect to nuclearity. We discuss the split
  property for the restriction of 
  $\lok{A}$ to its vacuum subrepresentation $\lok{A}e_\lok{A}$.
  According to arguments in  \cite{dB74} we only need to prove a
  faithful normal product state $\phi$ to exist on 
  ${\lok{A}}(I_1)e_\lok{A}\vee{\lok{A}}(I_2')e_\lok{A}$, if
  $\overline{I_1}\subset I_2$. 

From $\overline{I_1}\subset I_2$ we conclude that there is  $I_3\Subset\Seins$
satisfying $I_1\cup I_2' \subset I_3$, and because of modular
covariance $\eta: 
\lok{A}(I_3)\rightarrow \lok{A}(I_3)e_\lok{A}$, given by
$A\mapsto Ae_\lok{A}$,  defines an isomorphism of \Name{v.Neumann}
algebras. Upon restriction, $\eta$ becomes an isomorphism from
${\lok{A}}(I_1)\vee{\lok{A}}(I_2')$ onto
${\lok{A}}(I_1)e_\lok{A}\vee{\lok{A}}(I_2')e_\lok{A}$.
Hence, we get the desired state $\phi$ by taking a normal, faithful
product state $\psi$ on 
${\lok{B}}(I_1)\vee{\lok{B}}(I_2')$ (which exists since $\lok{B}$ is
split) and setting: $\phi:=\psi\circ\eta^{-1}$. 
\end{pf}

\subsubsection{On the isotony problem of chiral subnets}

A chiral subnet $\lok{A}\subset\lok{B}$ is called {\em
  cofinite} (cf \cite{fX00}), if the inclusion
$\lok{A}\coset\lok{C}_{max}\subset\lok{B}$ is of finite index, which
  is equivalent to the requirement that the \Name{DHR} 
  endomorphism of $\lok{A}\otimes \lok{C}_{max}$ which induces the
  representation $\lok{A}\coset\lok{C}_{max}\subset\lok{B}$ has finite
  statistics\footnote{More details on inclusions with
  finite index in chapter \ref{cha:finind}.}. 
In general it is hard to prove that a subnet $\lok{A}\subset\lok{B}$
is cofinite. However, \Name{Xu} \cite{fX00} was able to prove a large
  class of current subalgebras being cofinite.
 Cofiniteness of $\lok{A}\subset\lok{B}$
has remarkable consequences, like: $\lok{B}$ is strongly additive, if 
  and only if both  $\lok{A}$ and $\lok{C}_{max}$ are
  \cite{rL01}. The current subalgebras considered by \Name{Xu} are
  strongly additive \cite[corollary IV.1.3.3.]{vL97}.

Yet, finiteness of index is not a general assumption. For example,
\Name{Rehren} studied the chiral subnet induced by the inclusion
of the stress-energy tensor of
central charge $c=1$ in the current algebra $ \Loop{SU(2)}_1$ and found that
this inclusion  does not have a finite index
\cite{khR94}. Furthermore, \Name{Carpi} \cite{sC02}  has shown that
sectors of the theory generated by this stress-energy tensor typically
do not have finite statistical dimension. 

%%%%%%%%%%%%%%%%%%%%%%%%%%%%%%%%%

So, even if both $\lok{A}$ and $\lok{B}$ are strongly additive, it is
unclear, in general, whether $\lok{C}_{max}$ or any other non-trivial
\Name{Coset} model is strongly additive\footnote{Not even for current
  subalgebras the situation has been 
  clarified. Relying on results of \Name{Wassermann} \cite[theorem
  E]{aW98}, \Name{Xu} claimed strong additivity to hold for the
  maximal \Name{Coset} model of current subalgebras
  $\lok{A}\subset\lok{B}$, $\lok{B}$ a current algebra associated with
  $SU(n)$ \cite{fX00}, but the method of proof for \cite[theorem
  E]{aW98} is not valid, actually, and thus the claim was withdrawn
  \cite{fX01}.}. 
If we now look at the subnet
$\lok{C}_{max}\subset\lok{B}$ and consider its
\Name{Coset} models, we arrive at the isotony problem for this inclusion. 
In case 
$\lok{C}_{max}$ is strongly additive as well, it is obvious that
$\lok{C}_{max}\subset\lok{B}$ and its maximal \Name{Coset} model  are
locally their mutual 
relative commutants. Inclusions of this type are of particular
interest; \Name{Rehren} called them {\em normal pairs of subtheories}
\cite{khR00}. Hence, we have reasons to take the isotony problem
serious, even if we start with an inclusion  of strongly additive
theories. 

%%%%%%%%%%%%%%%%%%%%%%%%%%%

We would like to have a simple and applicable characterisation
of local observables in $\lok{B}$ which belong to a \Name{Coset} model
associated with a subtheory $\lok{A}$, and we would like this
characterisation to involve only {\em local data} following the
conviction that every observation is of finite extension and of finite
duration. Of course, it is possible to make this decision
 simply by taking all operators from $\lok{C}_I$ and discarding all
 operators which do not commute with all operators belonging to an
 algebra $\lok{A}(J)$, $J$ slightly enlarged. But such a method will
 in most cases not prove useful when looking at a particular model
 and, furthermore, chiral conformal
 quantum field theories usually behave well when taking the limit
 $J\rightarrow I$. 
So, we are led to the conjecture that equality $\lok{C}_I=\lok{C}_{max}(I)$ 
 should hold in very general circumstances.

As it stands at the moment, the maximal \Name{Coset} model is
determined by {\em global data}, the inner-implementing representation
$U^\lok{A}$, and establishing the equality
$\lok{C}_I=\lok{C}_{max}(I)$ would prove that all \Name{Coset} models
are of a {\em local nature}, their local operators being singled out by a
simple algebraic relation only involving local data associated with
the very same localisation region, namely the commutativity with the
operators of $\lok{A}(I)$. 

For dealing with the isotony problem in our context\footnote{Apparently,
  \Name{Carpi and Conti} encountered the same problem while
  generalising their analysis \cite{CC01} to general field algebras
  and solved it by methods quite different from the ones applied here
  \cite{CC03}.}, we look at the
action of   $Ad_{U^\lok{A}}$ on the local observables of $\lok{B}$. 
Because the  %\Name{Borchers-Sugawara}
 construction of $U^\lok{A}$ does not refer to the local structure  of
 $\lok{A}$ at all (see section \ref{sec:bosug}), we need some more
 information on the way this 
 representation is generated by local observables. 

In chiral conformal
 field theory it is natural to assume that $U^\lok{A}$  is generated
 by integrals of a stress-energy tensor 
 affiliated with $\lok{A}$. This assumption does not imply strong
 additivity \cite{BS90} and concerning the models known today seems
 more general, because all strongly additive 
 models do contain a stress-energy tensor. 
Because of the special features of stress-energy tensors in chiral
(and $\opo$-dimensional) conformal field theory, mainly due to the
\Name{L\"uscher-Mack} theorem \cite{FST89,gM88,LM76}, this assumption
 admits a successful discussion of the isotony problem, see chapter
 \ref{cha:netend}. 
But the presence of a stress-energy tensor does not 
trivialise the problem at all. In fact, one is led to pinpoint the
problem using general arguments, before the
stress-energy tensor actually is needed to prove two crucial
lemmas. Our discussion should, therefore, serve well as a 
setup for further generalisations.

Even for current subalgebras, which always contain a stress-energy tensor
by the \Name{Sugawara} construction, the action of the  stress-energy
tensor $\Theta^\lok{A}$ of a current subalgebra on general currents
in the larger current algebra $\lok{B}$ has not been studied as such,
yet. Only in connection with the classification of {\em conformal
inclusions}, ie the case that the stress-energy tensor
$\Theta^\lok{B}$ coincides with that of $\lok{A}$
\cite{SW86,AGO87,BB87}, this action has
been object of research, however, by quite indirect methods. The new
perspective of analysing the action of $U^\lok{A}$ on $\lok{B}$ (in
this context: of $\Theta^\lok{A}$ on $\lok{B}$) 
 has led to a simple  characterisation of
conformal inclusions by methods familiar in (axiomatic) quantum field
theory, see section \ref{cha:currsub}.\ref{sec:cocosub}. There is a natural
notion of conformal inclusion of chiral nets in our broader approach
(definition \ref{def:confinc}). 

Our analysis directly applies to the maximal \Name{Coset} models
of current subalgebras, because these contain the \Name{Coset}
stress-energy tensor 
$\Theta^\lok{B}-\Theta^\lok{A}$. This way we extend the finding on
normal pairs for cofinite current subalgebras to all inclusions  
$\lok{A}\subset\lok{B}$ where both $\lok{B}$ 
and $\lok{A}$ contain a stress-energy tensor, independent
of strong additivity or the index of the inclusion
$\lok{A}\coset\lok{C}_{max}\subset\lok{B}$. 

We will give a summary on current subalgebras in chapter
\ref{cha:currsub} and show them to have all the features which we
assume to hold for our analysis. The isotony problem
will  be solved step by step in chapter \ref{cha:netend}. But, first 
we have to show that our notion of maximal \Name{Coset} model applies
to all chiral subnets by proving the inner-implementing representation
associated with them to exist and to have  properties required later on.

%%%%%%%%%%%%%%%%%%%%%%%%%%%%%%%%%%%%%%%
%%%%% BORCHERS SUGAWARA CONSTRUCTION %%
%%%%%%%%%%%%%%%%%%%%%%%%%%%%%%%%%%%%%%%

\sekt{Conformal transformations as observables}{Conformal
  transformations as observables}{sec:bosug} 
%\section{\Name{Borchers-Sugawara} construction}
%\label{sec:bosug}

Spacetime symmetries are of paramount importance to relativistic
quantum field theory. Intuitively we expect such coordinate
transformations to be connected to observables. Time translations, for
example, should be observable due to their connection with the energy
operator. If we have a stress-energy tensor in the theory, as it is
often the case in models, the energy operator itself is given
as an integral of this local quantum field. Yet, the implementation of
covariance may be given in abstract terms or may stem
from a larger theory into which the theory of interest is embedded,
and it is not always manifest how covariance 
may be implemented by observables of the subtheory.

More specifically, as a fact of life any observation is of finite
extension in space and time  and thus we regard the {\em local}
observables as {\em the} constituting objects in quantum field theory. 
For this reason we shall work with the
\Name{v.\-Neu\-mann} algebra $\lok{A}$ which is generated by all local
observables. 
Thereby our setting includes quantum field theories  which are
not necessarily 
described completely by covariant quantum {\em fields}, and which might not
possess a stress-energy tensor. In fact, the main result  of this
section (theorem
\ref{hauptsatz}) is an abstract
statement about  \Name{v.Neumann} algebras, without reference to the local
structure of a quantum field theory.

We consider  
representations of such theories which admit a unitary
implementation 
of covariance, $V$,  and the task thus amounts to a search for observable, unitary,
implementing operators.  Quite obviously these operators can not
be {\em local} 
observables, since  locality implies that adjoint action of these
operators 
is trivial on algebras which are associated with causally disconnected
regions. On the
other hand we believe any observation has to 
be local in nature and we conclude: spacetime transformations should
be non-local limits of local observables, ie genuine global observables.

The problem of identifying spacetime symmetry transformations as
global observables is of interest only, if the given
representation is reducible. In irreducible representations, such as
the vacuum representation, {\em any} bounded
operator can be represented as a weak limit of local operators. The
representations induced by non-trivial chiral subnets
$\lok{A}\subset\lok{B}$ are 
manifestly reducible,  as we have for the cyclic projection:
$e_\lok{A}\neq \Einsop, 0$.

To our knowledge this problem so far has been dealt with only in the case of
abelian groups of translations satisfying the spectrum condition
(positivity of energy). \Name{Borchers} \cite{hjB66} has solved this
problem relying almost entirely on the spectrum condition and using a
deep result on the innerness of norm-continuous connected
automorphism groups  of \Name{v.Neumann} algebras \cite[corollary 8]{KR67}. His result is the key building block in our construction. 

 In the abelian case there are many inner-implementing representations
with different spectral properties. It was a
challenging task to ensure 
existence of an inner-implementing representation satisfying the
spectrum condition. 
\Name{Arveson} \cite{wA74} gave a proof for a
one-parameter group, \Name{Borchers} and \Name{Buchholz}
\cite{hjB84,BB85} succeeded in
solving this problem in general; see \cite{hjB87} for a summary.

The situation for an inner-implementing representation
of $\PSL(2,\dopp{R})^\sim$ is different. Because
$\PSL(2,\dopp{R})^\sim$ is
 identical with its commutator subgroup,
the result of our construction is unique  and validity of the
spectrum condition follows. We show as well that $V$-invariant vectors
are left invariant by the action of the inner-implementing
representation $V^{\lok{A}}$. Another result is the proof of
complete reducibility of $V^{\lok{A}}$ under weak assumptions on the
original representation $V$.

In the course of our argument we will construct an inner-implementing
representation  $V^{\lok{A}'}$ for the commutant of the
\Name{v.Neumann} algebra $\lok{A}$ as well. We have the following relation:
$ V(\tilde{g}) \,= \,\,V^\lok{A}(\tilde{g}) \,V^{\lok{A}'}(\tilde{g})\, , \quad \forall
\,\tilde{g}\in\PSL(2,\dopp{R})^\sim$.
This equation reminds of the \Name{Coset} construction \cite{GKO86}
involving stress-energy tensors of chiral current
algebras, which are given by the
\Name{Sugawara} construction. It is not difficult to show
that our result agrees with the 
outcome of integrating the respective stress-energy tensors (see
chapter \ref{cha:currsub}).

 Although the 
relation to \Name{Coset} constructions as considered by \Name{Goddard,
  Kent and Olive} \cite{GKO86} motivates the enterprise undertaken
here, its result is independent of the existence of a stress-energy
tensor. We have made use of this, already,  and connected it to a
generalised notion of \Name{Coset} construction (lemma
\ref{lem:cosmax}). In subsection \ref{subsec:bosappl} we will 
give some more details on this and other applications to chiral subnets.

Although there are special features of an
inner-implementing representation connected to a stress-energy
tensor, the result of our construction serves well as a substitute for
the \Name{Sugawara} stress-energy 
tensor in many respects. On the other hand we believe our construction
to be somewhat 
special to (chiral) conformal field theories as we argue in the
discussion concluding this section, and we know that the deeper part
of it is due to \Name{Borchers}. Summing up these thoughts we consider
the term {\em Borchers-Sugawara construction} appropriate.

Most of this section has been published already \cite{sK02}, but there
are some  differences which we have indicated in the text. Moreover,
 we have included  some  additional material, in particular proposition
    \ref{prop:JonUA}.

\subsection{\Name{Borchers-Sugawara} construction}
\label{subsec:prepbosug}

In the following $\Hilb{H}$ stands for a separable
\Name{Hilbert} space and $V$ is a unitary, strongly continuous
representation of $\PSL(2,\dopp{R})^\sim$ on $\Hilb{H}$.  If not
stated otherwise, $\lok{A}$ 
stands for a \Name{v.Neumann} algebra of operators on $\Hilb{H}$,
$\lok{A}'$ for its commutant and $\alpha$, $\alpha'$ for automorphic
actions of $\PSL(2,\dopp{R})^\sim$ on $\lok{A}$,  $\lok{A}'$ respectively. We
note that any spatial automorphism of $\lok{A}$, given by the adjoint
action of a unitary operator, induces a spatial
automorphism of $\lok{A}'$ as well.

We first prove a lemma on the spectrum condition. The
result is well known and our proof
is not new, presumably; its
second part is adapted from \cite{gM77}. 

\begin{prop}\label{prop:poengcond}
  If any one of the
  operators $L_0$, $P$, $-K$ has positive spectrum, then all three of
  them. In this case we say that $V$ satisfies the spectrum condition.
\end{prop}

\begin{pf}
Assume $L_0$ is positive. Take any vector $\phi$
analytic for the
representation $V$ (cf eg \cite{BR77}). We have:
\begin{equation}
  \label{eq:scalpos}
  0 \leqslant 2 \langle\phi, V(\widetilde{D}(\tau))L_0
    V(\widetilde{D}(-\tau))\phi\rangle = e^{\tau}
    \langle\phi, P\phi\rangle +
  e^{-\tau} \langle\phi, -K\phi\rangle \zendot 
\end{equation}
Multiplying by $e^{\pm \tau}$ and taking the appropriate 
limits $\tau\rightarrow \mp\infty$ we deduce  $\omega_\phi(P)\geq 0$ and
$\omega_\phi(-K)\geq 0$. Since the analytic vectors for the
representation $V$ form a core for all generators we may apply
criterion 5.6.21 of \cite{KR83}.

Now assume $P$ or $-K$ is positive. Special conformal transformations
and translations are conjugate in $\PSL(2,\dopp{R})$: $S(-n)=
R(\pi) T(n) R(-\pi)$. Defining $g_t := S(n) R(t) T(n) R(-t)$ this identity
becomes: $\lim_{t \nearrow \pi} g_t = id$. Now we see that the corresponding
relation holds true in $\PSL(2,\dopp{R})^\sim$, since we know it for
$\PSL(2,\dopp{R})$, the relation is continuous in $n$ and the 
covering projection is continuous as well. Because conjugation by a unitary
operator does not change the spectrum, positivity of $P$ follows from
positivity of $-K$ and vice versa. Positivity of $L_0$ follows from
equation (\ref{eq:L0PK}) by criterion 5.6.21 of \cite{KR83} applied as
before.
\end{pf}

Alternatively, one may prove proposition \ref{prop:poengcond} by
decomposing any unitary representation satisfying the spectrum
condition into a direct integral of irreducible representations (eg
\cite[chapter 5, \S 6, theorem 3]{BR77}). The latter are known
explicitly (eg \cite{dG93}) and the ones fulfilling the spectrum
condition on $L_0$ have positive spectrum for the translations as
well. This procedure is followed in the literature (eg \cite{GL96,PS86,FG93}). 

\begin{prop}\label{prop:inneruni}
  Assume  $Ad_{V}$ induces an automorphism
  group $\alpha$ on $\lok{A}$. If there exists a representation $V^\lok{A}$ of
  $\PSL(2,\dopp{R})^\sim$ by unitary operators in  $\lok{A}$ implementing
  $\alpha$ by its adjoint action on $\lok{A}$, then 
  this representation is unique.
\end{prop}

\begin{pf} Assume there are two such representations, $V_1^\lok{A}$
and $V_2^\lok{A}$. Then the operators
$V_1^\lok{A}(\tilde{g})V_2^\lok{A}(\tilde{g})^*$, $\tilde{g}\in\PSL(2,\dopp{R})^\sim$, implement the
trivial automorphism. For this reason these operators belong to the
centre of $\lok{A}$. Using this fact it is straightforward to show
that the operators  $V_1^\lok{A}(\tilde{g})V_2^\lok{A}(\tilde{g})^*$ form a
representation of $\PSL(2,\dopp{R})^\sim$. This representation is abelian and
its kernel contains all elements of the form
$\tilde{g}_1\tilde{g}_2\tilde{g}_1^{-1}\tilde{g}_2^{-1}$. 

Now these elements generate the whole of 
$\PSL(2,\dopp{R})^\sim$ since it is a {\em perfect group}, ie it
coincides with its commutator subgroup. One can reduce this statement
on the \Name{Lie} group to the structure of the corresponding
\Name{Lie} algebra by standard arguments (eg \cite{HN91}, lemmas
III.3.19, III.3.20, definition I.5.1 and remarks nearby this
definition). So, 
$\PSL(2,\dopp{R})^\sim$ is perfect since it has a simple \Name{Lie}
algebra and is connected.  Thereby
$V_1^\lok{A}(\tilde{g})V_2^\lok{A}(\tilde{g})^*=\Einsop$ $\forall 
\tilde{g}\in\PSL(2,\dopp{R})^\sim$.\end{pf}

We call a representation $V^\lok{A}$ in the sense of the proposition
above an {\em inner-implementing representation} (corresponding to the
pair $(V,\lok{A})$). We immediately have:

\begin{prop}\label{prop:uastrich} 
 Assume the unique inner-implementing representation 
 $V^\lok{A}$ to exist. Then $V^{\lok{A}'}\equiv V
 (V^\lok{A})^*$ is the unique  inner-implementing representation
 corresponding to $(V,\lok{A}')$. If $V^\lok{A}$ is  strongly
 continuous, then so is $V^{\lok{A}'}$. 
\end{prop}

\begin{pf} First we prove innerness of the operators 
$V(\tilde{g}) V^\lok{A}(\tilde{g})^*$ by recognising that their
adjoint action on 
$\lok{A}$ implements the trivial automorphism. Making use of this
it is straightforward to show that these operators do in
fact define a representation. The implementation property and
unitarity are trivial.  Uniqueness follows from proposition
\ref{prop:inneruni} directly. Continuity is fulfilled, since we are multiplying
continuous functions.
\end{pf}    

We now come to the derivation of the main result of this section
(theorem \ref{hauptsatz}). It depends on the following statement:
\begin{lemma}\label{th:borchers}
  Let $V$ satisfy the spectrum
  condition and let $Ad_V$ induce an automorphism group $\alpha$ of
  $\lok{A}$. Then there  
  are strongly continuous, unitary, inner-implementing  representations
  $\widetilde{T}^\lok{A}$, $\widetilde{S}^\lok{A}$,
  $\widetilde{R}^\lok{A}$ for the restrictions of
  $\alpha$ to the one-parameter subgroups of translations, special
  conformal transformations and rotations, respectively.
\end{lemma}

\begin{pf} This is an application of
\Name{Borchers}' theorem \cite{hjB66} and proposition \ref{prop:poengcond}.\end{pf}

At this point we stress that it is not clear at all
whether these restricted inner-implementing groups form a
representation of $\PSL(2,\dopp{R})^\sim$. We will show that the
inner-implementing representation may be constructed from any given  
pair $\widetilde{T}^\lok{A}$, $\widetilde{S}^\lok{A}$.
 The fact that there
are sufficiently many subgroups satisfying the spectrum condition
to generate the whole group seems to be special.

According to the \Name{Iwasawa} decomposition of $\PSL(2,\dopp{R})$ \cite{FG93} (appendix
I) we can write every $g\in \PSL(2,\dopp{R})$ in the form 
$g = {T}(p_g) {D}(\tau_g) {R}(t_g)$. Each term in this decomposition
depends continuously on $g$. By a 
short consideration on the covering projection from $\PSL(2,\dopp{R})^\sim$ to
$\PSL(2,\dopp{R})$ we readily see that the same decomposition works for
$\PSL(2,\dopp{R})^\sim$ as every element $\tilde{g}$ is of the form
$\tilde{R}(2\pi)^m g$, $m\in\dopp{Z}$ and $g$ from the first sheet of
the covering, which we may identify with $\PSL(2,\dopp{R})$.

Again in $\PSL(2,\dopp{R})$ we may check that every dilatation and every rotation may
be written as follows:
\begin{eqnarray}
  {D}(\tau) &=& {S}(-(e^{\frac{\tau}{2}}-1)e^{-\frac{\tau}{2}})\,
  {T}(1)\, {S}(e^{\frac{\tau}{2}}-1)\,
  {T}(-e^{-\frac{\tau}{2}}) \,\, ,\label{eq:stiwa1}\\
{R}(2t) &=& {S}((-1+\cos t)(\sin t)^{-1}) \,
{T}(\sin t) \, {S}((-1+\cos t)(\sin t)^{-1}).
\label{eq:stiwa2}
\end{eqnarray}
By looking at the curves in $\PSL(2,\dopp{R})^\sim$ defined by the left and
the right-hand sides and the action of the covering projection we
conclude that the corresponding equations hold true in $\PSL(2,\dopp{R})^\sim$.

Now we have found that any $\tilde{g}\in \PSL(2,\dopp{R})^\sim$ may be
written as a 
product of four translations and four special conformal
transformations each of them depending continuously on
$\tilde{g}$. Using the results of \Name{Borchers}' construction (lemma
\ref{th:borchers}), the \Name{Iwasawa} decomposition and (\ref{eq:stiwa1}), (\ref{eq:stiwa2}) we define for each
$\tilde{g}\in\PSL(2,\dopp{R})^\sim$: 
\begin{equation}
  \label{eq:pia}
  \pi^\lok{A}(\tilde{g}) := \prod_{i=1}^{4}
  \widetilde{T}^{\lok{A}}(p^{(i)}_{\tilde{g}})\widetilde{S}^{\lok{A}}(n^{(i)}_{\tilde{g}})\,
  ,\quad \tilde{g}\in\PSL(2,\dopp{R})^\sim \zendot
\end{equation}
We have $\pi^\lok{A}(id)=\Einsop$.
The following lemma asserts that the
$\pi^\lok{A}(\tilde{g})$ define an  inner-implementing
representation up to a cocycle in the centre of $\lok{A}$. To this end
we define operators sensitive to 
the violation of the group multiplication law:
$ z^\lok{A}({\tilde{g}},{\tilde{h}}):= \pi^\lok{A}({\tilde{g}}) \pi^\lok{A}({\tilde{h}}) \pi^\lok{A}({\tilde{g}}{\tilde{h}})^*,\,
{\tilde{g}}, {\tilde{h}}\in\PSL(2,\dopp{R})^\sim$. 

\begin{lemma}\label{lem:words}
  $\pi^\lok{A}: {\tilde{g}}\mapsto \pi^\lok{A}({\tilde{g}})$ defines a strongly continuous
  mapping with unitary values in $\lok{A}$. The adjoint action of
  $\pi^\lok{A}({\tilde{g}})$, ${\tilde{g}}\in\PSL(2,\dopp{R})^\sim$, on $\lok{A}$ implements the
  automorphism $\alpha_{\tilde{g}}$. $z^\lok{A}: ({\tilde{g}},{\tilde{h}})\mapsto z^\lok{A}({\tilde{g}},{\tilde{h}})$
  defines a strongly continuous 2-cocycle with unitary values in
  $\lok{A}'\cap\lok{A}$.
\end{lemma}

\begin{pf} 
Unitarity is obvious. Strong continuity follows in both cases from
continuity of products of continuous functions. The
implementing property of the $\pi^\lok{A}({\tilde{g}})$ follows
immediately by the 
decomposition of ${\tilde{g}}$ into a word containing four translations and
four special conformal transformations, the definition of
$\pi^\lok{A}({\tilde{g}})$ and lemma \ref{th:borchers} due to \Name{Borchers}. At
this point all
but the cocycle properties of $z^\lok{A}$ follow immediately from
its definition. If
we look at $\pi^\lok{A}({\tilde{f}})\pi^\lok{A}({\tilde{g}})\pi^\lok{A}({\tilde{h}})$, insert some
identities appropriately, we find:
$  z^\lok{A}({\tilde{f}},{\tilde{g}}{\tilde{h}})z^\lok{A}({\tilde{g}},{\tilde{h}})=z^\lok{A}({\tilde{f}},{\tilde{g}})z^\lok{A}({\tilde{f}}{\tilde{g}},{\tilde{h}})$. 
Even more immediate are the equalities 
$z^\lok{A}(id,{\tilde{g}})=z^\lok{A}({\tilde{g}}, id) = \Einsop$.
\end{pf}

We write the
abelian \Name{v.Neumann} algebra generated by the cocycle operators
$z^\lok{A}({\tilde{g}},{\tilde{h}})$ as follows:
$
  \lok{Z}^\lok{A} \equiv \{z^\lok{A}({\tilde{g}},{\tilde{h}}),
      z^\lok{A}({\tilde{g}},{\tilde{h}})^* | {\tilde{g}},{\tilde{h}} \in \PSL(2,\dopp{R})^\sim\}''
$. Obviously $\lok{Z}^\lok{A}$ is contained in the centre of $\lok{A}$.
Now we are prepared to realise the construction itself by proving that
the cocycle $z^\lok{A}$ is exact: 
\begin{lemma}\label{lem:constr}
  For every ${\tilde{g}}\in\PSL(2,\dopp{R})^\sim$ exists a unitary operator
  $z^\lok{A}({\tilde{g}})\in\lok{Z}^\lok{A}$ such that
  \begin{equation}
    \label{eq:defUA}
    V^\lok{A}({\tilde{g}}):= z^\lok{A}({\tilde{g}})\pi^\lok{A}({\tilde{g}})
  \end{equation}
defines a strongly continuous, inner-implementing
representation of $\PSL(2,\dopp{R})^\sim$.
\end{lemma}

\begin{pf} As 
$ \lok{Z}^\lok{A}\subset \lok{A}\cap \lok{A}'$ we may apply the direct
integral decomposition (cf eg \cite[chapter 14]{KR86}). This yields
a decomposition of $\Hilb{H}$ as a direct integral of \Name{Hilbert}
spaces $\Hilb{H}_x$ and it implies for the operators under
consideration: the action of 
$z^\lok{A}({\tilde{g}},{\tilde{h}})$ on $\Hilb{H}_x$, denoted by $z^\lok{A}({\tilde{g}},{\tilde{h}})(x)$, is a
multiple of the identity $\Einsop_x$ and thereby defines for
almost every $x$ a  continuous
2-cocycle $\omega({\tilde{g}},{\tilde{h}})_x\in\dopp{C}_1\Einsop_x$.
The action of the operators $\pi^\lok{A}({\tilde{g}})$ on 
$\Hilb{H}_x$, denoted by $\pi^\lok{A}({\tilde{g}})(x)$, defines for almost
every $x$ a unitary, strongly continuous, projective
representation of $\PSL(2,\dopp{R})^\sim$.

For \Name{Lie} groups with a simple \Name{Lie} algebra the lifting
criterion is valid \cite{dS68}, \cite[section III.10, theorem
10]{nJ62}. This ensures for almost every $x$ the 
existence of continuous phases $\omega({\tilde{g}})(x)$ such that
$\omega({\tilde{g}})(x) \pi^\lok{A}({\tilde{g}})(x)$ defines a representation of
$\PSL(2,\dopp{R})^\sim$. Integrating $\omega_x({\tilde{g}})$ over all $x$ yields a
unitary $z^\lok{A}({\tilde{g}})\in \lok{Z}^\lok{A}$, depending strongly continuously on
${\tilde{g}}$. Integrating the representations defined by the
$\omega({\tilde{g}})(x)\pi^\lok{A}({\tilde{g}})(x)$ yields a unitary, strongly continuous
representation $V^\lok{A}$ satisfying equation
(\ref{eq:defUA}). $V^\lok{A}({\tilde{g}})$ is an element of $\lok{A}$ for every
${\tilde{g}}$ and implements $\alpha_{\tilde{g}}$ by its adjoint action due to lemma
\ref{lem:words}.
\end{pf}

\noindent We summarise the discussions above:

\begin{theo}\label{hauptsatz} Let $\Hilb{H}$ be a
  separable \Name{Hilbert} space, and $V$ a representation of 
  $\PSL(2,\dopp{R})^\sim$ on $\Hilb{H}$, unitary, strongly
  continuous  and satisfying the spectrum
  condition. $\lok{A}$ is taken to be a \Name{v.Neumann} algebra
  of bounded operators 
  on $\Hilb{H}$. The adjoint actions of $V$ on $\lok{A}$,
  $\lok{A}'$ shall
  define groups $\alpha$, $\alpha'$ of automorphisms of $\lok{A}$,
  $\lok{A}'$, respectively. 

Then there exist unique unitary, strongly continuous,
inner-implementing representations $V^\lok{A}$, $V^{\lok{A}'}\equiv
V(V^\lok{A})^*$ of $\PSL(2,\dopp{R})^\sim$. 
\end{theo}

\begin{pf} Direct consequence of the
propositions and lemmas above.
\end{pf}

\noindent {\em Two remarks:} If we start with a proper representation  $V$ of
$\PSL(2,\dopp{R})$, then one arrives at representations $V^\lok{A}$,
$V^{\lok{A}'}$ 
which will be (generalized) ray representations of $\PSL(2,\dopp{R})$
and proper 
representations of 
$\PSL(2,\dopp{R})^\sim$. The cocycles of $V^\lok{A}$, $V^{\lok{A}'}$ have to
be mutually inverse, and common eigenvectors of $L_0^\lok{A}$,
$L_0^{\lok{A}'}$ have eigenvalues which sum up to integers.

In \cite{sK02} an alternative derivation is given for
$\PSL(2,\dopp{R})$ and $\PSO(4,2)$ (conformal group in $3+1$
dimensions) which uses an argument from \cite{BGL95} and should
generalise to all conformal groups. This
method does not extend to the (universal) covering groups, at least not
directly.  The approach presented here is close to the one of
\cite{BDF00} for deriving a representation of the \Name{Poincar{\'e}}
group from modular conjugations of wedge algebras. The corollaries
which follow in this section 
generalise directly to $\PSO(4,2)$, as shown in \cite{sK02}.

Now we derive three features of the inner-implementing
representations which they inherit from the original representation:
spectrum condition, invariant vectors, complete reducibility. 

\begin{cor}\label{cor:poseng}
  Both $V^\lok{A}$ and $V^{\lok{A}'}$
  satisfy the spectrum condition. 
\end{cor}

\begin{pf} The operators
$V^{\lok{A}\vee\lok{A}'}({\tilde{g}},{\tilde{h}}):=V^\lok{A}({\tilde{g}})V^{\lok{A}'}({\tilde{h}})$ define a
unitary, strongly continuous representation of $\PSL(2,\dopp{R})^\sim\times
\PSL(2,\dopp{R})^\sim$.  With respect to $V^{\lok{A}\vee\lok{A}'}$ we have a
dense domain of analytic vectors and we take an arbitrary vector
$\psi$ from it. The result follows now as in the
proof of proposition \ref{prop:poengcond} 
from the inequality
$$
  0 \leqslant
  \langle V^\lok{A}(\widetilde{D}(\tau))\psi,PV^\lok{A}(\widetilde{D}(\tau))\psi\rangle =
  \langle\psi,e^{- \tau}P^\lok{A}\psi\rangle +
  \langle\psi,P^{\lok{A}'}\psi\rangle
$$
by letting $\tau\rightarrow\infty$ and $\tau\rightarrow -\infty$, respectively.
\end{pf}

\begin{cor}\label{cor:Bosuginvvec}
Let $\Omega\in\Hilb{H}$ be a vector left invariant by $V$. Then
$V^\lok{A}$, $V^{\lok{A}'}$ both leave $\Omega$ invariant.   
\end{cor}

\begin{pf} 
Since translations and special conformal transformations generate the
whole of $\PSL(2,\dopp{R})^\sim$ it is sufficient to show invariance of
$\Omega$ for these two subgroups. We consider translations only; the
argument for special conformal transformations is the same.

Take arbitrary $\psi\in\Hilb{H}$. We have
$\langle\psi,V^\lok{A}({\tilde{g}})\Omega\rangle=\langle\psi,V^{\lok{A}'}({\tilde{g}})^*\Omega\rangle$
by assumption.
Set $f_\psi(p):=\langle\psi,V^\lok{A}(\widetilde{T}(p))\Omega\rangle$,
$g_\psi(p):=\langle\psi,V^{\lok{A}'}(\widetilde{T}(p))^*\Omega\rangle$.
Due to the spectrum condition (corollary \ref{cor:poseng}) $f_\psi$
may be extended to the upper half of the complex plane by means of the
\Name{Laplace} transform (cf eg \cite[chapter 2]{SW64}). This
continuation is analytic in the interior and of at most polynomial
growth for complex arguments. On the real line we have
$|f_\psi|\leq\|\Omega\|\,\|\psi\|$ and due to the theorem of
\Name{Phragmen-Lindel\"{o}f} \cite[section 5.62]{eT39} this bound
holds true for the continuation of $f_\psi$ as well. 

The same line of argument works for $g_\psi$ with respect to the lower
half of the complex plane. Since $f_\psi$ and $g_\psi$ coincide on the
real line both are restrictions of an entire function (reflection
principle). This entire function is bounded by
$\|\Omega\|\,\|\psi\|$, and due to \Name{Liouville}'s theorem
it is constant. Since the vectors $V^\lok{A}(\widetilde{T}(p))\Omega$,
$V^{\lok{A}'}(\widetilde{T}(p))^*\Omega$ are determined by the scalar
products $f_\psi(p)$ and $g_\psi(p)$, $\psi\in\Hilb{H}$, invariance
follows by taking $p=0$.
\end{pf} 

For the next corollary we prepare ourselves by a comment and a
lemma. In the corollary the representation $V$ is assumed completely  
reducible with finite multiplicities. Although this is a pretty strong
assumption in group theoretical terms, we consider this a rather
convincing 
assumption from the quantum field theoretical point of view. In this
context it is somewhat weaker than a common nuclearity condition
\cite{BGL93}. Nuclearity is desirable for quantum field theories
 and in our setting it corresponds to demanding
the $L_0$ eigenspaces to be finite-dimensional with
degeneracies growing at most exponentially. Typical (integrable)
chiral models such as current algebras exhibit this behaviour
\cite{FG93}.  This implies our
assumption as the following lemma clarifies.

The centre of $\PSL(2,\dopp{R})^\sim$ is an infinite cyclic group generated by
the rotation $\widetilde{R}(2\pi)$. The following lemma shows that
complete reducibility of a representation $V$ as in theorem
\ref{hauptsatz} is equivalent to requiring 
the representation space to have a decomposition
into a direct sum of eigenspaces of $\widetilde{R}(2\pi)$. Due to the
infinite order of the centre of $\PSL(2,\dopp{R})^\sim$ this is not obvious.

\begin{lemma}\label{lem:disspec}
  Assume $V(\widetilde{R}(2\pi))$ to have pure point spectrum. Then
  the spectrum of 
  $L_0$ is pure point and $V$ is completely reducible into a direct
  sum of irreducible representations.
\end{lemma}

\begin{pf}
Let $\Hilb{H}_i$ denote the eigenspace belonging to eigenvalue $e^{i2\pi
  h_i}$. The restriction of $V(\widetilde{R}(t))e^{-i h_i t}$
to $\Hilb{H}_i$ defines a representation of $U(1)$. This
representation is completely reducible due to the compactness of
$U(1)$ (cf eg \cite[ chapter 7, \S 7, theorem 4; chapter 5, \S
4, proposition 5]{BR77}). This proves the claim on the spectrum of $L_0$.

By the spectrum condition there are vectors of lowest
eigenvalue. Because of the complete analysis of lowest weights in
unitary representations of $\PSL(2,\dopp{R})^\sim$ \cite{dG93}, it is
known which lowest eigenvalues may occur and that the cyclic
representations generated from the lowest weight vectors are
irreducible. Taking such a lowest weight vector, applying to it the linear
span of the $V({\tilde{g}})$, ${\tilde{g}}\in\PSL(2,\dopp{R})^\sim$,
and taking the 
completion yields an irreducible representation space. We may reduce
 with respect to it because of
unitarity. We iterate this procedure and arrive at the second claim
since $\Hilb{H}$ is separable.
\end{pf}

\begin{cor}
  Assume $V$ to be completely reducible with finite
  multiplicities. Then $V^\lok{A}$ and  $V^{\lok{A}'}$ are completely
  reducible. 
\end{cor}

\begin{pf} Denote the lowest weight vectors by
$\varphi_{(d,i)}$, $i$ being the multiplicity index and $d$ the
eigenvalue of $L_0$. For any fixed $d$ the $\varphi_{(d,i)}$ span a
finite-dimensional \Name{Hilbert} space. This space is left invariant
by the operators $V^\lok{A}(\widetilde{R}(2\pi))$,
$V^{\lok{A}'}(\widetilde{R}(2\pi))$. Both operators may be diagonalised
on this space simultaneously, the result being a mere relabelling of
the irreducible 
subrepresentations of $V$. Now $V^\lok{A}(\widetilde{R}(2\pi))$,
$V^{\lok{A}'}(\widetilde{R}(2\pi))$ both are diagonal on the irreducible
subspaces generated from the ``new'' lowest weight vectors
$\varphi_{(d,i)}'$  and thus on the whole  of $\Hilb{H}$. Now the claim
follows as in the proof of lemma \ref{lem:disspec}.
\end{pf}
 
\noindent{ Remark:} Non-trivial unitary representations of
$\PSL(2,\dopp{R})^\sim$ are necessarily infinite-dimensional and the
multiplicity spaces of $V^\lok{A}$
serve as representation spaces for  $V^{\lok{A}'}$ and vice versa. The irreducible
representations of $V^\lok{A}$ and  $V^{\lok{A}'}$ will, therefore, not have
finite multiplicities in general.

%%%%%%%%%%%%%%%%%%%%%%%%%%%%%%%%%%%%
%%%% Subnet applications %%%%%%%%%%%
%%%%%%%%%%%%%%%%%%%%%%%%%%%%%%%%%%%%

\subsection{Direct applications to chiral subnets}
\label{subsec:bosappl}

We  collect a few immediate implications
of the \Name{Borchers-Sugawara} construction for chiral subnets
$\lok{A}\subset\lok{B}$.

\begin{prop}\label{prop:subnets}
  The inner-implementing unitaries $U^\lok{A}(g)\neq\Einsop$ are not
  elements of any local algebra. If $\lok{A}\neq\dopp{C}\Einsop$, then
  $\lok{A}$ contains non-trivial non-local operators,
  the vacuum is not faithful for $\lok{A}$, and the action of
  $Ad_{U^\lok{A}}$ on the local operators of the net $\lok{A}$ is
  ergodic. 
\end{prop}

\begin{pf} Suppose that for some ${\tilde{g}}\in\PSL(2,\dopp{R})^\sim$
  the unitary 
$U^\lok{A}({\tilde{g}})\neq\Einsop$ is contained in a local
algebra. By locality and 
invariance of the vacuum there is a
local algebra $\lok{B}(I)$ such that all vectors $B\Omega$,
$B\in\lok{B}(I)$, remain unchanged when acted upon by
$U^\lok{A}({\tilde{g}})$. Thus, by the \Name{Reeh-Schlieder} property of $\lok{B}$,
$U^\lok{A}({\tilde{g}})$ has to be trivial and the existence of such operators
is denied. 

The kernel of $U^\lok{A}$ has to be different from $\PSL(2,\dopp{R})^\sim$,
else $\lok{A}$ is left invariant 
point-wise by the covariance automorphisms and therefore must be abelian by
locality. But  local algebras of $\lok{A}$ have to be factors as
elements of a chiral subnet. So $\lok{A}\neq\dopp{C}\Einsop$ requires
the existence of operators $U^\lok{A}({\tilde{g}})\neq\Einsop$. These are not
local operators.

Any fixed point of the action
of $Ad_{U^\lok{A}}$ on a local algebra $\lok{A}(I)$ has to be
contained in its centre due to locality. This centre is trivial since
$\lok{A}(I)$ is a factor. $\Omega$ can not be separating, because we
have: $(U^\lok{A}({\tilde{g}})-\Einsop)\Omega=0$.
\end{pf} 

%%%%%

The inversion on the light-ray, $I_+: x\mapsto -x\in \dopp{R}$,
induces an outer automorphism of $\PSL(2,\dopp{R})$: $\eta(g)= I_+ g
I_+$. We have in particular:
$\eta(T(a)) = T(-a)$, $\eta(D(t))= D(t)$, $\eta(S(n))=S(-n)$,
$\eta(R(\varphi))= R(-\varphi)$.
Obviously, this outer automorphism extends to the
universal covering group, where we denote it by $\tilde{\eta}$. We
have for the covering projection $\symb{p}$: $\symb{p}\tilde{\eta}=
\eta\symb{p}$, which may be checked directly eg by looking at the
\Name{Iwasawa} decomposition.  

The \Name{Bisognano-Wichmann} property of $\lok{B}$ implies that the
modular conjugation of $\lok{B}(\Seins_+)$, namely $J$, implements
$I_+$. By this, the representation $U$ and the operator $J$ generate a
representation of $\PSL(2,\dopp{R})_\pm \equiv
\PSL(2,\dopp{R})\rtimes_\eta\dopp{Z}_2$, the group generated by $I_+$
and $\PSL(2,\dopp{R})$ modulo the relation $\eta(g)= I_+ g I_+$. The
following simple proposition shows, that $J$ and $U^\lok{A}$ together generate
a representation of $\PSL(2,\dopp{R})^\sim_\pm$.
\begin{prop}\label{prop:JonUA}
  The adjoint action of the modular conjugation $J$ of
  $\lok{B}(\Seins_+)$ on $U^\lok{A}$ implements the outer automorphism
  $\tilde{\eta}$, ie we have: $JU^\lok{A}(\tilde{g})J =
  U^\lok{A}(\tilde{\eta}(\tilde{g}))$. The same is true for
  $U^{\lok{A}'}$. 
\end{prop}

\begin{pf}
$JU^\lok{A}J$ defines a globally inner representation of
$\PSL(2,\dopp{R})^\sim$, since $Ad_J$ defines automorphisms of the
global algebra $\lok{A}$. For arbitrary $A\in\lok{A}(I)$,
$I\Subset\Seins$, we set $A':=JAJ$. This allows us to write:
\begin{equation}
  \label{eq:conimplUA}
  Ad_{JU^\lok{A}(\tilde{g})J}(A) = Ad_J(\alpha_{\symb{p}(\tilde{g})}(A')) =
  \alpha_{\eta(\symb{p}(\tilde{g}))}(A)  \zendot
\end{equation}
Thus, $JU^\lok{A}J$ implements an automorphic action $\alpha\circ\eta$
of $\PSL(2,\dopp{R})$ on $\lok{A}$. Conversely,
$JU^\lok{A}(\tilde{\eta}(.))J$ implements the automorphic action
$\alpha$, is a globally inner representation of
$\PSL(2,\dopp{R})^\sim$, and hence identical to the unique
inner-implementing representation $U^\lok{A}$ (proposition 
\ref{prop:inneruni}). This proves the statement for $U^\lok{A}$. A look
at the very definition of $U^{\lok{A}'}$ completes the proof.
\end{pf}

In a large class of chiral conformal models such as free fermions and chiral
current algebras there are explicit constructions for the
transformation operators as observables in terms of local quantum
fields (cf eg \cite{FST89} and chapter \ref{cha:currsub}). In both
cases the construction 
yields a representation of the whole \Name{Virasoro} algebra.  This
diffeomorphism invariance 
is necessarily broken in any positive-energy representation; it
remains a $\PSL(2,\dopp{R})^\sim$ symmetry only. 

We have constructed the inner implementation of this remaining
symmetry in a completely model independent way and used the result in
lemma \ref{lem:cosmax} for the definition of the maximal \Name{Coset} model
$\lok{C}_{max}$ associated with a given subnet
$\lok{A}\subset\lok{B}$. 
It might happen that a subnet $\lok{A}\subset\lok{B}$ admits no
\Name{Coset} theory at all:
\begin{defi}\label{def:confinc}
  A chiral subnet $\lok{A}\subset\lok{B}$ is called a {\bf conformal
  inclusion}, if $\lok{C}_{max}(I)=\dopp{C}\Einsop$, $I\Subset\Seins$.
\end{defi}
This term stems
from studies on chiral current algebras. Here we have for both nets
$\lok{A}\subset\lok{B}$ stress-energy tensors $\Theta^\lok{A}$,
$\Theta^\lok{B}$. A simple argument shows that their difference
$\Theta^\lok{B}-\Theta^\lok{A}\equiv \Theta^{coset}$ is a stress
energy tensor alike. By 
the \Name{Reeh-Schlieder} theorem and the \Name{L\"uscher-Mack} theorem
\cite{FST89,gM88,LM76} $\Theta^{coset}$ vanishes iff its central charge
vanishes. Its central charge  is completely determined by the
finite-dimensional \Name{Lie} algebras from which the  
current algebras are constructed and by the embedding of the smaller one
into the larger one. Its zeros, characterising the notion of conformal
embeddings for these models, have been 
classified \cite{SW86, AGO87, BB87}.

In section \ref{cha:currsub}.\ref{sec:cocosub} we give a
  characterisation of conformal  
  inclusions of current algebras by methods familiar in axiomatic
  quantum field theory and, thus, giving an  illustration
  of our perspective to look at such problems by analysing the action
  of the inner implementation of covariance of the subnet on the
  larger theory. This approach avoids arithmetic
  considerations on the central charge. 

The following proposition shows that our definition of conformal
inclusions covers conformal inclusions of current algebras (understood
as above) as special cases (cf theorem \ref{th:SegSugCon}). Chiral
subnets of finite index  are necessarily conformal inclusions in the
sense  of definition  \ref{def:confinc} (lemma \ref{lem:lokirredincl}).

\begin{prop}\label{prop:genconfincl}
  Suppose the inner-implementing representation of theorem
  \ref{hauptsatz} for a chiral subnet $\lok{A}\subset\lok{B}$
  satisfies $U=U^\lok{A}$. Then $\lok{A}\subset\lok{B}$ is conformal.
\end{prop}

\begin{pf} By assumption we have
$U^{\lok{A}'}=\Einsop$. Since $U^{\lok{A}'}$ implements covariance on
any \Name{Coset} theory, the local algebras of $\lok{C}_{max}$ have to
be trivial by the reasoning given in the proof of proposition
\ref{prop:subnets}.
\end{pf}

Looking at the factorisation $U\circ\symb{p}=U^\lok{A}U^{\lok{A}'}$
and the spectrum condition of all three of them (theorem
\ref{hauptsatz}, corollary \ref{cor:poseng}), this proposition says
that one can not have non-trivial \Name{Coset} models if all the
energy of $\lok{B}$,  represented in a global sense by $U$, already
belongs to $\lok{A}$, ie if $U^{\lok{A}'}=
\Einsop$. Hence, we are interested in situations where not all of the
``energy content'' of $\lok{B}$ is contained in $\lok{A}$, as
mentioned in the introduction (chapter \ref{cha:intro}).

%%%
It is not clear in
general  whether $U^{\lok{A}'}$ is globally inner in
$\lok{C}_{max}$. It is, of course, if both $\lok{B}$ and $\lok{A}$
possess a stress-energy tensor, in which case  $U^{\lok{A}'}$ is
generated by the \Name{Coset} stress-energy tensor (theorem
\ref{th:SegSugCon}). If $U^{\lok{A}'}$ 
is contained  in 
$\lok{C}_{max}$ and we apply the \Name{Borchers-Sugawara} construction
with respect to the action of $U$ on $\lok{C}_{max}$, yielding the
representation $U^{\lok{C}_{max}}$, then we have: $U=U^\lok{A}U^{\lok{C}_{max}}$. This
identity is the complete analogue of the decomposition of the  stress
energy tensor $\Theta^\lok{B}$ of an ambient theory into  the
stress-energy tensor $\Theta^\lok{A}$ of a
subtheory and the respective 
\Name{Coset} stress-energy tensor
$\Theta^\lok{B}-\Theta^\lok{A}$. Probably, this equality holds true 
in general, but we know of no proof to date. 

Taking this lack of understanding  seriously, one is led to the
following consideration: If 
we iterate the \Name{Borchers-Sugawara} construction with respect to
the action of $U^{\lok{A}'}$ on 
$\lok{C}_{max}$, and with respect to the action of $U^{\lok{C}_{max}'}$ on
$\lok{A}$, we get the inner-implementing representations $U^{\lok{C}_{max}}$
and $U^\lok{A}$ and the respective remainders
$U^{\lok{C}_{max}'}_{(\lok{A}')}$, $U^{\lok{A}'}_{(\lok{C}_{max}')}$
such that we 
have the identities: 
$$U^{\lok{A}'}= U^{\lok{C}} \, U^{\lok{C}_{max}'}_{(\lok{A}')}\, , \quad
U^{\lok{C}_{max}'} = U^{\lok{A}} \, U^{\lok{A}'}_{(\lok{C}_{max}')} \zendot$$  
Obviously,
$U^{\lok{A}}, U^{\lok{C}}$ commute 
with each other and with $U^{\lok{C}_{max}'}_{(\lok{A}')}$,
$U^{\lok{A}'}_{(\lok{C}_{max}')}$ and, moreover, the following holds:  
\begin{equation}
  \label{eq:UACdef}
  U\circ \symb{p} = U^{\lok{A}} \,  U^{\lok{C}} \, 
  U^{\lok{C}_{max}'}_{(\lok{A}')} = U^{\lok{C}}  \, U^{\lok{A}} \, 
  U^{\lok{A}'}_{(\lok{C}_{max}')} \zendot
\end{equation}
Thus, $U^{\lok{C}_{max}'}_{(\lok{A}')}$ and $U^{\lok{A}'}_{(\lok{C}_{max}')}$
coincide with the remainder of the \Name{Borchers-Sugawara}
construction on the \Name{Coset} pair
$\lok{A}\coset\lok{C}_{max}\subset\lok{B}$, namely:
\begin{equation}
  \label{eq:UACrem}
  U^{\lok{C}_{max}'}_{(\lok{A}')} = U^{\lok{A}'}_{(\lok{C}_{max}')} =
  U^{\lok{A}'\cap\lok{C}_{max}'} =  U\circ\symb{p}  \, (U^\lok{A})^* \, 
  (U^{\lok{C}_{max}})^* 
  \zendot
\end{equation}

As a next step one may observe that the inclusion
$\lok{A}\coset\lok{C}_{max}\subset\lok{B}$ is conformal: the local
operators of its maximal \Name{Coset} theory have to be central in
the respective local algebras of $\lok{C}_{max}$. In case
$U^{\lok{A}'\cap\lok{C}_{max}'}$ is non-trivial, the representation
$U$ can not be generated by local data according to the scheme
indicated  (\ref{eq:UACdef}), ie the result would exhibit a kind of
commutativity of its constituents which its local
approximations can not have. This appears unsatisfactory.

It is straightforward to check that in case $\lok{A}\subset\lok{B}$ is
cofinite and both $U^\lok{A}$ and $U^{\lok{C}_{max}}$ have the
{\em net-endomorphism property} (definition \ref{def:netend}) the
representation $U^{\lok{A}'\cap\lok{C}_{max}'}$ has to be trivial
(proposition \ref{prop:UACconf}). This result relies on a purely
representation theoretic argument concerning the action of  $\PSL(2,
\dopp{R})^\sim$  on finite-dimensional subspaces of a  
unitary representation, which, basically, trivialises the problem. A
general argument based on fundamental assumptions such as the
split property is highly desirable, but out of reach to date. 

Therefore, it is not known whether our definition of conformal
inclusion, referring to triviality of all \Name{Coset} models
associated with $\lok{A}\subset\lok{B}$, actually is equivalent to
$U=U^\lok{A}$. 
For this work there is no need to go into this problem in detail, but we
would like to state which extended \Name{Coset} pair might be attached to
the possibility $U^{\lok{A}'\cap\lok{C}_{max}'}\neq\Einsop$. 

The maximal iterated \Name{Coset} model, $\lok{A}_2$ %\lok{A}_{moritz}
, associated with
$\lok{C}_{max}\subset\lok{B}$ is given by the local algebras
$\lok{A}_2(I) := \{U^{\lok{C}_{max}}\}'\cap\lok{B}(I)$. It contains the subnet
$\lok{A}_{max}$ consisting of all observables which are {\em covariant with
respect to $U^\lok{A}$}, ie commute with $U^{\lok{A}'}$:
\begin{equation}\label{eq:Amax}
 \lok{A}_{max}(I) := \{U^{\lok{A}'}(g), \,\,
    g\in\PSL(2,\dopp{R})^\sim\}'\cap\lok{B}(I) \zendot
\end{equation}
 This is
obvious since $\lok{A}_{max}$ is a \Name{Coset} model associated with
$\lok{C}_{max}$ (lemma \ref{lem:cosmax}). Clearly, $\lok{A}_2$ and
$\lok{C}_{max}$ can not be extended without spoiling their character
as a \Name{Coset} pair. 

As argued above, we expect $U=U^\lok{A}U^{\lok{C}_{max}}$, which yields
$\lok{A}_2=\lok{A}_{max}$. We take $\lok{A}_{max}$ to be the other natural
object in our studies of chiral subnets and their \Name{Coset}
models. To put it  plainly: We do not {\em assume}
$U^{\lok{A}'}=U^{\lok{C}_{max}}$ to hold,
we rather {\em proceed without having an answer} to this problem. In all
what follows, our analysis will be done in terms of $U^\lok{A}$ and
$U^{\lok{A}'}$ and we will not have to face again the questions just
raised, because the forthcoming arguments will be  
independent.  The further development leads to satisfactory results,
which support our opinions on this issue.

We mention
another reason for considering $\lok{A}_{max}$ as a natural object of
investigation. While 
given $\lok{A}$ and $U$ the inner-implementing representation 
$U^\lok{A}$ is unique, $U^\lok{A}$ does not determine the subnet
$\lok{A}\subset\lok{B}$, as examples of conformal inclusions show. In
general there will be  subnets transforming covariantly under
the action of $U^\lok{A}$ (transformation property) and subnets
containing the operators of $U^\lok{A}$ as global 
observables (generating property). Generically  there will be no simple
relation such as inclusion 
or commutativity etc for any pair $\lok{A}_\alpha$, $\lok{A}_\beta$ of chiral
subnets having one or both properties. $\lok{A}_{max}$ is, of course,
the  maximal subnet
transforming covariantly and having the generating property. 
Any subnet $\lok{A}$ having both properties defines a conformal
inclusion
$\lok{A}e_{\lok{A}_{max}}\subset\lok{A}_{max}e_{\lok{A}_{max}}$. Since
studies on conformal 
inclusions form an area of research of their own, $\lok{A}_{max}$ should
be a generic object to explore.

\subsection{Discussion on  the \Name{Borchers-Sugawara} construction}
\label{sec:bosugsum}

We have presented a construction applying and
generalising the result 
of \Name{Borchers} \cite{hjB66}. Hence we obtained the unique
inner-implementing representation of $\PSL(2,\dopp{R})^\sim$. It
generalises, within its 
limits, the \Name{Sugawara} construction \cite{hS68}. We have proposed the name
{\em Borchers-Sugawara construction} because of these relations.
The construction is completely model independent and does not require
existence of a stress-energy tensor. Properties
connected with representations generated by a stress-energy tensor
 will be discussed in  \ref{cha:netend}.

It is natural to ask if this construction may be applied to other
spacetime symmetry groups. In our view the key tools in our
construction are the following: The original representation satisfies
the spectrum condition for some translation subgroups. There are
sufficiently many of them to generate the whole group and we have
an argument how to derive a representation of the covering group from
the unitary group generated by the operators constructed by means of
\Name{Borchers}' key result \cite{hjB66}.

In the case of the \Name{Poincar\'e} group the
translations usually satisfy the spectrum condition. Unfortunately, so
to say, they form an invariant subgroup and although one is tempted to
generate the group from $\PSL(2,\dopp{R})$ subgroups (as eg in
\cite{KW01}) this seems impossible with subgroups satisfying the
spectrum condition.

For conformal groups the situation is different, as shown in
\cite{sK02}. These groups are generated by their subgroups of
translations and special conformal transformations \cite{BGL93} and 
both subgroups
satisfy the spectrum condition, if
the conformal {Hamilton}ian has positive spectrum. 
The proof of main theorem in \cite{sK02} (corresponding to theorem
\ref{hauptsatz})  extends to all conformal
groups $\PSO(d,2)$, $d\geq 3$,  and the construction given here applies
directly to the conformal group in $\opo$ dimensions, since this is a
factor group of
$\PSL(2,\dopp{R})^\sim\times\PSL(2,\dopp{R})^\sim$. 

The construction
used in \cite{sK02}, however, does not extend to coverings of the
conformal groups, apparently. Applying the method used here requires
establishing an explicit, continuous decomposition
into translations and special conformal transformations, which ought
to be possible.

%%%%%%%%%%%%%%%%%%%%%%%%%%%%%%%%%%%%%%%%%%%%%%%%%%%%%%%%%%
%% ENDE KAPITEL COSETPAIRS %%%%%%%%%%%%%%%%%%%%%%%%%%%%%%%
%%%%%%%%%%%%%%%%%%%%%%%%%%%%%%%%%%%%%%%%%%%%%%%%%%%%%%%%%%

%\input{Currsub.tex}
%%%%%%%%%%%%%%%%%%%%%%%%%%%%%%%%%%%%%%%%%%%%%%%%%%%%%%%%%%
%% KAPITEL Stromunteralgebren %%%%%%%%%%%%%%%%%%%%%%%%%%%%
%%%%%%%%%%%%%%%%%%%%%%%%%%%%%%%%%%%%%%%%%%%%%%%%%%%%%%%%%%

\chap{Subnets of chiral current algebras}{Subnets of chiral current algebras}{cha:currsub} 
%%%%%%%%%%%%%%%%%%%%%%%%%%%%%%%%%%%%%%%%%%%%%%%%%%%%%%%%%%%%%%
%%%%%%%%%% CURRENT SUBALGEBRAS %%%%%%%%%%%%%%%%%%%%%%%%%%%%%%%
%%%%%%%%%%%%%%%%%%%%%%%%%%%%%%%%%%%%%%%%%%%%%%%%%%%%%%%%%%%%%%

%\sekt{Subnets of chiral current algebras}{Subnets of chiral current algebras}{sec:currsub}
% \section{Current subalgebras}
% \label{sec:currsub}

The purpose of this chapter is to introduce  the objects of most
investigations of chiral subnets and 
their \Name{Coset} models: the chiral conformal theories and
subtheories generated by chiral current algebras. 
These models are either formulated as theories of \Name{Wightman}
fields in $\opo$ dimensions, which factorise into independent chiral
parts, or more abstractly through unitarisable highest-weight
representations of {\em affine {Kac-Moody} algebras}, which then
are shown to integrate to chiral conformal nets by group theoretical
methods and can be seen to define chiral conformal quantum field
theories in the sense of \Name{Wightman}'s axioms as well.

If one wants to study model specific properties of current algebras,
their superselection structure in particular, other approaches 
to constructing chiral current algebras prove very useful. They may be
defined by the action of \Name{Bogoljubov}
(gauge) transformations on the \Name{CAR} algebra of chiral
fermions \cite{aW98} or one may use the basic construction of
\Name{Frenkel-Kac} \cite{FK80} for affine \Name{Kac-Moody} algebras
$A_n^{(1)}$, $D_n^{(1)}$, $E_{6-8}^{(1)}$ at level $1$, using 
vertex operators (see eg \cite{vL97}).  Closely related to the latter
point of view,  one may consider these models as local
extensions of the ${rank}_\lie{g}$-fold tensor product of the
$U(1)$-current algebra (eg \cite{cS95}); a variant is the approach of  
\cite{FG93}.      

Several authors have
dealt with specific properties of chiral current 
algebras, their inclusions and their \Name{Coset} models (eg
\cite{fX00, fX99, fX00a, fX01, fX02, KL02,  vL97, aW98, tL94}). 
We are primarily
interested in the archetypical, instructive character of these
models and for this reason we focus on 
establishing the properties needed in the definitions \ref{def:chcotheo} and 
  \ref{def:chsubnet} and additional properties connected with the
  {\em {Sugawara} construction}, which ensures presence of a
  stress-energy tensor in any chiral current algebra.

Therefore, we take a straightforward approach to these models
through a construction using free, chiral  fermion fields, the
{\em quarks}, take a direct route from the \Name{Wightman} fields to the
chiral conformal theory they generate, and discuss other aspects of
these models later on. The term {\em chiral current algebra} will be
used for all these models.

%%%%%%%%%%%%%%%%%%%%%%%%%%%%%%%%%%%%%%%%%%%%%%%%%%%
%%%%%% QUARK MODELS %%%%%%%%%%%%%%%%%%%%%%%%%%%%%%%
%%%%%%%%%%%%%%%%%%%%%%%%%%%%%%%%%%%%%%%%%%%%%%%%%%%

\sekt{Quark models of chiral current algebras}{Quark models of chiral current algebras}{sec:quarkm}
%\subsection{Quark models of chiral current algebras }
%\label{sec:quarkm}

First, we  describe how chiral current algebras may be
obtained directly as quantum fields in the sense of \Name{Wightman}'s
axioms  from free, massless fermion fields, the {\em quarks}, a method
introduced in \cite{BH71}. The \Name{Sugawara} construction for chiral
currents, which yields a stress-energy tensor for each chiral current
algebra, is discussed as the next step.

The fundamental objects of the framework laid down in
\Name{Wightman}'s axioms  are
{\em local quantum fields}, $\Phi$, which consist of maps 
from smooth test functions on spacetime, $f$, to (generically
unbounded) closable, linear operators, $\Phi(f)$, defined on a
dense subspace  of some \Name{Hilbert} space. The fields $\Phi$
usually carry multiple indices referring 
to their type, in particular to their transformation behaviour
under gauge and spacetime transformations. Covariance with 
respect to spacetime symmetries is required to be implemented by a
unitary, strongly continuous representation which satisfies the
spectrum condition and admits a unique (up to its phase)
normalised, invariant vector, the vacuum $\Omega$.  Multiple application
of smeared fields $\Phi(f)$ to the vacuum, $\Omega$, is assumed to
generate a dense subspace, the {\em {Wightman} domain}, which is common
to all smeared fields, invariant under their action and allows their
reconstruction, ie is a {\em core} for all fields. The matrix elements
of the fields $\Phi(f)$ have to 
depend continuously on the test function $f$, ie they shall be
{\em tempered distributions}. Locality is formulated as (graded)
commutativity of the smeared fields $\Phi_1(f_1)$, $\Phi_2(f_2)$ on
the \Name{Wightman} domain for test functions $f_1$, $f_2$ which have
spacelike separated supports. 
According to the  \Name{Wightman} reconstruction theorem \cite{aW56},
the whole theory including its covariance with respect to a unitary
implementation of  spacetime and gauge symmetry is encoded in the
properties of the hierarchy of {\em $n$-point functions}, namely in the
numerical, tempered distributions defined by the vacuum expectation
values of products of fields as eg $\langle\Omega,
\Phi_1(f_1)\ldots\Phi_n(f_n)\Omega\rangle$. 

Reviews and textbooks on this framework of relativistic quantum field
theory are eg \cite{GW65, SW64, rJ65, iT65}, a short summary is
contained in \Name{R. Haag}'s book  \cite[sections II.1,
II.2]{rH92}. The discussion in sections \ref{sec:currwick} and
\ref{sec:sugcurralg} is a summary 
of well-known results. 
The review of \Name{Furlan, Sotkov and Todorov} \cite{FST89} serves as
a source for the formulation of the general theory of quantised fields
with conformal symmetry in $\opo$ dimensions, and 
\cite{GO86,jF92} contain expositions of quark
models and compact \Name{Lie} algebras.

\subsection{Chiral currents as \Name{Wick} squares}
\label{sec:currwick}

Take $N$ independent, free, massless, complex fermion fields, the
{\em quarks} $\psi^i$, $i=1,\ldots,N$,  in 
$\opo$ dimensions. They satisfy the corresponding \Name{Dirac} equation:
\begin{displaymath}
  i\dolch{\partial} \psi^i (t,x)=
  i(\gamma^0\partial_0+\gamma^1\partial_1) \psi^i = 0 \zendot
\end{displaymath}
Choosing units in which the speed of light equals
$1$, $t=x^0$ stands for the time coordinate and $x=x^1$ for the space 
coordinate.  
The algebra of the $\gamma$-matrices is given by $(\gamma^0)^2 =
\Einsop = -(\gamma^1)^2$, $\antkomm{\gamma^0}{\gamma^1} =0$. With
$\gamma^5:=\gamma^0\gamma^1$ 
and $P_\pm:= \frac{1}{2}(1\pm\gamma^5)$ the quarks can be decomposed
into their  chiral components, each of which depends on
one light-cone 
coordinate only:
\begin{equation}\label{eq:quarchidec}
  (\partial_0\pm\partial_1) P_\pm \psi^i(t,x) = 0 \,\,\Leftrightarrow
  \,\,   \psi_\pm^i(t\mp x) := P_\pm \psi^i(t,x) \zendot
\end{equation}
For the time being we use the light-cone coordinates $x_\pm := t\mp x$.  

The quantum field theoretical nature of these fields is encoded in
canonical anti-commutation relations (\Name{CAR}):
\begin{displaymath}
  \begin{array}{ccccc}
\antkomm{\psi^i_+(x_+)}{\psi^j_-(y_-)}&=&
\antkomm{\psi^i_+(x_+)}{\psi^j_-(y_-)^*}&=&0 \zencom\\
\antkomm{\psi^i_+(x_+)}{\psi^j_+(y_+)}&=&
\antkomm{\psi^i_-(x_-)}{\psi^j_-(y_-)}&=&0 \zencom\\
\antkomm{\psi^i_+(x_+)}{\psi^j_+(y_+)^*} &=&\delta(x_+-y_+)
\delta^{ij} \zencom\\
\antkomm{\psi^i_-(x_-)}{\psi^j_-(y_-)^*} &=&\delta(x_--y_-)
\delta^{ij} \zendot\\
  \end{array}
\end{displaymath}
The non-vanishing two-point functions of these fields are given by:
\begin{displaymath}
  \begin{array}{ccccl}
  \skalar{\Omega}{\psi^i_\pm(x_\pm)\psi^j_\pm(y_\pm)^*\Omega} &=&
  \skalar{\Omega}{\psi^i_\pm(x_\pm)^*\psi^j_\pm(y_\pm)\Omega} &=&
  \frac{\delta^{ij}}{2\pi} \Delta(x_\pm-y_\pm) \zendot
  \end{array}
\end{displaymath}
The distribution $\Delta(t)$ is given as boundary value of the  analytic
function $\frac{-i}{t-i\varepsilon}$ in
the limit  
$\varepsilon\searrow 0$. 

The \Name{CAR} determine the $n$-point functions completely in terms
of the two-point 
functions and it is well-known that fermion fields as above define 
quantised fields satisfying \Name{Wightman}'s axioms on the fermionic
\Name{Fock} space (see eg \cite[section II.5]{rJ65}). 
The transformation behaviour of $\Delta$ with respect to
coordinate transformations shows that conformal symmetry is 
present, implemented by a unitary representation $U_\psi$ of the
twofold covering of 
$(\PSL(2,\dopp{R})^\sim \times \PSL(2,\dopp{R})^\sim)/\dopp{Z}$, where
the cyclic group generated by the {\em simultaneous rigid conformal
rotations} by $2\pi$, 
$\tilde{R}(2\pi)\times \tilde{R}(2\pi)$, is factored out. Moreover,
the decomposition into chiral components 
(\ref{eq:quarchidec}) yields a decomposition of the \Name{Wightman}
theory of quarks $\psi^i$ on fermionic \Name{Fock} space,
$\Hilb{H}_\psi$, into the 
tensor product of the \Name{Wightman} theories of the chiral 
quarks $\psi^i_\pm$: $\Hilb{H}_\psi= \Hilb{H}_{\psi_+}\otimes 
\Hilb{H}_{\psi_-}$.       

The quarks $\psi^i$ are found to be  {\em
  quasi-primary} fields of 
{\em scaling dimension} $\frac{1}{2}$, ie they have a definite
transformation behaviour determined by the scaling dimension, which we
  state for one chiral component: 
\begin{equation}
  \label{eq:transquasiprim}
 U_\psi(\tilde{g})\psi^i_+(x_+)U_\psi(\tilde{g})^* =
(d(\symb{p}(\tilde{g})(x))/dx)^{\frac{1}{2}}\psi^i_+(\symb{p}(\tilde{g})(x_+)) \,,\,\, \tilde{g}\in
\PSL(2,\dopp{R})^\sim.
\end{equation}
For the transformation
law of a general quasi-primary field
of scaling dimension $h$ replace $\frac{1}{2}$ by $h$. 

The normal ordering of free complex fermions is given by the usual
point-splitting procedure:
\begin{equation}
  \label{eq:normordferm}
:\psi^i_+\psi^j_+{}^*:(x_+) = \lim_{x_+'\rightarrow x_+} \left(\psi^i_+(x_+)
\psi^j_+(x_+')^* -\frac{\delta^{ij}}{2\pi} \Delta(x_+-x_+') \right) \zendot
\end{equation}
It is common to call this \Name{Wightman} field the {\em {Wick}
square} of the free complex fermions.

Now, we assume that there is an $N$-dimensional, non-trivial
representation of a compact, simple  \Name{Lie} algebra $\lie{g}$ by
Hermitian matrices 
$\{(M^a_{ij})_{i,j=1,\ldots,N}\}_{a=1,\ldots,d_\lie{g}}$, where
$d_\lie{g}$ denotes the dimension of $\lie{g}$. This means that we
have: $\overline{M^a_{ij}} = M^a_{ji}$ and $\komm{M^a}{M^b} = 
  i f^{ab}{}_c M^c$. The real numbers $f^{ab}{}_c$ are the structure
  constants of $\lie{g}$ defined with respect to some basis
  $\{T^a\}_{a=1,\ldots,d_\lie{g}}$. The $\opo$-dimensional {\em
   complex quark model} of the current algebra associated with $\lie{g}$ is
  given as follows:
  \begin{equation}
    \label{eq:quarkmodeldef}
    j^a{}^\mu(t,x) := M^a_{ij} :\overline{\psi^i}\gamma^\mu
    \psi^j:(t,x) = M^a_{ij} :\psi^i{}^*\gamma^0\gamma^\mu
    \psi^j:(t,x) \zendot
  \end{equation}
It is straightforward to check that this defines a Hermitian field of
scaling dimension $1$ and that it is a conserved current:
$\partial_\mu j^a{}^\mu=0$.

$j^a{}^\mu$ decomposes into two independent chiral parts, the {\em
  chiral currents} $j^a_\pm$:
\begin{displaymath}
  j^a_\pm := \frac{1}{2}(j^a{}^0\pm j^a{}^1) =
\, M^a_{ij} \, :\psi^i{}^* P_\pm \psi^j:\quad = \, M^a_{ij}
\, :\psi^i_\pm{}^*\psi^j_\pm: \zendot
\end{displaymath}
 The
decomposition of the fermionic \Name{Fock} space into the
tensor product of chiral fermionic \Name{Fock} spaces induces a
tensor-product decomposition of the \Name{Hilbert} space which the
currents generate from the vacuum, $\Hilb{H}_j$, into the
tensor product of the vacuum \Name{Hilbert} spaces of  the chiral currents: 
$\Hilb{H}_j = \Hilb{H}_{j_+}\otimes \Hilb{H}_{j_-}$. The chiral
currents act as $j^a_+ = j^a_+ \otimes \Einsop_{j_-}$, $j^a_- =
\Einsop_{j_+}\otimes j^a_- $.

The commutation relations between chiral currents read:
\begin{equation}
  \label{eq:quacurcom}
  \komm{j^a_\pm(x_\pm)}{j^b_\pm(y_\pm)} = i f^{ab}{}_c j^c_\pm(y_\pm)
  \delta(x_\pm-y_\pm) +   tr(M^aM^b)   \frac{i}{2\pi}
  \delta'(x_\pm-y_\pm)\, .
\end{equation}
The currents are  examples of {\em {Lie} fields}, as their
commutator is linear in the field itself. In higher dimensions the
possibilities of \Name{Lie} fields are very restricted (see eg
\cite{kB76} and references therein). 
The $c$-number part in (\ref{eq:quacurcom}) is called the 
{\em {Schwinger} term}.

The trace $tr(M^aM^b)$ is invariant under
cyclic permutations and hence induces an invariant symmetric bilinear
form on $\lie{g}$. Since $\lie{g}$ is simple, there is up to
normalisation only one such form and thus $tr(M^aM^b)$ has to be
proportional to the trace in the adjoint representation of
$\lie{g}$. The latter is usually normalised such that the highest
weight of the adjoint representation has length 
$2$ in the scalar product which is induced by the form. Because 
$\lie{g}$ is compact, the result is a {Euclid}ean scalar product which
we will call the {\em {Killing} form} of $\lie{g}$. 

The
\Name{Killing} form will be denoted $\skalar{.}{.}_\lie{g}$ and  the
matrix which induces it will be written $g^{ab}_\lie{g}$. The inverse
of $g^{ab}_\lie{g}$ is written with two lower indices:
$g_{ab}^\lie{g}$.
With these conventions, there exists a positive integer
$k$ such that $tr(M^aM^b) = k
g^{ab}_\lie{g}$. $k$ is the {\em second {Dynkin} index} of the
representation of $\lie{g}$ by the matrices $M^a$ and in relation to
the commutation relations  (\ref{eq:quacurcom}) it is the {\em level}
of the current algebra 
associated with $\lie{g}$. 

The constant functions on the light-cone are among the admissible
test functions for currents (cf below). If one smears out currents with
constant test functions, one reads off the current algebra
(\ref{eq:quacurcom}) that such current operators form a representation
of $\lie{g}$, which is contained in the current algebra and is
commonly called the {\em horizontal subalgebra}. The adjoint action of
the horizontal subalgebra on quarks and on currents generates global
gauge transformations of both fields and thus the currents may be
viewed as conserved currents associated with a gauge symmetry. For
this reason we regard the label ``$a$'' of the currents $j^a_\pm$ as
their {\em colour} and will call the algebra of the $T^a$, the
\Name{Lie} algebra $\lie{g}$,  the {\em colour algebra} of our
current algebra. 
The colour algebra and the level characterise a
current algebra completely. 

For orthogonal representations, there is the {\em real quark
  model}. We assume there are $N$-dimensional, skew-symmetric, real matrices
  $M^a$ forming a non-trivial, orthogonal representation of
  $\lie{g}$. Then we 
  take $N$ real fermions $\psi^i_r(t,x) :=
  \frac{1}{2}(\psi^i(t,x)+\psi^i(t,x)^*)$. It is readily checked that
  the following definition yields a current algebra associated with
  $\lie{g}$ at the level $k$ satisfying $kg^{ab}_\lie{g}=
  \frac{1}{2}tr(M^aM^b)$:    
  \begin{equation}
    \label{eq:realquark}
    j^a{}^\mu(t,x) = (iM^a_{ij}) \,\,
    :\psi^i_r\gamma^0\gamma^\mu\psi^j_r: (t,x)\zendot
  \end{equation}
These models factorise into  their chiral parts as above.

An interesting question is: For which level $k$ exist current
algebras as given above?  If one takes the direct sum of
representations as above, ie by matrices $M^a_1\oplus M^a_2$, it
is trivial to check that the level of the corresponding current
algebra is the sum of the levels determined by the $M^a_1$ and
$M^a_2$, respectively. Therefore, we have primarily to look for level
$1$ representations. 

As a well known
fact, there are representations of \Name{Dynkin} index $1$ by
Hermitian matrices for the compact real \Name{Lie} algebras of type 
$A_n$, the \Name{Lie} algebras of the special unitary groups
$SU(n+1)$, 
and of type $C_n$, the \Name{Lie} algebras of the symplectic groups
$Sp(n)$. 
Orthogonal representations of
\Name{Dynkin} index $2$ exist for $B_n$ ($SO(2n+1)$), $D_n$ ($SO(2n)$)
and the exceptional \Name{Lie} algebra $G_2$ (for both facts see eg
 \cite[(1.6.78)]{jF92}). Therefore, for all but the  colour 
algebras $E_{6-8}, F_4$ there are quark models with level $1$ and
hence at all levels\footnote{
For the remaining four simple colour algebras, $E_{6-8}, F_4$, there are
quark models for some
levels (see eg \cite{MP81}), but not for level $1$. Yet, the
corresponding 
chiral conformal nets 
are available at any level by different means (see section
\ref{sec:currloop}).}.

There are obvious adaptations of quark models to abelian, semi-simple
and reductive colour algebras. As reductive \Name{Lie} algebras
decompose into a direct sum of simple ideals and a $d$-dimensional ideal
isomorphic to $\dopp{R}^d$, we take this as the general case. Assume
now that $\lie{g}$ is a reductive \Name{Lie} algebra with the
decomposition $\lie{g} = \bigoplus_\alpha \lie{g}_\alpha \oplus
\dopp{R}^d$, where the $\lie{g}_\alpha$ stand for simple ideals. For
the semi-simple \Name{Lie} algebra $\bigoplus_\alpha \lie{g}_\alpha$
we simply take direct sums of the current algebras above. 

The {\em abelian current algebra}, ie the current algebra associated with an
 abelian colour algebra of dimension $d$,  is constructed by taking $d$
independent, 
complex quarks and setting with respect to some basis
$\{T^k\}_{k=1,\ldots,d}$ of $\dopp{R}^d$:
\begin{equation}
  \label{eq:defabcurr}
  j^k{}^\mu(t,x) := \,\, :\overline{\psi^k}\gamma^\mu\psi^k:(t,x)
  \zendot
\end{equation}
These currents decompose into independent, chiral components as above and their
commutation relations are given in terms of these chiral components as:
\begin{equation}\label{eq:abcurralgkron}
  \komm{j^k_\pm(x_\pm)}{j^l_\pm(y_\pm)} = \frac{i}{2\pi}\delta^{kl}
  \delta'(x_\pm-y_\pm)  \zendot
\end{equation}

Simply by changing the basis of $\dopp{R}^d$, one may
replace in (\ref{eq:abcurralgkron}) the \Name{Kronecker} symbol
$\delta^{kl}$ by any other {Euclid}ean scalar product on 
$\dopp{R}^d$. To capture the general commutation relation of an
abelian current algebra, we take an arbitrary {Euclid}ean scalar
product, $g_{\dopp{R}^d}$, and admit an additional
multiplicative constant, $\kappa$:
\begin{equation}
    \label{eq:abcurralg}
\komm{j^{k'}_\pm(x_\pm)}{j^{l'}_\pm(y_\pm)} = \frac{i}{2\pi} \,\,\kappa
\,\,g^{k'l'}_{\dopp{R}^d}\,\,\delta'(x_\pm-y_\pm)  \zendot
\end{equation}
 Because of the indicated arbitrariness, $\kappa$
should not be viewed as a ``level'' like for current algebras associated
with simple colour algebras. The general formulation of abelian
current algebras is the one needed for current subalgebras in current
algebras of simple colours, where 
$\kappa$ and $g^{kl}_{\dopp{R}^d}$ are given by the
embedding. The current algebra for $d=1$ will be called the
{\em $U(1)$-current algebra}. 

%%%%%%%%%%%%%%%%%%%%%%%%%%%%%%%%%%%%%%%%%%%%%%%%%%%
%%%%%%% SUGAWARA SUBSECTION %%%%%%%%%%%%%%%%%%%%%%%
%%%%%%%%%%%%%%%%%%%%%%%%%%%%%%%%%%%%%%%%%%%%%%%%%%%

\subsection{\Name{Sugawara} construction for chiral current algebras}
\label{sec:sugcurralg}

For a general quantum field theory it is not clear, whether
and how the coordinate and gauge transformations are
generated by local quantum fields which are intrinsic to the theory and may be
viewed as densities of energy, 
momentum, or charges. While in classical {Lagrange}an
field theory the explicitly known \Name{Noether} currents and their
integrals over 
space yield such objects of a local nature, an analogous result for a
general quantum field theory in \Name{Wightman}'s framework is not
known% \footnote{For local quantum
%   theories having the split property there is a very general quantum
%   version of \Name{Noether}'s theorem \cite{BDL86,ADF87}, but the local
%   objects associated with unitarily implementable symmetries, the
%   ``local implementers'' have a somewhat different character: they
%   form unitary representations of the symmetry group by local
%   operators with the same
%   spectral properties as the original unitary implementation of the
%   symmetry. We will  come back to this in chapter \ref{cha:noset}.}
.    

In $\opo$-dimensional conformal quantum field theory the situation is
different: under weak assumptions, \Name{L\"uscher and Mack}
found that conformal stress-energy tensors always decompose into
independent chiral parts and that these yield a local
formulation of the \Name{Virasoro} algebra, ie an
infinitesimal, 
projective representation of the group of orientation preserving
diffeomorphisms of the circle, $\Diff_+(\Seins)$.  Such stress-energy
tensors serve  well as  densities for 
the conformal transformations and
their specific properties allow to apply powerful
tools connected to structure 
and representation theory of the  \Name{Virasoro}
algebra and of $\Diff_+(\Seins)$.  

It is a remarkable feature of
current algebras that we have a simple method for constructing their
stress-energy tensor, the {\em {Sugawara} construction}. We come
back  to the general  result, the {\em theorem of
{L\"uscher and Mack}},  after we have sketched the
\Name{Sugawara} construction and hence given a non-trivial example. 

The \Name{Sugawara} construction applies independently to both chiral
parts of a $\opo$-dimensional current algebra. It was introduced by
\Name{Sugawara} \cite{hS68} for currents in $\opd$-dimensional
\Name{Minkowski} space; for an account on the history of the \Name{Sugawara}
construction in the context of $\opo$-dimensional and chiral current
algebras see eg \cite{CHK96}.  We will deal with chiral currents only
and we  drop the suffices $+$, $-$. 

The discussion for a chiral current algebra associated with a 
simple colour algebra, $\lie{g}$, covers the general case of reductive colour
algebras with straightforward modifications, so we deal with the
former in detail. The normal ordering of currents will be used: 
\begin{eqnarray}
   && :j^aj^b:(x)\nonumber\\
&=& \lim_{x'\rightarrow x} \left({j^a(x)j^b(x') -
  \frac{\Delta^2(x-x')}{(2\pi)^2}k g^{ab}_\lie{g}-if^{ab}{}_c j^c (x)
  \frac{\Delta(x-x')}{2\pi}}\right) .  \label{eq:defnormordcurr}
\end{eqnarray}

We define {\em the {Sugawara} stress-energy tensor},
$\Theta^\lie{g}$, of the current algebra of colours in $\lie{g}$ at
level $k$ by:
\begin{equation}
  \label{eq:defsugset}
  \Theta^\lie{g}(x):=\frac{\pi}{k+\symb{g}^\vee_\lie{g}} \,\,
  g_{ab}^\lie{g} \, :j^aj^b:(x) \zendot 
\end{equation}
The symbol $\symb{g}^\vee_\lie{g}$ stands for the {\em dual
{Coxeter} number} which amounts to half of the {\em second
{Casimir} operator}, $C_2^\lie{g} := g_{ab}^\lie{g} T^aT^b$, in
the adjoint representation of $\lie{g}$.

From the following commutation relation of $\Theta^\lie{g}$ with the
currents it can be readily deduced that
integrals of $\Theta^\lie{g}$ indeed generate the  conformal
transformations on the currents:
\begin{equation}
  \label{eq:setoncurr}
  \komm{\Theta^\lie{g}(x)}{j^c(y)} = ij^c(x)\delta'(x-y) \zendot
\end{equation}

The commutation relation of $\Theta^\lie{g}$ with itself reads: 
\begin{eqnarray}
  \label{eq:LueMackalg}
  \komm{\Theta^\lie{g}(x)}{\Theta^\lie{g}(y)}&=&i2\Theta^\lie{g}(x)\delta'(x-y)-i\klammer{\frac{d}{dy}\Theta^\lie{g}(y)}\delta(x-y)\nonumber\\
&&-i {\frac{c_\lie{g}(k)}{24\pi}} \delta'''(x-y) \zendot
\end{eqnarray}
The number $c_\lie{g}(k)$, which determines the central extension of this
algebra, is  called {\em central charge} and has for the
\Name{Sugawara} stress-energy tensor the following value:
\begin{equation}
  \label{eq:sugcenchar}
  c_\lie{g}(k)=\frac{k \, d_\lie{g}}{k+\symb{g}^\vee_\lie{g}} \zendot
\end{equation}
It is not difficult to show that $c_\lie{g}(k)$ lies between the { rank
of} $\lie{g}$, $r_{\lie{g}}$,  and its dimension, $d_{\lie{g}}$. 

%%%%%%%%%%%%%%%%%%%%%%%%%%%%%%%%%%%%%%%%%%%%
%%  Sugawara for Abelian CURRALG %%%%%%%%%%%
%%%%%%%%%%%%%%%%%%%%%%%%%%%%%%%%%%%%%%%%%%%%

The \Name{Sugawara} stress-energy tensor for the current algebras of
semi-simple colour algebras, $\bigoplus_\alpha \lie{g}_\alpha$, is
simply given by $\Theta^{\oplus_\alpha} = \sum_\alpha
\Theta^{\lie{g}_\alpha}$. For a current algebra associated with an
abelian colour algebra of dimension $d$, metric tensor
$g^{kl}_{\dopp{R}^d}$ and ``level'' $\kappa$ as discussed above we
set (normal ordering as in (\ref{eq:defnormordcurr}) with obvious
alterations): 
\begin{equation}
  \label{eq:defabsugset}
  \Theta^{\dopp{R}^d}(x) := \frac{\pi}{\kappa} \,\, g_{kl}^{\dopp{R}^d}
  \,\,:j^kj^l:(x) \zendot
\end{equation}
This stress-energy tensor has central charge
$c_{\dopp{R}^d}=d$. Concerning a
current algebra of a reductive colour algebra one has to add this
contribution for the abelian ideal.

%%%%%%%%%%%%%%%%%%%%%%%%%%%%%%%%%%%%%%%%%%%%%%%%%%%%%%
%%%%%%% DISCUSSION L-M THEOREM %%%%%%%%%%%%%%%%%%%%%%%
%%%%%%%%%%%%%%%%%%%%%%%%%%%%%%%%%%%%%%%%%%%%%%%%%%%%%%

The theorem of \Name{L\"uscher and Mack}  determines
the commutation 
relations of a chiral stress-energy tensor of a conformally invariant theory
to  be of the form (\ref{eq:LueMackalg}) where the central charge is
the only free parameter \cite{FST89,gM88,LM76}. The prerequisites are
the following:  
It is assumed
that the stress-energy tensor $\Theta^{\mu\nu}$ is a symmetric-tensor
($\Theta^{\mu\nu}=\Theta^{\nu\mu}$), Hermitian
($\Theta^{\mu\nu}{}^\dagger=\Theta^{\mu\nu}$), 
conserved ($\partial_\mu  \Theta^{\mu\nu}=0$) local \Name{Wightman}
field, which is relatively local to all the fields of a
\Name{Wightman} theory and which generates the 
translations as $P^\mu = \int dx^1 \Theta^{\mu 0}$. The
\Name{Wightman} theory itself shall be scale invariant, ie there shall
be a unitary implementation $V(D(t))$ of the scale transformations
$x^\mu\mapsto e^t x^\mu$ which leaves the vacuum invariant. All
translationally invariant vectors are assumed to be scalar multiples of
the vacuum, $\Omega$.
$\Theta$ is required to have scale dimension $2$, ie we have
$V(D(t))\Theta^{\mu\nu}(t,x)V(D(t))^* = e^{2t}
\Theta^{\mu\nu}(e^tt,e^tx)$. 

Then $\Theta$ is traceless ($\Theta_\mu{}^\mu=0$), takes in light-cone
coordinates the 
form $\Theta(x_+,x_-)= diag(\Theta_{++}(x_+), \Theta_{--}(x_-))$,
ie $\Theta$ decomposes into its independent chiral parts
$\Theta_{++}$ and $\Theta_{--}$ and its chiral components obey the
commutation relation 
(\ref{eq:LueMackalg}) above with some central charges $c_\pm$. The
two-point  function of the chiral components is given by:
\begin{equation}
  \label{eq:set2pt}
  \skalar{\Omega}{\Theta_{\pm\pm}(x_\pm)\Theta_{\pm\pm}(y_\pm)\Omega}
  = \frac{c_\pm}{8\pi^2} \Delta(x_\pm-y_\pm)^4 \zendot
\end{equation}
Hence, we have $c_\pm\geq 0$ and 
$\Theta_\pm$ vanishes if and only if  $c_\pm$
vanishes\footnote{Follows from the
\Name{Reeh-Schlieder}  theorem \cite{RS61}; alternatively, one may
apply \cite[lemma 2, section  V.3.B]{rJ65}. For the \Name{Virasoro}
algebra there are other  proofs \cite[lemma 1.1]{GW85},
\cite{fG86a}, too.}. If parity is 
conserved, then both central charges coincide. 

The assumptions of the \Name{L\"uscher Mack} theorem are comparatively
weak and natural in the setting of conformal quantum field
theory in $\opo$  dimensions; they should be fulfilled in a large class
of models. The most remarkable outcome of the theorem is the
determination of the commutation relations (\ref{eq:LueMackalg})
which, on a ``decent'' \Name{Hilbert} space, exponentiate to a
projective representation of $\Diff_+(\Seins)^\sim$, the universal
enveloping group of $\Diff_+(\Seins)$ \cite{GW85,vL99a}. This does
not mean, however, that the theory itself has a diffeomorphism symmetry. 

The \Name{Sugawara} stress-energy tensor 
fulfills all the properties of a stress-energy tensor in the sense of
\Name{L\"uscher and Mack} with respect to the currents. In
addition, it is relatively local to the quarks by its very
definition and the quark model construction. In most cases, however, the
\Name{Sugawara} stress-energy tensor does not implement the conformal
transformations on the quarks and differs from the stress-energy
tensor which may be constructed from the quarks directly.

Of course, the \Name{Sugawara} construction yields stress-energy
tensors for rational $c\geq 1$
only. \Name{Buchholz and Schulz-Mirbach} \cite{BS90} have given a
construction for all $c>1$. As results of \Name{Friedan, Qiu and Shenker}
\cite{FQS84,FQS84a,FQS85} (cf \cite{rL88}) show, all values of $c$
below $1$ which are 
compatible with unitarity and positivity of energy lie in a discrete
series: $c(m) = 1-6/(m+2)(m+3)$, $m\in \dopp{N}$. The
stress-energy tensors having these central charges were obtained by
\Name{Goddard, Kent and Olive} (\Name{GKO}) \cite{GKO85,GKO86} by a
\Name{Coset}  construction, see
section \ref{cha:finind}.\ref{sec:GKOcospa}.

%%%%%%%%%%%%%%%%%%%%%%%%%%%%%%%%%%%%%%%%%%%%%%%%%%%%%
%%%%%%%%% HAAG-KASTLER-AXIOMS FOR CURRENT ALGEBRAS %%
%%%%%%%%%%%%%%%%%%%%%%%%%%%%%%%%%%%%%%%%%%%%%%%%%%%%%

\sekt{Chiral conformal nets of current algebras}{Chiral conformal nets of current algebras}{sec:hakacurr}
%\subsection{Chiral conformal nets of current algebras}
%\label{sec:hakacurr}

This section establishes the connection between chiral
current algebras and chiral conformal nets. 
As a first step we need to formulate the conformal extension of the
current algebras on the conformal covering of the
light-ray. From there we come to the  description of chiral conformal
fields in terms of non-local operators, their  {\em modes}. The mode
picture is  particularly suitable for  controlling the action of the
unbounded current operators on vectors in the \Name{Wightman} domain
and enables us to show how current algebras generate chiral conformal  
nets. It identifies chiral  current algebras as   highest-weight
representations of affine \Name{Kac-Moody} algebras as 
well. This connection makes it possible to establish chiral current
algebras on abstract grounds, ie beyond the quark model construction.

\subsection{Compact picture of chiral current algebras}

Fields which are defined for test functions on the light-ray will be
denoted with a hat, eg $\widehat{\Phi}$, such that we can clearly
distinguish between fields on the light-ray and on its conformal
compactification.  
The natural definition of conformal fields extended to $\Seins$
referring to the \Name{Cayley} transformation (\ref{eq:cayley}) is given
for an arbitrary chiral, quasi-primary field $\widehat{\Phi}$ of scaling
dimension $h\in\dopp{N}$ as follows: 
\begin{equation}
  \label{eq:comppich}
\widetilde{\Phi}(z) = N_\Phi \klammer{i\frac{dx}{dz}}^h \widehat{\Phi} (x(z))
\zendot
\end{equation}
The normalisation constant $N_\Phi$ is fixed such that the
commutation relations take a simple form; we use $N_\Phi= 2\pi$ in the
following.

\clearpage
The operator-valued distributions $\widetilde{\Phi}$ are defined
with respect to the integration measure $\oint dz/2\pi i$ on $\Seins$. We
write $f_\wedge$ for a test function on the light-cone:
\begin{eqnarray}  
\widetilde{\Phi} (\widetilde{f_\wedge})
&:=& \int_\dopp{R} dt \widehat{\Phi} (t) f_\wedge (t)\nonumber\\
&=& \oint \frac{dz}{2\pi i} \widetilde{\Phi}(z) \underbrace{
  (1+z)^{2(h-1)}  2^{-(h-1)} f_\wedge (t(z))}_{=:
  \widetilde{f_\wedge}(z)}
\zendot
 \label{eq:licocompchii}
\end{eqnarray}
The fields $\widehat{\Phi}$ are declared on smooth test functions on
$\dopp{R}$ with compact support. Conformal symmetry permits to extend
the fields $\widetilde{\Phi}$ from the space of test functions
$\widetilde{f_\wedge}$ to all $f_\sim \in C^\infty(\Seins)$, which in
turn may be used to define extensions of the fields $\widehat{\Phi}$ on all
test functions $\widehat{f_\sim}$; the introduction of the compact
picture is not necessary for extending the fields, of course, but it
is convenient:
\begin{eqnarray}  
  \widehat{\Phi}(\widehat{f_\sim})
&:=& \oint \frac{dz}{2\pi i} \widetilde{\Phi}(z) f_\sim (z)\nonumber\\
&=& \int_\dopp{R} dt \widehat{\Phi} (t)
  \underbrace{\klammer{1-it}^{2(h-1)} 2^{-(h-1)}
  f_\sim(z(t))}_{=: \widehat{f_\sim} (t)}
 \zendot
\label{eq:licocompchi}
\end{eqnarray}
The change between the
{\em compact picture} (fields on $\Seins$) and the {\em light-cone
picture} (fields on $\dopp{R}$) induces mutually inverse, linear
transformations $f_\sim\mapsto
\widehat{f_\sim}$ and 
$f_\wedge\mapsto \widetilde{f_\wedge}$, which depend on
the scaling dimension of the field under consideration.

The smooth functions on the circle form the test-function space of a
conformally covariant chiral theory. Their images on the light-ray,
$\widehat{f_\sim}$, $f_\sim\in C^\infty(\Seins)$, are smooth functions
which grow at most as 
$|x|^{2(h-1)}$ and have the same asymptotic behaviour for 
$x\rightarrow +\infty$ as for $x\rightarrow -\infty$. This allows us
to integrate currents ($h=1$) with  
constant test functions which yields the horizontal subalgebra. The
stress-energy tensor ($h=2$) may be integrated with test functions of
growth up to $x^2$. This allows us to define the generators of global
conformal symmetry as integrals of the stress-energy tensor; we come
to this in section \ref{sec:curexp}.

For a Hermitian field on the light-ray we have:
$\widehat{\Phi}(f_\wedge)^\dagger =
\widehat{\Phi}(\overline{f_\wedge})$. Hence, for real $f_\wedge$, the
smeared field $\widehat{\Phi}(f_\wedge)$ is a Hermitian operator. The
corresponding condition on a test function $f_\sim$ reads:
\begin{equation}
  \label{eq:realitycond}
  \widetilde{\Phi}(f_\sim)^\dagger = \widetilde{\Phi}(f_\sim)
  \quad\Leftrightarrow\quad f_\sim =
  \widetilde{\overline{\widehat{f_\sim}}} \zendot
\end{equation}

Now that we have given the rules for switching between the compact and
the light-cone picture, it is straightforward to calculate the
commutation relations in the compact picture from the relations
between fields on the light-cone. For the currents we have (cf
equation \ref{eq:quacurcom}): 
\begin{equation}\label{eq:compcurralg}
  \komm{\widetilde{j^a}(f_\sim)}{\widetilde{j^b}(g_\sim)} = i
  f^{ab}{}_c  \widetilde{j^c}(f_\sim g_\sim) + k g^{ab}_\lie{g} 
\oint \frac{dz}{2\pi i}  f_\sim'(z) g_\sim (z) \zendot
\end{equation}
$f_\sim'$ stands for $d/dz f_\sim$. 

The \Name{L\"uscher-Mack} algebra for a stress-energy tensor of
central charge $c$ (cf equation \ref{eq:LueMackalg}) reads:
\begin{equation}
  \label{eq:lumackcomp}
  \komm{\widetilde{\Theta}(f_\sim)}{\widetilde{\Theta}(g_\sim)} =
  \widetilde{\Theta} (f_\sim' g_\sim -f_\sim g_\sim') + 
  \frac{c}{12} \oint \frac{dz}{2\pi i} f_\sim''' (z) g_\sim (z) \zendot
\end{equation}
The action of the \Name{Sugawara} stress-energy  tensor on the
currents takes the following form (cf equation \ref{eq:setoncurr}):
\begin{equation}
  \label{eq:setoncurrcomp}
   \komm{\widetilde{T}(f_\sim)}{\widetilde{j}^a(g_\sim)} =
    \widetilde{j}^a(-f_\sim g_\sim') \zendot
\end{equation}
All these relations are mere rewritings of those on the light-ray.

Test functions on the circle possess a \Name{Fourier}
expansion: $f_\sim = \sum_{n\in\dopp{Z}} f_n z^{-n}$. 
We write the test function defined by $z\mapsto z^m$ by $[z^m]$ and
define for a quasi-primary chiral field $\widetilde{\Phi}$ of scaling
dimension $h$ its {\em modes} as: 
\begin{equation}
  \label{eq:defmodes}
  \Phi_n := \widetilde{\Phi}([z^{h-1+n}])\, , \,\, n\in\dopp{Z} \quad. 
\end{equation}
If $\widehat{\Phi}$ is Hermitian, we have for the modes
$\Phi_n{}^\dagger = \Phi_{-n}$. 
The modes of currents and stress-energy tensors are given by:
\begin{eqnarray}
  \label{eq:defmodcurrset}
  L_n := \widetilde{\Theta} ([z^{n+1}]) \,\,  , \quad 
j^a_n := \widetilde{j^a} ([z^n]) \zendot
\end{eqnarray}

It is a matter of simple arithmetics to calculate the commutation
relations in terms of the modes:
\begin{eqnarray}
  \label{eq:moderelacurr}
  \komm{j_m^a}{j_n^b}&=&i f^{ab}{}_c j_{m+n}^c + k g^{ab}_\lie{g} m
  \delta_{m,-n} \zencom\\
\label{eq:virasoroalg}
\komm{L_m}{L_n} &=& (m-n) L_{m+n} + \frac{c}{12} \delta_{-n,m}
m(m^2-1) \zencom\\
\label{eq:vironcurr}
\komm{L_m}{j^a_n} &=& -n j^a_{m+n} \zendot
\end{eqnarray}
The commutation relations (\ref{eq:virasoroalg}) are known as
{\em {Virasoro} algebra} at central charge $c$. 
The commutation relations (\ref{eq:moderelacurr}) are manifestations
of the relations in an {\em affine {Kac-Moody} algebra} or {\em affine
{Lie} algebra} (see eg \cite{vKbook}, \cite{jF92}). 

The relations (\ref{eq:vironcurr}) and (\ref{eq:virasoroalg}) show
that the zeroth mode of 
$\Theta$, the conformal {Hamilton}ian $L_0$, defines a natural
energy grading on the modes and hence on  the vacuum \Name{Hilbert}
spaces of
both current algebras and stress-energy tensors. Since the spectrum of
$L_0$ is positive and pure-point on these spaces, it is a positive,
essentially self-adjoint operator. Moreover, the eigenspaces are
finite dimensional.  These facts are founding elements
of the next section.

%%%%%%%%%%%%%%%%%%%%%%%%%%%%%%%%%%%%%%%%%%%%%%%%%%%%%%%%%
%%%%%%%%%% INTEGRATING CURRENT ALGEBRAS %%%%%%%%%%%%%%%%%
%%%%%%%%%%%%%%%%%%%%%%%%%%%%%%%%%%%%%%%%%%%%%%%%%%%%%%%%%

\subsection{Integrating chiral current algebras}

In spite of their strong conceptual and structural similarities, the
relation between the formulations of quantum field theory in terms of
tempered operator-valued distributions (\Name{Wightman}'s axioms)
and in terms of local algebras 
of bounded operators (\Name{Haag-Kastler} axioms) is not that of
a straightforward equivalence. We shall not
elaborate on the remarkable general achievements in establishing relations in
both directions (see eg 
\cite{kF91} for a short summary and a list of references). In the
following we discuss this relation for current algebras. The
technical problems present in general surface here as well, but in a
 mild and manageable form. 

As we will see in a moment, the
energy grading with respect of $L_0$, the \Name{Sugawara} construction
and the \Name{Lie} algebra of the modes allow us to establish the
following properties for the smeared currents:  For real
test functions\footnote{Currents have scaling dimension $1$ which
  leads to $\widetilde{\overline{\widehat{f_\sim}}} =
  \overline{f_\sim}$, $f_\sim\in C^\infty(\Seins)$; see 
  (\ref{eq:licocompchi}), (\ref{eq:licocompchii}). Hence,
  real test functions $f_\sim$ lead to symmetric smeared current
  operators, see equation (\ref{eq:realitycond}).} $f_\sim, g_\sim $ 
 the current 
operators $j^a(f_\sim)$, $j^b(g_\sim)$ are essentially self-adjoint on
the \Name{Wightman} domain and all bounded functions of
their self-adjoint closures commute, ie their closures
$\closure{j^a(f_\sim)}$, $\closure{j^b(g_\sim)}$ commute in the
sense of self-adjoint operators  \cite[VIII.5]{RS72}, if $f_\sim$,
$g_\sim$ have disjoint supports. Both properties
are not obvious due to the unbounded character of the smeared fields
(cf eg \cite[VIII.5]{RS72}, \cite{hR88}), but when they hold one may define the
local algebras of the chiral net generated by the current algebra as:
\begin{equation}
  \label{eq:deflokcuralg}
  \lok{B} (I) := \menge{\closure{\widetilde{j^a}(f_\sim)},
  \,\, supp(f_\sim) \subset I, f_\sim = \overline{f_\sim} , \,\,
  T^a \in \lie{g}}'', \,\, I\Subset \Seins \zendot
\end{equation}
Covariance and isotony are direct consequences of this definition and the
properties of the current algebras as \Name{Wightman} theories. The
cyclicity of the vacuum may be established by a simple argument based
on the spectral resolution of the closures of the smeared fields
\cite[theorem 4.1]{DF77}. In general, \Name{Wightman} fields do not
behave that well and one has to introduce more refined definitions
for the local algebras (cf \cite{DSW86}). 

\Name{Buchholz and Schulz-Mirbach} \cite{BS90} introduced a method for
establishing essential self-adjointness and locality for the closures
of symmetric smeared fields based on proving linear energy bounds with
respect to the conformal {Hamilton}ian. This approach is
particularly well-suited for chiral conformal models for which the
commutation relations are explicitly known in terms of their
modes. 

In \cite{BS90}, the method is applied to the \Name{Virasoro} algebras
and this scheme covers, as stated by \Name{Buchholz and
  Schulz-Mirbach}, the $U(1)$-current algebra as well. Hence it is
known already that both \Name{Wightman} fields generate chiral
conformal nets.
The argument on the $U(1)$-current algebra is extended
to another class of conformally covariant fields in chapter
\ref{cha:noset}. If one is interested in current algebras with simple
colour algebras, one needs to apply another argument, which uses the
connection between $L_0$ and the current algebra via the
\Name{Sugawara} construction, because the zero modes of in these
algebras are not scalars. Since the $U(1)$-current algebra
possesses a \Name{Sugawara} stress-energy tensor as well, the
arguments below apply to this case as well.

To the author's knowledge, a proof of the following result is not
yet available:
\begin{theo}\label{th:chcurrchlok}
Assume $\lie{g}$ is a reductive colour algebra for which
the current algebra is available as a chiral conformal \Name{Wightman}
theory on the conformally compactified light-ray, $\Seins$. Then 
equation (\ref{eq:deflokcuralg}) defines a chiral conformal theory.   
\end{theo}
\begin{pf}
We establish linear energy bounds for the currents $\widetilde{j^a}$ of
a simple colour algebra ${\lie{g}_\alpha}\subset\lie{g}$; the argument
relies on the \Name{Sugawara} 
construction, the energy grading on \Name{Hilbert} space and positivity
of energy.  

We choose a basis $\{T^a\}$ in colour space which diagonalises the
\Name{Killing} metric on $\lie{g}_\alpha$ such that:
$2\symb{g}^\vee_{\lie{g}_\alpha}\skalar{T^a}{T^b}_{\lie{g}_\alpha} =
\delta^{ab}$. Then the conformal
{Hamilton}ian $L_0^{\lie{g}_\alpha}$ of colours in
$\lie{g}_\alpha$ reads:
\begin{equation}\label{eq:L0galpha}
  L_0^{\lie{g}_\alpha} = \frac{\symb{g}^\vee_{\lie{g}_\alpha}}{k_{\lie{g}_\alpha}+\symb{g}^\vee_{\lie{g}_\alpha}}
  \left(\sum_{i=1}^{d_{\lie{g}_\alpha}} j^a_0j^a_0 + 2
    \sum_{i=1}^{d_{\lie{g}_\alpha}} \sum_{n>0} j^a_{-n}j^a_n\right)
  \zendot
\end{equation}
This positive operator is dominated by the conformal
{Hamilton}ian of the whole current algebra, $L_0^\lie{g}$.

We take a vector $\phi_N$ of energy $N$ with respect to $L_0^\lie{g}$
and of unit length. There is the following bound for the action of the
annihilation modes $j^a_n$, $n\geq 1$, on $\phi_N$ (using
(\ref{eq:L0galpha})):
\begin{equation}
  \label{eq:annbound}
  \norm{j^a_n\phi_N}^2 \leqslant
  \frac{k_{\lie{g}_\alpha}+\symb{g}^\vee_{\lie{g}_\alpha}}{2\symb{g}^\vee_{\lie{g}_\alpha}}
  \skalar{\phi_N}{L_0^{\lie{g}_\alpha}\phi_N} \leqslant
  \frac{k_{\lie{g}_\alpha}+\symb{g}^\vee_{\lie{g}_\alpha}}{2\symb{g}^\vee_{\lie{g}_\alpha}}
  N \zendot
\end{equation} 
(\ref{eq:L0galpha}) yields a similar bound for the zero modes $j^a_0$, 
and for the generating modes $j^a_{-n}$, $n\geq 1$, we use the bound
(\ref{eq:annbound}) and the commutation relations (\ref{eq:moderelacurr}).
Thus, we have for all $n\in\dopp{Z}$:
\begin{equation}
  \label{eq:allbound}
  \norm{j^a_{n}\phi_N}^2 \leqslant
  \frac{k_{\lie{g}_\alpha}+\symb{g}^\vee_{\lie{g}_\alpha}}{\symb{g}^\vee_{\lie{g}_\alpha}}
  N + k_{\lie{g}_\alpha} 2 \symb{g}^\vee_{\lie{g}_\alpha} |n| \leqslant
  d_{\lie{g}_\alpha} k_{\lie{g}_\alpha} N + k_{\lie{g}_\alpha} 2
  \symb{g}^\vee_{\lie{g}_\alpha} |n| \zendot
\end{equation}
The second inequality follows from $c_{\lie{g}_\alpha}=
d_{\lie{g}_\alpha} k_{\lie{g}_\alpha}/
(k_{\lie{g}_\alpha}+\symb{g}^\vee_{\lie{g}_\alpha}) \geq
r_{\lie{g}_\alpha}\geq 1$ and
$\symb{g}^\vee_{\lie{g}_\alpha}{}\geq 1$. We note that the
right-hand side is greater than $1$ for $N\neq 0$; the case $N=0$, ie 
$\phi_0\sim \Omega$, does not pose any problem in the following.

In order to obtain a bound for a general $\phi$ from the
\Name{Wightman} domain, we proceed as follows: The symmetric operator
$j^{a}_{-n}j^{a}_{n}$ leaves invariant the finite-dimensional
energy eigenspace of energy $N$ and we 
may choose an orthonormal basis of vectors $\phi_N^\iota$
diagonalising the action 
of $j^{a}_{-n}j^{a}_{n}$. Expanding $\phi$ in terms of
this basis yields:
\begin{eqnarray}
  \label{eq:wightmanbound}
  \norm{j^{a}_{n}\phi}^2  
&\leqslant& \sum_{N,\iota} \overline{c_{N,\iota}} c_{N,\iota}
  \norm{(d_{\lie{g}_\alpha} k_{\lie{g}_\alpha} N + k_{\lie{g}_\alpha} 2 \symb{g}^\vee_{\lie{g}_\alpha} |n|)\phi_{N}^{\iota}}^2
\nonumber\\
&=& \norm{(d_{\lie{g}_\alpha} k_{\lie{g}_\alpha} L_0^\lie{g} +
  k_{\lie{g}_\alpha} 2 \symb{g}^\vee_{\lie{g}_\alpha} |n|)\phi}^2
\zendot
\end{eqnarray}
This argument gives a linear energy bound for the action of the modes
on the whole \Name{Wightman} domain.

Our interest is in bounds for smeared currents. If we take a
test function $f_\sim\in C^\infty(\Seins)$, then the \Name{Fourier}
coefficients decrease fast enough to make any series $\sum_n |f_n|
|n|^l$, $l\in\dopp{N}$, converge. Hence one gets for an
arbitrary smeared current $\widetilde{j^a}(f_\sim)$ and an arbitrary
vector $\phi$ from the \Name{Wightman} domain:
\begin{equation}
  \label{eq:hboundcurr}
    \norm{j^a(f_\sim)\phi} \leqslant c_{f_\sim} \norm{(L_0+\Einsop)\phi} \zendot
\end{equation}
The constant $c_{f_\sim}$ depends on $f_\sim$ only (and not on
$\phi$). In fact, one may choose: $  c_{f_\sim} = \sum_{n\in\dopp{Z}}
\betrag{f_{n}} k_{\lie{g}_\alpha} (d_{\lie{g}_\alpha}+2
\symb{g}^\vee_{\lie{g}_\alpha}|n|) $.  
By applying the arguments above for all simple ideals of
$\lie{g}$ and (in a simplified version) for the abelian ideal, we get
linear energy bounds for all smeared currents. 

$L_0$ has a total set of eigenvectors and it is straightforward to
check that $L_0$ is essentially self-adjoint on the \Name{Wightman}
domain. By \Name{Nelson}'s analytic vector theorem \cite{eN59}, $L_0$ has a dense set 
of {\em analytic vectors} $\phi$. A vector $\phi$ is analytic for an
operator $A$ if for any $n=0,1, \ldots$ the $n$th power of $A$ is
declared on $\phi$ and the following  holds
for some $0<s$:
\begin{equation}
  \label{eq:anavec}
  \sum_{n=0}^\infty \frac{\norm{A^n\phi}}{n!}  s^n  < \infty \zendot
\end{equation}
The symmetric smeared fields $\widetilde{j^a}(f_\sim)$  may be
extended to the space of finite energy vectors, the domain
$D(\closure{L_0})$ of the closure of $L_0$, because of the bound
(\ref{eq:hboundcurr}). Any analytic vector of $\closure{L_0}+\Einsop$
is an analytic vector 
for the extended smeared currents. Again by \Name{Nelson}'s theorem,
the extended smeared currents are essentially self-adjoint and hence
the \Name{Wightman} domain is a core for the self-adjoint closures
$\closure{\widetilde{j^a}(f_\sim)}$. 

As indicated by \Name{Buchholz and Schulz-Mirbach} \cite{BS90},
finally, we make contact with a result of \Name{Driessler and
Fr\"ohlich} \cite{DF77}, which requires to establish linear
energy bounds on the 
smeared fields $\widetilde{j^a}(f_\sim)$, for
$Ad_{L_0}(\widetilde{j^a}(f_\sim))$ and for
$Ad_{L_0}^2(\widetilde{j^a}(f_\sim))$ as quadratic forms on
the \Name{Wightman} domain. Because  we argued for arbitrary test functions,
we need not worry about the action of $Ad_{L_0}$ and  the
bounds for the quadratic forms follow from (\ref{eq:hboundcurr}) by
\cite[theorem X.18]{RS75}. The result of \Name{Driessler and
  Fr\"ohlich} now says: if 
smeared currents commute on the core, then all spectral projections of
their self-adjoint closures commute. This ensures locality for the
algebras defined above.  

Cyclicity of the vacuum follows by the argument in \cite[theorem
4.1]{DF77}. Since the other properties of definition \ref{def:chcotheo} follow
directly, we completed the proof..
\end{pf}

 The argument above works for test functions with support in some
 $I\Subset\Seins$ as
 well as for test functions supported in all of $\Seins$. For
 test functions of support in some $I\Subset\Seins$ one could 
have proceeded along the lines of 
\Name{Borchers and Zimmermann} \cite{BZ63} by recognising that the linear
energy bounds prove the vacuum to be an analytic vector for all
currents. 

% Moreover, this analyticity proves that the generating
% functionals for the \Name{Wightman} functions actually depend
% analytically on $z$:
% \begin{equation}\label{eq:analytgenfcts}
%   \mathcal{W}[j^a(f)](z) \equiv  \skalar{\Omega}{e^{zj^a(f)}\Omega}\,,\,\,
%  \mathcal{W}[\Theta(g)](z) \equiv \skalar{\Omega}{e^{z\Theta(g)}\Omega}
% \end{equation}
% Hence, one may expand these vacuum expectation values into the
% corresponding power series. \Name{Baumann} \cite{kB94} gave a closed
% formula for 
% $\mathcal{W}[\Theta(g)]$ for central charge $c=1$ and a procedure for
% obtaining the functions for other $c$ which is an alternative for the
% usual recursive calculation (cf eg \cite{FST89}). 

The results of \cite{BS90} on the \Name{Virasoro}
algebra show that the \Name{Sugawara} stress-energy tensor generates a
local and relatively local subnet of the chiral theory generated by
the currents.  The energy  bounds for the modes $L_0, L_{\pm1}$ of
such a tensor ensure that their symmetric linear combinations generate
a unitary representation  of $\PSL(2,\dopp{R})^\sim$ \cite{jF77}.

%%%%%%%%%%%%%%%%%%%%%%%%%%%%%%%%%%%%
%%%%% SECTION LOOP TO CURRENT %%%%%%
%%%%%%%%%%%%%%%%%%%%%%%%%%%%%%%%%%%%

\subsection{From loop algebras to chiral current algebras}
\label{sec:currloop}

Both chiral current algebras and chiral stress-energy tensors carry
the structures of  infinite-dimensional
\Name{Lie} algebras. The
chiral current algebra of colours in a simple, compact $\lie{g}$ represents
a central extension of $\luup{g}\equiv C^\infty (\Seins,\lie{g})$, the
{\em loop algebra} 
of smooth mappings from $\Seins$ into $\lie{g}$. The commutation
relations of stress-energy tensors represent central extensions of the
\Name{Lie} algebra of smooth vector fields on the circle,
$\diff(\Seins)$, known as {\em {Witt} algebra} (see eg \cite{mS95}). 
$\luup{g}$ is the \Name{Lie} algebra of the {\em loop group}
$\Loop{G}\equiv C^\infty (\Seins,G)$ of smooth mappings from $\Seins$
into $G$, the compact, connected and simply 
 connected \Name{Lie} group having $\lie{g}$ as its \Name{Lie}
 algebra. And $\diff(\Seins)$ is the \Name{Lie} algebra of the group
 of orientation preserving diffeomorphisms on 
 the circle, $\Diff_+(\Seins)$.

Multiplication in $\Loop{G}$ is declared point wise: $g_1\cdot g_2 (z) =
g_1(z)\cdot g_2(z)$, $z\in\Seins$, and in $\Diff_+(\Seins)$ by successive
evaluation: $\varphi_1\cdot\varphi_2 (z)= \varphi_1(\varphi_2(z))$.
$\Diff_+(\Seins)$ acts on $ \Loop{G}$ through reparametrisations of
$\Seins$. The group multiplication law in the
semi-direct product induced by this action,
$\Loop{G}\rtimes\Diff_+(\Seins)$, reads: $$(g_1,\varphi_1)\cdot
(g_2,\varphi_2) = (g_1\circ\varphi_2 \cdot g_2, \varphi_1\cdot
\varphi_2) \zendot$$ 
 General references on these structures
 are: \cite{jM83}, \cite{PS86}.

These relations may be used to establish current algebras for all
simple colour algebras and at all positive integer levels as
chiral conformal theories both in accordance with \Name{Wightman}'s axioms
and the \Name{Haag-Kastler} axioms. The starting point are
unitarisable, highest-weight representations of affine
\Name{Kac-Moody} algebras with the vacuum, $\Omega$, as highest-weight vector.

These representations are given in terms of modes $j^a_n$,
$a=1,\ldots,d_\lie{g}$, $n\in\dopp{Z}$, and their complex linear
combinations. One defines $j^a_n \Omega := 0$, $n\geq 0$. On the
vector space of linear combinations of vectors of the form
$j^{a_1}_{-n_1}\ldots j^{a_m}_{-n_m}\Omega$, 
$n_1,\ldots, n_m >0$, called $\Hilb{H}^{fin}$, there exists a 
scalar product such that the modes satisfy the
commutation relations (\ref{eq:moderelacurr}), are Hermitian
($j^a_{n}{}^\dagger=j^a_{-n}$), and  we have
and $\|\Omega\| =1$ \cite[theorem 11.7]{vKbook}. For this it is
necessary and sufficient that the level $k$ is a positive integer.

The \Name{Sugawara} stress-energy tensor finds its counterpart in this
context by
the {\em {Segal-Sugawara} formula}:
\begin{equation}
  \label{eq:defsegsug}
  L_n := \frac{1}{2(k+\symb{g}^\vee_\lie{g})} \,\, g^\lie{g}_{ab}
  \sum_{n_1+n_2=n} :j^a_{n_1}j^b_{n_2}: \zendot
\end{equation}
Here, $g^\lie{g}_{ab}:\!j^a_kj^b_l\!: \, = \,
g^\lie{g}_{ab}j^a_kj^b_l$ for $l>0$, and
$g^\lie{g}_{ab}:\!j^a_kj^b_l\!: \, = \, g^\lie{g}_{ab}j^b_lj^a_k$ for
$l\leq 0$. The $L_n$ satisfy the 
relations (\ref{eq:virasoroalg}), (\ref{eq:vironcurr}) on $\Hilb{H}^{fin}$.

The unitarisable, highest-weight representations of affine
\Name{Kac-Moody} algebras (cf eg \cite{vKbook})  are known to
exponentiate uniquely to 
projective, unitary representations of $\Loop{G}$ \cite{GW84,vL99a}.  
In  $\Loop{G}$ a group element is {\em localised} in some
$I\Subset\Seins$, if it coincides with the neutral element
of the group outside $I$. This intrinsic locality structure makes it
possible to prove, by group 
theoretical methods, that the
exponentiated projective representations with the vacuum as highest
weight define chiral conformal nets  \cite{vL97,aW98,FG93}. Since the
level may identified in any locally normal represention within local
algebras, it is a constituting element of the theory. We  denote
the local 
quantum theory connected with an exponentiated projective representation
of $ \Loop{G}$ at level $k$ by
$\Loop{G}_k$, in marginal deviation from the 
common practice which leaves out ``$\Loop{}$''. \label{ind:loop}  

For $\Diff_+(\Seins)$ one can follow the same
program\footnote{The relevant results are contained in \cite[theorem
11.12]{vKbook} (unitarity), \cite{GW85,vL99a} (integrability), 
\cite{tL94} (locality etc).}, but \Name{Wightman}'s axioms are established for
stress-energy tensors of all central charges $c$ (cf section
\ref{sec:sugcurralg}), hence 
these models were established as chiral nets in \cite{BS90} already.
  The chiral net
generated by a stress-energy tensor with central charge $c$ will be
called the {\em {Virasoro} model} $\Vir_c$. \label{ind:virc}       

Intermediate results which were obtained in order to establish the
$\Loop{G}_k$ models can be used to show that for every simple colour 
algebra $\lie{g}$ the action of the modes on $\Hilb{H}^{fin}$ extends
to a \Name{Wightman} theory of chiral currents. This covers in
particular the four cases
$E_{6-8}$, $F_4$
for which there are no quark models at level $1$, and it  will be
applied in the context of current subalgebras 
(see section \ref{sec:curexp}). The proof of the following proposition
relies mainly on results of \Name{Toledano-Laredo} \cite{vL97,vL99a}, but
the statement has  not appeared explicitly, yet: 

\begin{prop}
  For every compact, simple colour algebra $\lie{g}$ and at every
  positive, integer level $k$ the corresponding current algebra exists
  as a conformally covariant chiral quantum field theory in agreement
  with \Name{Wightman}'s axioms.
\end{prop}
\begin{pf}
  We start from the action of the Hermitian modes $j^a_n$ on the
  pre-\Name{Hilbert} space $\Hilb{H}^{fin}$, which we introduced
  above. The closure of $\Hilb{H}^{fin}$, the \Name{Hilbert} space
  $\Hilb{H}$, is separable by construction.

One controls the action of the modes $j^a_n$ in terms of {\em
  {Sobolev} norms}:
\begin{equation}
  \label{eq:sobolev}
  \norm{\xi}_s := \norm{(1+L_0)^s\xi}\, , \quad \xi\in
  \Hilb{H}^{fin}\,,\,\, s\in\dopp{R} \, .
\end{equation}
The right-hand side may be calculated by expansion into energy
eigenvectors.
The closure of $\Hilb{H}^{fin}$ with respect to the norm $\norm{.}_s$
is called the  
{\em scale} $\Hilb{H}_s$,  and the space
$\Hilb{H}^\infty=\bigcap_s\Hilb{H}_s$ is the space of {\em smooth
vectors} for the action of the conformal {Hamilton}ian
$L_0$. Since $L_0$ has a total set of eigenvectors, its closure  is 
essentially self-adjoint on $\Hilb{H}^{\infty}$.

The commutation relations of the modes $j^a_n$ yield bounds for their
action on $\Hilb{H}^{fin}$ in terms of the norms $\norm{.}_s$ 
\cite{GW84}. These allow to prove that this action on
$\Hilb{H}^{fin}$ extends to a jointly continuous mapping
$\luup{g}\times \Hilb{H}^\infty\rightarrow
\Hilb{H}^\infty$ such that the commutation relations
(\ref{eq:compcurralg}) are satisfied, ie we have a projective
representation of  
$\luup{g}$ on $\Hilb{H}^\infty$ \cite[corollary
1.3.1]{vL97}. Hence, the currents $\widetilde{j^a}$ are local. 

The matrix elements of currents, namely $\langle v,
\widetilde{j^a}(f_\sim) w\rangle$, $v,w\in\Hilb{H}^\infty$, define
tempered distributions on $C^\infty(\Seins)$ by the definition of
the topology on $\luup{g}$. The currents $\widetilde{j^a}(f_\sim)$ for
real $f_\sim$ are essentially self-adjoint on $\Hilb{H}^\infty$, which
means that all currents are closable operators \cite[corollary
1.3.1]{vL97}. 

$\Hilb{H}^\infty$ is the \Name{Wightman} domain of the currents. It is
clear that the vacuum is cyclic for the currents, since
$\Hilb{H}=\Hilb{H}_0$. Arguments given in \cite{BS90} show that the
Hermitian linear combinations of $L_{0,\pm 1}$ satisfy linear energy
bounds with respect to $L_0$ and hence exponentiate to a
positive-energy representation
$U$ of $\PSL(2,\dopp{R})$ \cite{jF77}. Obviously, $\dopp{C}\Omega$
is the space of $U$-invariant vectors. 

The action of the current modes and of the $L_n$ on
$\Hilb{H}^{fin}$ exponentiates uniquely to a projective 
representation of $ \Loop{G}\rtimes\Diff_+(\Seins)$
\cite{GW84,vL99a}. This shows that the currents are  
covariant with respect to $U$. 
\end{pf}

Irreducible, locally
normal representations $\pi$ of $ \Loop{G}_k$ on a \Name{Hilbert} space
$\Hilb{H}_\pi$ yield  unitary, strongly
continuous, projective representations of $ \Loop{G}$. % , ie a strongly
% continuous homomorphism
% \begin{displaymath}
%    \Loop{G} \longrightarrow PU(\Hilb{H}_\pi) = U(\Hilb{H}_\pi)/
%    \dopp{C}_1\Einsop_\pi 
% \end{displaymath}
This can be seen looking at the one-parameter groups in
$\Loop{G}$. Referring to a basis $\{T^a\}$ of $\lie{g}$, a general
element in $\luup{g}$ is given in terms of an $d_\lie{g}$-tuple
$f_\sim=(f_\sim^{(a)})_a$ of smooth functions,  and the one-parameter
subgroups are of the form $g_{f_\sim}(t)=exp(-it\sum_af_\sim^{(a)}T^a)$. With
respect to a partition of unity on $\Seins$ by two smooth,
non-negative functions $\chi_{1,2}$, $supp(\chi_{1,2})\Subset\Seins$,
we  see that a locally normal representation $ \Loop{G}_k$
defines a strongly continuous mapping from the one-parameter subgroups
of $\Loop{G}$ to the unitaries on $\Hilb{H}_\pi$. In a neighbourhood
of the identity the exponential 
map $exp: \luup{g} \rightarrow \Loop{G}$ is a homeomorphism, and thus
we have\footnote{For an analogous argument on the $\Vir_c$ models one
  may use lemma \ref{locmap}.} a unitary, strongly continuous,
projective representation of $\Loop{G}$. 

The diffeomorphism symmetry
of loop groups can be used to show that every locally normal
representation is automatically covariant with respect to
$\PSL(2,\dopp{R})$ (see section \ref{cha:app}.\ref{sec:locnorrep}), but it is not
clear {\em a priori} that the representation is of positive
energy. Likewise, there is no 
argument yet which ensures an action of
$\luup{g}$ on a dense subspace of smooth vectors in representation space.

Therefore, it is not sure whether the locally normal representations
are covered by the classification of unitarisable highest-weight
representations of $\luup{g}$ at level $k$ in terms of the {\em {Weyl}-alcove
  condition} (eg \cite[(2.4.26)]{jF92}).  These representations
are completely determined  by the action of the horizontal subalgebra
on the lowest energy eigenspace, that is by the  highest weight 
$\lambda$ of this irreducible, unitary representation of $\lie{g}$. We
discuss the \Name{Weyl}-alcove condition for $\luup{s\lie{u}(n)}$ in
detail.

$\lambda$ is of the form $\lambda=\sum_{i=1}^{n-1} \lambda_i
\lambda^{(i)}$, where the $\lambda^{(i)}$ are the fundamental weights of
$\lie{s}\lie{u}(n)$ and the $\lambda_i$ are non-negative integers.
The \Name{Weyl}-alcove condition for $\luup{s\lie{u}(n)}$ at level $k$
reads:
$
%\begin{displaymath}
  \sum_{i=1}^{n-1} \lambda_i \leq k
%\end{displaymath}
$.
We denote the set of highest weights $\lambda$ satisfying this
condition by $P_+^k(SU(n))$. The corresponding (projective) highest-weight
representations of $\luup{g}$ are known to exponentiate to unitary,
projective representations of 
$\Loop{SU(n)}$ with positive energy \cite{GW84}, which form locally
normal representations of $\Loop{SU(n)}_k$ (covariant with positive
energy) \cite{aW98}. 

The fusion rules of these positive energy representations of
$\Loop{SU(n)}_k$ are known to
coincide with the ones expected from operator product expansions
\cite{aW98} which means that their statistical dimensions coincide
with the asymptotic dimensions of the corresponding highest weights
(eg \cite{jF94,jF92}), which shall be denoted $d_\lambda$, $\lambda\in
P_+^k(SU(n))$.  \Name{Xu} has calculated the {\em $\mu$-index} \cite{KLM01} of
these models explicitly as \cite[theorem 4.1]{fX00a}:
\begin{equation}
  \label{eq:muLSUnk}
  \mu( \Loop{SU(n)}_k) = \sum_{\lambda\in P_+^k(SU(n))} d_\lambda^2
  \zendot
\end{equation}
Because the $\mu$-index is finite and since the models are strongly additive
\cite{vL97} and split \cite{FG93}, the chiral nets $\Loop{SU(n)}_k$ are
{\em completely rational} \cite{KLM01}, and by the value of the
$\mu$-index  the 
unitarisable, highest-weight representations are known to exhaust all
sectors \cite[theorem 33]{KLM01}. 

This result opens the door to the classification of 
locally normal representations of many other chiral nets; in this work
it will be needed only in the proof of proposition
\ref{prop:E81trivsss} where it is shown that $\Loop{E(8)}_1$ has only
one sector. The identification of
all sectors of the $\Vir_{c<1}$ models \cite{KL02} is another achievement
of this sort, cf section \ref{cha:finind}.\ref{sec:GKOcospa}. 
  
%%%%%%%%%%%%%%%%%%%%%%%%%%%%%%%%%%%%%%%%%%%%%%%%%%%%%%%%%%%%%%%
%%%%%% CURRENT SUBALGEBRAS AS INSTRUCTIVE EXAMPLES %%%%%%%%%%%%
%%%%%%%%%%%%%%%%%%%%%%%%%%%%%%%%%%%%%%%%%%%%%%%%%%%%%%%%%%%%%%%

\sekt{Current subalgebras as instructive examples}{Current subalgebras as instructive examples}{sec:curexp}
%\subsection{Current subalgebras as instructive examples}
%\label{sec:curexp}

An injective \Name{Lie} algebra homomorphism $\iota$ from a reductive
colour algebra $\lie{h}$ into a reductive colour algebra $\lie{g}$,
denoted $\iota: \lie{h}\hookrightarrow \lie{g}$, is regarded as an
inclusion of \Name{Lie} algebras $\iota(\lie{h})\subset\lie{g}$. It
clear from the current algebra (\ref{eq:quacurcom}) that the inclusion
of colour algebras
induces an inclusion of the corresponding current algebras.
By theorem
\ref{th:chcurrchlok}, the inclusion of current algebras 
yields a chiral subnet, the {\em current subalgebra}, which we denote by
$\lok{A}\subset\lok{B}$.

The \Name{Sugawara} construction endows the current subalgebra
associated with $\iota(\lie{h})$ with its own stress-energy tensor
$\Theta^{\iota(\lie{h})}$. Since $\Theta^{\iota(\lie{h})}$ and
$\Theta^\lie{g}$ obey the same commutation relations with currents
of colour in $\iota(\lie{h})$, the tensor $ \Theta^\lie{g}-
\Theta^{\iota(\lie{h})}$ commutes with all these currents and hence
with $\Theta^{\iota(\lie{h})}$. It follows that $ \Theta^\lie{g}-
\Theta^{\iota(\lie{h})}$ satisfies the \Name{L\"uscher-Mack}
commutation relations (\ref{eq:LueMackalg}) with central charge
$c_{\lie{g}}-c_{{\iota(\lie{h})}}$ and, therefore, it defines a
stress-energy tensor. $ \Theta^\lie{g}-
\Theta^{\iota(\lie{h})}$ is called the {\em {Coset} stress-energy
tensor} (associated with the respective  current
subalgebra). 

It is clear
from the arguments of \Name{Buchholz and Schulz-Mirbach} \cite{BS90}
that the \Name{Coset} stress-energy tensor generates a chiral subnet
of $\lok{B}$ which is contained in 
$\lok{C}_{max}$. As arguments below show, the Hermitian linear
combinations of modes $L_{0,\pm 1}^{\iota(\lie{h})}$ of
$\Theta^{\iota(\lie{h})}$ generate the \Name{Borchers-Sugawara}
representation $U^{\lok{A}}$ of
$\lok{A}\subset\lok{B}$ 
and   the corresponding  modes of the \Name{Coset} stress-energy tensor
  generate $U^{\lok{A}'}$, which coincides with
  the \Name{Borchers-Sugawara} representation $U^{\lok{C}_{max}}$ of
  the maximal \Name{Coset} model $\lok{C}_{max}$ associated with
$\lok{A}\subset\lok{B}$ (cf section
\ref{cha:cospa}.\ref{sec:bosug}).  

If we have $\Theta^\lie{g}=\Theta^{\iota(\lie{h})}$, ie
$c_{\Theta^\lie{g}}=c_{\Theta^{\iota(\lie{h})}}$, then the inclusion
$\iota(\lie{h})$ is called a {\em conformal inclusion} or {\em conformal
embedding}.  The following theorem \ref{th:SegSugCon} shows that
this notion of conformal inclusion coincides, for current subalgebras,
with the one given in definition \ref{def:confinc}.

We now establish that the {\em Additional Assumption}  on 
$U^\lok{A}$ needed in chapter \ref{cha:netend} is satisfied for
current subalgebras (cf page
\pageref{ass}). The covering projection from
$\Diff_+(\Seins)^\sim$ onto $\Diff_+(\Seins)$ is denoted by
$\symb{p}$. The subgroup of diffeomorphisms $\varphi$ which are
localised in some $I\Subset\Seins$, ie which satisfy
$\varphi\restriction{I'} = id\restriction{I'}$, is denoted as
$\Diff_I(\Seins)$. Localised diffeomorphisms $\varphi$ will be identified with
their preimage $\symb{p}^{-1}(\varphi)$ in the first
sheet of the covering: \label{ind:locdiff}

\begin{theo}\label{th:SegSugCon}
  Let the chiral subnet $\lok{A}\subset\lok{B}$ stem from an embedding
  of chiral current algebras and let $\lok{C}_{max}$ denote the maximal
  \Name{Coset} model associated with this inclusion. Then there are
  unitary, projective representations $\Upsilon^\lok{B}$ of
  $\Diff_+(\Seins)$ and 
  $\Upsilon^\lok{A}$, $\Upsilon^\lok{C}$ of $\Diff_+(\Seins)^\sim$ having
  the following properties:
  \begin{itemize}
\item It holds true: $\Upsilon^\lok{B}\circ\symb{p}(\tilde{\varphi}) = \Upsilon^\lok{A}(\tilde{\varphi})
  \Upsilon^\lok{C}(\tilde{\varphi})$,
  $\tilde{\varphi}\in\Diff_+(\Seins)^\sim$.  
  \item For $\varphi\in\Diff_I(\Seins)$, $I\Subset \Seins$, we have:
  \begin{equation}
    \label{eq:intsuginner}
  \Upsilon^\lok{B}(\varphi)\in\lok{B}(I) \, , \,\,
  \Upsilon^\lok{A}(\symb{p}^{-1}(\varphi))\in\lok{A}(I) \, , \,\,
  \Upsilon^\lok{C}(\symb{p}^{-1}(\varphi))\in\lok{C}_{max}(I).    
  \end{equation}
 \item For all elements $\tilde{g}$ of the universal covering group of
  global conformal transformations, $\PSL(2,\dopp{R})^\sim$, holds true:
\begin{equation}
    \label{eq:setbosug}
    \begin{array}[b]{ccccccccc}
    \Upsilon^\lok{B}(\symb{p}(\tilde{g}))&\!=\!& U(\symb{p}(\tilde{g})) &\!\!, \,&
    \Upsilon^\lok{A}(\tilde{g}) &\!=\!&  U^\lok{A}(\tilde{g}) & , \,\\
    &&&&\Upsilon^\lok{C}(\tilde{g}) &\!=\!& U^{\lok{A}'}(\tilde{g})&\!=\!&
    U^{\lok{C}_{max}}(\tilde{g}) \zendot  
    \end{array}
  \end{equation}
  \end{itemize}
\end{theo}

\begin{pf}
The representations  of current algebras induced by embeddings
$\iota(\lie{h})\subset\lie{g}$, $\lie{h}$, $\lie{g}$ reductive colour
algebras, are known to be completely reducible into irreducible
highest-weight representations of the current algebra associated with 
$\iota(\lie{h})$; this yields branchings of the vacuum representation 
space $\Hilb{H}$ of the current 
algebra associated with $\lie{g}$ of the form \cite[\S
12.12]{vKbook} \cite[\S 4.9]{KP84} \cite{KW88}:
\begin{equation}
  \label{eq:branchcurralg}
  \Hilb{H} = \bigoplus_{\Lambda}
  \Hilb{H}_{(\Lambda)}\otimes\Hilb{H}_\Lambda \zendot
\end{equation}
Here $\Hilb{H}_{(\Lambda)}$ is the multiplicity space of the
  highest-weight representation $\pi_\Lambda$ living on
  $\Hilb{H}_\Lambda$; the corresponding representations are of
  the form $\Einsop_{(\Lambda)}\otimes\pi_\Lambda(.)$. 

The \Name{Hilbert} space $\Hilb{H}$ is completely reducible into
irreducible highest-weight representations with respect to the action
of the \Name{Sugawara} stress-energy tensor $\Theta^\lok{B}$ of
$\lok{B}$ and the representation spaces 
$\Hilb{H}_\Lambda$ are completely reducible into  highest-weight
representations with respect to the stress-energy tensor
$\Theta^\lok{A}$ of $\lok{A}$
\cite[\S 11.12]{vKbook}. These representations exponentiate to
projective representations of $\Diff_+(\Seins)^\sim$
\cite{GW85,vL99a}. In the latter reference it is 
  shown that the cocycles of the irreducible, exponentiated
  highest-weight representations stemming from $\Theta^\lok{B}$ all
  coincide. Therefore, we can take their direct sum which defines a
  projective representation  of $\Diff_+(\Seins)^\sim$; we denote this
  representation by $\Upsilon^\lok{B}$. Mutatis mutandis, the same is
  true for the direct  sum of irreducible, projective representations of
  $\Diff_+(\Seins)^\sim$  stemming  from $\Theta^\lok{A}$, which we
  write as $\Upsilon^\lok{A}$. 

Because the spectrum of the
  conformal {Hamilton}ian, the zeroth mode of $\Theta^\lok{B}$,
  is integer, 
  it is clear that the kernel of $\symb{p}$, the cyclic group
  generated by $\tilde{R}(2\pi)$, is represented trivially in
  $\Upsilon^\lok{B}$. Thus,  $\Upsilon^\lok{B}$ is  a projective 
  representation of $\Diff_+(\Seins)$.    
%%%%%%%%%%%%%%%%%%%%%%%%

The exponentiated self-adjoint integrals of the stress-energy tensor
represent the one-parameter groups in $\Upsilon^\lok{B}$ and
$\Upsilon^\lok{A}$, respectively \cite[proposition
I.1.13]{tL94}. \Name{Haag} duality 
of $\lok{B}$ and linear energy bounds (theorem \ref{th:chcurrchlok},
\cite{BS90}) imply  that 
localised one-parameter groups of diffeomorphisms are represented in
$\Upsilon^\lok{B}$, $\Upsilon^\lok{A}$ by
local operators. Moreover, $\Upsilon^\lok{A}$ commutes with the cyclic
projection $e_\lok{A}$ onto $\overline{\lok{A}\Omega}$ because of the
\Name{Sugawara} construction; modular
covariance of the subnet $\lok{A}\subset\lok{B}$ implies that
localised one-parameter groups of diffeomorphisms are represented in
$\Upsilon^\lok{A}$ by local observables in $\lok{A}$. According to
\cite[proposition V.2.1]{tL94}, products of one-parameter groups in
$\Diff_I(\Seins)$ are dense in $\Diff_I(\Seins)$, which proves the
statement in (\ref{eq:intsuginner}) for $\Upsilon^\lok{B}$ and
$\Upsilon^\lok{A}$.

 The Hermitian linear combinations of
the modes $L^{\Theta^\lok{A}}_{0,\pm 1}$ of $\Theta^\lok{A}$ 
satisfy linear energy bounds by arguments as in \cite{BS90} and
integrate to a representation of $\PSL(2,\dopp{R})^\sim$ \cite{jF77},
which we know to be globally inner in $\lok{A}$
 and which has the same
infinitesimal action as $U^\lok{A}$ on the currents in
$\lok{A}$. Denoting the corresponding linear combinations of
generators of $U^\lok{A}$ by $L^\lok{A}_{0,\pm 1}$, we recognise that
$L_i^\lok{A}-L^{\Theta^\lok{A}}_i$, $i=0, \pm 1$, defines an abelian
and hence trivial representation of $\lie{s}\lie{l}(2)$.  
This yields the statement on $\Upsilon^\lok{A}$ in
(\ref{eq:setbosug}). By the same  
line of argument we see that $U$ and $\Upsilon^\lok{B}$ coincide for
global conformal transformations.  

The \Name{Coset} {Hamilton}ian, $L_0-L_0^\lok{A}$, defines an
energy grading on the multiplicity spaces $\Hilb{H}_{(\Lambda)}$ with
finite-dimensional energy eigenspaces. Hence we know that the
$\Hilb{H}_{(\Lambda)}$ are completely reducible into highest-weight
representations with respect to the action of the \Name{Coset}
stress-energy tensor \cite[\S 11.12]{vKbook}. These representations
yield a projective, unitary representation $\Upsilon^\lok{C}$
of $\Diff_+(\Seins)^\infty$ (see arguments above on
$\Upsilon^\lok{A}$). In order to show the statement in
(\ref{eq:intsuginner}), it is sufficient to look at localised
one-parameter groups for which it is seen to hold immediately.  The
statement concerning $\Upsilon^\lok{C}$ in 
(\ref{eq:setbosug})  is obvious. The
identity $\Upsilon^\lok{B}\circ\symb{p}= \Upsilon^\lok{A}
\Upsilon^\lok{C}$ follows from the uniqueness of the exponentiation
and from $\Theta^\lok{B}= \Theta^\lok{A} +
(\Theta^\lok{B}-\Theta^\lok{A})$. 
\end{pf}

\noindent { Remarks:}  
Chiral current algebras are known to be strongly additive
\cite[corollary IV.1.3.3]{vL97}, \cite{HL82, BS90} which means that the
local algebras of 
the maximal  \Name{Coset} model $\lok{C}_{max}(I)$ associated with a
current subalgebra coincide with the
respective local relative commutants $\lok{C}_I$. If the \Name{Coset}
stress-energy 
tensor has central charge $c\leq 1$,  the \Name{Coset} model generated
by $ \Upsilon^\lok{C}$ is strongly additive \cite{KL02,fX03}. In this case,
the subnet $\lok{A}_{max}$ (see equation \ref{eq:Amax})
is automatically given by the local relative commutants of
$\lok{C}_{max}$ (lemma \ref{lem:cosmax}). 

 If the \Name{Coset} central charge is greater than $1$, the subnet
generated  by $\Upsilon^\lok{C}$ is not strongly additive
\cite{BS90}. The discussion of the following chapter applies directly to all
\Name{Coset} models of current subalgebras containing
$\Upsilon^\lok{C}$, but  the arguments indicate that our analysis
probably admits 
extensions to subnets which do not possess a stress-energy tensor at
all (cf discussion in chapter \ref{cha:disc}). 
 
%  \Name{Carpi} has shown that the \Name{Coset} model
%  generated by $\Upsilon^\lok{C}$ is minimal, ie it has no non-trivial chiral
%  subnet \cite{sC98}.

%%%%%%%%%%%%%%%%%%%%%%%%%%%%%%%%%%%%%%%%%%%%%%%%%%%%%%%%%%%%%%%%%%
%%%%%%%%%%%% Conformal covariance subalgebras %%%%%%%%%%%%%%%%%%%%
%%%%%%%%%%%%%%%%%%%%%%%%%%%%%%%%%%%%%%%%%%%%%%%%%%%%%%%%%%%%%%%%%%

\sekt{Conformal covariance subalgebras}{Conformal covariance
  subalgebras}{sec:cocosub}
%\section{Conformal covariance subalgebras}
%\label{sec:cocosub}

Conformal inclusions of chiral current algebras are of interest for a
large variety of reasons. Their classification was undertaken some time
ago, because they are particularly relevant to string theory: they
make string compactification possible without altering  
conformal covariance. Using general
arguments this task was transferred to checking maximal inclusions of reductive
\Name{Lie} algebras in simple \Name{Lie} algebras, for which a
classification was available already, mainly due to the work of
\Name{Dynkin} \cite{eD57a,eD57b}. The classification of conformal
inclusions was thus achieved, looking at the central charge of the
respective stress-energy tensors, by several authors \cite{AGO87,
  BB87, SW86}.

Many of the conformal inclusions were found to correspond to
{\em symmetric spaces} (cf \cite{GNO85, cD96} in particular), and
{\em isotropy irreducibility} of the coset space proved a useful yet
neither necessary nor 
sufficient criterion for an inclusion being conformal. We undertake a complete
characterisation of conformal inclusions by means of straightforward
arguments familiar in (axiomatic) quantum field theory. On the course
we prove a longstanding\footnote{To the author's surprise, there
  does not seem to be a proof available yet.}
conjecture of \Name{Schellekens and Warner} 
\cite{SW86}. 

We use properties of any  \Name{Wightman} quantum field
theory: positivity of energy, separating property of
the vacuum for local quantum fields, and unitarity. 
Our analysis clarifies the situation in natural
group theoretical terms and in direct correspondence to quantum field
theoretical notions. In addition, there is no need to specialise in
maximal subalgebras and our approach is rather direct in that respect.

The methods applied here arise  from a more general question:
how does the inner-implementing representation $U^\lok{A}$, uniquely
associated with every 
covariant subtheory $\lok{A}$ of a chiral conformal theory
$\lok{B}$ by means of the \Name{Borchers-Sugawara} construction
 (section \ref{cha:cospa}.\ref{sec:bosug}), act on the
observables of the ambient theory $\lok{B}$? While detailed knowledge
of the action of the \Name{Sugawara} stress-energy tensor of a current
subalgebra on the
currents of the larger current algebra (equation \ref{geneitcomm}) does not directly lead to an
understanding of the geometric character of this action, it is
helpful for characterising the currents on which this action
implements conformal covariance. The geometrical impact is
resolved in chapter \ref{cha:netend} in a broader setting.

We proceed as follows: In the following subsection we introduce notations
and conventions, prove the conjecture of \Name{Schellekens and Warner}
and provide a direct argument for conformal inclusions being
necessarily restricted to level $1$. The second subsection is about
studying {\em conformal covariance subalgebras} associated to
\Name{Lie} algebra inclusions, these being intermediate to the
original inclusion, if not trivial. The section will be closed by a
simple characterisation of {\em {Coset} currents}, ie current
subalgebras commuting with  the given current subalgebra. The contents
of this section are available as \cite{sK03a}.

%%%%%%%%%%%%%%%%%%%%%%%%%%%%%%%%%%%%%%%%%%%%%%%%%%%%%%%%%%%%%%%
%%%% Characterisation of conformal inclusions %%%%%%%%%%%%%%%%%
%%%%%%%%%%%%%%%%%%%%%%%%%%%%%%%%%%%%%%%%%%%%%%%%%%%%%%%%%%%%%%%

\subsection{Characterisation of conformal inclusions}
\label{sec:char}

We study a current algebra with colours in a  simple\footnote{General
  reasoning leads to an extension of the following discussion to 
inclusions of reductive subalgebras in reductive \Name{Lie}
algebras, cf eg \cite{AGO87}.}, compact \Name{Lie}
algebra $\lie{g}$ as quantum fields on the chiral light-ray. 
Basis elements of $\lie{g}$ will
be denoted by $T^a$; they give the colour of the corresponding
current $j^a$. The current algebra is given by the following
commutation relations: 
\begin{displaymath}
  \komm{j^a(x)}{j^b(y)} = i f^{ab}{}_c j^c(x) \delta(x-y) + k
  g^{ab}_\lie{g} \frac{i}{2\pi} \delta' (x-y) \zendot
\end{displaymath}
$g_\lie{g}$ denotes the \Name{Killing} metric of $\lie{g}$,
$f^{ab}{}_c$ its structure constants and $k$ the current algebra's
level; $k$ is a positive integer.

By embedding a reductive \Name{Lie} subalgebra $\lie{h}$ into
$\lie{g}$ via an injective homomorphism $\iota: \lie{h}
\hookrightarrow \lie{g}$ we have an associated current
subalgebra. $\lie{h}$ consists of 
several simple ideals, denoted (for the time being) by
$\lie{h}_\alpha$, and an abelian 
ideal of dimension $n\geq 0$. The inclusions $\iota(\lie{h}_\alpha)
\subset \lie{g}$ are partly characterised by their
{\em {Dynkin} index} $I_\alpha$, which is defined through the relation
$I_\alpha g^{ab}_{\alpha} =
g^{\iota(a)\iota(b)}_\lie{g}$ between the \Name{Killing} metric of
$\lie{h}_\alpha$, denoted $g_\alpha$, and the restriction of the
\Name{Killing} metric of 
$\lie{g}$ to $\iota(\lie{h}_\alpha)$. The commutation relations
(\ref{eq:quacurcom}) take the following form for currents associated
with colours in $\iota(\lie{h}_\alpha)$: 
\begin{displaymath}
  \komm{j^{\iota(a)}(x)}{j^{\iota(b)}(y)} = i f^{\iota(a)\iota(b)}{}_{\iota(c)}
    j^{\iota(c)}(x) \delta (x-y) + I_\alpha k g^{ab}_\alpha
    \frac{i}{2\pi} \delta' (x-y) \zendot 
\end{displaymath}

%%%%%%%%%%%%%%%%%%%%%%%%%%%%%%%%%%%%%%%%%%%%%%%%%%%%%%%%%%%%

The infinitesimal conformal transformations are implemented by the
adjoint action of the \Name{Sugawara} stress-energy tensor
$\Theta^\lie{g}$, see equation (\ref{eq:defsugset}). %It is given by:
% \begin{displaymath}
%    \Theta^\lie{g} (x) = \frac{\pi}{k+\symb{g}^\vee_\lie{g}}
%   g^\lie{g}_{ab} :j^{a} j^{b}:(x)
% \end{displaymath}
% $\symb{g}^\vee_\lie{g}$ is the dual \Name{Coxeter} number of $\lie{g}$. The
% commutation relation of $\Theta^\lie{g}$ with a current reads as:
% \begin{displaymath}
%   \komm{\Theta^\lie{g}(x)}{j^c(y)} = i j^c(x) \delta'(x-y)
% \end{displaymath}
Restricting to colours in $\iota(\lie{h}_\alpha)$ the \Name{Sugawara}
stress-energy 
tensor $\Theta^\alpha$ has the same commutation relations with
currents associated with colours in $\iota(\lie{h}_\alpha)$ as
$\Theta^\lie{g}$: 
\begin{displaymath}
    \Theta^\alpha (x) =\frac{\pi}{I_\alpha k+\symb{g}^\vee_\alpha}
  \,\, g^\alpha_{ab} :j^{\iota(a)} j^{\iota(b)}:(x) \zendot
\end{displaymath}

For the abelian ideal we adopt the following conventions:
$I_{\dopp{R}^n}:=1$, $\symb{g}^\vee_{\dopp{R}^n} := 0$, $g_{\dopp{R}^n}^{ij} :=
g^{\iota(i)\iota(j)}_{\lie{g}}$. Using these as input all the formulas
above apply to currents associated with colours in
$\iota(\dopp{R}^n)$. We shall, 
therefore, drop the distinction between simple and abelian ideals of $\lie{h}$
and use the symbol $\lie{h}_\alpha$ for any simple or the abelian
ideal from now on.

With this general notation the action of a stress-energy tensor
$\Theta^\alpha$ on an arbitrary current $j^c$ reads:
\begin{eqnarray}
\komm{\Theta^\alpha(x)}{j^c(y)}\nonumber
&=&\frac{\pi}{I_\alpha k+ \symb{g}^\vee_\alpha}
\,  g^\alpha_{ab} \, i f^{\iota(b)c}{}_d
:j^{\iota(a)}j^d+j^dj^{\iota(a)}:(x)\delta(x-y)\nonumber\\
&&+i\frac{k}{I_\alpha k+ \symb{g}^\vee_\alpha}  \, j^{\iota(a)}(x)
g^\alpha_{ab} g_\lie{g}^{\iota(b)c} \,\, \delta'(x-y)
\nonumber\\&&
+i \frac{1}{2(I_\alpha k+ \symb{g}^\vee_\alpha)} \,  j^d(x) 
\klammer{C_2^{\alpha}}_d{}^c \,\, \delta'(x-y) \zendot\label{geneitcomm}
\end{eqnarray}
 This equation is obtained by applying the current algebra
 and the normal ordering prescription for currents (see equation
 \ref{eq:defnormordcurr})% (cf treatment of
                         % \Name{Sugawara}-construction in
                         % \cite{khR98}) 
. The
 matrix $C_2^{\alpha}$ stands for the second \Name{Casimir} element of
 $\lie{h}_\alpha$ in the representation $Ad_\lie{g}\circ
 \iota\restriction{}{\lie{h}_\alpha}$, if $\lie{h}_\alpha$ is a simple ideal. In
 any case we have:
\begin{displaymath}
  \klammer{C_2^{\alpha}}_d{}^c = g^\alpha_{ab}
  \,\, if^{\iota(b)e}{}_d \,\, if^{\iota(a)c}{}_e = g^\alpha_{ab} \klammer{
  Ad_{T^{\iota(b)}} Ad_{T^{\iota(a)}} }_d{}^c \zendot
\end{displaymath}
Taking the trace of this matrix one may readily see that it does
not vanish for the abelian ideal.

Now we are prepared to state and prove our main
result. \Name{Schellekens and Warner} conjectured it in their
discussion closing \cite{SW86}.
\begin{theo}\label{schwacon}
  The following holds true for the {\bf weighted {Casimir} element}
  $\widetilde{C}_2^{\iota(\lie{h})}$ 
  of $\iota(\lie{h})$ ($P_\alpha$ stands for the projection onto
  $\iota(\lie{h}_\alpha)$): 
  \begin{equation}\label{schwainq}
    \widetilde{C}_2^{\iota(\lie{h})} := \sum_\alpha \frac{2 I_\alpha k P_\alpha + C_2^{\alpha}}{2(I_\alpha
      k+ \symb{g}^\vee_\alpha)} \leqslant \Einsop \zendot
  \end{equation}

This inequality is saturated if and only if
$\iota(\lie{h})\subset\lie{g}$ yields a 
conformal inclusion, ie $\sum_\alpha \Theta^\alpha =:
\Theta^{\iota(\lie{h})} = \Theta^\lie{g}$. 
\end{theo}

\begin{pf}
By invariance of $g^\lie{g}$ the orthocomplementation $\lie{g}=
\iota(\lie{h}) + \iota(\lie{h})^\perp$ provides a reduction of the
representation $Ad_\lie{g}\circ\iota$. We
have  $C^\alpha_2\restriction{}{\iota(\lie{h})} = 2 \symb{g}^\vee_\alpha P_\alpha$,
ie $\widetilde{C}_2^{\iota(\lie{h})}\restriction{}{\iota(\lie{h})}=\Einsop$, and
the inequality only remains to be proven for colours orthogonal to
$\iota(\lie{h})$, where $P_\alpha\restriction{}{\iota(\lie{h})^\perp}=0$. Because
all \Name{Casimir} elements commute and all are 
positive operators, we assume as well that $T^c$ is a common
eigenvector for all linear mappings $C^\alpha_2$.

We prove the inequality by looking at specific expectation values of the
\Name{Coset}  {Hamilton}ian 
$L_0^\lie{g}-L_0^\lie{h}$. This is a positive operator, which is given
by the \Name{Coset} stress-energy tensor
$\Theta^\lie{g}-\Theta^{\iota(\lie{h})}$ smeared with the test function
$\xi_{L_0}(x)= \frac{1}{2}(x^2+1)$. The infinitesimal action of a
conformal {Hamilton} operator on the test function of a smeared
field covariant with respect 
to it shall be abbreviated by $l_0$, ie we have
$$\komm{L_0^\lie{g}}{j^c(g)} = i \int dx g'(x) \xi_{L_0}(x) j^c(x)\equiv i \int
dx (l_0 g) (x)j^c(x) = i\, j^c(l_0 g) \zendot$$

Using the general commutation relation (\ref{geneitcomm}), calculating
two and three point functions of currents (cf \cite{FST89}), observing
that some group-theoretical tensors involved are null for reasons of
permutation symmetry/ antisymmetry and carefully taking into
account the normal ordering of currents    \cite{FST89} one arrives at
the following formula:
\begin{eqnarray}
  0 &\leqslant& \skalar{\Omega}{j^c(g)^\dagger (L_0^\lie{g}-L_0^\lie{h})
    j^c(g)\Omega}\nonumber\\
&=&i \klammer{1- \sum_\alpha \frac{C^\alpha_2[T^c]}{2(I_\alpha k+
    \symb{g}^\vee_\alpha)}} \skalar{\Omega}{j^c(g)^\dagger
    j^c(l_0 g)\Omega}\label{star} \zendot
\end{eqnarray}
The desired inequality may be established through
  division by $i\langle\Omega,j^c(g)^\dagger
    j^c(l_0 g)\Omega\rangle$, which does not vanish for generic $g$ and is
  positive as an expectation value of $L_0\geq 0$. 

If we have
$\Theta^\lie{g}=\Theta^{\iota(\lie{h})}$,  (\ref{schwainq}) is
saturated on $\iota(\lie{h})$ trivially ($C_2^\alpha P_\beta =
2\symb{g}^\vee_\alpha \delta^\alpha{}_\beta$) and because of
(\ref{star}) on 
$\iota(\lie{h})^\perp$ as well, hence on all of $\lie{g}$. The
conclusion in the opposite direction  
is, actually, a consequence of equation (\ref{eq:geneittwo}) in
proposition \ref{crueqns}% (cf eg \cite{khR98})
: This leads to trivial commutation relations for
$\Theta^\lie{g}-\Theta^{\iota(\lie{h})}$, especially to $c_\lie{g}=c_\lie{h}$,
which yields, by the
 \Name{Reeh-Schlieder} theorem% \footnote{See for example \cite{rJ65}(lemma 2,
%  section V.3.B); for an argument directly referring to the
%  \Name{Virasoro} algebra see \cite{GW85,fG86a}.}
,
 $\Theta^\lie{g}-\Theta^{\iota(\lie{h})} = 0$. 
\end{pf} % Fuer erweiterte Version
         % dieses auskommentieren.
% But we have an alternative argument as well. We
% assume the above inequality (\ref{schwainq}) to hold true as an equality on
% all of $\lie{g}$. Again
% we may extend a basis of $\iota(\lie{h})$ to a basis of $\lie{g}$ by
% adding colour elements which are common eigenvectors for all operators
% $C^\alpha_2$ and are othogonal to $\iota(\lie{h})$. For such a colour
% $T^c\in\iota(\lie{h})^\perp$ it is straightforward to calculate the
% following commutation relation for the associated current:
% \begin{displaymath}
%   \komm{\klammer{\Theta^\lie{g}-\Theta^{\iota(\lie{h})}}(x)}{j^c(y)} = -
%   \sum_\alpha \frac{\pi}{I_\alpha k+\symb{g}^\vee_\alpha} g^\alpha_{ab} i
%   f^{\iota(a)c}{}_d 
% :j^{\iota(a)}j^d+j^dj^{\iota(a)}(x)\delta(x-y)
% \end{displaymath}

% Next one has this to calculate the following vacuum expectation value,
% which is proportional to the central charge of the \Name{Coset} stress-energy
% tensor following the general commutation relation, but which is found
% to be null because of the symmetry of its group theoretical
% coefficients: 
% \begin{displaymath}
%   -i (c_\lie{g}-c_\lie{h}) \delta'''(x-y) =
%   \skalar{\Omega}{\komm{\klammer{\Theta^\lie{g}-\Theta^{\iota(\lie{h})}}(x)}{\Theta^\lie{g}(y)}\Omega} = 0
% \end{displaymath}
% By the the direct analysis of \cite{fG86a} or by a variant of the
% \Name{Reeh-Schlieder} theorem (eg. \cite[lemma 2, sect. V.3.B]{rJ65})
% we conclude that the \Name{Coset} stress-energy tensor has to vanish- its two
% point function is proportional to its central charge. 
%\end{pf}

\begin{cor}\label{kone}
  An embedding $\iota(\lie{h})\subset \lie{g}$ can give rise to a
  conformal inclusion of the associated current algebras only, if the
  current algebra associated with $\lie{g}$ has level $k=1$.
\end{cor}

\begin{pf}
Highest-weight representations of current algebras may be characterised
uniquely by a vector of lowest energy which is a highest-weight vector
with respect to the horizontal subalgebra. 
We look at the representation defined by the highest weight
$\psi_\lie{g}$ of the adjoint representation of $\lie{g}$. Since
$\psi_\lie{g}$ has, by the usual convention, length 2, this
representation is in accordance with the \Name{Weyl}-alcove condition
\cite[(4.51)]{FST89} for unitary representations of current algebras
for $k\geq 2$. The following argument applies, therefore, to all but
level $1$. 

Actually, we may restrict attention to the action of
$L_0^\lie{g}-L_0^\lie{h}$ on $\lie{g} \psi_\lie{g}$, the highest
weight module of $\lie{g}$ generated from the vector with lowest
energy and highest weight $\psi_\lie{g}$. Here we have (cf equation
\ref{eq:L0galpha}): 
\begin{displaymath}
  0 \leqslant (L_0^\lie{g}-L_0^\lie{h})\restriction{}{\lie{g}
    \psi_\lie{g}} = \frac{\symb{g}^\vee_\lie{g}}{k+\symb{g}^\vee_\lie{g}} \Einsop - \sum_\alpha
  \frac{C^\alpha_2}{2(I_\alpha k+ \symb{g}^\vee_\alpha)} \zendot
\end{displaymath}
This implies a strictly sharper bound than (\ref{schwainq}) and by theorem
\ref{schwacon} this immediately yields the desired result.
\end{pf}

%%%%%%%%%%%%%%%%%%%%%%%%%%%%%%%%%%%%%%%%%%%%%%
%%%% Conformal covariance subalgebras %%%%%%%%
%%%%%%%%%%%%%%%%%%%%%%%%%%%%%%%%%%%%%%%%%%%%%%

\subsection{Covariant and invariant colours}
\label{sec:cocoalg}

After we have given a characterisation of conformal inclusions
$\iota(\lie{h})\subset\lie{g}$, we now pursue further the structures in
colour space which are associated with the action of $\Theta^{\iota(\lie{h})}$
on currents with colours in $\lie{g}$. We find that {\em covariant} and
{\em invariant} colours form reductive \Name{Lie} algebras,
the first being intermediate to the original embedding
$\iota(\lie{h})\subset\lie{g}$, the second being orthogonal to and
commuting with it. All these results are in terms on the weighted
\Name{Casimir} element $\widetilde{C}_2^{\iota(\lie{h})}$ of the
\Name{Lie} algebra $\iota(\lie{h})$.

The following is the main ingredient of the results in this section:
\begin{prop} \label{crueqns}
For an arbitrary colour $T^c\in\lie{g}$ we have:
  \begin{eqnarray}
    \label{eq:geneittwo}
&&\norm{\komm{\klammer{\Theta^\lie{g}-\Theta^{\iota(\lie{h})}}(f)}{j^c(g)}\Omega}^2
\nonumber\\ 
&=&8k \pi^2 \skalar{(\Einsop - \widetilde{C}_2^\lie{h})
  T^c}{\widetilde{C}_2^\lie{h} T^c}{}_\lie{g}
\,\,\widetilde{\Delta}^4(\overline{f\cdot g}, f\cdot g) \nonumber\\
&& + 
k
\skalar{(\Einsop-\widetilde{C}^\lie{h}_2)T^c}{(\Einsop-\widetilde{C}^\lie{h}_2)T^c}_\lie{g} 
\,\,\widetilde{\Delta}^2 (\overline{f\cdot g'}, f\cdot g') \zendot 
  \end{eqnarray}
Here $\skalar{.}{.}_\lie{g}$ stands for the scalar product on
$\lie{g}$ induced by the \Name{Killing} form. We define: 
$$\Phi^c(x) := \sum_\alpha 
\frac{1}{2(I_\alpha k + \symb{g}^\vee_\alpha)} g^\alpha_{ab} f^{\iota(b)c}{}_d
:j^{\iota(a)}j^d+j^dj^{\iota(a)}:(x)$$
The two-point function of $\Phi^c$ is given by:
\begin{equation}
  \label{eq:covfietwo}
  \skalar{\Omega}{\Phi^c(x)\Phi^c(y)\Omega} = 2 k
  \widetilde{\Delta}^4 (x-y) \skalar{(\Einsop - \widetilde{C}_2^\lie{h})
  T^c}{\widetilde{C}_2^\lie{h} T^c}{}_\lie{g} \zendot
\end{equation}

The numerical distributions in these formulae are given by:
\begin{eqnarray*}
  %\label{eq:numdist}
  \widetilde{\Delta}^4(\overline{f\cdot g}, f\cdot g) &=&(2\pi)^{-4}
  \int\!\!\!\int dx \,dy \,\,\klammer{i[(x-y)-i\varepsilon]}^{-4}
  \overline{f\cdot g}(x)f\cdot g(y) \, ,\\
 \widetilde{\Delta}^2(\overline{f\cdot g'}, f\cdot g') &=&(2\pi)^{-2}
  \int\!\!\!\int dx \, dy\,\, \klammer{i[(x-y)-i\varepsilon]}^{-2}
  \overline{f\cdot g'}(x)f\cdot g'(y) \, .
\end{eqnarray*}
\end{prop}

\begin{pf}
  We will not give the derivation of these formulae in detail. We
  rather indicate their verification. First, one may restrict
  attention to colours $T^c\in\iota(\lie{h})^\perp$ since the weighted
  \Name{Casimir} respects the orthogonal decomposition
  $\lie{g}=\iota(\lie{h})\oplus \iota(\lie{h})^\perp$ with respect to
  $Ad\circ\iota$ and $\Theta^\lie{g}-\Theta^{\iota(\lie{h})}$ commutes with all
  currents whose colours are in $\iota(\lie{h})$. ``All'' that one has
  to do  is to   
  apply the general commutation relation (\ref{geneitcomm}) restricted
  to colours from $\iota(\lie{h})^\perp$, follow carefully the normal
  ordering of currents, observe symmetries of group theoretical
  coefficients, keep in mind $T^c\in\iota(\lie{h})^\perp$, calculate
  some $n$-point functions of currents 
  following the scheme in \cite{FST89}, use \Name{Jacobi}'s identity a
  few times and recognise the second \Name{Casimir} element in the
  adjoint representation, which amounts to twice the dual
  \Name{Coxeter} number. With all that, it is a straightforward
  algebraic exercise.
\end{pf}

Taking  $g$ as the test function of
constant value $1$, equation (\ref{eq:geneittwo}) implies
$\widetilde{C}_2^\lie{h} (\Einsop -\widetilde{C}_2^\lie{h})\geq 0$,
from which we immediately get inequality (\ref{schwainq}), and the second
statement in  theorem \ref{schwacon} follows from
(\ref{eq:geneittwo}), too.

\begin{defi}
  A current $j^c$ is said to {\bf transform covariantly}  with respect to
  $\Theta^{\iota(\lie{h})}$, if and only if $\komm{\Theta^\lie{g}(f)}{j^c(g)} =
  \komm{\Theta^{\iota(\lie{h})}(f)}{j^c(g)}$ $\forall f,g$.
\end{defi}
The proof of the following corollary shows that, in fact, it is
sufficient to require the equality to hold for a few, but sufficiently
many test functions: two suitable pairs $(f_i,g_i)$, $i=1,2$, are enough. 

\begin{cor}\label{confcoval}
  A current $j^c$ transforms covariantly  with respect to
  $\Theta^{\iota(\lie{h})}$, if and only if its
  colour fulfills: 
  $\widetilde{C}_2^{\iota(\lie{h})} T^c = T^c$. These {\bf covariant
  colours} form a reductive \Name{Lie}
  algebra, $\lie{k}$, containing $\iota(\lie{h})$ as a subalgebra. The
  currents with colours in $\lie{k}$ form the {\bf conformal
  covariance subalgebra}. If
  $\lie{k}\neq \iota(\lie{h})$, then the level of the current algebra
  associated with $\lie{g}$ has to be $k=1$.  
\end{cor}

\begin{pf}
If we have $\widetilde{C}_2^{\iota(\lie{h})} T^c = T^c$, we know from
the variant of the \Name{Reeh-Schlieder} theorem (eg \cite[lemma 2,
section V.3.B]{rJ65}) and proposition \ref{crueqns} above, that $j^c$ and
the \Name{Coset} stress-energy tensor commute. This is another way of saying:
$j^c$ transforms covariantly with respect to 
  $\Theta^{\iota(\lie{h})}$. 

Conversely: If $j^c$ is covariant with respect to
$\Theta^{\iota(\lie{h})}$, the group theoretical scalar products  in
equation (\ref{eq:geneittwo}) have to be zero, since the numerical
distributions involved are linearly independent. The second one of
these is the norm of $(\Einsop -\widetilde{C}_2^{\iota(\lie{h})})T^c$,
which makes the equation $\widetilde{C}_2^{\iota(\lie{h})} T^c =T^c$
valid.

Now, if $T^a$ and $T^b$ are covariant colours, then so is $-i
\komm{T^a}{T^b}$. This becomes  clear, if one observes
$f^{ab}{}_c j^c(g) = -i \komm{j^a([1])}{j^b(g)}$, where $[1]$ is a
constant test function: $[1](x) = 1$. 

The reductivity of $\lie{k}$ is not difficult to prove, either. 
%Because 
$\lie{k}$
%is an eigenspace of a symmetric operator, namely of
%$\widetilde{C}_2^{\iota(\lie{h})}$, it 
is a subspace
of $\lie{g}$, endowed with an invariant scalar product, which is given by the
restriction of the \Name{Killing} form on $\lie{g}$. By invariance of
this scalar product on $\lie{k}$ with respect to $Ad_\lie{k}$ (this
being a mere restriction of invariance under $Ad_\lie{g}$) any
invariant subspace of $\lie{k}$ has an invariant orthogonal
complement. Now this is complete reducibility of $Ad_\lie{k}$ and by
\cite[25.3.a]{jC89} $\lie{k}$ is reductive.

Since one can reduce the problem of understanding all conformal
inclusions to the studies of reductive inclusions in simple \Name{Lie}
algebras (cf eg \cite{AGO87}) the last part of the claim follows
immediately from 
corollary  
\ref{kone}, as $\iota(\lie{h})\subset \lie{k}$ {is}, by construction of
$\lie{k}$, a conformal inclusion and the \Name{Dynkin} indices of the
simple ideals in $\lie{k}$ are greater than or equal to $1$.
\end{pf}

\begin{cor}
  A current $j^c$, whose colour $T^c$ lies in $\iota(\lie{h})^\perp$
  and fulfills $\widetilde{C}_2^{\iota(\lie{h})} T^c = 0$, commutes
  with the entire current algebra associated with
  $\iota(\lie{h})$. These colours form a reductive \Name{Lie} algebra,
  the algebra of {\bf  invariant colours}; we call their
  current algebra {\bf {Coset} current algebra}.  
\end{cor}

\begin{pf}
  % We know from proposition \ref{crueqns} that such a $j^c$ commutes
%   with $\Theta^{\iota(\lie{h})}$. It is possible to give a rigorous proof of
%   the commutativity of $j^c$ (in the sense of essentially self adjoint
%   operators) with $\iota(\lie{h})$'s current algebra
%   on grounds of covariance and locality (cf \cite[lemma 11]{sK02}),
%   but 
 We set
  $V_0:=  ker(\widetilde{C}_2^{\iota(\lie{h})}) \cap
  \iota(\lie{h})^\perp$. $V_0$  is an invariant  subspace with respect
  to  the action of $\lie{h}$ on $\lie{g}$ via  
$Ad_\lie{g}\circ\iota$. In fact, it is the representation 
space for the trivial subrepresentation on $\iota(\lie{h})^\perp$: For
any simple ideal 
$\lie{h}_\alpha$ we have by complete reducibility $C_2^\alpha\restriction{}{V_0} =
\sum_\Lambda \skalar{\Lambda+ 2 \rho}{\Lambda}_{\lie{h}_\alpha} =
0$. Since both the \Name{Weyl} vector $\rho$ and the contributing highest
weight vectors $\Lambda$ are dominant, we have $\Lambda = 0$. For the
abelian ideal the irreducible subrepresentations on $V_0$ are given by
common eigenvectors, such that $C_2^{\dopp{R}^n} v =
g^\lie{g}_{\iota(i)\iota(j)} \lambda^i\lambda^j v = 0$. This gives the
same result. This means, that all of $V_0$ commutes with
$\iota(\lie{h})$, ie $V_0\subset \iota(\lie{h})'$. We gain directly:
$V_0 = \iota(\lie{h})' \cap \iota(\lie{h})^\perp$.

By \Name{Jacobi}'s identity and invariance of the \Name{Killing}
metric, $\iota(\lie{h})' \cap\iota(\lie{h})^\perp$ forms a \Name{Lie}
subalgebra of $\lie{g}$. This is
reductive by the same argument as in the proof to corollary
\ref{confcoval}. 
\end{pf}

 Generically, the \Name{Coset} theory is not
generated by \Name{Coset} currents, although there are examples of
this structure \cite{KS88}. Obviously $\iota(\lie{h})\oplus(\iota(\lie{h})'
\cap \iota(\lie{h})^\perp)\subset \lie{g}$ has to be a conformal
inclusion for that to be the case, since the \Name{Coset} stress-energy
tensor has to be the \Name{Sugawara} stress-energy tensor of the
current algebra associated with $\iota(\lie{h})'
\cap \iota(\lie{h})^\perp$. \Name{Casimir} elements of $\iota(\lie{h})'
\cap \iota(\lie{h})^\perp$ give, when transferred to the corresponding
horizontal subalgebra,  charge operators of the \Name{Coset}
theory. These will, in general, fail to separate the representations
of the \Name{Coset} theory. The same goes for the \Name{Cartan} subalgebra of
$\iota(\lie{h})'\cap \iota(\lie{h})^\perp$, whose spectrum defines
characters of the representations of the \Name{Coset} theory.  The
\Name{Coset} current algebra is trivial for all inclusions with {\em minimal} \Name{Coset}
theory: Here the \Name{Coset} theory is generated by the \Name{Coset} stress-energy
tensor and this theory contains nothing but this field
\cite{sC98}.  Triviality of \Name{Coset} current algebra ought to be regarded
as the generic situation. 

Currents $j^c$ leading to vanishing $\Phi^c$ are linear
combinations of covariant and \Name{Coset} currents (proposition \ref{crueqns}). This is obvious, since
a decomposition of $T^c$ into eigenvectors
of $\widetilde{C}_2^\lie{h}$ with distinct eigenvalues $\lambda$
yields: $$\skalar{(\Einsop - \widetilde{C}_2^\lie{h}) 
  T^c}{\widetilde{C}_2^\lie{h} T^c}{}_\lie{g} = \sum_\lambda \lambda
(1-\lambda) \skalar{
  T^c_\lambda}{T^c_\lambda}{}_\lie{g} \zendot$$
 As $0\leq \lambda \leq 1$ this scalar 
product vanishes, if and only if just $0$ and $1$
contribute. This means, that there are no currents with a ``simple''
intermediate transformation behaviour with respect to the action of
$\Theta^{\iota(\lie{h})}$. Typically, a current $j^c$ has $\Phi^c\neq 0$, ie
a ``complicated'' transformation behaviour. By the analysis in
chapter \ref{cha:netend} this behaviour will be seen to be physically
satisfactory, still.   

%%%%%%%%%%%%%%%%%%%%%%%%%%%%%%%%%%%%%%%%%%%%%%%%%%%%%%%%%%
%% ENDE KAPITEL Stromunteralgebren %%%%%%%%%%%%%%%%%%%%%%%
%%%%%%%%%%%%%%%%%%%%%%%%%%%%%%%%%%%%%%%%%%%%%%%%%%%%%%%%%%

%\input{NetEnd.tex}
%%%%%%%%%%%%%%%%%%%%%%%%%%%%%%%%%%%%%%%%%%%%%%%%%%%%%%%%%%
%% KAPITEL Netzendomorphismen %%%%%%%%%%%%%%%%%%%%%%%%%%%%
%%%%%%%%%%%%%%%%%%%%%%%%%%%%%%%%%%%%%%%%%%%%%%%%%%%%%%%%%%

\chap{Local Nature of {Coset} Models}{Local Nature of
\Name{Coset} Models}{cha:netend}
%\chapter{Net-endomorphism property \& Chiral holography}
%\label{cha:netend}

In this chapter we discuss the action of the inner-implementing
representation $U^\lok{A}$ of a chiral subnet
$\lok{A}\subset\lok{B}$ on general local observables in $\lok{B}$. The
first section studies the ``geometric 
impact'' of $U^\lok{A}$, ie the localisation of
$Ad_{U^\lok{A}(\tilde{g})}\lok{B}(I)$ depending on $I\Subset\Seins$
and $\tilde{g}\in\PSL(2,\dopp{R})^\sim$.  Intuitively, we do not expect
an observable 
of $\lok{B}$ to be more sensitive to the action of $Ad_{U^\lok{A}}$
than to that of $Ad_{U}$: the generator of translations, $P$, is known
to decompose into two commuting positive parts,
$P=P^\lok{A}+P^{\lok{A}'}$, and regarding them as chiral analogues of
{Hamilton}ians leads us to the expectation that $P^\lok{A}$
should not transport observables of $\lok{B}$ ``faster'' than $P$
itself. A typical observable $B$ in $\lok{B}$ should exhibit a
behaviour interpolating between invariance ($B$ in $\lok{C}_{max}$)
and covariance ($B$ in $\lok{A}_{max}$). 

For this behaviour to be ensured we have, as it turns out, only to
show that scale transformations represented through $U^\lok{A}$
respect the two fixed points of scale transformations, namely $0$ and
$\infty$, when acting on $\lok{B}$. We can prove this to be the case
in presence of a stress-energy tensor and it seems natural in any
case. The sub-geometrical
transformation behaviour for translations, which we expect, then
follows by results of \Name{Borchers} \cite{hjB97,hjB97a} using the
spectrum condition and modular theory. We collected, rearranged and
reformulated 
results of \Name{Borchers} and \Name{Wiesbrock} in order to provide a
natural {\em converse of {Borchers}' theorem on half-sided
translations}, which was not yet available in the literature. By
extending the analysis to general conformal transformations we arrive
at the notion 
of {\em net-endomorphism property} for the action of $U^\lok{A}$ on
$\lok{B}$.

In the second section we use the net-endomorphic action of $U^\lok{A}$
to construct from the chiral conformal theory $\lok{B}$ a conformal
net in $\opo$ dimensions which contains the chiral algebras as
{\em time-zero algebras}. The result satisfies all axioms of a
$\opo$-dimensional conformal quantum theory, except that not
translations in 
futurelike directions have positive spectrum but rather translations
in right spacelike directions. While this prohibits interpreting the
picture of {\em chiral holography}
as genuinely  physical, it provides a helpful
geometrical framework of a {\em quasi-theory} in $\opo$ dimensions. The
maximal \Name{Coset} 
model appears as a subtheory of {\em chiral observables}
and hence we make contact with results of \Name{Rehren}
\cite{khR00}, which have interesting consequences for known examples.

In the last section we provide our solution to the isotony problem
(main theorem \ref{th:main}),
ie we establish the local nature of the maximal \Name{Coset}
model. We start by giving a new characterisation of $\lok{C}_{max}$
making use of the particular structure of the group of chiral
conformal transformations. And then, again, the presence of a
stress-energy tensor for $\lok{A}$ is  
only needed in order to establish a rather natural, but crucial lemma
on the representation of scale transformations through
$U^\lok{A}$.  Most of this chapter is available as \cite{sK03c}.

%%%%%%%%%%%%%%%%%%%%%%%%%%%%%%%%%%%%%%%%%%%%%%%%%%%%%%%%%%%
%%%%%%% NET-ENDOMORPHISM PROPERTY %%%%%%%%%%%%%%%%%%%%%%%%%
%%%%%%%%%%%%%%%%%%%%%%%%%%%%%%%%%%%%%%%%%%%%%%%%%%%%%%%%%%%

\sekt{Net-endomorphism Property}{Net-endomorphism
Property}{sec:netend}

In the following we deduce, step by step, the sub-geometric character
of the adjoint action of $U^\lok{A}$ (and of $U^{\lok{A}'}$) on
$\lok{B}$. The analysis relies 
 on a single property of the dilatations in $U^\lok{A}$. The notion of
 {\em net-endomorphisms} arises naturally in the course of the argument
 and will be discussed at the end of this section. We, therefore,
 define:

\begin{defi}\label{def:netend}
  $U^\lok{A}$ is said to have the {\bf net-endomorphism property}, if
  the adjoint action of ${U^\lok{A}(\tilde{D}(t))}$, $t\in\dopp{R}$,
  defines a group of automorphisms of  $\lok{B}(\Seins_+)$. 
\end{defi}

The net-endomorphism  property holds making the following

\noindent{\bf Additional Assumption:}\label{ass}{\em
  There is a unitary, strongly continuous, projective representation
  $\Upsilon^\lok{A}$ of the universal covering group of orientation preserving
  diffeomorphisms of the circle, $\Diff_+(\Seins)^\sim$, on $\Hilb{H}$ such
  that:
  \begin{itemize}
  \item If a diffeomorphism $\varphi\in\Diff_+(\Seins)$ is localised in
    $I\Subset\Seins$, ie
    $\varphi\restriction{}{I'}=id\restriction{}{I'}$, it is
    represented 
    by a local observable of $\lok{A}$, namely:
    $\Upsilon^\lok{A}(\symb{p}^{-1}(\varphi))\in\lok{A}(I)$.
  \item
    $\Upsilon^\lok{A}(\tilde{D}(t))U^\lok{A}(\tilde{D}(t))^*
    \in\dopp{C}\Einsop$  
    for all $t\in\dopp{R}$. 
  \end{itemize}
} %end of \em
\noindent Here, the covering projection from
$\Diff_+(\Seins)^\sim$ onto $\Diff_+(\Seins)$ is denoted by
$\symb{p}$.  Localised diffeomorphisms $\varphi$ are identified
with their preimage $\symb{p}^{-1}(\varphi)$ in the first
sheet of the covering.

The {\sc Additional Assumption} only
enters through the lemmas \ref{lem:autUA} and \ref{lem:isoUA}, which
we believe to hold true in a lot more general circumstances. It was
verified in presence of an {\em integrable stress-energy tensor in 
$\lok{A}$} for current subalgebras in theorem \ref{th:SegSugCon}. In
this case the  
representations $\Upsilon^\lok{A}\restriction {\PSL(2,\dopp{R})^\sim}$ and
$U^\lok{A}$ coincide, whereas we have only assumed
that the respective generators agree up to a multiple of $\Einsop$. At
this point we want to stress: We do not assume $\lok{A}$ to be
diffeomorphism covariant, ie the adjoint action of $\Upsilon^\lok{A}$
on $\lok{A}$ to implement a geometric, automorphic action of
$\Diff_+(\Seins)$ on $\lok{A}$.

\begin{lemma}\label{lem:autUA}
  $U^\lok{A}$ has the
  net-endomorphism property, if the {\sc Additional Assumption}  holds.
\end{lemma}

\begin{pf}
  By lemma \ref{lem:diffD} there exist, for small $t\in\dopp{R}$,
  diffeomorphisms 
  $g_{\varepsilon}$, $g_{\delta}$ localised in arbitrarily small
  neighbourhoods of $-1$ and  $1$, respectively, and diffeomorphisms
  $g_+$, $g_-$ localised in $\Seins_+$ and $\Seins_-$, respectively,
  such that we have: $D(t)= g_+g_-g_\delta 
  g_\varepsilon$. If the closure of a proper interval $I$ is
  contained in $\Seins_+$,
  we have with an appropriate 
  choice of $g_\delta$, $g_\varepsilon$ by the {\sc Additional Assumption} :
  \begin{equation}\label{eq:autUA}
%$
    U^\lok{A}(\tilde{D}(t)) \lok{B}(I) U^\lok{A}(\tilde{D}(t))^* =
    \Upsilon^\lok{A}(\symb{p}^{-1}(g_+)) \lok{B}(I)
    \Upsilon^\lok{A}(\symb{p}^{-1}(g_+))^*\subset 
    \lok{B}(\Seins_+) \zendot
%$.
  \end{equation}
Because $\lok{B}(\Seins_+)$ is continuous from the inside, we see that
$Ad_{U^\lok{A}(\tilde{D}(t))}$ induces an endomorphism of
$\lok{B}(\Seins_+)$. The same holds true for
$U^\lok{A}(\tilde{D}(-t))$ and, therefore, these endomorphisms are
automorphisms.
\end{pf}

Remark: From formula (\ref {eq:SETnetend}) we
readily see that a stress-energy tensor
 yields lemma \ref{lem:autUA} without any direct reference to the
structure of $\Diff_+(\Seins)$ or, indeed, the \Name{L\"uscher-Mack}
algebra (\ref{eq:LueMackalg}).

The  next step is to give a characterisation of one-parameter groups of
unitary operators
which define, by their adjoint action, endomorphism semigroups of a
standard \Name{v.Neumann} algebra. 
 The following theorem is mainly a new
 formulation of results by \Name{Borchers} and 
 \Name{Wiesbrock}. It appears to be a natural {\em converse of 
{Borchers}' theorem on half-sided translations}. 
The  
methods involved in the  proof are completely standard, but the result
ought to be 
made available\footnote{Compare \cite{dD96} for another
characterisation of endomorphism semigroups related to
\Name{Borchers}' theorem.}. 

\begin{theo}\label{bowiesatz}
  Assume $\lok{M}\subset\lok{B}(\Hilb{H})$ to be a
  \Name{v.Neumann} algebra having a cyclic and separating vector
  $\Omega$ in the separable \Name{Hilbert} space $\Hilb{H}$.
  $J,\Delta$ shall stand for the modular data of this pair. Let
  $V(t)$, $t\in\dopp{R}$, be a strongly continuous one-parameter
  group. Then any two from $\{\ref{BWspec},\ref{BWvac},\ref{BWscal}\}$
  imply the remaining two in the list below; $\ref{BWcone}$ yields
  $\ref{BWspec}$, $\ref{BWvac}$, $\ref{BWscal}$.
%%%%%%%%%%%%%%%%%%%%%%%%%%%%%%%%
  \begin{enumerate}
  \item \label{BWspec}
    \begin{enumerate}
    \item \label{BWspecspec}$V(s)=e^{iHs}$, $H\geqslant 0$,
    \item $V(s)\lok{M}V(s)^*\subset\lok{M}$, $s\geqslant 0$.
    \end{enumerate}
  \item \label{BWvac}
    \begin{enumerate}
    \item $V(s)\Omega=\Omega$, $ s\in\dopp{R}$,
    \item $V(s)\lok{M}V(s)^*\subset\lok{M}$, $s\geqslant 0$.
    \end{enumerate}
   \item \label{BWscal}
    \begin{enumerate}
    \item \label{BWscalscal}$\Delta^{it}V(s)\Delta^{-it} = V(e^{-2\pi
      t}s)$, $JV(s)J=V(-s)$, 
      $ t,s\in\dopp{R}$,
    \item $V(s)\lok{M}V(s)^*\subset\lok{M}$, $s\geqslant 0$.
    \end{enumerate}
   \item \label{BWcone}
    \begin{enumerate}
    \item \label{BWcsp}$V(s)=e^{iHs}$, $H\geqslant 0$,
    \item \label{BWcs}$\Delta^{it}V(s)\Delta^{-it} = V(e^{-2\pi t}s)$,
      $t,s\in\dopp{R}$,
    \item \label{BWcc}$\skalar{m_+'\Omega}{V(s)m_+\Omega}\geqslant 0$,
      $s\geqslant 0$, 
      $m_+\in\lok{M}_+$, $m_+'\in\lok{M}'{}_+$.
    \end{enumerate}
   \end{enumerate}
%%%%%%%%%%%%%%%
\end{theo}
$\lok{M}_+$ denotes the cone of positive elements in $\lok{M}$,
$\lok{M}_+'$ the cone of positive elements in its commutant $\lok{M}'$.

\begin{pf}
Most of the implications were proved by \Name{Borchers}
and  \Name{Wiesbrock}, respectively:
$\ref{BWspec}\wedge \ref{BWvac} \Rightarrow \ref{BWscal}$:
\cite{hjB92} (cf \cite{mF98}); 
$\ref{BWvac}\wedge \ref{BWscal} \Rightarrow \ref{BWspec}$:
\cite{hwW92}; 
$\ref{BWspec}\wedge \ref{BWscal} \Rightarrow \ref{BWvac}$:
\cite{hjB98a}; 
$\ref{BWspec}\wedge \ref{BWvac} \wedge  \ref{BWscal}\Rightarrow
\ref{BWcone}$: \cite[proposition 2.5.27]{BR87}.  

We prove the remaining statement, namely $\ref{BWcone} \Rightarrow
\ref{BWspec}\wedge \ref{BWvac} \wedge  \ref{BWscal}$, by reduction to 
\cite[theorem 1.1]{hjB97}\footnote{Alternatively, one may use the same
  statement in \cite[theorem 2.5]{hjB97a}.}.  As a first step we look
at the domain of entire analytic vectors with respect to
$\Delta^{iz}$, which we denote by $D_\Delta$, and derive an analytic
continuation of relation $\ref{BWcs}$ as a quadratic form on $D_\Delta$. 
We define: 
\begin{displaymath}
 F(z,w) := 
  \langle\Delta^{i\overline{z}}\psi, e^{ie^{2\pi
        w}H}\Delta^{iz}\phi\rangle \zendot
\end{displaymath}
According to the spectrum condition on $H$,  $F$ is analytic in $w$ for
$0< Im(w) <\frac{1}{2}$, and this function is bounded and continuous
for the closure of this region; the region itself shall be denoted by
$\dopp{S}$. In fact, by \Name{Hartog}'s theorem,  $F$ is analytic on
$\dopp{C}\times \dopp{S}$ as a function in two complex variables. We
make full  
use of relation $\ref{BWcs}$ by looking at another function $G$, which
agrees with $F$ for $0<
Im(w) + Im(z)< \frac{1}{2}$:  
\begin{displaymath}
  G(z,w) := \langle\psi, e^{ie^{2\pi (w+z)}H}\phi\rangle \zendot
\end{displaymath}
Evaluating at $w\in\dopp{R}$ and $z=\frac{i}{4}$ we get:
\begin{equation}\label{eq:analcontscaltrans}
   \langle\Delta^{\frac{1}{4}}\psi, e^{ie^{2\pi
        w}H}\Delta^{-\frac{1}{4}}\phi\rangle = \langle\psi,
    e^{-e^{2\pi w}H}\phi\rangle \zendot
\end{equation}

Both $\psi, \phi$ are of the form
$\psi=\Delta^{-\frac{1}{4}}\psi'$, 
$\phi=\Delta^{\frac{1}{4}}\phi'$, $\psi', \phi'\in D_\Delta$. Since
the set of such $\psi', \phi'$ is dense in $\Hilb{H}$, the equation
above becomes an equation for bounded operators, which yields:
\begin{equation}\label{eq:boskal}
  e^{isH} = \Delta^{-\frac{1}{4}}e^{-sH}\Delta^{\frac{1}{4}}\, , \quad
  s\geqslant 0 \zendot
\end{equation}

Next, we show invariance of $\Omega$ following arguments from the
proof of \cite[lemma 2.3.c]{hjB98a}: let $E$ be the projection onto
the eigenvectors of
$\Delta$ having eigenvalue $1$. Multiplying the identity
(\ref{eq:boskal}) from both
sides by $E$ leads to:
\begin{displaymath}
  Ee^{isH}E = Ee^{-sH}E\, , \quad  s\geqslant 0\,.
\end{displaymath}
Here, the right hand side is a positive operator and thus we have as well:
\begin{displaymath}
  \klammer{Ee^{isH}E}^* = Ee^{-isH}E = Ee^{-sH}E = Ee^{isH}E\, , \quad
  s\geqslant 0 \,.
\end{displaymath}
According to a standard argument\footnote{Such an argument is given,
  for example, in the proof of corollary \ref{cor:Bosuginvvec} and uses the
  spectrum condition, the \Name{Phragmen-Lindel\"of} theorem,
  \Name{Schwarz}' reflection principle and \Name{Liouville}'s 
  theorem.}, this invariance with respect to conjugation yields:
%\begin{displaymath}
 $ Ee^{isH}E = Ee^{i0H}E = E$. 
%\end{displaymath}
Therefore, all vectors $\xi$ satisfying $\xi= E\xi$ are 
invariant under the action of $V$ and this means in particular:
$V(s)\Omega = \Omega$, $\forall s\in\dopp{R}$. 

It now follows from $\ref{BWcc}$ and \cite[proposition 2.5.28]{BR87} that 
$e^{-sH}$, $s\geq 0$, leaves the natural cone of
$\klammer{\lok{M},\Omega}$ globally fixed. The other assumptions of
\cite[theorem 1.1]{hjB97} are the identities:
\begin{eqnarray*}
  \Delta^{it}e^{-Hs}\Delta^{-it}&=& e^{-se^{-2\pi t} H}\, , \quad
  s\geqslant 0\,\, ,\\
e^{-Hs} \Omega &=& \Omega\, , \quad s\geqslant 0\,\, .
\end{eqnarray*}
These relations are obvious by analytic continuation of results
derived above. By \cite[theorem 1.1]{hjB97} the adjoint action of
$V(s)$, $s\geq 0$, does indeed induce 
endomorphisms of $\lok{M}$ and we have completed the proof. 
\end{pf}

%%%%

The analytic continuation of the dilatation-translation relation 
$\ref{BWcs}$ to imaginary arguments as in 
(\ref{eq:analcontscaltrans}) is a consequence of the spectrum
condition and does not follow from general group
theoretical bounds (see eg \cite[chapter 11, \S3, theorem
4]{BR77}). It does not appear to be possible to drop  the assumption
$\ref{BWcsp}$ on the spectrum condition and to deduce it in the course
of the argument, since one inevitably runs into domain problems (cf
\cite[theorem 1]{dD96}, \cite[corollary 2.8]{rL97}, \cite[proposition
2.4]{BCL98}). 

%%%%%

The arguments in the proof of theorem \ref{bowiesatz} apply, with
minor alterations, to translation groups with negative generator, as eg the
special conformal transformations $U(S(.))$. While $J$ has
the same action, $JU(S(n))J = U(S(-n))$, the scaling behaviour is
opposite:
\begin{equation}
  \label{eq:specconfscal}
  \Delta^{it} U(S(n)) \Delta^{-it} = U(S(e^{2\pi t}n)) \zendot
\end{equation}
The negative spectrum together with the opposite scaling law
(\ref{eq:specconfscal}) shows that the condition characterising
endomorphism semi-groups is again given by the one in 
$\ref{BWcc}$ Since the arguments are the same as for the case of
positive spectrum and scaling law $\ref{BWscalscal}$, $\ref{BWcs}$ we
state the following corollary without proof:
\begin{cor}\label{cor:bowie}
  The statements in theorem \ref{bowiesatz} still hold, if one
  uses  $V(s)= e^{iKs}$, $K\leq 0$, instead of $\ref{BWspecspec}$,
  $\ref{BWcsp}$ and replaces the scaling law in $\ref{BWscalscal}$,
  $\ref{BWcs}$ by 
  $\Delta^{it} V(s) \Delta^{-it} = V(e^{2\pi t}s)$, 
  $s,t\in\dopp{R}$.
\end{cor}

%%%%%

At this stage our intuition about the geometric impact of
$U^\lok{A}$ 
on $\lok{B}$ can be verified. We will discuss the general situation
after the following corollary:

\begin{cor}\label{cor:netend}
  Assume ${U^\lok{A}}$ to have the net-endomorphism property. Then 
  the adjoint action of ${U^{\lok{A}'}(\tilde{D}(.))}$ on
  $\lok{B}(\Seins_+)$ defines a group of automorphisms. 

For $s\geq 0$
  the adjoint action of $U^\lok{A}(\tilde{T}(s))$ 
  induces  endomorphisms of $\lok{B}(\Seins_+)$ and the adjoint action
  of $U^\lok{A}(\tilde{T}(-s))$ maps $\lok{B}(\Seins_+)$ into
  $\lok{B}(T(-s)\Seins_+)$. The corresponding statements hold true, if
  one replaces $\lok{A}$ by $\lok{A}'$ or $\tilde{T}(.)$
  by $\tilde{S}(.)$.
\end{cor}

\begin{pf}
The statement on $Ad_{U^{\lok{A}'}(\tilde{D}(.))}$  follows from
$U^{\lok{A}'}= U\!\circ\!\symb{p} \,U^\lok{A}{}^*$ and covariance of
$\lok{B}$. Using the factorisation $U(T(s))=
U^\lok{A}(\tilde{T}(s))U^{\lok{A}'}(\tilde{T}(s))$, covariance and  
isotony of $\lok{B}$,  the
statement on $Ad_{U^{\lok{A}'}(\tilde{D}(.))}$ and invariance of
$\Omega$ with respect to $U^{\lok{A}'}$, we have the following
inequality for all $t\in\dopp{R}$, $s\geq 0$,
$B_+\in\lok{B}(\Seins_+)_+$, $B'_+\in\lok{B}(\Seins_-)_+$:
\begin{eqnarray*}
0 &\leqslant&
\skalar{U^{\lok{A}'}(\tilde{D}(t))^*B_+'U^{\lok{A}'}(\tilde{D}(t))\Omega}{
  U(T(s))
  U^{\lok{A}'}(\tilde{D}(t))^*B_+U^{\lok{A}'}(\tilde{D}(t))\Omega}\\ 
&=& \skalar{B_+'\Omega}{
  U^\lok{A}(\tilde{T}(s))U^{\lok{A}'}(\tilde{T}(e^{t}s))
 B_+\Omega} \zendot
\end{eqnarray*}
In the limit $t\rightarrow -\infty$ strong continuity of
$U^{\lok{A}'}$ implies $\langle B_+'\Omega, 
  U^\lok{A}(\tilde{T}(s)) B_+\Omega\rangle \geq 0$, which in turn yields the
statement on $U^\lok{A}(\tilde{T}(s))$, $s\geq 0$, by theorem
\ref{bowiesatz}, because of the \Name{Bisognano-Wichmann} property of
$\lok{B}$ and general results on $U^\lok{A}$ (section
\ref{cha:cospa}.\ref{sec:bosug}). Following
the same argument 
with $\lok{A}$ instead of $\lok{A}'$ and vice versa leads to the
corresponding statement on  $U^{\lok{A}'}(\tilde{T}(s))$, $s\geq
0$. If one replaces in both statements $\tilde{T}(s)$ by
$\tilde{S}(s)$, one may apply the argument as well, but using the
limit $t\rightarrow \infty$ and corollary \ref{cor:bowie}.

The remainder follows immediately from the following argument, which
we indicate for the translations represented through $U^\lok{A}$:
\begin{eqnarray*}
  Ad_{U^\lok{A}(\tilde{T}(-s))} \lok{B}(\Seins_+) &=& Ad_{U(T(-s))}
Ad_{U^{\lok{A}'}(\tilde{T}(s))}\lok{B}(\Seins_+)\\
& \subset& Ad_{U(T(-s))}
\lok{B}(\Seins_+) = \lok{B}(T(-s)\Seins_+) \zendot
\end{eqnarray*}
\end{pf}

%%%%%%%%%%%%%%%%%%%% From here

The geometric impact of a general $U^\lok{A}(\tilde{g})$,
$\tilde{g}\in \PSL(2,\dopp{R})^\sim$, on an arbitrary local algebra
$\lok{B}(I)$ is discussed easily. We may restrict our attention to
group elements $\tilde{g}$ for which there is a single sheet of the
covering projection $\symb{p}$ containing both $\tilde{g}$ and the
identity, as the following discussion indicates. 

Every element $g$ in $\PSL(2,\dopp{R})$ is contained in (at least) one
one-parameter group\footnote{I am indebted to D. Guido for providing
  the reference. In the particular case of $\PSL(2,\dopp{R})$  this
  fact may be checked directly (proposition \ref{prop:PSLexp}).}
\cite{mM94,mM97}.  We use the  local identification of
one-parameter subgroups in $\PSL(2,\dopp{R})$ and in
$\PSL(2,\dopp{R})^\sim$, choose a parametrisation such 
that $\tilde{g}=\tilde{g}(1)$, $id= \tilde{g}(0)$, and we set
$\gamma_{\tilde{g}}(I) := \bigcup_{\tau=0}^1 \symb{p}(\tilde{g}(\tau))
I$. For $\tilde{g}$ further away from the identity we set
$\gamma_{\tilde{g}}(I) := \Seins$ and take $\lok{B}(\Seins)$ to be the
algebra of all bounded operators on $\Hilb{H}$.  Then we have:
  
\begin{prop}\label{prop:netend}
Assume $U^\lok{A}$ to have the net-endomorphism property. Then    
for any $\tilde{g}\in \PSL(2,\dopp{R})^\sim$ and any $I\Subset
  \Seins$: $$Ad_{U^\lok{A}(\tilde{g})}\lok{B}(I)\subset
  \lok{B}(\gamma_{\tilde{g}}(I))\, , \quad 
  Ad_{U^{\lok{A}'}(\tilde{g})}\lok{B}(I)\subset 
  \lok{B}(\gamma_{\tilde{g}}(I)) \zendot$$
\end{prop}

\begin{pf}
Each proper interval $I$ in $\Seins$ may be
identified by the ordered 
pair consisting of its boundary points,  $z_+$
and $z_-$.
%%%%%%%%%%%%%%%%%%%%%%%%%%%
We define three one-parameter subgroups in $\PSL(2,\dopp{R})$
referring to each $I\Subset\Seins$ with
respect to a particular choice $h\in\PSL(2,\dopp{R})$ satisfying
$h\Seins_+ =I$: $D_I(.) = h D(.) h^{-1}$, $T_I(.) = h T(.)
h^{-1}$, $S_I(.) = h S(.) h^{-1}$.
%%%%%%%%%%%%%%%%%%%%%%%%%%%%
% There are three distinguished 
% one-parameter groups in $\PSL(2,\dopp{R})$ referring to $I$: the group
% $D_I$ of dilatations associated with $I$ (leaves both $z_+$ and $z_-$
% fixed), the group of special conformal transformations $S_I$
% associated with $I$ (leaves $z_+$ fixed only), and the group of
% translations $T_I$ associated with $I$ (leaves $z_-$ fixed only). The
% one-parameter groups associated with $\Seins_+$ (or equivalently with
% $\Seins_-$) are the 
% well-known subgroups which we introduced above: $D_{\Seins_+}(.)=
% D(.)$ and so forth. All three groups $D_I$, $T_I$, $S_I$ are conjugate
% to the corresponding one-parameter subgroups associated with
% $\Seins_+$, ie we may
% choose parameters on them such that with $h\Seins_+ =I$ we have: $D_I(.)
% = h D(.) h^{-1}$ and so forth.   

Each element $g$ in $\PSL(2,\dopp{R})$ is fixed, up to a dilatation
$D_I(t)$, by its action on $\{z_+,z_-\}$. Under the action of elements
$g(\tau)$, $\tau=0,\ldots, 1$, interpolating in the one-parameter
group associated with $g$ between the identity and $g=g(1)$, the orbits
of $z_\pm$ are given by monotonous functions $z_+(\tau)$,
$z_-(\tau)$.   Demanding $s$, $n$,
$t$ to depend continuously on $\tau$ and to take value $0$ at $\tau=0$,
every $g(\tau)$ may be represented as
 $g(\tau)=
S_{T_I(s(\tau))I}(n(\tau)) T_I(s(\tau)) D_I(t(\tau))$ or as $g(\tau)=
T_{S_I(n(\tau))I}(s(\tau)) S_I(n(\tau)) D_I(t(\tau))$. We choose one
form which works for all interpolating elements. By the requirements we
have made it is ensured that the representation works (after obvious
identifications) in $\PSL(2,\dopp{R})^\sim$ as well.  Corollary
\ref{cor:netend} implies the claim of the proposition now.
\end{pf}
%%%%%%%%%%%%%%%%%%%% to here

This proves in particular: For every $I\Subset\Seins$ there is a
neighbourhood of the identity in $\PSL(2,\dopp{R})^\sim$ for which
the action of $Ad_{U^\lok{A}(.)}$ on $\lok{B}(I)$   delivers local
observables. The result of this proposition can be improved on grounds
of some further assumptions; see proposition \ref{prop:AutchirRot}.

We have found $Ad_{U^\lok{A}}$ to induce homomorphisms from local
algebras of $\lok{B}$ into algebras associated with an enlarged localisation
region. This
sub-geometrical action respects isotony, ie the net-structure. The
adjoint action of $U$ induces the covariance isomorphisms of local
algebras and one 
usually regards these as {\em automorphisms of the net $\lok{B}$}. We
consider, therefore, the term {\em net-endomorphisms} appropriate. The
automorphic action of $Ad_{U^\lok{A}(\tilde{D}(.))}$ on 
$\lok{B}(\Seins_+)$ which we proved in lemma \ref{lem:autUA} does not,
apparently, follow from the endomorphism property for translations
in corollary \ref{cor:netend}. This motivated definition
\ref{def:netend} above.

%%%%%%%%%%%%%%%%%%%%%%%%%%%%%%%%%%%%%%%%%%%%%%%%%%%%%%%%%%%
%%%%%%% CHIRAL HOLOGRAPHY %%%%%%%%%%%%%%%%%%%%%%%%%%%%%%%%%
%%%%%%%%%%%%%%%%%%%%%%%%%%%%%%%%%%%%%%%%%%%%%%%%%%%%%%%%%%%

\sekt{Chiral holography}{Chiral holography}{sec:chihol} 
%\section{Chiral holography}
%\label{sec:chihol}

We give a {\em holographic} interpretation
of the  net-endomorphism property. This shows that the results
achieved so far are satisfactory, and it yields new insights into
structures associated with chiral conformal subnets 
and their \Name{Coset} models. 

If we define $U^\lok{A}\times U^{\lok{A}'}$ as representation of
$\PSL(2,\dopp{R})^\sim\times \PSL(2,\dopp{R})^\sim$ through
$(\tilde{g},\tilde{h})\mapsto U^\lok{A}(\tilde{g})
U^{\lok{A}'}(\tilde{h})$, this is, in fact, a representation
of the conformal symmetry group of a local conformal quantum theory in
$\opo$ dimensions, which is isomorphic to
$(\PSL(2,\dopp{R})^\sim\times \PSL(2,\dopp{R})^\sim)/\dopp{Z}$. This
factor group arises, if one identifies the simultaneous rigid
conformal rotation by $2\pi$, namely
$(\tilde{R}(2\pi),\tilde{R}(2\pi))$, with the trivial
transformation. The last section taught us a lot about the
sub-geometrical action of $U^\lok{A}$, $U^{\lok{A}'}$ on
the local observables in $\lok{B}$. So, it is natural to look for a
relation between the geometrical character of this action and structures in
$\opo$ dimensions.

This relation turns out to be a complete correspondence: We construct
a $\opo$-dimensional, local, conformal theory from the 
original chiral theory $\lok{B}$ applying the net-endomorphism
property of $U^\lok{A}$. In
order to prove locality in $\opo$ dimensions we are led to a
particular choice of light-cone coordinates, by which  the
original local algebras $\lok{B}(I)$, $I\Subset\Seins$, are included
in the $\opo$-dimensional picture as {\em time zero algebras}. This
choice of coordinates yields an unphysical spectrum condition:
translations in the right  spacelike wedge 
have positive spectrum. Whereas this prohibits an interpretation of
the new theory as a genuinely physical one, where we would have
positivity of the spectrum in future-like directions, the construction
does provide  a useful geometrical picture for
questions concerned with chiral subnets and their \Name{Coset} models.
For this reason we have to regard the result of our construction as a
local, conformal {\em quasi-theory} in $\opo$ dimensions.

If, on the opposite, one takes a  conformal local quantum theory
in $\opo$ dimensions and defines a chiral conformal net by restriction
to time zero algebras, a similar phenomenon arises (cf
\cite{KLM01,rL01}): the spectrum condition disappears altogether, but
powerful tools of local quantum theory are available still, because
the \Name{Reeh-Schlieder} property survives. In our case there remains
a spectrum condition from 
which one can still derive the \Name{Reeh-Schlieder} property. In
this sense we find a natural ``converse'' of the restriction process
which  justifies the term {\em chiral holography} for our
construction. 

The main result of this section will be proved by making contact with
the analysis of \Name{Brunetti, Guido and Longo} \cite{BGL93} who
discussed conformal quantum field theories in general spacetime
dimensions 
as local quantum theories on the conformal covering of the respective
\Name{Minkowski} space given by  extensions of local nets living on
\Name{Minkowski} space itself.

\subsection{The holographic quasi-theory in $\opo$ dimensions}
\label{sec:bopo}

In $\opo$ dimensions, \Name{Minkowski} space $\dopp{M}$ is the
Cartesian product of two chiral light-rays, which we take as
light-cone coordinates of  $\dopp{M}=\dopp{R}\times\dopp{R}$. One
arrives at the (physical) conformal covering $\widetilde{\dopp{M}}$ of
$\dopp{M}$, if one compactifies both light-rays adding the points at
infinity, takes the infinite, simply connected covering of the compactification
$\Seins\times\Seins$, which yields $\dopp{R}\times\dopp{R}$,
and, finally, one identifies all points which are connected by the
action of simultaneous rigid conformal rotations by $2\pi$. The result
has the shape of a cylinder having infinite timelike
extension: $\widetilde{\dopp{M}}= \Seins\times\dopp{R}$. Without the
final identification we would have 
spacelike separated copies of $\dopp{M}$ in covering space, which we
consider unphysical; conformally covariant quantum fields can be
proven to live on this (physical) conformal covering of
\Name{Minkowski} space, see \cite{LM75}. 

Light-rays in $\widetilde{\dopp{M}}$ are infinitely extended, universal
coverings of the compactified light-rays and serve well as light-cone
coordinates of $\widetilde{\dopp{M}}$. The localisation regions, which
we will consider, are 
{\em $\opo$-dimensional double cones} given as Cartesian products
of two intervals, $I\times J$, where $I$, $J$ are properly contained
in a single copy of $\Seins$ on the left and right light-rays in
$\widetilde{\dopp{M}}$, respectively.  

$\PSL(2,\dopp{R})^\sim$, the universal covering group of the chiral conformal
transformations,  has an action on the infinite covering $\dopp{R}$ of
$\Seins$ which is transitive
for the intervals which are properly contained in a single copy of
$\Seins$. We exclude the point of
infinity from $\Seins$ and choose a fixed interval $I$ which is
properly contained in the remainder. This interval is identified with
its first pre-image in covering space. For intervals $J_{L}$, $J_R$
which are properly included in a single copy of 
$\Seins$ we choose group elements  
$\tilde{g}_{L,R}\in \PSL(2,\dopp{R})^\sim$ satisfying
$J_{L}=\tilde{g}_{L}I$, 
$J_R=\tilde{g}_RI$. Making use of this choice we  define a set of
(local) algebras indexed by $\opo$-dimensional double cones:
\begin{equation}
  \label{eq:oodef}
  \lok{B}^{\opo}(J_L\times J_R) := U^\lok{A}(\tilde{g}_{L})
  U^{\lok{A}'}(\tilde{g}_{R}) \lok{B}(I) U^\lok{A}(\tilde{g}_{L})^*
  U^{\lok{A}'}(\tilde{g}_{R})^*  \zendot
\end{equation}
By covariance of $\lok{B}$, the resulting algebra $\lok{B}^{\opo}(J_L\times
J_R)$ is uniquely determined by $J_L\times J_R$.   

Furthermore, we define a covering projection $\symb{p}$ from
$\dopp{R}$ onto $\Seins$ referring to the covering projection
$\symb{p}: \PSL(2,\dopp{R})^\sim\rightarrow \PSL(2,\dopp{R})$ such that
we have: $\symb{p} J_{L,R} := \symb{p}(\tilde{g}_{L,R}) I$. Because of
the close relation between both covering projections and the fact that 
the respective argument distinguishes between them, we did not
introduce a new 
symbol. This definition enables us to state two identities for the
algebras defined in equation (\ref{eq:oodef}):
\begin{eqnarray*}
  \lok{B}^{\opo}(J_L\times J_R) &=&
  U^\lok{A}(\tilde{g}_{L}\tilde{g}_{R}{}^{-1}) 
  \lok{B}(\symb{p} J_{R})
  U^\lok{A}(\tilde{g}_{L}\tilde{g}_{R}{}^{-1})^*\\
&=&  U^{\lok{A}'}(\tilde{g}_{R}\tilde{g}_{L}{}^{-1}) \lok{B}(\symb{p} J_{L}) 
  U^{\lok{A}'}(\tilde{g}_{R}\tilde{g}_{L}{}^{-1})^* \zendot 
\end{eqnarray*}

Double cones $J\times J$, which are centered at the time zero axis, are
called {\em time zero double cones} and we get for the corresponding
{\em time zero algebras}: 
\begin{displaymath}
  \lok{B}^{\opo}(J\times J) = \lok{B}(\symb{p} J) \zendot
\end{displaymath}
Thus, the local algebras of the original chiral conformal theory
$\lok{B}$ are included into the new quasi-theory $\lok{B}^{\opo}$ as time zero
algebras:
\begin{theo}\label{th:chihol}
  If $\lok{A} \subset\lok{B}$ is an inclusion of chiral conformal
  theories and if the unique inner-implementing representation
  $U^\lok{A}$ associated with this inclusion has the net-endomorphism
  property, then equation (\ref{eq:oodef}) defines a set
  $\lok{B}^{\opo}$ of local 
  algebras assigned to double cones in $\opo$-dimensional conformal
  space time, $\widetilde{\dopp{M}}$, having all but one of the usual
  properties of a local, conformal, weakly additive quantum theory
  in $\opo$ dimensions (see \cite{BGL93}): the spectrum condition holds
  for translations in the right spacelike wedge.   
\end{theo}

\begin{pf}
Obviously, the set $\lok{B}^{\opo}$ of local algebras is covariant with
respect to the representation $U^\lok{A}\times U^{\lok{A}'}$. Because
of the identity
$U^\lok{A}(\tilde{R}(2\pi))U^{\lok{A}'}(\tilde{R}(2\pi))=\Einsop$
the set $\lok{B}^{\opo}$ is in fact labelled by the double cones in
$\widetilde{\dopp{M}}$, and $U^\lok{A}\times U^{\lok{A}'}$ is a
representation of the conformal group in $\opo$ dimensions, namely the
group $(\PSL(2,\dopp{R})^\sim\times
\PSL(2,\dopp{R})^\sim)/\dopp{Z}$. The spectrum condition for
$U^\lok{A}\times U^{\lok{A}'}$ was proved in corollary \ref{cor:poseng}.

The vacuum vector is invariant with respect to $U^\lok{A}\times
U^{\lok{A}'}$ (corollary \ref{cor:Bosuginvvec}) and it is a basis for
the space of vectors with this property, because the space of
$U$-invariant vectors is one-dimensional. $\Omega$ is cyclic for all
local algebras in $\lok{B}^{\opo}$ because of the
\Name{Reeh-Schlieder} property of $\lok{B}$. 

Isotony follows directly from the net-endomorphism property. An
inclusion of $\opo$-dimensional double cones $\tilde{g}_LI\times
\tilde{g}_RI\subset \tilde{h}_LI\times
\tilde{h}_RI$ contained in \Name{Minkowski} space $\dopp{M}$, yields
the relations: $\tilde{h}_{L,R}{}^{-1}\tilde{g}_{L,R}I\subset
I$. Applying proposition \ref{prop:netend} we get:
$Ad_{U^{\lok{A}'}(\tilde{h}_{R}{}^{-1}\tilde{g}_{R})U^\lok{A}(\tilde{h}_{L}{}^{-1}\tilde{g}_{L})}\lok{B(I)}\subset
\lok{B}(I)$. This is equivalent to $\lok{B}^{\opo}(\tilde{g}_LI\times
\tilde{g}_RI)\subset \lok{B}^{\opo}(\tilde{h}_LI\times
\tilde{h}_RI)$. 

Locality for double cones in $\dopp{M}$ is shown easily as well. We
can reduce the  discussion to  the situation where there is a double
cone $J_1\times J_2$ spacelike to our basic time zero double cone $I\times I$
simply by applying an appropriate transformation. There is a time zero
double cone $J\times J$ which contains $J_1\times J_2$ and is spacelike to
$I\times I$. Since we have shown isotony for $\lok{B}^{\opo}$, locality for
this set follows from locality of $\lok{B}$.    

Weak additivity may be proved as in the chiral case. By scale
covariance the local algebras of $\lok{B}^{\opo}$ are continuous from
the inside as well as from the outside \cite{LRT78}. Because we can
restrict the discussion to time zero algebras and since the argument of
\Name{J\"{o}r\ss} \cite{mJ96} for the corresponding chiral situation
may be extended directly, we have weak additivity for
$\lok{B}^{\opo}$. 

The proof is complete, if one recognises that the proof of
\cite[proposition 1.9]{BGL93}, which establishes  the unique extendibility of
$\lok{B}^{\opo}$ from $\dopp{M}$ to all of $\widetilde{\dopp{M}}$, only
requires the prerequisites 
established so far. In particular, not the spectrum condition itself is
needed, but only its consequence, the \Name{Reeh-Schlieder} property. 
\end{pf}

In light of this theorem we obtain a straightforward interpretation
of the sub-geometrical action of $U^\lok{A}$ on $\lok{B}$. If we apply
a chiral coordinate transformation $\tilde{g}_R$ to a time zero double 
cone $J\times J$ and if we test the localisation of the correspondingly
transformed local algebra of $\lok{B}^{\opo}$ only by looking at time
zero algebras, then we find that the result commutes just with
 time zero algebras $\lok{B}(K)$ assigned to
proper intervals $K$ contained in the causal complement of
$\gamma_{\tilde{g}_R}J$. The statement of proposition
\ref{prop:netend} follows from \Name{Haag} duality of $\lok{B}$.
(Compare figure \ref{fig:netend}.) 
{\par
 \begin{figure} %Abb.pos: normal:[h]
\begin{minipage}{0cm} \end{minipage}\hfill
\begin{minipage}{8.0cm}
\epsfig{file=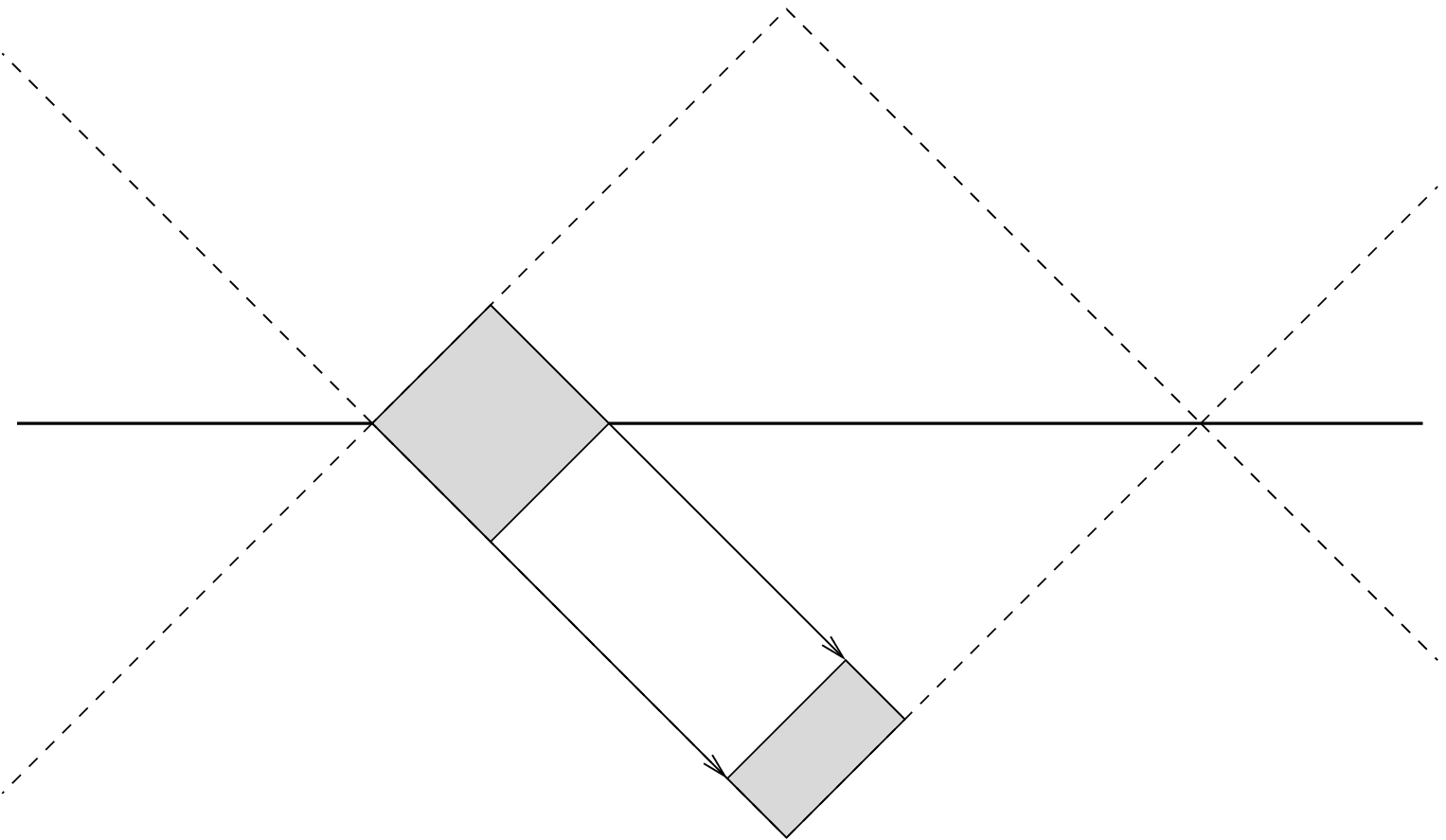,width=6cm} 
\refstepcounter{figure}\label{fig:netend}%Zeigername f"ur die Abb.
\end{minipage}
\hfill\begin{minipage}{0cm} \end{minipage}
 \begin{minipage}{0cm} \end{minipage}
 \parbox{14cm}{\vspace{2ex} \centerline{\small {\bf \figurename{} 
       \ref{fig:netend}:} 
     Chiral transformations and locality in $d=\opo$.}} 
%#5: Bildunterschrift
 \end{figure}}

The theorem has some direct applications to chiral subtheories and
their \Name{Coset} models: We have found that the maximal \Name{Coset}
model $\lok{C}_{max}$ associated with a subtheory
$\lok{A}\subset\lok{B}$ may be regarded as the chiral conformal theory
of all right chiral observables in  $\lok{B}^{\opo}$ in the sense of
\Name{Rehren} \cite{khR00}, ie the local observables of
$\lok{B}^{\opo}$ which are invariant under the action of
transformations on the left light-cone coordinate only.

The observables of $\lok{A}$ may be viewed as left chiral
observables and the chiral conformal subnet
$\lok{A}_{max}\subset\lok{B}$ consisting of local observables
invariant with respect to the action of $U^{\lok{A}'}$ (and
accordingly covariant with respect to the action of $U^\lok{A}$) is to
be identified with the chiral theory of all left chiral observables in
$\lok{B}^{\opo}$. 

Thus, we have identified $\lok{A}_{max}$ and $\lok{C}_{max}$ as
fixed-points of a spacetime symmetry acting on a
suitably extended theory, namely $\lok{B}^{\opo}$. In presence of the
net-endomorphism property it is not necessary to extend the
``classical''  symmetry concept (see eg \cite{hA92}), if one wants to
interpret the chiral subtheories $\lok{A}_{max}$ and $\lok{C}_{max}$
as fixed-points of a symmetry; all one has to do is to extend the
theory $\lok{B}$ to its holographic image
$\lok{B}^\opo$. Generalisations of the 
symmetry concept are necessary for a large class of chiral conformal
subtheories \cite{LR95, khR94a}. 
\label{cospairsymm} %Symmetry interpretation of Coset pairs

Another interesting, direct
consequence of theorem 
\ref{th:chihol} is the following: The cyclic subspaces of
$\lok{C}_{max}$ and $\lok{A}_{max}$, namely
$\overline{\lok{C}_{max}(I)\Omega}$ and
$\overline{\lok{A}_{max}(I)\Omega}$, coincide
with the spaces of $U^\lok{A}$- and $U^{\lok{A}'}$-invariant vectors,
respectively, as was shown by \Name{Rehren} \cite[lemma
2.3]{khR00}. 
%%%%%%%%%%%%%%%%%%%%%%%%%%%%
The proof of the following proposition includes an alternative proof
of this statement; 
together with the 
other parts, this proposition may be viewed as a generalised version
of \cite[theorem 2.4]{fX00}, which applies to a particular
class of chiral subnets.
%%%%%%%%%%%%%%%%%%%%%%%%%%%
\begin{prop}\label{prop:lemma23}
  Assume $U^\lok{A}$ to have the net-endomorphism property and denote
  the projections onto the subspaces of $U^\lok{A}$- and
  $U^{\lok{A}'}$-invariant vectors by $E_{U^\lok{A}}$ and
  $E_{U^{\lok{A}'}}$, 
  respectively. Then we have for the maximal $U^\lok{A}$-covariant
  extension of $\lok{A}$,  given by
  $\lok{A}_{max}(I) = \{U^{\lok{A}'}\}'\cap\lok{B}(I)$, and  the
  maximal \Name{Coset} model associated with $\lok{A}\subset\lok{B}$,
  given by $\lok{C}_{max}(I) = \{U^{\lok{A}}\}'\cap\lok{B}(I)$, for
  arbitrary $I \Subset\Seins$: 
  \begin{equation}
    \label{eq:lemma23}
  \overline{\lok{A}_{max}(I)\Omega}=  E_{U^{\lok{A}'}}\Hilb{H}\, , \quad
  \overline{\lok{C}_{max}(I)\Omega}= 
  E_{U^\lok{A}}\Hilb{H}  \zendot
  \end{equation}
For any \Name{Coset} model $\lok{C}$ associated with
  $\lok{A}\subset\lok{B}$ we have a unitary equivalence of chiral
  conformal theories:
  $\lok{A}\coset\lok{C}e_{\lok{A}\smcoset\lok{C}}  \cong
  \lok{A}e_\lok{A}\otimes\lok{C}e_\lok{C}$. $E_{U^\lok{A}}\Hilb{H}$
  has a direct 
  interpretation as multiplicity space of the vacuum subrepresentation
  of $\lok{A}\subset\lok{B}$.  
\end{prop}
\begin{pf}
  Concerning the proof of (\ref{eq:lemma23}) it suffices to deal with
  the case $I=\Seins_+$ (because of the \Name{Reeh-Schlieder}
  theorem). By lemma \ref{lem:invvec} the spaces of vectors which 
  are invariant with respect to translations are identical with
  $E_{U^\lok{A}}\Hilb{H}$ and $E_{U^{\lok{A}'}}\Hilb{H}$, respectively. Taking
  into account corollary \ref{cor:netend} above the statement
  (\ref{eq:lemma23}) was  proved by \Name{Borchers} \cite[theorem
  2.6.3]{hjB98a}.

The statement on the tensor-product character of vacuum
representations of \Name{Coset} pairs was proved for
proposition \ref{prop:cospatenpos} already.  

%%%%%%%%%%%%%%%%%%%%%%%%%%%%%%%%%%%%%%%%%%%%%5
In the following, $A$ denotes local observables in
$\lok{A}\subset\lok{B}$ and $\pi_0(A)= A e_\lok{A}$ its representative
in the 
vacuum representation on $e_\lok{A}\Hilb{H}=:\Hilb{H}_0$. The
implementation of conformal covariance in $\pi_0$ shall be written $U_0$. 
For every vacuum subrepresentation in
$\lok{A}\subset\lok{B}$ there is a partial isometry $R: \Hilb{H}
\rightarrow \Hilb{H}_0$ satisfying  $RA = \pi_0(A)R$ for all local $A$ in
$\lok{A}\subset\lok{B}$.  

The projection $e_R:=R^*R$ commutes with all of
$\lok{A}$. $RU^\lok{A}(.)R^*$ is a unitary strongly continuous
representation of $\PSL(2,\dopp{R})^\sim$ which implements global
conformal covariance in $\pi_0$, thus: $RU^\lok{A}(.)R^*= U_0(.)$. It
follows directly that $\Phi_\Omega:=R^*\Omega$, the vacuum of the
subrepresentation associated with $R$, is invariant with
respect to $U^\lok{A}$, ie $\Phi_\Omega\in E_{U^\lok{A}}\Hilb{H}$. This
completes the proof of the last statement.     
%%%%%%%%%%%%%%%%%%%%%%%%%%%%%%%%%
 \end{pf}

It is not clear in general that the representation
$\lok{A}\coset\lok{C}_{max}\subset\lok{B}$ of the tensor-product theory
defined by the vacuum  representation of a chiral subnet
$\lok{A}\subset\lok{B}$ and the vacuum representation of its maximal
\Name{Coset} model has a (spatial) tensor-product decomposition. This
is known under certain conditions \cite{KLM01}.   

Proposition \ref{prop:lemma23} gives results from 
character arguments on inclusions of current algebras a direct
and rigorous meaning in the context of the analysis of the respective
inclusions of 
chiral conformal theories and \Name{Coset} models. We will discuss
this in section \ref{cha:finind}.\ref{sec:cospactps}.

%%%%%%%%%%%%%%%%%%%%%%%%%%%%%%%%%%%%%%%%%%%%%%%%%%
%%%% Sharp geometrical action %%%%%%%%%%%%%%%%%%%%
%%%%%%%%%%%%%%%%%%%%%%%%%%%%%%%%%%%%%%%%%%%%%%%%%%

\subsection{Sharp geometrical action and time-like commutativity}
\label{sec:sharp}

It is interesting to look at two properties which are closely related
to but stronger than the net-endomorphism property. The first of
these, the {\em sharp geometrical action}, is sufficient for
reinterpreting the holographic quasi-theory as a physically sensible model:
\begin{defi}\label{def:shageoact}
  We say that the unique inner-implementing representation $U^\lok{A}$
  of a chiral subnet $\lok{A}\subset\lok{B}$ has {\bf sharp geometrical
  action} (on $\lok{B}$) if for every $I\Subset\Seins$ and for every
  $\tilde{g}$ of $\PSL(2,\dopp{R})^\sim$ we have:
  \begin{equation}
    \label{eq:sharpgeoact}
    Ad_{U^\lok{A}(\tilde{g})} \lok{B}(I) \subset
    \lok{B}(\gamma_{\tilde{g}}(I))\cap
    \lok{B}(\gamma_{\tilde{g}}'(I)') \zendot
  \end{equation}
\end{defi}
Here, we have introduced a new notation:
$\gamma_{\tilde{g}}'(I) := \gamma_{\tilde{g}}(I)\setminus(\overline{I}\cup
\symb{p}(\tilde{g})\overline{I})$. For the case
$\gamma_{\tilde{g}}'(I)=\varnothing$ we set $\varnothing' := \Seins$,
$\lok{B}(\varnothing) := \dopp{C}\Einsop$ and $\lok{B}(\varnothing') :=
\lok{B}(\varnothing)'$.  

One recognises immediately that sharp
geometrical action of $U^\lok{A}$ implies the same property for
$U^{\lok{A}'}$:
\begin{eqnarray}
  \label{eq:sharpUAprime}
  Ad_{U^{\lok{A}'}(\tilde{g}^{-1})} \lok{B}(I) &\subset&
  \lok{B}(\gamma_{\tilde{g}}(\symb{p}(\tilde{g}^{-1})I))\cap
  \lok{B}(\gamma_{\tilde{g}}'(\symb{p}(\tilde{g}^{-1})I)')\nonumber\\
&=&\lok{B}(\gamma_{\tilde{g}^{-1}}(I))\cap
  \lok{B}(\gamma_{\tilde{g}^{-1}}'(I)') \zendot
\end{eqnarray}
Here we have used  the elementary identity
$\gamma_{\tilde{g}}(\symb{p}(\tilde{g}^{-1})I)=
\gamma_{\tilde{g}^{-1}}(I)$. 

For $\tilde{g}$ close to $id$ we have  $I\cup\symb{p}(\tilde{g})I =
\gamma_{\tilde{g}}(I)\Subset\Seins$ and the inclusion
(\ref{eq:sharpgeoact}) is clear, if we have the net-endomorphism
property. In fact, sharp geometrical action implies the
net-endomorphism property. When $I$ and $\symb{p}(\tilde{g})I$
become disjoint, then 
(generically) the  complement of  $I$ and $\symb{p}(\tilde{g})I$ in
$\Seins$ consists of two proper, disjoint intervals, $\gamma_{\tilde{g}}(I)'$
and $\gamma_{\tilde{g}}'(I)$; sharp geometrical action says in this
case, that $Ad_{U^\lok{A}(\tilde{g})} \lok{B}(I)$ commutes with all
observables associated with either localisation  region.

\begin{prop}\label{prop:quasisharp}
  Assume $U^\lok{A}$ to have the net-endomorphism property. Then sharp
  geometrical action is equivalent to commutativity for timelike
  separation in the quasi-theory $\lok{B}^{\opo}$, which thus may be
  reinterpreted as a physical model.
\end{prop}

\begin{pf}
  We take a   $\tilde{g}\subset\PSL(2,\dopp{R})^\sim$ such that $I$ and
  $\tilde{g}I$ 
   both lie in a single copy of $\Seins$ in
  $\dopp{R}$ and are disconnected: $\overline{I} \cap
  \tilde{g}\overline{I}=\varnothing$. Assuming 
  sharp geometrical action for $U^\lok{A}$ we have:
  \begin{displaymath}
    Ad_{U^\lok{A}(\tilde{g})}\lok{B}(\symb{p}I) = \lok{B}^{\opo}
    (\tilde{g}I\times I) \subset
    \lok{B}(\symb{p}\gamma_{\tilde{g}}'(I))' \zendot
  \end{displaymath}

Repeating the argument with $U^{\lok{A}'}$, varying $I$ and
$\tilde{g}$ it is clear that sharp geometrical action implies
timelike commutativity of $\lok{B}^{\opo}$. The opposite direction is
obvious. (Compare figure \ref{fig:shageo}).
{\par
 \begin{figure} %Abb.pos: normal:[h]
\psfrag{1}{{$I\times I$}}
\psfrag{2}{{$\gamma_{\tilde{g}}'(I)\times \gamma_{\tilde{g}}'(I)$}}
\psfrag{3}{{$\tilde{g}I\times \tilde{g}I$}}
\psfrag{4}{{$\tilde{g}I\times I$}}
\begin{minipage}{0cm} \end{minipage}\hfill
\begin{minipage}{10.0cm}
\epsfig{file=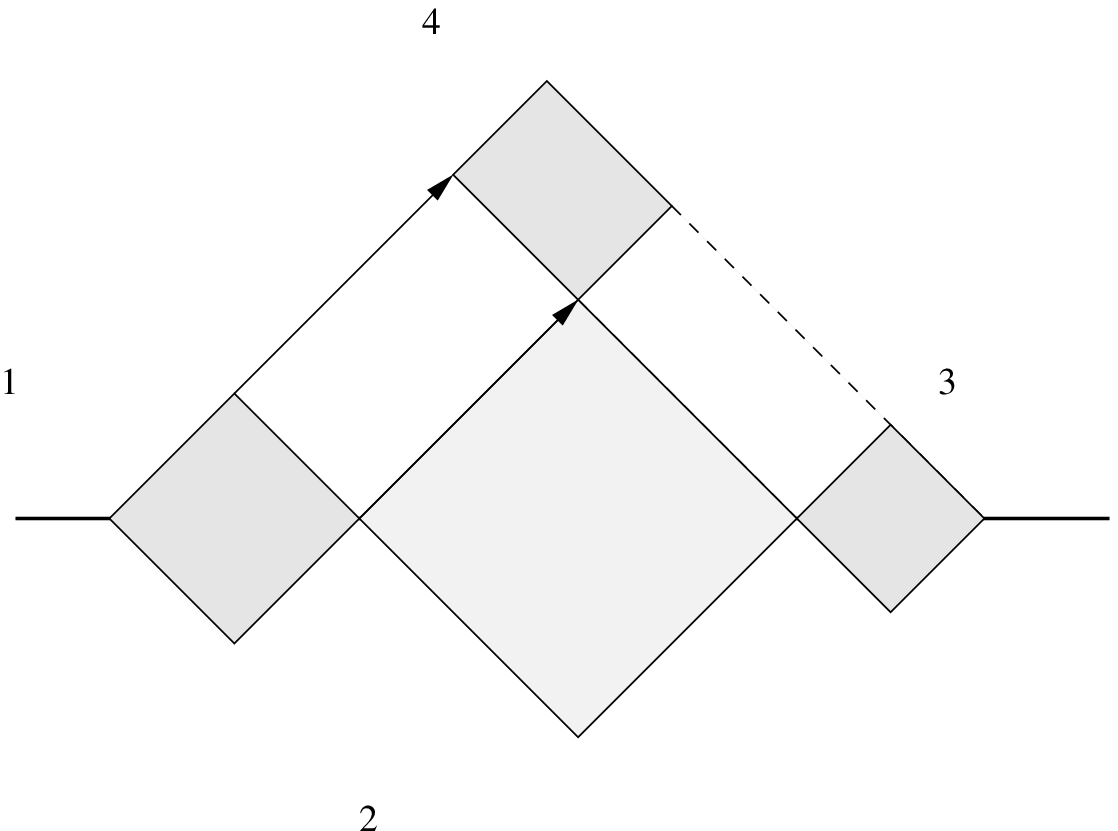,width=6cm} 
\refstepcounter{figure}\label{fig:shageo}%Zeigername f"ur die Abb.
\end{minipage}
\hfill\begin{minipage}{0cm} \end{minipage}
 \begin{minipage}{0cm} \end{minipage}
 \parbox{14cm}{\vspace{2ex} \centerline{\small {\bf \figurename{}
    \ref{fig:shageo}:} 
    Sharp geometrical action and timelike commutativity.}} 
\begin{minipage}{1cm} $\quad$\end{minipage}
%#5: Bildunterschrift
 \end{figure}}

Hence, in presence of sharp geometrical  action the distinction between
timelike and spacelike directions from locality disappears. We may
swap the time axis with the space axis. Then the original chiral local
algebras become {\em $x$-zero 
algebras} associated with intervals on the time axis. With this
definition the usual spectrum condition holds.
\end{pf}

Making some assumptions on the inclusion $\lok{A}_{max}\coset\lok{C}_
{max}\subset\lok{B}$  we give a straightforward
analysis in chapter \ref{cha:finind} which yields
a necessary condition on a possible sharp geometrical action: the
spectrum of $U^\lok{A}(\tilde{R}(2\pi))$ has to be 
contained in $\{\pm 1\}$. This condition excludes sharp geometrical
action for all current subalgebras known to the author, except the ones
one can make up trivially; section
\ref{cha:finind}.\ref{sec:cospactps} contains some examples of
current subalgebras.

A very special sub-geometrical action arises, if $\lok{B}$ is actually
generated by a \Name{Coset} pair $\lok{A}_{max}\coset\lok{C}_{max}$, ie
we have $e_{\lok{A}_{max}\smcoset\lok{C}_{max}} = \Einsop$. This we call a
{\em conormal} chiral subnet $\lok{A}_{max}\subset\lok{B}$. In this
case, obviously, we have a stronger version of 
sharp geometrical action which we shall call {\em completely sharp
geometrical action}:
\begin{equation}
  \label{eq:comsharpact}
  Ad_{U^\lok{A}(\tilde{g})} \lok{B}(I) \subset \lok{B}(I)\vee
  \lok{B}(\symb{p}(\tilde{g})I) \zendot
\end{equation}
It is not known, if or under which additional conditions sharp
geometrical action actually implies conormality of the subnet
$\lok{A}_{max}\subset\lok{B}$ and thus the character of $\lok{B}$ as a
tensor product of chiral conformal theories. But we have some
immediate remarks to make.

Obviously, completely sharp geometrical action implies sharp geometrical
action. Both are equivalent if  for every pair of proper
intervals $I_0,I_2$ for which
$\Seins\setminus\overline{I_0}\cup\overline{I_2}$ is the union of two
disjoint proper intervals $I_1,I_3$ the inclusion 
\begin{eqnarray}
  \label{eq:jowassub}
  \lok{B}(I_0)\vee\lok{B}(I_2)\subset\klammer{\lok{B}(I_1)\vee\lok{B}(I_3)}'
\end{eqnarray}
is actually an equality. Generically, the inclusion is non-trivial and
for completely rational $\lok{B}$ equality implies that all locally
normal representations of $\lok{B}$ are unitarily equivalent to its
vacuum representation \cite{KLM01}. This is seldom the case; however,
 $ \Loop{E(8)}_1$ is an example of this, 
cf section \ref{cha:finind}.\ref{sec:e81ctps}. Inclusions of the type
(\ref{eq:jowassub}) have been studied eg in
\cite{cS95,fX00a,KLM01}. The {\em index} of the inclusion in
(\ref{eq:jowassub})  is the {\em
  $\mu$-index} of $\lok{B}$ \cite{KLM01} (cf chapter \ref{cha:finind}).

Sharp geometrical action is weaker than conormality of
$\lok{A}_{max}\subset\lok{B}$. The $\opo$-dimensional conformal theory
defined by tensor products of free, massless, chiral fermions
$\psi_{L,R}$  
({\em bounded {Bose} fields}, cf \cite{kB97, kB99, GR00, khR97a})
fulfills timelike commutativity, but 
is manifestly not generated by the subtheories defined by its maximal chiral
observables, $\lok{A}_{L,R}^{max}$, in the sense of \Name{Rehren}
\cite{khR00}. Since the chiral fermion fields possess a stress-energy
tensor, this model may be reinterpreted in the sense of chiral subnets
and their \Name{Coset} models: we restrict it to the time axis and
look at the subnets induced by the inclusions of the maximal chiral
observables in the whole theory. 
This gives subnets with sharp
geometrical action of the respective inner-implementing
representations, but the inclusions are not conormal.
 It is 
doubtful, whether the restriction to $x$-zero 
algebras does yield a subnet $\lok{A}_{L}^{max}\subset\lok{B}$ with a
completely sharp geometrical action for $U^{\lok{A}^{max}_L}$, as
products $\psi_L(f)\otimes\psi_R(g)$, $supp(f)\subset I$,
$supp(g)\subset J$ do not seem to be approximated by
operators from $\lok{B}(I\times I)$ and
$\lok{B}(J\times J)$.

It is clear that the adjoint action of
$U^\lok{A}(\tilde{R}(2\pi))$ induces an automorphism of each local
algebra $\lok{B}(I)$, $I\Subset\Seins$, if $U^\lok{A}$ has sharp
geometrical action (see definition \ref{def:shageoact}). This improves
our knowledge on 
the net-endomorphic character of the action of $U^\lok{A}$ on
$\lok{B}$ (proposition \ref{prop:netend}), but it does not apply to
many subnets. A more relevant statement of the same type can be 
verified for a broad class of \Name{Coset} pairs (proposition
\ref{prop:AutchirRot}). 
  
%%%%%%%%%%%%%%%%%%%%%%%%%%%%%%%%%%%%%%%%%%%%%%%%%%%
%%%% ISOTONY PROBLEM %%%%%%%%%%%%%%%%%%%%%%%%%%%%%%
%%%%%%%%%%%%%%%%%%%%%%%%%%%%%%%%%%%%%%%%%%%%%%%%%%
\sekt{Solving the isotony problem}{Solving the isotony problem}{sec:isoprob}

%\section{Isotony problem}
%\label{sec:isoprob}

In this section we use the  {\sc Additional Assumption}   
to  solve the isotony  problem for the local relative
commutants $\lok{C}_I$ of an inclusion  of chiral conformal theories,
$\lok{A}\subset\lok{B}$. Once their isotony is proved, they are known
to coincide with the  local
algebras of the maximal \Name{Coset} model $\lok{C}_{max}$ associated
with $\lok{A}\subset \lok{B}$. This way, we
reach one of the  main goals of this work: the maximal \Name{Coset} model is
found to be of a local nature, ie it is determined completely by local
data. 

The isotony problem is solved in two steps: 
A crucial, purely group theoretical lemma (lemma \ref{lem:invvec})
admits a simple, accessible characterisation of the isotony problem,
which we give in proposition \ref{prop:chariso}. Aside of
being an intermediate step of our analysis, it illustrates the
character of the isotony problem.  The argument is 
completed by an application  of the {\sc Additional  Assumption}
(lemma \ref{lem:isoUA}) and
summarised in the main theorem  of this work (theorem \ref{th:main}).

\begin{prop}\label{prop:chariso}
  Assume  the unique inner-implementing representation $U^\lok{A}$
  associated with a chiral subnet 
  $\lok{A}\subset\lok{B}$ to have the net-endomorphism
  property. Referring to $I\Subset\Seins$, $e^c_I$ shall denote the projection
  onto the \Name{Hilbert} subspace which the local relative
  commutant $\lok{C}_I = \lok{A}(I)'\cap\lok{B}(I)$ generates from the
  vacuum. The     following are equivalent: 
  \begin{enumerate}
  \item \label{chisoi}For some pair $I,K$ of intervals satisfying
    $K\subsetneq I\Subset\Seins$ holds: $e^c_K\subset  e^c_I$. 
  \item \label{chisoii} $\lok{C}_{\Seins_+}   \subset
  \{U^\lok{A}(\tilde{D}(t)), \, t\in\dopp{R}\}'$. 
  \item \label{chisoiii} $\lok{C}_{max}(I) = \{U^\lok{A}(\tilde{g}),
  \, \tilde{g}\in\PSL(2,\dopp{R})^\sim\}'\cap
  \lok{B}(I) =  \lok{C}_I$, $I\Subset\Seins$.
  \end{enumerate} 
\end{prop}
Remark: The statement on the  cyclic projections is non-trivial
since, although the local relative commutants are manifestly covariant
with respect to $U$, the \Name{Reeh-Schlieder} theorem does not apply
due to the unclear status of isotony (cf eg \cite{hjB68}).

\begin{pf}
  The implications \ref{chisoiii} $\Rightarrow$  \ref{chisoi},
  \ref{chisoii} are obvious.  We start the proof proper with a
  discussion on \ref{chisoi} $\Rightarrow$ \ref{chisoiii}  and here we
  look at the case $I=\Seins_+$ (general case by covariance). We set
  $e^c_{\Seins_+}= e^c_+$. The inclusion $e^c_K\subset  e^c_+$ yields
  by the separating property of the vacuum and modular covariance of
  $\lok{C}_{\Seins_+}\subset \lok{B}(\Seins_+)$:  
  $\lok{C}_K\subset\lok{C}_{\Seins_+}$. Thus, any $g\in
  \PSL(2,\dopp{R})$ satisfying $g\Seins_+ =K$ leads to an operator
  $U(g)$  which leaves $e^c_+\Hilb{H}$ globally invariant. $g$ has the
  form $g=  S(n)T(s)D(t)$, $n,s\geq 0$. $g$ may be chosen such that  $t=0$.

By modular covariance $J$, the modular conjugation of
$\lok{B}(\Seins_+)$, and $e^c_+$ commute and, by covariance and the
\Name{Bisognano-Wichmann} property of $\lok{B}$,  $Ad_{JU(R(\pi))}$ induces
an automorphism of $\lok{C}_{\Seins_+}$, so $e^c_+$ commutes with
$U(R(\pi))$, too. The relations $JT(s)J=T(-s)$, $JS(n)J=S(-n)$ lead to
$U(S(-n))U(T(-s))e^c_+\Hilb{H}\subset e^c_+\Hilb{H}$. 
We assume $n,s>0$ and define 
\begin{displaymath}
  g(n,s) := S\klammer{-n \, \frac{ns+(1+ns)^2}{2+ns}} \,\,
  T\klammer{-s \, \frac{2+ns}{ns+(1+ns)^2}} \,\, \klammer{S(n)T(s)}^2
  \zendot 
\end{displaymath}
Applying scale covariance we arrive at: $U(g(n,s))e^c_+\Hilb{H}\subset
e^c_+\Hilb{H}$. The group element $g(n,s)$ leaves the point
$1\in\Seins$ invariant and is not a pure scale transformation. 
This proves that
all special conformal 
transformations leave $e^+_c$ invariant. The same follows for the
translations because of $R(\pi)S(n)R(\pi)=T(-n)$, which proves
$\komm{U(g)}{e^c_+}=0$    for all $g\in \PSL(2,\dopp{R})$ recognising
that translations and special conformal transformations generate the
whole group (see section \ref{cha:cospa}.\ref{sec:bosug}). For  $n=0$
or $s=0$ the last part 
applies directly. This proves: $e^c_K=e^c_+$ for all
$K\Subset\Seins$. By modular covariance of the inclusions
$\lok{C}_K\subset \lok{B}(K)$ we have $\lok{C}_K =
\{e^c_K\}'\cap\lok{B}(K)$ and this yields isotony for the local
relative commutants. The remainder follows by maximality of
$\lok{C}_{max}$ (lemma \ref{lem:cosmax}).

Finally we discuss the implication \ref{chisoii} $\Rightarrow$
\ref{chisoiii}. If $B\in \lok{B}(\Seins_+)$ commutes with
$U^\lok{A}(\tilde{D}(t))$, $t\in\dopp{R}$, then $B\Omega$ is invariant
under the action of all of 
$U^\lok{A}$ (lemma \ref{lem:invvec}). If $\tilde{g}$ is sufficiently
close to the identity, $Ad_{U^\lok{A}(\tilde{g})}(B)$ is a local
operator (proposition \ref{prop:netend}),
and the separating property of the vacuum proves that $B$ commutes
with all of $U^\lok{A}$. Thereby, we arrive at 
$\lok{C}_{\Seins_+}\subset \lok{C}_{max}(\Seins_+)$, provided the
assumption in \ref{chisoii} holds. The other inclusion is trivial.  
\end{pf}

Remark: The  dilatations $U^\lok{A}(\tilde{D}(t))$, $t\in\dopp{R}$, induce
automorphisms of $\lok{B}(\Seins_+)$ and the last part of the proof
shows  $\lok{C}_{max}(\Seins_+)$ to  be the fixed-point subalgebra with
respect to this automorphism group. Covariance leads to a
corresponding identification of  every $\lok{C}_{max}(I)$,
$I\Subset\Seins$. This may be regarded as an alternative ``local''
characterisation of $\lok{C}_{max}$, but since the automorphism groups
are determined by global observables, namely non-trivial unitaries
from $U^\lok{A}$, this is not satisfactory.

 Only for the final step of our analysis we need to invoke the
 {\sc Additional Assumption}  once again:
\begin{lemma}\label{lem:isoUA}
  Assume the {\sc Additional Assumption}  to hold. Then we have:\newline
  $U^\lok{A}(\tilde{D}(t))\in \lok{A}(\Seins_+)\vee
  \lok{A}(\Seins_-)$, $t\in\dopp{R}$, and $U^\lok{A}$ has the
  net-endomorphism property.   
\end{lemma}
\begin{pf}
 %  By lemma \ref{lem:diffD}, for small, fixed $t$ there exist diffeomorphisms
%   $g_\delta$, $g_\varepsilon$ localised in arbitrarily small
%   neighbourhoods of $+1\in\Seins$ and $-1\in\Seins$, respectively, and
%   diffeomorphisms $g_+$, $g_-$ which are localised in $\Seins_+$ and
%   $\Seins_-$, respectively, such that there
%   holds: $D(t)= g_+g_-g_\varepsilon g_\delta$.
%
% Setting $g_\varepsilon^{\tau_1}:= D(\tau_1) g_\varepsilon D(-\tau_1)$,
% $g_\delta^{\tau_2}:= D(-\tau_2) g_\delta D(\tau_2)$ we can maintain
% this product representation for $\tau_{1,2}$ ranging from $0$ to
% $\infty$; in the limit $\tau_{1,2}\rightarrow \infty$ the supports of
% $g_\varepsilon$, $g_\delta$ shrink to the points $-1$ and $+1$,
% respectively.
%
According to the {\sc Additional Assumption}  and lemma
\ref{lem:diffD} there exist, for small, fixed $t$, diffeomorphisms
  $g_\delta$, $g_\varepsilon$ localised in arbitrarily small
  neighbourhoods of $+1\in\Seins$ and $-1\in\Seins$, respectively, and
  diffeomorphisms $g_+^{\tau_1,\tau_2}$, $g_-^{\tau_1,\tau_2}$ which
  are localised in $\Seins_+$ and $\Seins_-$, respectively, and  phases
$\varphi(\tau_1,\tau_2)$ such that for $\tau_{1,2}\in\dopp{R}_+$:  
\begin{eqnarray*}
  U^\lok{A}(\tilde{D}(t))&=& \varphi(\tau_1, \tau_2) \, 
  \Upsilon^\lok{A}(\symb{p}^{-1}(g_+^{\tau_1,\tau_2})) \, 
  \Upsilon^\lok{A}(\symb{p}^{-1}(g_-^{\tau_1,\tau_2}))\\
  && \cdot \, Ad_{U^\lok{A}(\tilde{D}(\tau_1))}
    (\Upsilon^\lok{A}(\symb{p}^{-1}(g_\varepsilon)))  \,   
  Ad_{U^\lok{A}(\tilde{D}(-\tau_2))}
    (\Upsilon^\lok{A}(\symb{p}^{-1}(g_\delta))) \,\, .  
\end{eqnarray*}

Following \Name{Roberts} \cite[corollary 2.5]{jR74a}, dilatation
invariance of the vacuum and the shrinking supports ensure that the last
two operators converge weakly to their vacuum expectation values in the
limit $\tau_{1,2}\rightarrow \infty$. We rewrite the equation above:
%%%%
\begin{eqnarray}
&&
  Ad_{U^\lok{A}(\tilde{D}(\tau_1))}(\Upsilon^\lok{A}(\symb{p}^{-1}(g_\varepsilon)))  
  Ad_{U^\lok{A}(\tilde{D}(-\tau_2))}(\Upsilon^\lok{A}(\symb{p}^{-1}(g_\delta)))
  U^\lok{A}(\tilde{D}(t))^* \nonumber\\
\label{eq:weakscalconv}
&=&\overline{\varphi(\tau_1, \tau_2)}
\Upsilon^\lok{A}(\symb{p}^{-1}(g_+^{\tau_1,\tau_2}))^*
  \Upsilon^\lok{A}(\symb{p}^{-1}(g_-^{\tau_1,\tau_2}))^*  \zendot
\end{eqnarray}
The operators to the right converge weakly by this equation in the
limit $\tau_1, \tau_2 \rightarrow \infty$. For small
$t$, $g_\varepsilon$ and $g_\delta$ may be chosen 
close to the identity, $\omega(.)$ is continuous and normalised, which
means that for  $g_\varepsilon, g_\delta \approx id$ we have
$\omega(\Upsilon^\lok{A}(\symb{p}^{-1}(g_\varepsilon)))\neq 0$, 
$\omega(\Upsilon^\lok{A}(\symb{p}^{-1}(g_\delta)))\neq 0$.  
This implies  $U^\lok{A}(\tilde{D}(t))\in \lok{A}(\Seins_+)\vee
  \lok{A}(\Seins_-)$ for small and hence for all $t$.

Because $\Upsilon^\lok{A}(\symb{p}^{-1}(g_+^{\tau_1,\tau_2}))$ and
$\Upsilon^\lok{A}(\symb{p}^{-1}(g_-^{\tau_1,\tau_2}))$ are unitary
operators, the right-hand side of equation (\ref{eq:weakscalconv})
converges, up to a phase, strongly against $U^\lok{A}(\tilde{D}(t))$
for small $t$. This strong convergence proves that for
$B\in\lok{B}(\Seins_+)$ and small $t$ holds true in the weak topology:
\begin{equation}
  \label{eq:denetend}
  U^\lok{A}(\tilde{D}(t)) B U^\lok{A}(\tilde{D}(t))^* =
  \lim_{\tau_1,\tau_2\rightarrow \infty}
  Ad_{\Upsilon^\lok{A}(\symb{p}^{-1}(g_+^{\tau_1,\tau_2}))} (B) \in
  \lok{B}(\Seins_+) \,\, .
\end{equation}
This establishes the net-endomorphism property (definition \ref{def:netend}).
\end{pf}

The first statement of this lemma holds trivially, if the global
algebra $\lok{A}$      coincides with
$\lok{A}(\Seins_+)\vee\lok{A}(\Seins_-)$. This is a desirable
property, eg for the {\em {Connes}' fusion} approach to superselection
structure (cf eg \cite{aW98}), and it holds true in presence of strong
additivity, but a proof of it relying on general properties of chiral
conformal subtheories seems out of reach.

However, it seems  natural for the  representatives
$U^\lok{A}(\tilde{D}(t))$, $t\in\dopp{R}$, to be contained in
$\lok{A}(\Seins_+)\vee\lok{A}(\Seins_-)$. One may  compare this with  the
construction of the inner implementation 
of translations in $\bigvee_{I\Subset\dopp{R}} \lok{A}(I)$ using the
spectrum condition \cite{hjB66}. Here, the  translations operate as
geometrical automorphism
group on the net of local \Name{v.Neumann} algebras associated with
localisation regions which do not contain the fixed-point of translations and
they are implementable by a unitary group contained in the
\Name{v.Neumann} algebra which is generated by this net. The action of
$U^\lok{A}(\tilde{D}(.))$ on $\lok{A}(\Seins_+)\vee\lok{A}(\Seins_-)$
has the same character in this respect.

The proof of lemma \ref{lem:isoUA} relies on the structure of $\Diff_+(\Seins)$
in order to establish convergence by equation 
(\ref{eq:weakscalconv}). This procedure does not apply to the
representation of dilatations through products of an exponentiated
smeared stress-energy tensor as in  equation (\ref{eq:SETnetend}).
If one  seeks  for nets of test functions  $f_\delta$,
$f_\varepsilon$ whose supports shrink to the  points $+1$ and  $-1$,
respectively, while the  properties needed for establishing equation
(\ref{eq:SETnetend})  are upheld, one lacks  control  
over the vacuum expectation values of $exp(-it
  \closure{\Theta^\lok{A}(f_\varepsilon)})$, $exp(-it
  \closure{\Theta^\lok{A}(f_\delta)})$. Such control could allow to
  establish 
  non-zero weak limit points for this procedure\footnote{Compare discussion in
  \cite[section 4]{BS90} on    strong additivity of the $U(1)$-current
  algebra.} 
  and thus to give an alternative proof for lemma \ref{lem:isoUA}
  doing without 
  the  structure of $\Diff_+(\Seins)$.

We summarise and state the main result of this work, which
proves that the maximal \Name{Coset}  models are of a {\em local nature}:
\begin{theo}\label{th:main} Let 
  $\lok{A}\subset\lok{B}$ be a chiral conformal subtheory and suppose
  the {\sc Additional Assumption}  to hold. Then the 
  unique  inner-implementing representation $U^\lok{A}$ has the
  net-endomorphism property and for all $I\Subset\Seins$ there holds:
  \begin{displaymath}
    \lok{C}_{max} (I) = \{U^\lok{A}(\tilde{g}), \,
    \tilde{g}\in\PSL(2,\dopp{R})^\sim \}' \cap \lok{B}(I) = 
    \lok{A}(I)' \cap \lok{B}(I) = \lok{C}_I \,\,. 
  \end{displaymath}
\end{theo}

\begin{pf}
  The net-endomorphism property of $U^\lok{A}$ holds by lemma
  \ref{lem:autUA} (and lemma \ref{lem:isoUA}), and $\ref{chisoii}$ in
  proposition \ref{prop:chariso} 
  is fulfilled  because of lemma \ref{lem:isoUA}.
\end{pf}

 In the cases where both $\lok{A}$ and $\lok{B}$ possess
  an integrable stress-energy tensor, and hence $\lok{C}_{max}$ alike,
  the theorem means in particular: $\lok{A}_{max}(I)$ and $\lok{C}_
  {max}(I)$, $I\Subset\Seins$, are their mutual relative
  commutants in $\lok{B}(I)$. We have learned that  
  the \Name{Coset} pair
  $\lok{A}_{max}\coset\lok{C}_{max}\subset\lok{B}$ is a  typical object
  for studies on the structures related to chiral subnets an their
  \Name{Coset} models. 
For these typical \Name{Coset} pairs  we
  introduce a new term reflecting the absence of the isotony problem
  both for the subnet and for the 
  \Name{Coset} model:
  \begin{defi}\label{def:normcos}
    A \Name{Coset} pair $\lok{A}\coset\lok{C}\subset \lok{B}$ is
    called {\bf normal} if for all $I\Subset\Seins$ the algebras
    $\lok{A}(I)$ and  $\lok{C}(I)$ are their mutual relative
    commutants in $\lok{B}(I)$. 
  \end{defi}
For normal \Name{Coset} pairs $\lok{A}\coset\lok{C}\subset\lok{B}$ the
  local inclusions are
  automatically {\em irreducible}, ie the 
  relative commutant of $\lok{A}\coset\lok{C}(I)$ in $\lok{B}(I)$ is
  $\dopp{C}\Einsop$, because the local algebras
  $\lok{A}(I)$ are factors.

The main theorem proves the conclusions of \Name{Rehren} \cite{khR00} to
hold true which rely on the {\em generating property} of nets of chiral
observables, if the $\opo$-dimensional theory contains a stress-energy
tensor in the sense of the \Name{L\"uscher-Mack} theorem
\cite{FST89}. Since such a stress-energy tensor factorises into its 
independent chiral components, our analysis applies directly. The
{\em generating property} introduced  in \cite{khR00} resisted 
attempts of proof even in presence of a stress-energy tensor,
unfortunately.

\chap{{Coset} pairs of finite Index}{{Coset} pairs of finite
Index}{cha:finind} 
%\chapter[On cases of finite Index]{On cases of finite Index}
%\label{cha:finind}

In this chapter we want to make contact with the frameworks of {\em nets of
subfactors} 
\cite{LR95} and of normal {\em canonical tensor product subfactors}
\cite{khR00}.   
The characteristic assumption for this purpose will be that
the inclusion $\lok{A}\coset\lok{C}\subset \lok{B}$ is of finite
index; this property is discussed in the first section, and some
simple statements on the local inclusions
$\lok{A}\coset\lok{C}(I)\subset \lok{B}(I)$ are derived. We fix some
notation on the way.    

In the second section, we look at the net of inclusions in the
$\opo$-dimensional quasi-theory $\lok{B}^{\opo}$ and derive their
character 
as covariant, 
localisable representations of $\lok{A}\otimes\lok{C}$ along the lines
of \cite{rL01}, where 
the corresponding situation for chiral subnets is
discussed. We exploit well-known structures of
the theory of superselection sectors in low dimensions
\cite{FRS89,FRS92}  in order to improve our
knowledge of the net-endomorphic action of $U^\lok{A}$ and we
determine the possible spectrum of $U^\lok{A}(\tilde{R}(2\pi))$ in case
$U^\lok{A}$ has sharp geometrical action.

In the third section we discuss some examples of \Name{Coset} pairs,
illustrating the structures developed more abstractly in the 
previous chapters and sections, and we interpret results of  studies on current
subalgebras \cite{GKO86,KW88,KS88} such that we get examples of normal
canonical tensor product subfactors which are \Name{Coset} pairs. 

%%%%%%%%%%%%%%%%%%%%%%%%%%%%%%%%%%%%%%%%%%%%%%%%%%%%
%%%%%% LOCAL CosPa INCLUSIONS OF FINITE INDEX %%%%%%
%%%%%%%%%%%%%%%%%%%%%%%%%%%%%%%%%%%%%%%%%%%%%%%%%%%%

\sekt{On local inclusions of \Name{Coset} pairs}{On local inclusions
  of {Coset} pairs}{sec:loccospa} 

As discussed in section
\ref{cha:cospa}.\ref{sec:disass} finiteness of index is presumably not
a generic 
feature of chiral subnets. Nevertheless, there are rich structures
associated with inclusions of finite index, and the theory 
on these is advanced and promises to teach us something regarding
the general situation as well.  In particular, a large number of
current subalgebras $\lok{A}\subset\lok{B}$ are known to be {\em
  cofinite}, ie the \Name{Coset} pair
$\lok{A}\coset\lok{C}_{max}\subset\lok{B}$ is known to have finite
index, and hence the methods related to finite index inclusions yield  
interesting results on \Name{Coset} models.

While there was not much need to introduce the machinery on modular
covariant subalgebras to the full in the previous chapters, we now
have to give the related structures in more detail. Rather
than duplicating existing summaries, we just state the relevant facts
in a form needed here and fix our notation on the way. The general
reference on what is to come is the seminal work  of \Name{Longo and
  Rehren} on {\em nets of subfactors} \cite{LR95}.  In
this section we will be concerned with local inclusions of
\Name{Coset} pairs, $\lok{A}\coset\lok{C}(I)\subset\lok{B}(I)$. 

Let $\lok{N}\subset\lok{M}$ be an inclusion of \Name{v.Neumann}
algebras with common unit $\Einsop$, $\Omega$ a cyclic and separating
vector for $\lok{M}$. $\lok{N}$ is a {\em modular
  covariant subalgebra} of $\lok{M}$, if it is globally invariant with
respect to the modular group $\sigma$ of $(\lok{M},\Omega)$:
\begin{equation}
  \label{eq:modinvalg}
  \sigma_t(\lok{N})= \Delta^{it}_\lok{M}\lok{N}\Delta^{-it}_\lok{M} =
  \lok{N}\, , 
  \quad \forall t\in\dopp{R} \zendot
\end{equation}
$\Delta_\lok{M}$ stands for the modular operator of $(\lok{M},\Omega)$; the
corresponding modular conjugation will be denoted $J_\lok{M}$.
The assumption of conformal covariance of chiral subnets in
definition \ref{def:chsubnet} implies that the local subalgebras are
modular covariant because of the \Name{Bisognano-Wichmann} property
 of $\lok{B}$. 

Modular covariance has remarkable consequences, mostly due to a
theorem of \Name{Takesaki} \cite{mT72} which states that there is a
normal, faithful conditional expectation
$\mu:\lok{M}\rightarrow\lok{N}$ leaving invariant the state
$\omega(.)=\skalar{\Omega}{.\Omega}$, ie $\omega\circ\mu= \omega$. The
latter property determines $\mu$ uniquely (see below 
\ref{implpropmu} on page \pageref{implpropmu}).

There are equivalent ways to define conditional expectations
on \Name{v.\-Neu\-mann} algebras (see eg \cite[chapter II, \S
9]{sS81}). We prefer to state one definition with a list of useful
properties of conditional expectations taken from \cite{KK92}. A
{\em conditional expectation} $\mu:\lok{M}\rightarrow\lok{N}$ is a
{\em projection} from 
$\lok{M}$ onto $\lok{N}$ of norm $1$, ie a 
linear mapping satisfying $\mu^2=\mu$ and $\|\mu\|=1$. Such $\mu$
has the following properties:
\begin{enumerate}
\item Idempotency: $\mu^2=\mu$,
\item Bimodule property: $\mu(n_1mn_2) = n_1 \mu(m) n_2 \, , \,\,
  m\in\lok{M}\, , \,\,   n_1, n_2\in\lok{N}$,
\item Positivity: $\mu(m^*m)\geq 0$,   $m\in\lok{M}$,
\item $*$-property: $\mu(m)^* = \mu(m^*)$, $m\in\lok{M}$,
\item Normalisation: $\mu(\Einsop)= \Einsop$,
\item \Name{Schwarz} property: $\mu(m^*)\mu(m)\leq\mu(m^*m)$, $m\in\lok{M}$.
\end{enumerate}
% The bimodule property may be formulated equivalently as: $\mu(n^*mn)=
% n^* \mu(m) n$; this is readily verified by the familiar polarisation
% identity for conjugate-bilinear mappings.  

$\mu$ is called {\em faithful} if $\mu(m^*m)=0$ is possible for
$m=0$ only. $\mu$ is a {\em normal} conditional expectation if it is normal
%(weak-$*$ continuous)
 as a linear map. 

Examples of conditional expectations are group means for fixed-point
inclusions with respect to a finite or compact automorphism group. If
the inner-implementing representation $U^\lok{A}$ associated with a
chiral subnet $\lok{A}\subset\lok{B}$ has the net-endomorphism
property, a result of \Name{Frigerio}  \cite{aF78} gives explicit formulae
for the faithful, normal conditional expectations of the inclusions
$\lok{A}_{max}(I)\subset\lok{B}(I)$, 
$\lok{C}_{max}(I)\subset\lok{B}(I)$ which  leave
invariant the vacuum. We state the corresponding formula for the
inclusion $\lok{C}_{max}(\Seins_+)\subset\lok{B}(\Seins_+)$:
\begin{equation}
  \label{eq:frigerio}
  \mu_{\lok{C}_{max}(\Seins_+)\subset\lok{B}(\Seins_+)} (b) = w^*-\lim_{\lambda\searrow 0} \lambda
  \int_0^\infty \!\!dt \,\, e^{-\lambda t} \, Ad_{U^\lok{A}(T(t))}(b)\, , \quad
  b\in \lok{B}(\Seins_+)\,\,.   
\end{equation}
One may use the action of dilatations in a similar manner
({\em mean ergodic theorem}, cf eg \cite{eL76,dP83}).

Normal, faithful conditional expectations $\mu:\lok{M}\rightarrow\lok{N}$
which leave the state $\omega$ invariant
have remarkable properties (see eg \cite{KK92},
\cite[theorem 6.4]{hjB97}, \cite[lemma VI.1.2]{hjB00}). We assume in
addition 
$\lok{M}, \lok{N}$ to be factors and $\Hilb{H}$ to be separable. If we
denote 
the projection onto $\Hilb{H}_\lok{N}= \overline{\lok{N}\Omega}$ by
$e_\lok{N}$, then we have:
\begin{enumerate}
\item $e_\lok{N}$ commutes with $\Delta_\lok{M}^{it}$, $t\in\dopp{R}$, and $J_\lok{M}$.
\item $\Delta_\lok{M} e_\lok{N}$ and $J_\lok{M}e_\lok{N}$ are the modular data of
  $\lok{N}e_\lok{N}$ on $\Hilb{H}_\lok{N}$.
\item We have $e_\lok{N}\lok{N}'e_\lok{N} = J_\lok{M}\lok{N}J_\lok{M}e_\lok{N}$.
\item It holds true: $\lok{N}= \{e_\lok{N}\}'\cap\lok{M}$.
\item \label{lokiso} The map $\lok{N}\ni n\mapsto ne_\lok{N}$ defines an isomorphism
  of \Name{v.Neumann} algebras. We have $\norm{n}=\norm{ne_\lok{N}}$.
\item \label{implpropmu} For given $m\in\lok{M}$, the unique solution
  of the equation  $e_\lok{N}me_\lok{N}=ne_\lok{N}$, $n\in\lok{N}$, is
  $n=\mu(m)$.
\end{enumerate}
%\ref{implpropmu} follows from an argument of \Name{Jones} \cite{vJ83}
%as well. 

 The index of a normal, faithful conditional
expectation $\mu:\lok{M}\rightarrow\lok{N}$, $\lok{M},\lok{N}$
infinite-dimensional factors, is defined as follows 
%(cf eg \cite[section 2]{LR95})
:
\begin{equation}
  \label{eq:condexind}
I_\mu := \inf\menge{\lambda: \lambda\geqslant 1\,, \,\,
\mu(m^*m)\geq \lambda^{-1} m^*m\,\, \forall m\in\lok{M}}
  \zendot
\end{equation}
If $I_\mu<\infty$ for some $\mu$, then there is a unique normal,
faithful conditional 
expectation $\mu_0:\lok{M}\rightarrow\lok{N}$ which has a minimal
index $I_{\mu_0}$. This is used to define the index of the subfactor
$\lok{N}\subset \lok{M}$: 
\begin{equation}
  \label{eq:defindsubf}
  [\lok{M}:\lok{N}]:=I_{\mu_0} = \inf_\mu I_\mu \zendot
\end{equation}

If $\lok{N}\subset\lok{M}$ is {\em irreducible}, ie the {\em relative
  commutant} $\lok{N}^c:=\lok{N}'\cap\lok{M}$ of $\lok{N}$ in
$\lok{M}$ is trivial,  $\lok{N}^c=\dopp{C}\Einsop$, then there is at most
one normal, faithful conditional expectation
$\mu:\lok{M}\rightarrow\lok{N}$ \cite[theorem 1.5.5]{aC73}. Local
  inclusions of chiral subnets of finite index are irreducible, as we 
shall see below (lemma \ref{lem:lokirredincl}), and for this reason
  their respective unique 
normal, faithful conditional expectation is automatically minimal.

Central for the studies on (nets of) subfactors is the {\em dual
  canonical endomorphism} $\rho: \lok{N}\rightarrow \lok{N}$
% (see \cite{rL89,rL90,LR95})
. If 
$\lok{N}$ and $\lok{M}$ are both properly infinite \Name{v.Neumann}
algebras on a separable \Name{Hilbert} space, there are
vectors $\phi$ which are cyclic and separating for both $\lok{N}$ and
$\lok{M}$ \cite{DL84}. Taking the modular conjugations
$J_{(\lok{N},\phi)}$ and $J_{(\lok{M},\phi)}$ corresponding to the
pairs $(\lok{N},\phi)$ and $(\lok{M},\phi)$, respectively, one defines
the {\em canonical endomorphism} $\gamma: \lok{M}\rightarrow
\lok{N}\subset\lok{M}$ by: 
\begin{equation}
  \label{eq:canenddef}
  \gamma(m) :=
  J_{(\lok{N},\phi)}J_{(\lok{M},\phi)}mJ_{(\lok{M},\phi)}J_{(\lok{N},\phi)}\,
  , \,\, m\in\lok{M} \zendot
\end{equation}
$\gamma$ depends only through a unitary conjugation in $\lok{N}$ on
the choice of $\phi$ \cite{rL87}. The restriction of $\gamma$ to
$\lok{N}$ defines the dual canonical endomorphism $\rho$. For
inclusions of infinite factors $\lok{N}\subset\lok{M}$ with finite
index $[\lok{M}:\lok{N}]$ one defines the {\em dimensions} $d_\rho$,
$d_\gamma$ of $\rho$, $\gamma$ to be:
\begin{equation}
  \label{eq:defdimrho}
  d_\rho := [\lok{N}:\rho(\lok{N})]^{\frac{1}{2}}\, ,\,\, d_\gamma :=
  [\lok{M}:\gamma(\lok{M})]^{\frac{1}{2}}  \zendot
\end{equation}
%For a subfactor $\lok{N}\subset\lok{M}$ 
We have: $d_\rho=d_\gamma=[\lok{M}:\lok{N}]$. 

Regarding a chiral subnet $\lok{A}\subset\lok{B}$ one finds that the
dual canonical endomorphism $\rho_I:\lok{A}(I)\rightarrow
\rho_I(\lok{A}(I))$, taken as an endomorphism of the isomorphic
algebra $\lok{A}(I)e_\lok{A}$,  may be extended to a \Name{DHR}
endomorphism $\rho$ of 
$\lok{A}e_\lok{A}$ localised in $I$. The \Name{DHR}
endomorphism $\rho$ induces a localised representation of
$\lok{A}e_\lok{A}$ which is unitarily equivalent to the
representation of $\lok{A}e_\lok{A}$ by the embedding
$\lok{A}\subset\lok{B}$ and denoted $\rho$ as well. This is shown in
\cite{rL01} (mostly relying on results in \cite{LR95}) and we will
give corresponding arguments in $\opo$ dimensions in the next
section.  

The dimension of the 
representation $\rho$ is defined as:
\begin{equation}
  \label{eq:defindrep}
  d(\rho):=
  [\rho_{I'}(\lok{A}(I')e_\lok{A})':
  \rho_I(\lok{A}(I)e_\lok{A})]^{\frac{1}{2}}   
  = [\lok{A}(I):\rho_I(\lok{A}(I))]^{\frac{1}{2}} = d_{\rho_I}  \zendot
\end{equation}
By the index-statistics theorem \cite{GL96} this dimension coincides with
the statistical dimension of the \Name{DHR} endomorphism
$\rho$. The index  
$[\lok{B}(I):\lok{A}(I)]$ is found to be constant for all
$I\Subset\Seins$ \cite{rL01}. The same arguments apply, with minor
alterations, to the $\opo$-dimensional situation and we refrain from
giving a proof that the index is constant in this setting as well.

The following lemma is a preparation for a simple statement on
\Name{Coset} pairs of 
finite index (proposition \ref{prop:UACconf}); it shows in
particular that a chiral subnet 
$\lok{A}\subset\lok{B}$ with finite index is a conformal inclusion in
the sense of definition \ref{def:confinc}. 

\begin{lemma}\label{lem:lokirredincl} 
  Assume $\lok{A}\coset\lok{C}\subset\lok{B}$ to be a \Name{Coset}
  pair of finite index. Then the local inclusions
  $\lok{A}\coset\lok{C}(I)\subset\lok{B}(I)$, $I\Subset\Seins$, are
  irreducible, as are 
  the inclusions $\lok{A}(I)\subset\lok{A}(I)^{cc}$,
  $\lok{C}(I)\subset\lok{A}(I)^c$. 
\end{lemma}

\begin{pf}
  Using formulae on the indices of  a sequence of inclusions and for
  tensor products of such we have \cite{rL90, rL89}:
  \begin{eqnarray}
    \label{eq:indexseqtensori}
    [\lok{B}(I):\lok{A}\coset\lok{C}(I)] &\leqslant&
    [\lok{B}(I):\lok{A}(I)^{cc}\otimes\lok{A}(I)^{c}]\cdot\nonumber\\
    &&\cdot \, 
    [\lok{A}(I)^{cc}\otimes\lok{A}(I)^{c}:\lok{A}\coset\lok{C}(I)] \,\,
    ,\\
    \label{eq:indexseqtensorii}
    [\lok{A}(I)^{cc}\otimes\lok{A}(I)^{c}:\lok{A}\coset\lok{C}(I)]&=&
    [\lok{A}(I)^{cc}:\lok{A}(I)] [\lok{A}(I)^{c}:\lok{C}(I)] \,\, .
  \end{eqnarray}
The inequality (\ref{eq:indexseqtensori}) is saturated, if
$\lok{A}\coset\lok{C}(I)\subset\lok{B}(I)$ is irreducible. This is the
case: the index $[\lok{B}(I):\lok{A}\coset\lok{C}(I)]$ is finite and
hence the relative commutant of $\lok{A}\coset\lok{C}(I)$ in
$\lok{B}(I)$ is finite dimensional \cite{rL90}. This
finite-dimensional algebra is modular covariant and hence, according
to lemma \ref{lem:diltransspace}, all vectors in
$\lok{A}\coset\lok{C}(I)^c\Omega$ are invariant with 
respect to the translation group $U(T_I(.))$ (cf proof of proposition
\ref{prop:netend}), but this means that they are fixed by all of $U$
(lemma \ref{lem:invvec}). Uniqueness of the vacuum and its separating
property show that the 
relative commutant is trivial. The other inclusions are
irreducible by the same argument. 
\end{pf}

%%%%%%%%%%%%%%%%%%%%%%%%%%%%%%%%%%%%%%%%%%%%%%%%%%%%%%%%%%%
%%%%%% harmonic analysis of local inclusions %%%%%%%%%%%%%%%
%%%%%%%%%%%%%%%%%%%%%%%%%%%%%%%%%%%%%%%%%%%%%%%%%%%%%%%%%%%

Inclusions of infinite subfactors $\lok{N}\subset\lok{M}$ with finite
index may be characterised in terms of intertwiners between
endomorphisms. 
  If $\rho, \sigma$ are endomorphisms of  $\lok{M}$, then an
  intertwiner from $\rho$ to $\sigma$ is a bounded
operator $T$ which satisfies for all $m\in\lok{M}$: $T\rho(m)=
\sigma(m) T$. This property is indicated by the notation  $T:\rho
\rightarrow  \sigma$.
The following may be found eg in \cite{LR95,RST96,khR94a}, where some
more useful identities on 
the intertwiners are stated as well. We take $\lok{N}\subset\lok{M}$ to be
irreducible \label{vwchar}:

\noindent The subfactor $\lok{N}\subset\lok{M}$ is completely
determined by the triple 
$(\gamma,v,w)$ where $\gamma$ is an endomorphism of $\lok{M}$,
$v,w\in\lok{M}$ are isometries, $v: id 
\rightarrow \gamma$, $w: \gamma \rightarrow \gamma^2$
and the following identities are satisfied:
\begin{enumerate}
\item \label{vwidi} $w^*v=[\lok{M}:\lok{N}]^{-\frac{1}{2}} \Einsop = w^*\gamma(v)$,
\item \label{vwidii}$ww^*=\gamma(w^*)w$,
\item \label{vwidiii}$ww=\gamma(w)w$.
\end{enumerate}
One may reconstruct $\lok{N}$ as the image of $\mu(.)=
w^*\gamma(.)w$. Every $m\in\lok{M}$ is of  the form:
\begin{equation}
  \label{eq:harmanalok}
  m= [\lok{M}:\lok{N}]\,\, \mu(mv^*)\,v = [\lok{M}:\lok{N}]\,\, v^*\,\mu(vm)\,.
\end{equation}

The dual canonical endomorphism $\rho$ has a finite decomposition into
irreducible, inequivalent endomorphisms $\rho_s$,
$\rho\cong\bigoplus_sN_s\rho_s$. For every $\rho_s$ there is a
complete orthonormal set of intertwiners
$\lok{N}\ni w^i_s:\rho_s\rightarrow\rho$, $i=1,\ldots, N_s$,
$\sum_{is}w^i_sw^i_s{}^*=\Einsop$, 
$w^i_s{}^*w^j_t=\delta_{st}\delta^{ij}\Einsop$. The second equation
defines a scalar product on the intertwiner spaces. 

The definition $\psi^i_s:=w^i_s{}^*v$ yields an anti-isomorphism
between the intertwiner spaces of the $w^i_s$ and the space of {\em
  charged intertwiners}, spanned by the $\psi^i_s:id \rightarrow
\rho_s$ (\cite[lemma 4.5]{LR95}, \cite[proposition
2]{khR94a}). By irreducibility of $\lok{N}\subset \lok{M}$, there is a
 scalar product on the space of charged intertwiners:
\begin{equation}
  \label{eq:defscalpchinter}
  \,\,\skalar{\psi^{i_1}_{s_1}}{\psi^{i_2}_{s_2}}\Einsop \equiv
\psi^{i_1}_{s_1}{}^*\psi^{i_2}_{s_2}\,\,
\in\lok{N}'\cap\lok{M}=\dopp{C}\Einsop\,. 
\end{equation}
 Equation (\ref{eq:harmanalok})
leads to a decomposition 
of $\lok{M}$ with respect to the $\rho_s$:
\begin{equation}
  \label{eq:harmanalokpsi}
  m= \sum_{s,i} \,\, [\lok{M}:\lok{N}] \, \mu(m\psi^i_s{}^*) \,\, \psi^i_s \zendot
 % \equiv \sum_{s,i} n_{s,i}(m) \psi^i_s
\end{equation}
This formula is the ground for the following proposition.

\begin{prop}\label{prop:UACconf} Let
  $\lok{A}\coset\lok{C}\subset\lok{B}$ be a \Name{Coset} pair of finite
  index. Assume the corresponding inner-implementing 
  representations $U^\lok{A}$, $U^\lok{C}$ to have the
  net-endomorphism property. Then the local inclusions
  $\lok{A}\coset\lok{C}(I)\subset\lok{B}(I)$ are irreducible,
  $\lok{A}_{max}(I)$ and $\lok{C}_{max}(I)$ are their mutual relative
  commutants and we have: $U\circ\symb{p}= U^\lok{A}U^\lok{C}$. 
\end{prop}
\begin{pf}
The irreducibility of $\lok{A}\coset\lok{C}(I)\subset\lok{B}(I)$ was
shown in lemma \ref{lem:lokirredincl}.

Exploiting covariance, we restrict our attention to the case
$I=\Seins_+$. 
We denote elements in $\lok{A}\coset\lok{C}(\Seins_+)$ by $[ac]$.  
Each $b\in \lok{B}(\Seins_+)$ has the form
$b=\sum_{s,i}[ac]_{s,i}\psi^i_s$. By assumption,
$Ad_{U^{\lok{A}'\cap\lok{C}'}}(\tilde{D}(\tau))$ (see equation
(\ref{eq:UACrem}) for 
the definition of $U^{\lok{A}'\cap\lok{C}'}$) defines an automorphism
of $\lok{B}(\Seins_+)$, which acts trivial on
$\lok{A}\coset\lok{C}(\Seins_+)$ and leaves globally invariant the
finite-dimensional spaces of charged intertwiners
$\psi^i_s$. According to the lemmas \ref{lem:diltransspace} and
\ref{lem:invvec} the vectors 
$\psi^i_s\Omega$ are invariant with respect to
$U^{\lok{A}'\cap\lok{C}'}$.  The \Name{Reeh-Schlieder} theorem on the
chiral theory $\lok{B}$ and the decomposition $b=\sum_{s,i}[ac]_{s,i}\psi^i_s$
for each $b\in \lok{B}(\Seins_+)$ yields
$U^{\lok{A}'\cap\lok{C}'}=\Einsop\,\Leftrightarrow
U\circ\symb{p}=U^\lok{A}U^\lok{C}$. 

The same line of argument works for the inclusion
$\lok{A}_{max}(\Seins_+)\subset\lok{A}(\Seins_+)^{cc}$, where it proves
every element in $\lok{A}(\Seins_+)^{cc}$ to be invariant with respect
to $Ad_{U^\lok{C}}$ and hence to be  contained in
$\lok{A}_{max}(\Seins_+)$ (invoking proposition \ref{prop:netend}). The proof
is complete, if we apply the same routine to
$\lok{C}_{max}(\Seins_+)\subset\lok{A}(\Seins_+)^{c}$. 
\end{pf}

Proposition \ref{prop:UACconf} establishes a complete analogy
between  the \Name{Coset} construction of
stress-energy tensors for current subalgebras of \Name{Goddard, Kent,
  Olive} \cite{GKO86} and the \Name{Borchers-Sugawara} representations
of a \Name{Coset} 
pair of finite index: The identity $U\circ\symb{p}=
U^\lok{A}U^\lok{C}$ shows that $U$ possesses a factorisation  with
respect to the \Name{Coset} pair $\lok{A}\coset\lok{C}\subset\lok{B}$
which is generated by commuting local observables (cf discussion in
section \ref{cha:cospa}.\ref{subsec:bosappl}).
Furthermore, it shows that the net-endomorphism property is sufficient
to solve the isotony problem for \Name{Coset} pairs of finite index
and to prove that the maximal \Name{Coset} pair is normal.

The proof of proposition \ref{prop:UACconf} is similar to the
ones in the
well-known situation of current subalgebras for which the vacuum
representation of the larger one decomposes into finitely many
irreducible representations when restricted to the subalgebra (see eg
\cite[corollary 3.2.1]{KW88}): loosely speaking, the assumption of
 finite index trivialises the problem, since there
are no non-trivial finite-dimensional, unitary representations of
$\PSL(2,\dopp{R})^\sim$. 

%%%%%%%%%%%%%%%%%%%%%%%%%%%%%%%%%%%%%%%%%%%%%%%%%%%%%
%%%%%% NETS OF FINITE INDEX CosPas %%%%%%%%%%%%%%%%%%
%%%%%%%%%%%%%%%%%%%%%%%%%%%%%%%%%%%%%%%%%%%%%%%%%%%%%

\sekt{\Name{Coset} pairs as nets of subfactors}{{Coset} pairs as
  nets of subfactors}{sec:netcospa}  

In this section we study some aspects of the inclusion of \Name{Coset}
pairs as left and right chiral observables in the $\opo$-dimensional
quasi-theory $\lok{B}^{\opo}$ 
constructed from the subnet $\lok{A}\subset\lok{B}$ by means of chiral
holography 
(section \ref{cha:netend}.\ref{sec:chihol}). Our arguments
naturally apply to inclusions of left and right chiral
observables in a $\opo$-dimensional conformal theory as well, and we
try to make this analogy manifest in the following. Hence, we make
contact with structures discussed by  
\Name{Rehren} \cite{khR00}, who has analysed the inclusion of chiral
observables in a (genuine) $\opo$-dimensional conformal theory. 

In particular, we investigate the character of the inclusion
$\lok{A}\coset\lok{C}\subset\lok{B}^{\opo}$ as a localisable
representation of $\lok{A}\otimes\lok{C}$, which is considered as a
$\opo$-dimensional theory. The assumption of finite
index for the \Name{Coset} pair forms the ground for most of the
following. If the \Name{Coset} pair is {\em spatial} (see below) in
addition, then we can improve easily our knowledge on the
sub-geometrical action of $U^\lok{A}$ on $\lok{B}$ following from the
net-endomorphism property (proposition \ref{prop:netend}) and show
that  the spectrum of  
$U^\lok{A}(\tilde{R}(2\pi))$ has to be contained in $\menge{\pm 1}$,  if
$U^\lok{A}$ has sharp geometrical action.

We proceed along the lines of \cite{LR95,RST96,khR94a,GL96} and, in
particular, \cite{rL01}. Some of
the discussions below are given merely for the convenience of the
reader, as most of the arguments in \cite{rL01} are adapted easily
from the chiral setting to
the $\opo$-dimensional situation, or essentially stem from \cite{LR95}
where the formulation does not refer to a specific spacetime at
all.

\subsection{The localised representation of a \Name{Coset} pair}
\label{sec:locrepcos}

In  section \ref{sec:loccospa} we considered inclusions of local
algebras in the chiral case; now, 
we want to take the net-structure of inclusions into account and we do
that for the inclusion of the \Name{Coset} pair $\lok{A}\coset\lok{C}$
in the $\opo$-dimensional quasi-theory $\lok{B}^\opo$.  In the
terminology of  \Name{Longo and Rehren}, who developed the
general, abstract framework of {\em nets of subfactors} \cite{LR95},
our subnets are particular examples of 
(irreducible) {\em quantum field theoretical nets of subfactors}. We
 have to extend the arguments of \cite{LR95} from directed nets of
subfactors to the situation in the $\opo$-dimensional conformal
covering space.  

The assumption of conformal covariance on chiral subnets (definition
\ref{def:chsubnet}) implies that we have a conditional
expectation for each local inclusion which is normal, faithful and
leaves invariant the vacuum.  These conditional expectations reflect
the  net-structure:

The reconstruction of each conditional expectation
$\mu_I:\lok{B}(I)\rightarrow \lok{A}(I)$ through the cyclic projection
$e_\lok{A}$ (see \ref{implpropmu} on page \pageref{implpropmu}) shows
that the conditional expectations are {\em consistent} with the net of
inclusions: $\mu_I\restriction\lok{B}(J) = \mu_J$, $J\subset
I$. Namely, for $b\in \lok{B}(J)\subset\lok{B}(I)$ the equation
$e_\lok{A}be_\lok{A}= ae_\lok{A}$ has only one solution
$a\in\lok{A}(J)\subset\lok{A}(I)$, namely $a=\mu_I(b)=\mu_J(b)$. The
same line of argument proves the conditional expectations to be
covariant:
\begin{displaymath}
  \alpha_g(\mu_I(b))e_\lok{A} = \alpha_g(e_\lok{A}be_\lok{A}) =
  e_\lok{A}\alpha_g(b)e_\lok{A} = \mu_{gI}(\alpha_g(b))e_\lok{A} \zendot
\end{displaymath}
These structures carry over to the $\opo$-dimensional situation, which
we study now.
%These proofs are founded essentially on the separating property of the
%vacuum and the \Name{Reeh-Schlieder} property.

Let $\lok{B}^{\opo}$ be a $\opo$-dimensional conformal (quasi-) theory
(cf section \ref{cha:netend}.\ref{sec:chihol})
and denote its restriction to \Name{Minkowski} space by
$\lok{B}^{\opo}_0$, ie we look at the (directed) net of local algebras
$\lok{B}^{\opo}(\Geb{O})$, $\Geb{O}$ a bounded double cone in
$\dopp{M}$, denoted $\Geb{O}\Subset\dopp{M}$. Bounded double
cones are Cartesian products $I\times J$ 
of bounded intervals on the chiral light-rays. The localisation
regions of $\lok{B}^{\opo}$ are double cones $\Geb{O}$ which are
properly contained a single copy of $\dopp{M}$; such double cones will
be indicated as $\Geb{O}\Subset\widetilde{\dopp{M}}$.

Suppose, we have a
localised representation $\rho$ of $\lok{B}^{\opo}$, that is: for each
double cone $\Geb{O}\Subset\widetilde{\dopp{M}}$  the algebra
$\rho_\Geb{O}(\lok{B}^{\opo}(\Geb{O}))$ forms a normal representation 
of the respective local algebra $\lok{B}^{\opo}(\Geb{O})$ on the
vacuum \Name{Hilbert} space 
$\Hilb{H}$ of $\lok{B}^{\opo}$, and the $\rho_\Geb{O}$ fulfill:
\begin{itemize}
\item consistency: $\Geb{O}_1\subset\Geb{O}_2$ $\Rightarrow$
  $\rho_{\Geb{O}_2}\restriction\lok{B}^{\opo}(\Geb{O}_1) =
  \rho_{\Geb{O}_1}$,
\item locality: $\Geb{O}_1\subset\Geb{O}_2'$  $\Rightarrow$
  $\rho_{\Geb{O}_1}(\lok{B}^{\opo}(\Geb{O}_1))\subset
  \rho_{\Geb{O}_2}(\lok{B}^{\opo}(\Geb{O}_2))'$,
\item localisation: for some $\Geb{O}_0'\Subset\widetilde{\dopp{M}}$ we
  have $\rho_\Geb{O}=id_\Geb{O}$ for all $\Geb{O}\subset \Geb{O}_0'$.
\end{itemize}

\noindent By \Name{Haag} duality of $\lok{B}^{\opo}$ \cite{BGL93}, $\rho$
satisfies the following properties:
\begin{enumerate}
\item \label{DHRdefi} If $\Geb{O}\Subset\widetilde{\dopp{M}}$,
  $\Geb{O}_0\subset \Geb{O}$, then 
  $\rho_{\Geb{O}}(\lok{B}^{\opo}(\Geb{O}))
  \subset\lok{B}^{\opo}(\Geb{O})$, and for all 
  $\Geb{O}_1\Subset\widetilde{\dopp{M}}$ containing $\Geb{O}$ we have:
  $\rho_{\Geb{O}_1}\restriction \lok{B}^{\opo}(\Geb{O}) =
  \rho_{\Geb{O}}$.
\item \label{DHRdefii} If $\Geb{O}\subset \Geb{O}_0'$, we have $\rho_\Geb{O}=id_\Geb{O}$.
\item \label{DHRdefiii} For $\Geb{O}_1,\Geb{O}_2\Subset\widetilde{\dopp{M}}$,
  $\Geb{O}_0\cup\Geb{O}_1\subset\Geb{O}_2$, there is a unitary
  $u\in\lok{B}^{\opo}(\Geb{O}_2)$ such that $\Geb{O}\mapsto
  \tilde{\rho}_\Geb{O}(.):= u\rho_\Geb{O}(.)u^*$ is a representation
  localised in $\lok{O}_1$.
\end{enumerate}

We call any map $\rho$ which associates with every
$\Geb{O}\Subset\dopp{M}$ a normal representation $\rho_{\Geb{O}}$ of
$\lok{B}^{\opo}(\Geb{O})$ and which satisfies \ref{DHRdefi},
\ref{DHRdefii}, \ref{DHRdefiii} with $\widetilde{\dopp{M}}$ replaced
by $\dopp{M}$ a {\em {DHR} endomorphism} (of $\lok{B}^{\opo}_0$),
in direct correspondence with the works of \Name{Doplicher, Haag,
  Roberts} \cite{DHR69a,DHR69b,DHR71,DHR74}. 

It will prove possible to
construct a \Name{DHR} endomorphism for $\lok{A}\otimes\lok{C}_0$
from the inclusion $\lok{A}\coset\lok{C}\subset\lok{B}^{\opo}$ and we
need to extend this to a representation of $\lok{A}\otimes\lok{C}$. In the
following lemma $\dopp{M}_{0,\pm}$ stands for the open interior of the
closure of the union of $\dopp{M}$ and the two neighbouring sheets of
the covering as indicated in figure \ref{m}; the restriction of
$\lok{B}^{\opo}$ to  $\dopp{M}_{0,\pm}$ is denoted by
$\lok{B}^{\opo}_{0,\pm}$. We state the lemma in some more generality
than actually needed in the following: 

{\par
 \begin{figure} %Abb.pos: normal:[h]
\begin{minipage}{0cm} \end{minipage}\hfill
\psfrag{1}{{$\Geb{O}_0$}}
\psfrag{2}{{$\Geb{O}$}}
\psfrag{3}{{$\Geb{O}_1$}}
\psfrag{4}{{}}
\psfrag{5}{{}}
\psfrag{7}{{$\dopp{M}_-$}}
\psfrag{8}{{$\infty$}}
\psfrag{9}{{$\dopp{M}_+$}}
\psfrag{10}{{$\dopp{M}$}}
\begin{minipage}{6.0cm}
\epsfig{file=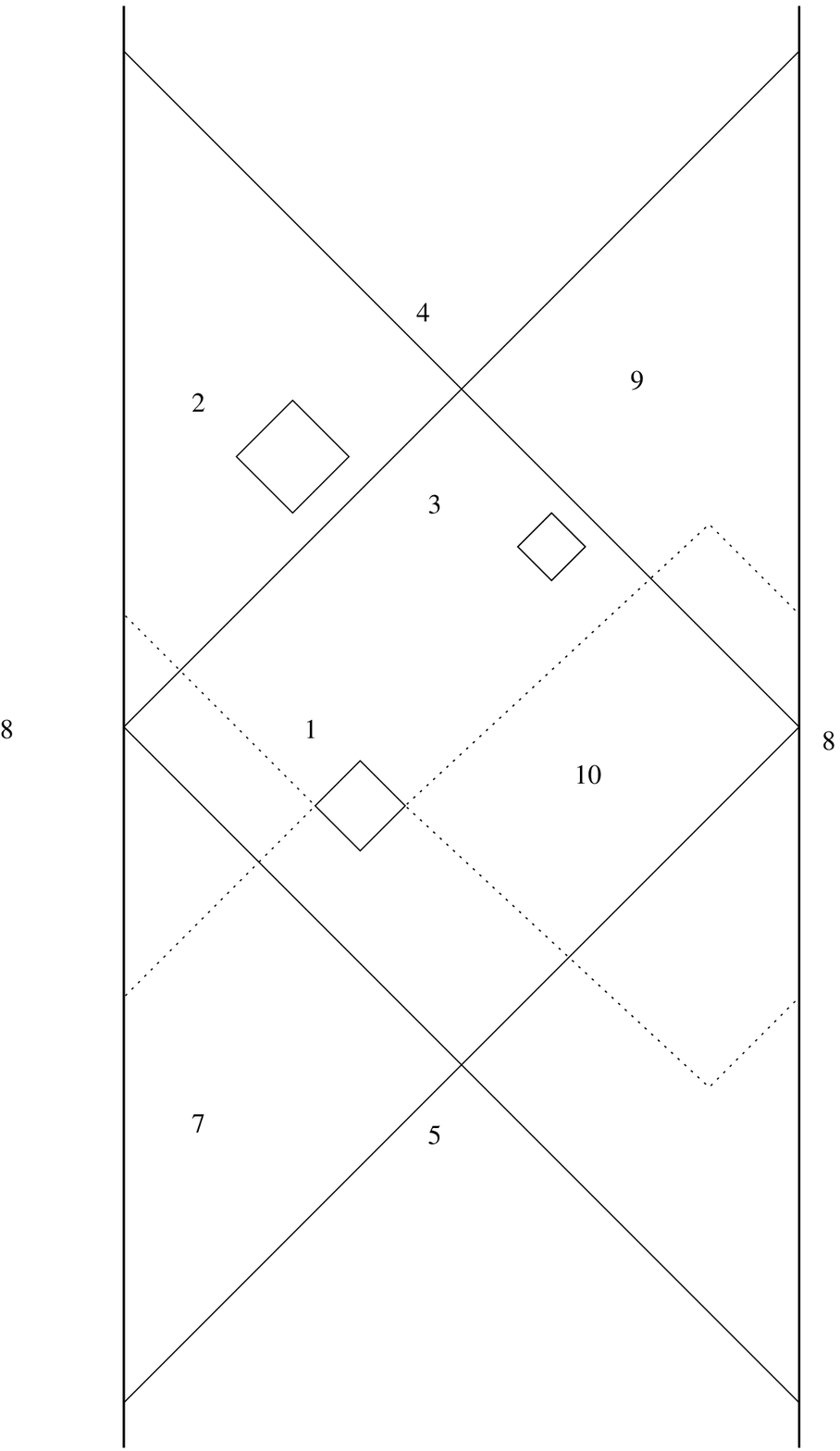,width=4cm} 
\refstepcounter{figure}\label{m}%Zeigername f"ur die Abb.
\end{minipage}
\hfill\begin{minipage}{0cm} \end{minipage}
 \begin{minipage}{0cm} \end{minipage}
 \parbox{14cm}{\vspace{2ex} \centerline{\small  {\bf \figurename{}
       \ref{m} \phantom{\hspace{2.5em}}}}} 
%#5: Bildunterschrift
 \end{figure}}

%%%%%%%%%%%%%%%%%%%%%%%%%%%%%%%%%%%%%%%%%%%%%%%%%%%%%%%%%%%%%%%
%%%%%%% LEMMA EXTENSION OF DHR END TO LOC REP %%%%%%%%%%%%%%%%%

\begin{lemma}\label{lem:DHRextend}
  Let $\rho$ be a \Name{DHR} endomorphism of $\lok{B}^{\opo}_0$ localised in
  some $\Geb{O}_0\Subset\dopp{M}$. There is a unique representation of
  $\lok{B}^{\opo}_{0,\pm}$ extending $\rho$ and localised in
  $\Geb{O}_0$.
\end{lemma}

\begin{pf}
The argument is straightforward and virtually identical to the chiral
case \cite[proposition 11]{rL01}. For geometrical details see figure
\ref{m}, where points 
on the left and right boundary are to be identified if they have the
same height. 

For each $\Geb{O}\Subset\dopp{M}_{0,\pm}$ there is a
$\dopp{M}\Supset\Geb{O}_1\subset\Geb{O}'$ and a
$\Geb{O}_2\Subset\dopp{M}$ containing 
both $\Geb{O}_0$ and $\Geb{O}_1$. Transporting $\rho$ from $\Geb{O}_0$
to $\Geb{O}_1$ we set with respect to a corresponding
$u\in\lok{B}^{\opo}(\Geb{O}_2)$: $\tilde{\rho}:= Ad_u\circ\rho$. The
extension of $\rho$ to $\lok{B}^{\opo}(\Geb{O})$,
$\Geb{O}\Subset\dopp{M}_{0,\pm}$, is 
defined by $\rho_\Geb{O}:= Ad_{u^*}$. 

Obviously, this definition 
extends $\rho$. It is well defined as a different
choice of $u$, $\Geb{O}_1$, $\Geb{O}_2$ for $\rho$'s transport
 results in adjoining a local unitary associated with the causal
complement of $\Geb{O}$. The extension is localised in $\Geb{O}_0$
since for $\Geb{O}\subset \Geb{O}_0'$ we may choose $u=\Einsop$,
$\Geb{O}_0=\Geb{O}_1=\Geb{O}_2$. The consistency of the extension for
$\Geb{O}_a\subset\Geb{O}_b$ is seen easily by a common choice $u$,
$\Geb{O}_1$, $\Geb{O}_2$. 

Locality of $\rho$ follows in this way: For any pair
$\Geb{O}_a, \Geb{O}_b\Subset\dopp{M}_{0,\pm}$,
$\Geb{O}_b\subset\Geb{O}_a'$ there is a covering $\menge{\Geb{O}^i_b}$
of $\Geb{O}_b$ by double cones $\Geb{O}^i_b$ such that
$\overline{\Geb{O}^i_b}\subset\Geb{O}_b$ and there is a
double cone $\Geb{O}_1^i\Subset\dopp{M}$  spacelike to both
$\Geb{O}_a$ and $\Geb{O}_b^i$. Transporting $\rho$ from $\Geb{O}_0$
into $\Geb{O}_1^i$ as above, we have: 
\begin{displaymath}
  \tilde{\rho}^i_{\Geb{O}^i_b} (\lok{B}^{\opo}(\Geb{O}^i_b)) =
  \lok{B}^{\opo}(\Geb{O}^i_b) \subset \lok{B}^{\opo}(\Geb{O}_a)' =
  \tilde{\rho}^i_{\Geb{O}_a} (\lok{B}^{\opo}(\Geb{O}_a))' \zendot
\end{displaymath}
Now one has to use the unitary equivalence of $\tilde{\rho}^i$ and
$\rho$ and weak additivity of $\lok{B}^{\opo}$, which yields:
$\rho_{\Geb{O}_b}(\lok{B}^{\opo}(\Geb{O}_b))\subset \rho_{\Geb{O}_a}(\lok{B}^{\opo}(\Geb{O}_a))'$.

Finally, we verify the uniqueness of the extension. For arbitrary
$\Geb{O}\subset\dopp{M}_{0,\pm}$, we transport $\rho$ into
$\Geb{O}_1\subset\Geb{O}'$ and define $\tilde{\rho}$ as above. If
$\rho^{(1)}$, $\rho^{(2)}$ are two extensions of $\rho$, we look at
$\tilde{\rho}^{(1),(2)}:= Ad_u\circ{\rho}^{(1),(2)}$. Since
$\tilde{\rho}^{(1),(2)}$ 
are localised in 
$\Geb{O}_1\subset\Geb{O}'$, they agree trivially on
$\lok{B}^{\opo}(\Geb{O})$, namely:
$\tilde{\rho}^{(1)}_{\Geb{O}}=id\restriction \lok{B}^{\opo}(\Geb{O}) =
\tilde{\rho}^{(2)}_{\Geb{O}}$. The remainder follows from the definition
of $\tilde{\rho}^{(1),(2)}$. 
\end{pf}

Remarks: Restricting to time-zero double cones yields the chiral case
\cite[proposition 11]{rL01}; here the complete conformal covering
space, $\Seins$, is reached by the argument. 

It is seen easily, that the 
extended $\rho$ may be transported into any $\Geb{O}_1$ for which
$\Geb{O}_1, \Geb{O}_1'\Subset \dopp{M}_{0,\pm}$ by  the usual
procedure relying on local normality.
% We take an isometry $V$ implementing the local
% equivalence $\rho_{\Geb{O}_1'}(\lok{B}^{\opo}(\Geb{O}_1'))\cong
% \lok{B}^{\opo}(\Geb{O}_1')$. If we set ${\rho}':= V\rho V^*$, we
% have a consistent set of representation of local algebras in
% $\lok{B}^{\opo}_{0,\pm}$ which is localised 
% in $\Geb{O}_1$. By \Name{Haag} duality, the unitary equivalence of
% ${\rho}'$ and $\rho$ is implemented by a unitary
% $u\in\lok{B}^{\opo}(\Geb{O}_2)$, $\Geb{O}_2\supset
% \Geb{O}_0\cup\Geb{O}_1$. 
This leads us to recognise that $\rho$ may not be extended beyond
$\dopp{M}_{0,\pm}$  by the procedure in the proof of lemma
\ref{lem:DHRextend}. Fortunately, this is not needed for our purposes
either. 

In the following, we extend the dual canonical endomorphism of a local
inclusion
$\lok{A}\coset\lok{C}(\Geb{O}_0)\subset\lok{B}^{\opo}(\Geb{O}_0)$ to a
\Name{DHR} endomorphism of $\lok{A}\otimes\lok{C}_0$. Lemma
\ref{lem:DHRextend} and the periodicity 
of $\lok{A}\otimes\lok{C}$ over $\widetilde{\dopp{M}}$ then tell us
that this \Name{DHR} endomorphism extends to a localised
representation.

%%%%%%%%%%%%%%%%%%%%%%%%%%%%%%%%%%%%%%%%%%%%%%%%%%%%%%%%%%%%%%%
%%%%%%% LEMMA COMMON EXTENSION OF CAN END %%%%%%%%%%%%%%%%%%%%%

\begin{lemma}\label{lem:commoncanend}
Choose a canonical endomorphism
$\gamma_{\Geb{O}_0}:\lok{B}^{\opo}(\Geb{O}_0)\rightarrow
\lok{A}\coset\lok{C}(\Geb{O}_0)$. Then for any
$\Geb{O}\supset\Geb{O}_0$ there is a canonical endomorphism
$\gamma_\Geb{O}$ satisfying: $\gamma_\Geb{O}\restriction
\lok{B}^{\opo}(\Geb{O}_0) = \gamma_{\Geb{O}_0}$. 

If the \Name{Coset} pair is of finite index, there are common
isometric intertwiners
\begin{displaymath}
  \lok{B}^{\opo}(\Geb{O}_0)\ni v: id \rightarrow \gamma_\Geb{O}\, ,
  \,\, \lok{A}\coset\lok{C}(\Geb{O}_0) \ni w: id \rightarrow
  \rho_\Geb{O}\,. 
\end{displaymath}
\end{lemma}

\begin{pf}
  (Along the lines of \cite[lemma 15]{rL01}) We set
  $\lok{M}:= \lok{B}^{\opo}(\Geb{O})$,
  $\lok{M}^0:= \lok{B}^{\opo}(\Geb{O}_0)$,
  $\lok{N}:= \lok{A}\coset\lok{C}(\Geb{O})$,
  $\lok{N}^0:= \lok{A}\coset\lok{C}(\Geb{O}_0)$. Moreover, we define
  $\lok{M}_1 := \lok{M}\vee\menge{e_\lok{N}}''$, $\lok{M}^0_1 :=
  \lok{M}^0\vee\menge{e_\lok{N}}''$. Let
  $\mu:\lok{M}\rightarrow\lok{N}$,
  $\mu^0:\lok{M}^0\rightarrow\lok{N}^0$ be the normal, faithful
  conditional expectations which leave the vacuum invariant.

There is an isometry $v_1\in\lok{M}^0_1$ satisfying $v_1v_1^* =
e_\lok{N}= e_{\lok{N}^0}$ which reconstructs $\gamma_{\Geb{O}_0}$ by
the unique solutions in
$\lok{A}\coset\lok{C}(\Geb{O}_0)$ of the following equation on elements
$b\in\lok{B}^{\opo}(\Geb{O}_0)$ \cite[proposition 2.9, theorem
3.2]{LR95}: 
\begin{equation}
  \label{eq:canenddefLR95}
  \gamma_{\Geb{O}_0}(b)e_\lok{N} \equiv v_1 b v_1^* \zendot
\end{equation}
The same equation for elements $b\in\lok{B}^{\opo}(\Geb{O})$
defines a canonical endomorphism $\gamma_{\Geb{O}}$ of
$\lok{B}^{\opo}(\Geb{O})$. Consistency of $\gamma_{\Geb{O}_0}$ and
$\gamma_{\Geb{O}}$ is obvious. 

In  case the index is finite, there exist normal, faithful conditional
expectations $\mu_1:\lok{M}_1\rightarrow\lok{M}$, $\mu_1^0:
\lok{M}_1^0\rightarrow \lok{M}^0$ with the same (minimal) index. These
are consistent: in $\lok{M}^0_1$ elements of the form
$\sum_ix_ie_\lok{N}y_i$, $x_i, y_i\in\lok{M}^0$ are dense
\cite{vJ83,PP86} and we have $\mu_1(e_\lok{N}) =
[\lok{M}:\lok{N}]^{-1} = \mu_1^0(e_\lok{N})$. Consistency follows as a
consequence of normality from the following identity:
\begin{displaymath}
  \mu_1(\sum_ix_ie_\lok{N}y_i) = \sum_ix_i\mu_1(e_\lok{N})y_i =
  \mu_1^0(\sum_ix_ie_\lok{N}y_i) \zendot
\end{displaymath}

The operators $v,w$ are obtained as follows (see \cite[(2.16)]{LR95}):
\begin{eqnarray}
  \label{eq:canendvdef}
  [\lok{M}:\lok{N}]^{-\frac{1}{2}}v &:=& \mu_1(v_1) = \mu_1^0(v_1)
  \zencom\\
  \label{eq:canendwdef}
  [\lok{M}:\lok{N}]^{-\frac{1}{2}}w &:=& \mu(v) = \mu^0(v) \zendot
\end{eqnarray}
They fulfill the relations \ref{vwidi}, \ref {vwidii}, \ref {vwidiii}
on page \pageref{vwidi}. 
\end{pf}

%%%%%%%%%%%%%%%%%%%%%%%%%%%%%%%%%%%%%%%%%%%%%%%%%%%%%%%%%%%%%%%
%%%%%%% LEMMA EXTENSION OF DUAL CAN END %%%%%%%%%%%%%%%%%%%%%%%
\clearpage
\begin{lemma}\label{lem:extendualcan} Let
  $v_1\in\lok{B}^{\opo}(\Geb{O}_0)\vee\menge{e_{\lok{A}\smcoset\lok{C}}}''$
  be an isometry satisfying $v_1v_1^*=e_{\lok{A}\smcoset\lok{C}}$. The
  dual canonical endomorphism $\rho_{\Geb{O}_0}$ defined by 
  \begin{equation}
    \label{eq:defdualcancos}
    \rho_{\Geb{O}_0}([ac])e_{\lok{A}\smcoset\lok{C}}   =
    v_1\,[ac]\,v_1^*\,\, , \quad [ac]\in
    \lok{A}\coset\lok{C}(\Geb{O}_0) \,\, ,
  \end{equation}
  has an extension defining a localised representation of
  $\lok{A}\otimes\lok{C}$ localised in $\Geb{O}_0$, which shall be
  called $\rho$. 
\end{lemma}

\begin{pf} (Along the lines of \cite[proposition 16]{rL01}.)
  We choose a sheet of the covering as \Name{Minkowski} space, $\dopp{M}$,
  containing $\Geb{O}_0$ as a bounded double cone. The extension of
  $\rho_{\Geb{O}_0}$ to all of $\lok{A}\coset\lok{C}_0$ is constructed
  as in the proof of lemma \ref{lem:commoncanend} by extending the
  canonical endomorphism $\gamma_{\Geb{O}_0}$. The isomorphisms
  between the local algebras of $\lok{A}\coset\lok{C}_0$ and
  $\lok{A}\otimes\lok{C}_0$ define the action of the extension of
  $\rho_{\Geb{O}_0}$ on $\lok{A}\otimes\lok{C}_0$ (see \ref{lokiso}, page
  \pageref{lokiso}) and hence of $\rho$ itself. 

We have to check
  the conditions 
  \ref{DHRdefi}, \ref{DHRdefii}, \ref{DHRdefiii} (page
  \pageref{DHRdefi}) for $\rho$. 
  \ref{DHRdefi} and \ref{DHRdefii} are contained in \cite[theorem
  3.2]{LR95}; condition \ref{DHRdefiii} follows from the uniqueness of
  the canonical endomorphism up to unitary equivalence with respect to
  a local unitary in $\lok{A}\coset\lok{C}_0$ \cite{rL87}. 

Finally, we
  apply lemma \ref{lem:DHRextend} and recognise the periodicity of
  $\lok{A}\otimes\lok{C}$ on conformal covering space which means that
  $\lok{A}\otimes\lok{C}$ effectively lives on $\Seins\times\Seins$ and
  hence extension to $\dopp{M}_{0,\pm}$ is sufficient.
\end{pf}

We have
$\rho_\Geb{O}(\lok{A}\otimes\lok{C}(\Geb{O})) \subset
\lok{A}\otimes\lok{C}(\Geb{O})$ for each $\Geb{O}\supset\Geb{O}_0$, 
because $\rho$ is localised and $\lok{A}\otimes\lok{C}$ satisfies \Name{Haag}
duality. Moreover,
$[ac] \mapsto [ac]e_{\lok{A}\smcoset\lok{C}}$,
$[ac]\in\lok{A}\coset\lok{C}(\Geb{O})$,
$\Geb{O}\Subset\widetilde{\dopp{M}}$, defines an isomorphism of
\Name{v.Neumann} algebras which can be used to define:
\begin{equation}
  \label{eq:defrhorho}
  \rho_\Geb{O} ([ac]) e_{\lok{A}\smcoset\lok{C}} \equiv \rho_\Geb{O}
  ([ac] e_{\lok{A}\smcoset\lok{C}}) \,, \,\, \Geb{O} \supset \Geb{O}_0 \zendot
\end{equation}
It is clear that this definition agrees with the dual canonical
endomorphisms for $\Geb{O}\Subset\dopp{M}$ which were used to
construct $\rho$ (see proof of
lemma \ref{lem:extendualcan}). 

%%%%%%%%%%%%%%%%%%%%%%%%%%%%%%%%%%%%%%%%%%%%%%%%%%%%%%%%%%%%
%%%%%%%%%% PROPOSITION DHR CHARACTER OF COSPAs %%%%%%%%%%%%%

\begin{prop}\label{prop:cospadhr}
  The representation $\rho$ of lemma
  \ref{lem:extendualcan} is 
  unitarily equivalent to the representation
  $\lok{A}\coset\lok{C}\subset\lok{B}^{\opo}$. In particular, it
  is covariant with the same spectrum condition on the
  translations. It holds true for $v_1$ as above and $\Geb{O}
  \Subset\widetilde{\dopp{M}}$: 
  \begin{equation}
    \label{eq:uniequivcanend}
    \rho_\Geb{O}([ac] e_{\lok{A}\smcoset\lok{C}}) = v_1\,[ac]\, v_1^*
    \,,\quad [ac]\in\lok{A}\coset\lok{C}(\Geb{O}) \,\, .
  \end{equation}
\end{prop}

\begin{pf} (Along the lines of \cite[proposition 17]{rL01}.)
  For $\Geb{O}_0\subset\Geb{O}\Subset\dopp{M}$ and for
  $\Geb{O}\subset\Geb{O}_0'$ there is nothing to show. For the general
  case we follow the same strategy as in the proof of lemma
  \ref{lem:DHRextend}. It suffices to look at proper double cones
  $\Geb{O}$ in $\dopp{M}_{0,\pm}$.

  We choose $\dopp{M}\Supset\Geb{O}_1\subset\Geb{O}'$ and
  $\Geb{O}_2\Subset\dopp{M}$ containing both $\Geb{O}_1$ and
  $\Geb{O}_0$. There is a  canonical endomorphism 
  $\gamma_{\Geb{O}_2}$ which extends $\gamma_{\Geb{O}_0}$ and yields
  through its restriction to $\lok{A}\coset\lok{C}(\Geb{O}_2)$, the
  dual canonical endomorphism 
  $\rho_{\Geb{O}_2}$,  the same localised representation $\rho$
  by the process described in the proof of lemma
  \ref{lem:extendualcan}.

Now choose a canonical endomorphism $\tilde{\gamma}_{\Geb{O}_1}$. According to
lemma  \ref{lem:commoncanend} there is a canonical endomorphism
$\tilde{\gamma}_{\Geb{O}_2}$ which extends
$\tilde{\gamma}_{\Geb{O}_1}$, and there is 
a unitary $u\in\lok{A}\coset\lok{C}(\lok{O}_2)$ such that
$\tilde{\gamma}_{\Geb{O}_2} = Ad_u\circ \gamma_{\Geb{O}_2}$
\cite{rL87}. The defining isometry $\tilde{v}_1\in
\lok{B}^\opo(\Geb{O}_1)\vee\{e_{\lok{A}\smcoset\lok{C}}\}''$ of 
$\tilde{\gamma}_{\Geb{O}_1}$  satisfies: $\tilde{v}_1 = u v_1$.

The dual canonical endomorphism $\tilde{\rho}_{\Geb{O}_1}$ extends to
a localised representation $\tilde{\rho}$ for which we have:
$\tilde{\rho} = Ad_u\circ\rho$. For
$[ac]\in\lok{A}\coset\lok{C}(\Geb{O})$ this leads to:
   \begin{displaymath}
u\rho_{\Geb{O}}([ac]e_{\lok{A}\smcoset\lok{C}})u^* =
\tilde{\rho}([ac]e_{\lok{A}\smcoset\lok{C}}) =
[ac]e_{\lok{A}\smcoset\lok{C}} = \tilde{v}_1 [ac] \tilde{v}_1^* =
uv_1[ac]v_1^* u^* \zendot 
   \end{displaymath}
\end{pf}

%%%%%%%%%%%%%%%%%%%%%%%%%%%%%%%%%%%%%%%%%%%%%%%%%%%%%%%%%%%%
%%%%%%%%%%%% LEMMA VW INTERTWINING RHO %%%%%%%%%%%%%%%%%%%%%

The following lemma determines to a large extent the transformation
behaviour of observables in $\lok{B}^{\opo}$ with respect to chiral
transformations, provided the \Name{Coset} pair has finite index:
\begin{lemma}\label{lem:vwrhointertw}
  Assume $[\lok{B}^{\opo}:\lok{A}\coset\lok{C}]<\infty$,
  and take $v\in\lok{B}^{\opo}(\Geb{O}_0)$,
  $w\in\lok{A}\coset\lok{C}(\Geb{O}_0)$ as in equations
  (\ref{eq:canendvdef}), 
  (\ref{eq:canendwdef}). If $\Geb{O}\Subset\widetilde{\dopp{M}}$
  contains $\Geb{O}_0$, we have for all $[ac]\in
  \lok{A}\coset\lok{C}(\Geb{O})$: 
  \begin{equation}
    \label{eq:vwrhointertw}
    \rho_\Geb{O}([ac])\,v_1 = v_1\,[ac] \,, \,\,   
    \rho_\Geb{O}([ac]) v = v [ac]\, , \,\,    
    \rho_\Geb{O}([ac]) w = w [ac] \zendot
  \end{equation}
\end{lemma}

\begin{pf}
  From equations (\ref{eq:defrhorho}) and (\ref{eq:uniequivcanend})
  follows immediately the first statement in
  (\ref{eq:vwrhointertw}). For the remainder apply the conditional
  expectations $\mu_1:
  \lok{B}^\opo(\Geb{O}_1)\vee\{e_{\lok{A}\smcoset\lok{C}}\}''
  \rightarrow \lok{B}^\opo(\Geb{O}_1)$ and $\mu \mu_1:
  \lok{B}^\opo(\Geb{O}_1)\vee\{e_{\lok{A}\smcoset\lok{C}}\}''
  \rightarrow \lok{A}\coset\lok{C}(\Geb{O}_1)$ (see proof of lemma 
  \ref{lem:commoncanend}).
\end{pf}

%%%%%%%%%%%%%%%%%%%%%%%%%%%%%%%%%%%%%%%%%%%%%%%%%%%%%%%%%%%%
%%%%%%%%%%%% PROP  ON FINITE DECOMPOSITION OF RHO %%%%%%%%%%

The following proposition is practically identical to \cite[corollary
18]{rL01}:
\begin{prop}
  Assume $[\lok{B}^{\opo}:\lok{A}\coset\lok{C}]<\infty$. $\rho$ shall
  denote a representation of $\lok{A}\otimes\lok{C}$ on
  $e_{\lok{A}\smcoset\lok{C}}\Hilb{H}$ which is unitarily equivalent to
  $\lok{A}\coset\lok{C}\subset\lok{B}^{\opo}$ and localised is some
  double cone $\Geb{O}_0$. Then $\rho$ has finite decomposition into
  irreducibles: $$\rho\cong\bigoplus_{i=0}^n N_i\rho^i \,\, .$$ The $\rho^i$
  are covariant representations of $\lok{A}\otimes\lok{C}$ with the
  same spectrum condition on the translations and are localised in
  $\Geb{O}_0$  alike. For $\Geb{O}\supset\Geb{O}_0$ the dual canonical
  endomorphism has the same decomposition: $$\rho_\Geb{O}
  \cong\bigoplus_{i=0}^n N_i\rho^i_\Geb{O}\,.$$    
\end{prop}

\begin{pf}
  Since the spacetime symmetry group of our problem is
  $(\PSL(2,\dopp{R})^\sim\times \PSL(2,\dopp{R})^\sim)/\dopp{Z}$ one may
  adapt the arguments of \cite{GL96} using the propositions and
  lemmas above. In particular, the identity of global and local
  intertwiner spaces may be proved in the same manner, replacing the
  choice of one point at infinity by two such points, one on each 
  chiral light-ray. We refrain from repeating the discussions in
  \cite{GL96}.
\end{pf}

Remark: 
Because $\lok{A}\coset\lok{C}\subset \lok{B}^{\opo}$ is {\em
  irreducible} for finite index, the vacuum sector of
  $\lok{A}\otimes\lok{C}$ appears 
{\em exactly once} in $\rho$; these two facts are equivalent (see
  \cite{LR95}). 

%  We have given a presentation of the interpretation of
% the dual canonical endomorphism of a net of quantum field theoretical
% subfactors as a localised representation of the subtheory in our
% particular setting. As hopefully has  become clear on the way, this
% scheme applies in many more contexts; the corresponding
% arguments for directed nets are contained in \cite{LR95} already.

%%%%%%%%%%%%%%%%%%%%%%%%%%%%%%%%%%%%%%%%%%%%%%%%%%%%%%%%%%%%%%%%%%%%%%
%%%%%%%%%%%%%%% LOCAL EXTENSION PROBLEM FOR COSPAS %%%%%%%%%%%%%%%%%%%

Before closing this section we only mention very briefly a  natural
question: For which pairs of chiral nets 
exist embeddings as \Name{Coset} pairs in another chiral conformal
theory and which are the
solutions to this problem? This problem will prove difficult to tackle in
general, but if one restricts attention  to  \Name{Coset}
pairs of finite index there is a detailed formulation.

 An
irreducible, infinite subfactor $\lok{N}\subset\lok{M}$ with
$[\lok{M}:\lok{N}]<\infty$ is completely determined by the following
data, the
{\em {DHR} triple $(\rho,w,w_1)$}, where  $\rho$ is an 
endomorphism of $\lok{N}$ with $d_\rho= [\lok{M}:\lok{N}]$ and
$\lok{N}\ni w:id\rightarrow\rho$, $\lok{N}\ni
w_1:\rho\rightarrow\rho^2$ are isometric intertwiners satisfying \cite{rL94}:
\begin{enumerate}
\item \label{DHRtripli} $w^*w_1 = [\lok{M}:\lok{N}]^{-\frac{1}{2}} \Einsop =
  \rho(w^*)w_1$,
\item \label{DHRtriplii} $w_1w_1^* = \rho(w_1^*)w_1$,
\item  \label{DHRtripliii} $w_1w_1= \rho(w_1)w_1$.
\end{enumerate}
This characterisation is equivalent to the one discussed on page
\pageref{vwchar}, where one has to take $\rho=\gamma\restriction\lok{N}$, 
$w_1=\gamma(v)$. Concerning chiral subnets and \Name{Coset} pairs this
 has the advantage of determining $\lok{B}$ entirely in terms of
 \Name{DHR} data of $\lok{A}$ or, respectively, of $\lok{A}\otimes\lok{C}$.

If we now combine some \Name{DHR} endomorphisms of
$\lok{A}\otimes\lok{C}$ of finite statistics to a reducible 
endomorphism $\rho$ it is known what is needed in addition to ensure
that the extension of $\lok{A}\otimes\lok{C}$ by means of \ref{DHRtripli},
\ref{DHRtriplii}, \ref{DHRtripliii} indeed defines a local theory
\cite[theorem 4.9]{LR95} (some more details in \cite{RST96}). We need
to have the following identity for the statistics operator of $\rho$,
$\varepsilon_\rho^<$, and $w_1$: 
\begin{equation}
  \label{eq:lokext}
  \varepsilon_\rho^< w_1 = w_1
\end{equation}
From \ref{DHRtripli},
\ref{DHRtriplii}, \ref{DHRtripliii} alone only follows relative locality of
the extension $\lok{B}$. Determining the possible solutions
$\lok{B}$ is a {\em local extension problem} for
$\lok{A}\otimes\lok{C}$. 
It is outside the purposes of this work
to give a more elaborate discussion on this issue. We only mention as
introductory references \cite{LR95,RST96,khR01} and for some examples
for solutions of such problems \cite{KL02}. %BE? further references?

%%%%%%%%%%%%%%%%%%%%%%%%%%%%%%%%%%%%%%%%%%%%%%%%%%%%%%%% 
%%% DEFINITION SPATIAL COSPAs %%%%%%%%%%%%%%%%%%%%%%%%%%

\subsection{On spatial \Name{Coset} pairs}
\label{sec:spatcospa}

It is the purpose of the following discussion to determine the
spectrum of the operator
$U^\lok{A}(\tilde{R}(2\pi))$ in  
presence of sharp  geometrical action under conditions which are
satisfied in a large set of examples. The spectrum of
$U^\lok{A}(\tilde{R}(2\pi))$ is known in many examples, and our result
excludes the possibility of sharp geometrical action in all these
cases (except, of course, the ones that one can make up trivially). On
the way we indicate how to improve our
knowledge about the geometrical impact of $U^\lok{A}$
(net-endomorphism property) for cofinite subnets
$\lok{A}\subset\lok{B}$. 

The key point is 
that the transformation behaviour of charged intertwiners is
completely given in terms of the \Name{Coset} pair. This is
true in a more general setting than the one we give below, but the
derivation of the general statement is a mere repetition of the
analysis in the chiral case (see \cite{rL01,GL96}) and appears
to be of little interest. So, we 
give a discussion aiming directly for our goals, which requires some
additional structure.
The starting point is
\begin{defi}\label{def:spatialcospa}
  $\lok{A}\coset\lok{C}\subset\lok{B}$ is called a {\bf spatial
  {Coset} pair}, if the localised representation
  $\rho$ of $\lok{A}\otimes\lok{C}$, which is unitarily equivalent to
  the \Name{Coset} pair representation, decomposes completely into
  tensor products, $\rho^\lok{A}_r\otimes\rho^\lok{C}_s$, of
  irreducible, localised representations $\rho^\lok{A}_r$ of $\lok{A}$
  and $\rho^\lok{C}_s$ of $\lok{C}$, respectively: 
  $$ \rho\cong \bigoplus Z_{rs}\rho^\lok{A}_r\otimes\rho^\lok{C}_s\,.$$
\end{defi}
Remark: Even for \Name{Coset} pairs of finite index it is not clear whether
they are automatically spatial. This is true under additional
conditions \cite[lemma 27]{KLM01}, which are known to be satisfied if
one of $\lok{A}$, $\lok{C}$ is {\em completely rational} \cite[corollary
14]{KLM01}.

%%%%%%%%%%%%%%%%%%%%%%%%%%%%%%%%%%%%%%%%%%%%%%%%%%%%%%
%%%%%% TRANSFORMATION BEHAVIOUR OF CHARGED INTERTW: %%

Covariance of a representation $\rho^\lok{A}_r$ localised in some
$I_0\Subset\Seins$ means that there are unitaries
$z_{\rho^\lok{A}_r}(\tilde{g})$ 
implementing the equivalence of $\rho^\lok{A}_r$ and
$\alpha_{\symb{p}(\tilde{g})}\circ\rho^\lok{A}_r\circ\alpha_{\symb{p}(\tilde{g})^{-1}}$:
\begin{equation}
  \label{eq:repcocycldef}
  Ad_{z_{\rho^\lok{A}_r}(\tilde{g})}\circ \rho^\lok{A}_r  =
  \alpha_{\symb{p}(\tilde{g})}\circ \rho^\lok{A}_r\circ
  \alpha_{\symb{p}(\tilde{g})^{-1}}    \zendot 
\end{equation}  
The map $g\mapsto z_{\rho^\lok{A}_r}(\tilde{g})$ may be chosen to be a
{\em localised $\alpha$-cocycle} with values in the universal
$C^*$-algebra $\lok{A}_{uni}$ such that:
\begin{enumerate}
\item \label{loccocycli}
    $z_{\rho^\lok{A}_r}(\tilde{g})\in\lok{A}(I_0\cup 
    \symb{p}(\tilde{g})I_0)e_\lok{A}$, for
    $\tilde{g}\in\PSL(2,\dopp{R})^\sim$ which are close to the
    identity and 
    satisfy $I_0\cup\symb{p}(\tilde{g})I_0\Subset\Seins$. %(cf
    %proposition \ref{prop:netend}).  
\item \label{loccocyclii} $z_{\rho^\lok{A}_r}(\tilde{g}\tilde{h}) =
  \alpha_{\symb{p}(\tilde{g})}(z_{\rho^\lok{A}_r}(\tilde{h}))z_{\rho^\lok{A}_r}(\tilde{g})$,
  $\tilde{g},\tilde{h}\in\PSL(2,\dopp{R})^\sim$.  
\end{enumerate}
If $U_{\rho^\lok{A}_r}$ implements covariance in the representation
$\rho^\lok{A}_r$, we have: 
 \begin{displaymath}
   z_{\rho^\lok{A}_r}(\tilde{g})=
 U(\symb{p}(\tilde{g}))e_\lok{A} U_{\rho^\lok{A}_r}(\tilde{g})^*\, , \,\,
 \tilde{g}\in\PSL(2,\dopp{R})^\sim\,.
 \end{displaymath}
%In fact, $z_{\rho^\lok{A}_r}$ allows to reconstruct ${\rho^\lok{A}_r}$;
More details on localised covariance cocycles may be found in  \cite{jR76},
  \cite{GL92}, \cite{GL96}, \cite[\S 27]{aS97}. 

For $\tilde{g}$ as in \ref{loccocycli} one may identify
$z_{\rho^\lok{A}_r}(\tilde{g})$ with its representative in the
inclusion $\lok{A}\subset\lok{B}$. By the very construction of the
covariance cocycles 
$z_{\rho^\lok{A}_r}(\tilde{g})$ for general $\tilde{g}$ (see eg
\cite[lemma 27.2]{aS97}), it is clear 
that $z_{\rho^\lok{A}_r}(\tilde{g})$  may be identified
with its representative in the inclusion $\lok{A}\subset\lok{B}$ as
well: The inclusion defines a locally normal representation of
$\lok{A}e_\lok{A}$ which lifts to a representation of
$\lok{A}_{uni}$. The universal $C^*$-algebra, in turn, contains the
covariance cocycle  $z_{\rho^\lok{A}_r}(\tilde{g})$.

All that has been said about the $\rho^\lok{A}_r$ holds, mutatis
mutandis, for the representations $\rho^\lok{C}_s$; these
representations shall be localised in $J_0\Subset\Seins$.  

We now choose complete sets of orthonormal intertwiners
$\lok{A}(I_0)\ni \,w^i_s:    \rho^\lok{A}_r\otimes id \rightarrow
\rho$ and $\lok{C}(J_0)\,  \ni w^j_s: id\otimes\rho^\lok{C}_s
\rightarrow \rho$ such 
that the $w^i_rw^j_s:  \rho^\lok{A}_r\otimes\rho^\lok{C}_s\rightarrow
\rho$ form a complete orthonormal set of intertwiners. These are
intertwiners for all 
$\lok{A}\otimes\lok{C}(I\times J)$, if $I\times J \supset I_0\times
J_0$ \cite[proof of theorem 2.3]{GL96}. 

Using the isometry $v$ from lemma \ref{lem:vwrhointertw} we define the
charged intertwiners for our spatial \Name{Coset} pair of finite index
as: 
\begin{equation}
  \label{eq:cospachargedef}
  \psi^{ij}_{rs}:= w^i_r{}^*w^j_s{}^* v\,
  \in\lok{B}^{\opo}(I_0\times J_0) \zendot
\end{equation}
These operators are charged intertwiners $id\rightarrow
\rho^\lok{A}_r\otimes\rho^\lok{C}_s$ for  all
$\lok{A}\otimes\lok{C}(I\times J)$, if $I\times J \supset I_0\times
J_0$. The map $w^i_rw^j_s\mapsto \psi^{ij}_{rs}$ defines an
anti-isomorphism of the respective intertwiner spaces \cite[lemma
4.5]{LR95}, \cite[proposition 2]{khR94a}.  Regarding the
transformation behaviour of 
the charged intertwiners we have:

\begin{lemma}\label{lem:covchargedinter}
  For
  $(\tilde{g},\tilde{h})\in\PSL(2,\dopp{R})^\sim\times\PSL(2,\dopp{R})^\sim$ 
  holds:
  \begin{equation}
    \label{eq:charinterwtransf}
    \alpha_{(\tilde{g},\tilde{h})} \psi^{ij}_{rs} =
    Ad_{U^\lok{A}(\tilde{g})}
    Ad_{U^{\lok{A}'}(\tilde{h})}(\psi^{ij}_{rs}) = 
    z_{\rho^\lok{A}_r}(\tilde{g})
    z_{\rho^\lok{C}_s}(\tilde{h}) 
    \psi^{ij}_{rs} \zendot
  \end{equation}
\end{lemma}

\begin{pf}
The method of proof is standard \cite[lemma
(3.6)]{khR94a},\cite[corollary 19]{rL01}. We look at some
neighbourhood $\mathcal{U}$ of the identity in
$\PSL(2,\dopp{R})^\sim\times\PSL(2,\dopp{R})^\sim$. It is easy to show
on grounds of the covariance of the endomorphisms $\rho^\lok{A}_r$,
$\rho^\lok{C}_s$ (see equation \ref{eq:repcocycldef}) that the
following mapping, defined for $(\tilde{g},\tilde{h})\in\mathcal{U}$,
leaves globally  invariant  the spaces of charged intertwiners:
\begin{equation}
  \label{eq:trivcovcharint}
   ((\tilde{g},\tilde{h}),\psi^{ij}_{rs}) \mapsto
  z_{\rho^\lok{A}_r}(\tilde{g})^* 
  z_{\rho^\lok{C}_s}(\tilde{h})^* 
  \alpha_{(\tilde{g},\tilde{h})}(\psi^{ij}_{rs})\, .
\end{equation}
Indeed, this map defines a unitary representation of
$\PSL(2,\dopp{R})^\sim\times\PSL(2,\dopp{R})^\sim$ on the
finite-dimensional \Name{Hilbert} spaces of charged intertwiners. This 
representation has
to be trivial, as there are no non-trivial unitary, finite-dimensional
representations of this group.
\end{pf}

Hence, the transformation behaviour
of $b\in\lok{B}^{\opo}(\Geb{O})$, $\Geb{O}\supset I_0\times J_0$ has
the following form:
\begin{equation}
  \label{eq:transflokobsB}
  \alpha_{(\tilde{g},\tilde{h})}(b) = 
  \sum_{i,j,r,s} [\lok{B}:\lok{A}\coset\lok{C}]\,\, 
  \alpha_{(\tilde{g},\tilde{h})}(\mu_\Geb{O}(b\psi^{ij}_{rs}{}^*))
  z_{\rho^\lok{A}_r}(\tilde{g})
  z_{\rho^\lok{C}_s}(\tilde{h})
  \psi^{ij}_{rs} \zendot
\end{equation}

For chiral rotations by $2\pi$ we have in terms of the lower bound
$h^\lok{A}_r$ of
the spectrum of the conformal {Hamilton}ian $L_0$ in the
representation $\rho^\lok{A}_r$ \cite{GL96}:
$z_{\rho^\lok{A}_r}(\tilde{R}(2\pi))= e^{-i2\pi h^\lok{A}_r}$.  This
proves, that the adjoint action of $Ad_{U^\lok{A}(\tilde{R}(2\pi))}$
on any $\lok{B}^{\opo}(\Geb{O})$, $\Geb{O}\supset I_0\times J_0$,
defines  an automorphism of this algebra. 
%%%%%%%%%%%%%%%%%%%%%%%%%%%%%%%%%%%%%%%%%%%%%%%%%%%%%%%%%%%%
%%%% PROPOSITION AUTOMORPHIC ACTION OF CHIRAL ROT BY 2Pi %%%
We state this result explicitly for the chiral situation:
\begin{prop}\label{prop:AutchirRot}
  Let $\lok{A}\coset\lok{C}\subset\lok{B}$ be a spatial \Name{Coset}
  pair with finite index and assume $U^\lok{A}$ to have the
  net-endomorphism property. Then  
  $Ad_{U^\lok{A}(\tilde{R}(2\pi))}(\lok{B}(I)) = \lok{B}(I)$ for all
  $I\Subset\Seins$. 
\end{prop}
\begin{pf}
  First, one goes into the chirally holographic picture, then one constructs the representation $\rho$ unitarily equivalent
  to $\lok{A}\coset\lok{C}\subset\lok{B}^{\opo}$ and localised in
  $I\times  I$, and
  one proceeds as above until one arrives at
  (\ref{eq:transflokobsB}). Now the 
  statement is obvious, if one restricts attention to the time-zero
  algebra $\lok{B}^{\opo}(I\times I)= \lok{B}(I)$.
\end{pf}

Hence we  improved proposition \ref{prop:netend} on the
net-endomorphic action of $U^\lok{A}$ on $\lok{B}$ under additional
assumptions.  
We expect this automorphic action of
$Ad_{U^\lok{A}(\tilde{R}(2\pi))}$ on each $\lok{B}(I)$,
$I\Subset\Seins$, in general.

We continue with an application:   
 the spectrum 
of $U^\lok{A}(\tilde{R}(2\pi))$ is determined (almost  completely),
provided $U^\lok{A}$ has sharp  
geometrical action. The argument applies to the corresponding 
situation of chiral observables in any $\opo$-dimensional conformal 
theory as well.   

%%%%%%%%%%%%%%%%%%%%%%%%%%%%%%%%%%%%%%%%%%%%%%%%%%%%%%%%%%
%%%%% PROPOSITION SPEC UAR2PI AND SHARP GEOM ACTION %%%%%%

\begin{prop}\label{prop:SpecUAR2pisharp}
Let $\lok{A}\coset\lok{C}\subset\lok{B}$ be a spatial \Name{Coset}
  pair with finite index and assume $U^\lok{A}$ to have sharp
  geometrical action on  $\lok{B}$. Then we have 
  $Spec( \, U^\lok{A}(\tilde{R}(2\pi)) \, )\subset\menge{\pm 1}$,
  and each $b\in\lok{B}(I)$, $I\Subset\Seins$, has a decomposition:
  \begin{equation}
    \label{eq:sharpgeodecomp}
    b=b_++b_-\, , \,\, b_\pm\in\lok{B}(I)\, , \,\,
    Ad_{U^\lok{A}(\tilde{R}(2\pi))}(b_\pm) = \pm b_\pm\,.
  \end{equation}
\end{prop}

\begin{pf}
We analyse this problem in the holographic picture in $\opo$
dimensions (see section \ref{cha:netend}.\ref{sec:chihol}). % The
% arguments are standard and mostly from \Name{DHR} theory in low
% dimensions and of chiral conformal theory \cite{FRS89,FRS92,GL96}
% There are good textbook-like reviews available: \cite{KMR90,aS97}. For
% the reader's convenience, we give references in the latter review
% for some standard identities.
%
The \Name{DHR} endomorphism $\rho^\lok{A}_r\otimes\rho^\lok{C}_s$ has
four statistics operators: left and right statistics operators for
spacelike directions, denoted $\varepsilon^>_{rs}$ and
$\varepsilon^<_{rs}$, and upper and lower statistics operators for
time-like directions, which we write as $\varepsilon^\vee_{rs}$ and
$\varepsilon^\wedge_{rs}$. All arguments on statistics have nothing to
do with positivity of energy, but  only with geometry and
commutativity. 

For the chiral endomorphisms there is a right and a
left statistics operator, namely $\varepsilon^+_r$ and
$\varepsilon^-_r$ for $\rho^\lok{A}_r$ and correspondingly
$\varepsilon^+_s$ and $\varepsilon^-_s$ for $\rho^\lok{C}_s$. We have:
\begin{equation}\label{eq:statchistat}
  \varepsilon^<_{rs} = \varepsilon^+_r\otimes\varepsilon^+_s\,,\,\,
  \varepsilon^\vee_{rs} = \varepsilon^+_r\otimes\varepsilon^-_s\,. 
\end{equation}
According to the general analysis of \Name{Rehren} \cite[proposition
4]{khR94}, \cite[lemma 3.1]{khR01} the corresponding statistical
phases satisfy: 
\begin{equation}\label{eq:statphasechistat}
  1=\, \kappa^<_{rs} = \kappa_r^+\kappa_s^+\,,\,\, 1=\,
  \kappa_{rs}^\vee = \kappa^+_r\kappa_s^- \,\, .
\end{equation}
The left identity follows from locality (commutativity for spacelike
separation) and the right one from commutativity for timelike separation. 

The left and right statistics operators are connected by the identity
$\varepsilon^+_{r,s} = \varepsilon^-_{r,s}{}^*$%\cite[theorem 16.5]{aS97}
, which yields directly $\kappa^+_{r,s} =
\overline{\kappa^-_{r,s}}$ 
%\cite[def. 16.8, (16.18), proof Bem. 16.11]{aS97}
. Combining this with the conformal spin and 
statistics theorem \cite{GL96} and equations (\ref{eq:statphasechistat})
leads to 
\begin{equation}
  \label{eq:specUAsharp}
  \kappa^+_r{}^2 = \klammer{e^{i2\pi h^\lok{A}_r}}^2 = 1
  \,\,\Leftrightarrow\,\, 
  h^\lok{A}_r  \in \frac{1}{2}\dopp{N} \zendot
\end{equation}

The remainder follows from the decomposition of local observables and
the transformation law of charged intertwiners, see equation
(\ref{eq:transflokobsB}). 
\end{pf}

If the decomposition of the vacuum representation of a chiral
conformal theory $\lok{B}$ considered as representation of a
spatial \Name{Coset} pair $\lok{A}\coset\lok{C}$ of  finite index is
known and $U^\lok{A}$ has the net-endomorphism property, it is very
simple to check whether the \Name{Borchers-Sugawara} 
representation $U^\lok{A}$ can have sharp geometrical action by means
of proposition \ref{prop:SpecUAR2pisharp}. In most
examples sharp geometrical action and hence time-like commutativity
of the quasi-theory in $\opo$ dimensions can be ruled out this way. In section
\ref{sec:cospactps} 
some typical branchings for current subalgebras and their \Name{Coset}
models are stated. Obviously, tensor products of chiral conformal
theories and tensor products of chiral fermions (analogous to bounded
\Name{Bose} fields) yield chiral subnets
with sharp geometrical action. It is unclear whether there are other
examples of this structure; compare remarks at the end of section
\ref{cha:netend}.\ref{sec:chihol}. 

The result of the spectrum of chiral conformal rotations by $2\pi$ in
proposition \ref{prop:SpecUAR2pisharp} is familiar in the context of
conformal quantum field theory in terms of operator-valued
distributions. Here, the two-point function of a quasi-primary field
$\Phi_{d_+d_-}$ of chiral scaling dimensions $d_+$ and $d_-$ is
determined by the transformation behaviour of the field to be:
\begin{equation}
  \label{eq:quaspri2pt}
  \left\langle\Omega, \, \Phi_{d_+ {d}_-} (x_+,x_-)
    \Phi_{d_+{d}_-}(y_+,y_-) \,\Omega\right\rangle = C_\Phi  
\klammer{\frac{-i}{\vartriangle_++i\varepsilon}}^{2d_+}
\klammer{\frac{-i}{\vartriangle_-+i\varepsilon}}^{2{d}_-} \zendot
\end{equation}
Here, $C_\Phi$ is some positive constant and $\vartriangle_+=x_+-y_+$,
$\vartriangle_-=x_--y_-$. 

Commutativity of the field for spacelike, $\vartriangle_+\vartriangle_-<0$,
and for time-like separation, $\vartriangle_+\vartriangle_->0$, yields a
symmetry of the two-point function in (\ref{eq:quaspri2pt}) with
respect to $\vartriangle_\pm\leftrightarrow-\vartriangle_\pm$. Taking this
together with the analyticity properties of the distributions on the
right-hand side of (\ref{eq:quaspri2pt})  results in restrictions on
the possible values of $d_\pm$:
%
% With a choice for the branch of the logarithm we have:
% \begin{equation}\label{eq:iepsbranch}
%   \klammer{\frac{-i}{- \vartriangle +i\varepsilon}}^{2d} =
%   \klammer{\frac{-i}{ \vartriangle -i\varepsilon}}^{2d} e^{-i2d \pi} 
% \end{equation}
% The $+i\varepsilon$ and the $-i\varepsilon$ prescriptions do not differ
% for $\vartriangle>0$ and the connection between both for $\vartriangle<0$ is
% known as well \cite[\S 8.3.4, \S 8.3.5]{fC74}: 
% \begin{equation}\label{eq:iepsrules}
%   \begin{array}{l@{=}l@{\,,\,\,}l}
%     (\vartriangle + i \varepsilon)^{2d}&
%     (\vartriangle - i \varepsilon)^{2d}&\vartriangle >0\\
%     (\vartriangle + i \varepsilon)^{2d}&
%     e^{i4\pi d}\,(\vartriangle - i \varepsilon)^{2d}&\vartriangle <0
%   \end{array}
% \end{equation}
% Combining (\ref{eq:iepsbranch}) and (\ref{eq:iepsrules}) leads to:
\begin{equation}
  \label{eq:iepsperm}
  \klammer{\frac{-i}{\vartriangle+i\varepsilon}}^{2d} = e^{\pm i 2\pi
    d} \klammer{\frac{-i}{-\vartriangle+i\varepsilon}}^{2d} \, , \quad
  \vartriangle \gtrless 0 \zendot
\end{equation}

Commutativity in spacelike directions results in
$e^{i2\pi(d_+-d_-)}{=} 1$ and commutativity for timelike
directions requires $e^{i2\pi(d_++d_-)}{=} 1$. Both
together imply $d_\pm\in\frac{1}{2}\dopp{N}$ and hence the spectrum of
chiral rotations by $2\pi$ is found again to be contained in
$\menge{\pm 1}$.

%%%%%%%%%%%%%%%%%%%%%%%%%%%%%%%%%%%%%%%%%%%%%%%%%%%%%
%%%%%% CosPas as normal, irred CTPSs %%%%%%%%%%%%%%%%
%%%%%%%%%%%%%%%%%%%%%%%%%%%%%%%%%%%%%%%%%%%%%%%%%%%%%
\clearpage
\sekt{\Name{Coset} construction of normal \Name{CTPS}}{Coset
  construction of normal {CTPS}}{sec:cospactps}
% \section[\Name{Coset} construction of normal \Name{CTPS}]{Coset
%   construction of normal {CTPS}}
% \label{sec:cospactps}    
% \fancyhead[CO]{\Name{Coset} construction of normal \Name{CTPS}}

The main goal in this section is to get examples of {\em normal
canonical tensor product subfactors} (normal \Name{CTPS}). This concept was
introduced by \Name{Rehren} \cite{khR00} in the context of inclusions
of tensor products of chiral observables in a $\opo$-dimensional
conformal theory. According to the definition of normal \Name{CTPS}
\cite{khR00},  \Name{Coset} pairs form another class of examples of
this structure, if they are spatial, normal and of finite index. Provided
 the net-endomorphism property holds for the respective
 inner-implementing representation, such \Name{Coset} pairs
have an interpretation directly analogous to that originally
considered in \cite{khR00} (by chiral holography, section 
\ref{cha:netend}.\ref{sec:chihol}).  

In the examples which we discuss below there is
a \Name{Coset} pair 
$\lok{A}\coset\lok{C}\subset\lok{B}$ which is spatial, the decomposition
({\em branching rules}) of this representation of
$\lok{A}\otimes\lok{C}$ available from group-theoretic
investigations. The branchings for
$\lok{A}\coset\lok{C}\subset\lok{B}$ are 
finite and, moreover, both $\lok{A}$ and 
$\lok{C}$ are {\em  completely rational}, a property which
consists of three parts: the models satisfy the split property and
strong additivity (cf section \ref{cha:cospa}.\ref{sec:disass}) and
their respective 
$\mu$-index is finite. The $\mu$-index is the index of the inclusion
described in equation (\ref{eq:jowassub}), and finite $\mu$-index
implies in particular that there are only finitely many sectors which
all have finite statistical dimension \cite[theorem 33]{KLM01}. For
this reason the branching 
tells us that the \Name{Coset} pair
$\lok{A}\coset\lok{C}\subset\lok{B}$ has finite index.

Because of strong additivity of $\lok{A}$ and $\lok{C}$ the \Name{Coset} 
pair $\lok{A}_{max}\coset\lok{C}_{max}\subset\lok{B}$ is known to be
normal, but the same follows from the presence of stress-energy
tensors in both subnets. We will
use our knowledge on $\lok{A}$, $\lok{C}$ and on
$\lok{A}\coset\lok{C}\subset\lok{B}$ in order to obtain the
branching rules for  $\lok{A}_{max}\coset\lok{C}_{max}\subset\lok{B}$.

%%%%%%%%%%%%%%%%%%%%%%%%%%%%%%%%%%%%%%%%%%%%%%%%%%%%%%%%%%%%%%%%%%%%%%%%%%%
%%%%%%%%% LEMMA SPATIALITY OF MAXIMAL COSET PAIR %%%%%%%%%%%%%%%%%%%%%%%%%%

\begin{lemma}\label{lem:spatcospamax}
  Assume $\lok{A}\coset\lok{C}\subset\lok{B}$ is a spatial
  \Name{Coset} pair of finite index.

  Then  $\lok{A}_{max}\coset\lok{C}_{max}\subset\lok{B}$ is spatial
  and of finite index as well. 
\end{lemma}

\begin{pf}
 The statement on the index follows from an identity contained in
 \cite{rL90}, compare the argument in the  proof of lemma
 \ref{lem:lokirredincl}. 

By assumption, $\lok{A}\subset\lok{B}$ is unitarily equivalent to a
direct sum of irreducible representations of
$\lok{A}$, the representation space decomposing as $\Hilb{H} = \sum_i
\Hilb{H}_{(i)}\otimes\Hilb{H}_i$, where the $\Hilb{H}_i$ are
the representation spaces of inequivalent, irreducible representations
$\pi_i$ and the $\Hilb{H}_{(i)}$ are multiplicity spaces. We denote the
projection onto $\Hilb{H}_{(i)}\otimes\Hilb{H}_i$ by $P_i$.

For an operator $V$ which commutes with all of $\lok{A}$, the operators
$P_i V P_j$ define intertwiners from $\Einsop_{(j)}\otimes \pi_j$ to
$\Einsop_{(i)}\otimes \pi_i$, which by inequivalence of the $\pi_i$
implies: $P_i V P_j = \delta_{ij} P_i V P_i$, ie $V = \sum_i
P_iVP_i$. Obviously, $P_iVP_i$ is of the form $P_iVP_i =:
V_{(i)}\otimes \Einsop_i$, and this means that the multiplicity spaces
$\Hilb{H}_{(i)}$ form representation spaces of $\lok{C}_{max}$.  
These representations are completely reducible into irreducibles by
 finiteness of index \cite{GL96}. The same argument works for
 $\lok{A}_{max}$ as well.
\end{pf}

For definiteness the following proposition summarises the relevant
results on spatial \Name{Coset} pairs of finite index obtained so far:
\begin{prop}\label{prop:spatialitymax}
  $\lok{A}\coset\lok{C}\subset\lok{B}$ a spatial \Name{Coset} pair
  with finite index. Assume $U^\lok{A}$, $U^\lok{C}$ to
  have the net-endomorphism property.

Then $\lok{A}_{max}\coset\lok{C}_{max}\subset\lok{B}$ is a normal,
irreducible, spatial \Name{Coset} pair with finite index, and thus
forms a normal \Name{CTPS}.
\end{prop}

\begin{pf}
The statement was proven with proposition \ref{prop:UACconf} and
lemma \ref{lem:spatcospamax}.   
\end{pf}

The {\em branchings} of the \Name{Coset} pairs
$\lok{A}\coset\lok{C}\subset\lok{B}$ and
$\lok{A}_{max}\coset\lok{C}_{max}\subset\lok{B}$ are written as
follows:
\begin{eqnarray}
  \label{eq:coupmatr}
  \rho&\cong&\bigoplus \,\, Z_{rs} \,\,
  \rho^\lok{A}_r\otimes\rho^\lok{C}_s \zencom\\
\label{eq:coupmatrmax}
  \rho_{max}&\cong&\bigoplus \,\, Z_{uv}^{max} \,\,
  \rho^{\lok{A}_{max}}_u\otimes\rho^{\lok{C}_{max}}_v \zendot
\end{eqnarray}
The finite  matrices $Z_{rs}$, $Z^{max}_{uv}$ are called the {\em coupling
  matrix} of the respective decomposition. Normality of a \Name{CTPS}
  is equivalent to $Z_{0s}=\delta_{0s}$, $Z_{r0}=\delta_{r0}$
  \cite[corollary 3.5]{khR00}.

The probably most remarkable result of \cite{khR00}, translated in our
setting, is the following:
If $\lok{A}_{max}\coset\lok{C}_{max}\subset\lok{B}$ forms a normal \Name{CTPS}, then the sets of superselection sectors
${[\rho^{\lok{A}_{max}}_u]}$,
${[\rho^{\lok{C}_{max}}_v]}$ appearing in (\ref{eq:coupmatrmax})
are invariant under conjugation and their direct sums span a subsystem
of sectors which is closed under fusion. The coupling matrix
$Z^{max}_{uv}$ is in fact a permutation matrix, ie for given $u$ there
is exactly one $v$ for which $Z^{max}_{uv}\neq 0$ and the map
$u\mapsto v(u)$ derived from this condition yields $Z^{max}_{uv} =
\delta_{uv(u)}$. In this way the coupling matrix induces an isomorphism
of fusion algebras for  
the subsystems of sectors of $\lok{A}_{max}$, $\lok{C}_{max}$
generated by the subsectors of $\rho_{max}$ through
$[\rho^{\lok{A}_{max}}_u]\mapsto [\rho^{\lok{C}_{max}}_{v(u)}]$. This
means in particular that the statistical dimensions of coupled sectors
have to coincide:
$d(\rho^{\lok{A}_{max}}_u)= d(\rho^{\lok{C}_{max}}_{v(u)})$. 

\Name{M\"uger} has announced an extension of these findings in case
$\lok{B}$ has trivial superselection structure, ie all locally normal
representations of $\lok{B}$ are equivalent to the vacuum
representation \cite{mM02}. In this case the isomorphism of fusion
rules extends to an isomorphism of the respective \Name{DHR}
subcategories. %  If all the sectors of $\lok{A}_{max}$ and
% $\lok{C}_{max}$ are contained in $\rho_{max}$, $\lok{B}$ has trivial
% superselection structure if and only if the coupling matrix is a
% {\em modular invariant}  with respect to the left and right
% statistics representation of $\SL(2,\dopp{Z})$ (cf \cite{khR01}).
These claims motivated our discussing the example contained in section
\ref{sec:e81ctps}. 

The results of \Name{Rehren} and \Name{M\"uger} show that branchings
of \Name{Coset} pairs establish direct links between the
superselection structure of $\lok{A}_{max}$ and
$\lok{C}_{max}$. Moreover, the \Name{DHR} triple of a chiral subnet of
finite index determines completely  the vacuum representation of the
larger theory and, in the opposite direction, the embedding itself
determines the vacuum representation of the subnet. In principle, the 
superselection structures of $\lok{A}$, $\lok{C}$, $\lok{A}_{max}$,
$\lok{C}_{max}$ all are connected by embedding, local extension and
coupling, provided the assumptions of proposition
\ref{prop:spatialitymax} hold (or, possibly, alterations of these). 

Of these connections we use only rudimentary details.  Let
$\iota^\lok{A}:\lok{A}(I)\hookrightarrow\lok{A}_{max}(I)$ denote the
embedding inducing the inclusion
$\lok{A}(I)\subset\lok{A}_{max}(I)$; this inclusion is irreducible by
the assumption of finite index (lemma \ref{lem:lokirredincl}). There
is an injective  homomorphism 
$\bar{\iota}^\lok{A}:\lok{A}_{max}(I)\rightarrow\lok{A}(I)$ such
that $\iota^\lok{A}\bar{\iota}^\lok{A}$ equals $\gamma^\lok{A}_I$, the
canonical endomorphism of $\lok{A}(I)\subset\lok{A}_{max}(I)$, and we have
$\bar{\iota}^\lok{A}\iota^\lok{A}=\rho_I^\lok{A}$ for the dual
canonical endomorphism. 
An endomorphism $\sigma$ of $\lok{A}_{max}(I)$
possesses $\bar{\iota}^\lok{A}\sigma\iota^\lok{A}$  as {\em
  restriction of $\sigma$}  to an endomorphism of $\lok{A}(I)$. For a
short  summary on these notions see, for example, 
\cite[section 2.8.C.]{LR95}. Since the decompositions of $\rho_{max}$,
$\rho^\lok{A}$ as localised representations and as dual canonical
endomorphisms are the same (as covariant representations of finite
index \cite{GL96,rL01}), we may study the fusion rules of
restricted representations by looking at the corresponding restricted
endomorphisms. 

 In the examples below, the fusion rules
of restricted endomorphisms are known, while their decomposition into
irreducibles is not (initially). But we do know the
following: The canonical endomorphism\footnote{It has to be emphasised that
  the canonical endomorphism does in general not have a continuation
  to a \Name{DHR} endomorphism and that thus not all irreducibles
  contained in it need
  to have an interpretation as a \Name{DHR} endomorphism. In this
  respect the canonical endomorphism is quite different from the dual
  canonical endomorphism.} $\gamma_I^\lok{A}$ contains the identity on
$\lok{A}_{max}(I)$ with multiplicity $1$. Since the dual canonical
endomorphism $\rho_{max}$ is self-conjugate, to 
  every $\rho^{\lok{A}_{max}}_u$ appearing in 
(\ref{eq:coupmatrmax}) the conjugate $\rho^{\lok{A}_{max}}_{\bar{u}}$
appears as well. The product of the corresponding restricted
endomorphisms must contain $\rho^\lok{A}$:
\begin{eqnarray}
    \bar{\iota}^\lok{A}\rho^{\lok{A}_{max}}_{Iu}\iota^\lok{A}\,\,\,
  \bar{\iota}^\lok{A}\rho^{\lok{A}_{max}}_{I\bar{u}}\iota^\lok{A}
  &=&
  \bar{\iota}^\lok{A}\rho^{\lok{A}_{max}}_{Iu}\gamma^\lok{A}_I 
  \rho^{\lok{A}_{max}}_{I\bar{u}}\iota^\lok{A}\nonumber\\
  &\succ&
  \bar{\iota}^\lok{A}\rho^{\lok{A}_{max}}_{Iu}
  id^{\lok{A}_{max}(I)}
  \rho^{\lok{A}_{max}}_{I\bar{u}}\iota^\lok{A}\nonumber\\
  \label{eq:restricfusion}
  &\succ&
  \bar{\iota}^\lok{A}id^{\lok{A}_{max}(I)}\iota^\lok{A} =
  \rho^\lok{A}_I \zendot
\end{eqnarray}
The same applies to $\lok{C}$ and $\lok{C}_{max}$, of course.

Below we will derive the branching of the extended \Name{Coset} pairs
$\lok{A}_{max}\coset\lok{C}_{max} \subset \lok{B}$, which define
normal \Name{CTPS}. The basis for the discussions are 
branching formulae for a \Name{Coset} pair
$\lok{A}\coset\lok{C}\subset\lok{B}$ which were derived by character
arguments in \cite{GKO86}, \cite{KW88}, \cite{KS88}. These (and other)
sources contain
branchings for a number of other \Name{Coset} pairs as well. We have
selected the three examples discussed in sections \ref{sec:GKOcospa},
\ref{sec:exepext}, \ref{sec:e81ctps} for their special interest.

The main tools of the arguments below are the present knowledge 
on current algebras and of $\Vir_{c<1}$ models, the spatial 
identification of $\lok{A}_{max}$ and $\lok{C}_{max}$ (proposition
\ref{prop:lemma23}), the results on normal \Name{CTPS} discussed above
\cite{khR00} and some general results on completely rational
models, mainly from \cite{KLM01}.
%%%%%%%%%%%%%%%%%%%%%%%%%%%%%%%%%%%%%%%%%%%%%%%%%%%%%%%%%%%%%%%%%%%%%%
%%%%%%%%% GKO COSET PAIRS %%%%%%%%%%%%%%%%%%%%%%%%%%%%%%%%%%%%%%%%%%%%
%%%%%%%%%%%%%%%%%%%%%%%%%%%%%%%%%%%%%%%%%%%%%%%%%%%%%%%%%%%%%%%%%%%%%%

\subsection{The \Name{Coset} pairs of \Name{Goddard, Kent and Olive}}
\label{sec:GKOcospa}

One achievement in $\opo$-dimensional  and chiral conformal quantum
field theory was the construction of the stress-energy tensors with
central charge less than $1$
as \Name{Coset} models by \Name{Goddard, 
  Kent and Olive} (\Name{GKO}) \cite{GKO85,GKO86}. The analysis of
\Name{Friedan, 
  Qiu and Shenker} \cite{FQS84,FQS84a,FQS86} (cf \cite{rL88}) had
shown that below $1$ the central charges had to be contained in the
{\em discrete series}:
\begin{equation}
  \label{eq:discserc}
c(m) = 1- \frac{6}{(m+2)(m+3)}\, , \quad m\in\dopp{N} \zendot
\end{equation}

The \Name{Coset} construction of \Name{GKO} proved these models to
exist as \Name{Coset} models associated with the chiral
subnet
$\Loop{SU(2)}_{m+1}\subset\Loop{SU(2)}_{1}\otimes\Loop{SU(2)}_{m}$,
the inclusion induced by the diagonal embedding of the colour algebras.
The \Name{Coset} stress-energy tensor, which generates the
$\Vir_{c(m)}$ model, is simply given by the difference of the
\Name{Sugawara} stress-energy tensors of
$\Loop{SU(2)}_{1}\otimes\Loop{SU(2)}_{m}$ and of $\Loop{SU(2)}_{m+1}$. 
There are various other ways to construct $\Vir_{c<1}$ models as
\Name{Coset} stress-energy tensors of current algebra inclusions
(classification: \cite{BG87}, examples eg in \cite{GKO86,KW88}). 

The branching of the vacuum  
representation of  $\Loop{SU(2)}_{1}\otimes\Loop{SU(2)}_{m}$ with
respect to the 
\Name{Coset} pair $\Loop{SU(2)}_{m+1}\coset\Vir_{c(m)}$ reads
\cite{GKO86}, \cite[4.1.a]{KW88}: 
\begin{equation}
  \label{eq:GKObranch}
   L(\Lambda_0)\otimes L(m\Lambda_0)\!\!\restriction \,\cong \sum_{l: 0\leq 2l \leq m+1}
  L((m+1-2l) \Lambda_0, 2l \Lambda_1)\otimes V_{1,2l+1}^{(m)} \,\, .
\end{equation}
By the symbol $\restriction$ we indicate that the formula gives the
decomposition of  $L(\Lambda_0)\otimes L(m\Lambda_0)$ as a
representation of $\Loop{SU(2)}_{m+1}\otimes\Vir_{c(m)}$.

The highest-weight representations of
$\Loop{SU(2)}_{k}$ are denoted as in \cite{KW88}
according to the classification of (unitarisable) highest-weight
representations of 
affine \Name{Kac-Moody} algebras (see \cite{vKbook}, summary in
\cite{KW88}). $L(k\Lambda_0)$ stands for the vacuum
representation of $\Loop{SU(2)}_{k}$ and $L((k-l) \Lambda_0, l
\Lambda_1)$ stands for the highest-weight representation of
$\Loop{SU(2)}_{k}$ in which the horizontal subalgebra acts as a spin
$l/2$ representation of $SU(2)$ on the vectors of lowest
energy.

The highest-weight vector of the representation $L((k-l) \Lambda_0, l
\Lambda_1)$ has conformal energy $l(l+2)/(4(2+k))$. Hence, not
all the lowest conformal energy eigenvalues occurring in
(\ref{eq:GKObranch}) can be half-integers; this prohibits sharp
geometrical action of the inner-implementing representation
$U^{\Loop{SU(2)}_{m+1}}$ (proposition \ref{prop:SpecUAR2pisharp}). 

The highest-weight representations $V_{r,s}^{(m)}$ of $\Vir_{c(m)}$
are identified by the energy $h^{(m)}_{r,s}$ of their lowest-energy
vector. There are only finitely many of such representations, namely:
\begin{equation}
  \label{eq:highvir}
  h^{(m)}_{r,s} = \frac{\left[(m+3)r-s(m+2)\right]^2-1}{4(m+2)(m+3)}\,
, \quad 1\leq s\leq r \leq m+1 \zendot
\end{equation}

Together with\footnote{Alternatively one may argue as in \cite[lemma 3.2,
  corollary 3.3]{KL02}} proposition
\ref{prop:lemma23} equation (\ref{eq:GKObranch}) shows that
$\Vir_{c(m)}$ is indeed the maximal
\Name{Coset} model; since there are no chiral subnets in a net generated
  by a stress-energy tensor \cite{sC98}, these \Name{Coset} models are minimal
  among the non-trivial ones as well.

\Name{Xu} \cite{fX00,fX99,fX01} studied the \Name{Coset} models of
the chiral subnets $
\Loop{SU(n)}_{k+l}\subset\Loop{SU(n)}_{k}\otimes\Loop{SU(n)}_{l}$
arising from the diagonal embedding of colour algebras. His results
cover, of course, the \Name{GKO} \Name{Cosets}\footnote{For an earlier
  investigation on the $\Vir_{c<1}$-models as \Name{GKO} \Name{Coset}
  models see \cite{tL94}.}. 
 \Name{Kawahigashi and Longo} \cite{KL02} completed the proof of the
$\Vir_{c<1}$ models being  completely rational. Since the
$\Loop{SU(n)}_{k}$ are completely rational as well
\cite{FG93,vL97,fX00a}, this tells us that 
$\Loop{SU(2)}_{m+1}\coset\Vir_{c(m)}\subset\Loop{SU(2)}_{1}
\otimes\Loop{SU(2)}_{m}$  
does indeed form a normal \Name{CTPS}.

Furthermore, \Name{Kawahigashi and Longo} \cite{KL02} completed the proof
of the $\Vir_{c(m)}$ fusion rules 
being identical with the ones expected from operator product
expansions of primary fields and the {\em statistical
representation} of $\SL(2,\dopp{Z})$ \cite{khR90b} being identical with the
{\em modular representation} (see eg \cite[section 10.6]{DMS96}). The latter
identity is induced by the diagonalisation of the fusion rules by both
representations (\Name{Verlinde} formulae, \cite{eV88} \cite{khR90b})
and the identification of the highest weights $h^{(m)}_{r,s}$. More
details on the problem of matching these two representations of
$SL(2,\dopp{Z})$ for general rational models in \cite{khR01}, \cite{jF94}. 

 For future
reference we state the fusion rules
\begin{equation}
  \label{eq:vircmfusion}
  \phi_{r_1,s_1}^{(m)}\phi_{r_2,s_2}^{(m)} \cong
\bigoplus_{{r=|r_1-r_2|+1, \atop r+r_1+r_2:{\rm odd}}}
^{r_{max}}
\bigoplus_{{s=|s_1-s_2|+1, \atop s+s_1+s_2:{\rm odd}}}
^{s_{max}}
\phi_{r,s}^{(m)}\, , 
\end{equation}
 where $r_{max}= \min(r_1+r_2-1, 2m+3-r_1-r_2)$,
 $s_{max}=\min(s_1+s_2-1, 2m+5-s_1-s_2)$. It easy to see that all
 $\phi_{r,s}^{(m)}$ are self-conjugate, ie the vacuum sector
 $\phi_{1,1}^{(m)}$ is contained only in the squares
 $(\phi_{r,s}^{(m)})^{2}$ and it appears in these precisely
 once. The  functions on the 
 right-hand side of  (\ref{eq:highvir}) are invariant 
with respect to the simultaneous replacement $r\rightarrow m+2-r$,
$s\rightarrow m+3-s$. This identification has to be taken into
account when calculating the fusion rules.

The statistical dimensions of the representations $\phi_{r,s}^{(m)}$ are:
 \begin{equation}
   \label{eq:virstatdim}
 d^{(m)}_{r,s} =
 \frac{\sin(r\pi/(m+2))}{\sin(\pi/(m+2))}\frac{\sin(s\pi/(m+3))}{\sin(\pi/(m+3))}
 \zendot
 \end{equation}
% In particular, these values cover almost all dimensions
% allowed for braided endomorphisms  below $\sqrt{6}$ \cite{khR95}; only
% $\sqrt{5}$ does not appear. 

%%%%%%%%%%%%%%%%%%%%%%%%%%%%%%%%%%%%%%%%%%%%%%%%%%%%%%%%%%%%%%%%%%%%%%
%%%%%%%%% EXCEPTIONAL EXTENSION OF VIR 21/22 %%%%%%%%%%%%%%%%%%%%%%%%%
%%%%%%%%%%%%%%%%%%%%%%%%%%%%%%%%%%%%%%%%%%%%%%%%%%%%%%%%%%%%%%%%%%%%%%

\subsection{Extension $(A_{10},E_6)$ of $\Vir_{\frac{21}{22}}$ as a \Name{Coset} model}
\label{sec:exepext}

\Name{Kawahigashi and Longo} \cite{KL02} completed the classification
of local extensions of the $\Vir_{c<1}$ models. Most of these are
given as {\em orbifolds}: the local extension contains the $\Vir_{c<1}$
model as a fixed-point subtheory with respect to a $\dopp{Z}_2$
symmetry; some of these are among the maximal \Name{Coset} models
associated with current subalgebras, as branchings contained in
\cite{KW88} show. Only four local extensions are of a different type
({\em exceptional cases}). For two of these \Name{Kawahigashi and Longo}
\cite{KL02} gave a rigorous interpretation as \Name{Coset} models of
current subalgebras following suggestions of \Name{B\"ockenhauer and
  Evans} \cite{BE99a}. These are the extension $(E_6,A_{12})$ of
$\Vir_{c(10)}$, associated with $ \Loop{SU(2)}_{11}\subset
\Loop{SO(5)}_1\otimes \Loop{SU(2)}_1$, and the 
extension $(E_8,A_{30})$ of $\Vir_{c(28)}$, associated with $
\Loop{SU(2)}_{29}\subset \Loop{G(2)}_1\otimes \Loop{SU(2)}_1$. 

A third of the exceptional local extensions is  identified through
 the branching rules for the \Name{Coset} pair
$\Loop{SU(9)_2}\coset\Vir_{c(9)}\subset \Loop{E(8)}_2$ \cite[4.3.a]{KW88}:
\begin{equation}
  \label{eq:su9e8br}
  L(2\Lambda_0)\!\!\restriction \,\cong \sum^4_{l=0}
  \left[L(\mu_{2l+1})+L(\sigma\mu_{2l+1})+L(\sigma^2\mu_{2l+1})\right]\otimes
  \left[V^{(9)}_{2l+1,1}+ V^{(9)}_{2l+1,7}\right] \, .
\end{equation}
By the symbol $\restriction$ we indicate that the formula gives the
decomposition of  $L(2\Lambda_0)$ as a
representation of $\Loop{SU(9)}_2\otimes\Vir_{c(9)}$. 

We have in (\ref{eq:su9e8br}): 
\begin{displaymath}
  \begin{array}{r@{\quad\vline\quad}c@{\quad\quad}c@{\quad\quad}c@{\quad\quad}c@{\quad\quad}c}
l&0&1&2&3&4\\
 \hline &&&&&\\[-2.5ex]
\mu_{2l+1}& 2\dot{\Lambda}_0 &\dot{\Lambda}_1+\dot{\Lambda}_8
&\dot{\Lambda}_2+\dot{\Lambda}_7 &\dot{\Lambda}_3+\dot{\Lambda}_6
&\dot{\Lambda}_4+\dot{\Lambda}_5\\  
\sigma\mu_{2l+1} &2\dot{\Lambda}_3  &\dot{\Lambda}_4+\dot{\Lambda}_2
&\dot{\Lambda}_5+\dot{\Lambda}_1 &\dot{\Lambda}_6+\dot{\Lambda}_0
&\dot{\Lambda}_7+\dot{\Lambda}_8\\ 
\sigma^2\mu_{2l+1} &2\dot{\Lambda}_6 &\dot{\Lambda}_7+\dot{\Lambda}_5
&\dot{\Lambda}_8+\dot{\Lambda}_4 &\dot{\Lambda}_0+\dot{\Lambda}_3
&\dot{\Lambda}_1+\dot{\Lambda}_2\\ 
  \end{array}
\end{displaymath}
As in \cite{KW88} we denote the fundamental weights of the affine
\Name{Kac-Moody} algebra with horizontal subalgebra isomorphic to the
\Name{Lie} algebra of $SU(9)$ as $\dot{\Lambda}_i$, $i=0,\ldots,8$. If
one is interested in the highest weight of the module of $SU(9)$ formed by the
vectors of lowest energy, one simply ignores the multiple of
$\dot{\Lambda}_0$ and takes $\dot{\Lambda}_i$, $i=1,\ldots,8$, to
stand for the corresponding 
fundamental weights of $SU(9)$. 
The notation for the representations of the $\Vir_{c<1}$ models was
introduced in the previous section already. 

The conformal energies of the highest-weight vectors for
$\Loop{SU(9)}_2$ occurring in (\ref{eq:su9e8br}) are given by:
\begin{displaymath}
  \begin{array}{c@{\quad\vline\quad}c@{\quad\quad}c@{\quad\quad}c@{\quad\quad}c@{\quad\quad}c}
l&0&1&2&3&4\\
\hline&&&&&\\[-2.5ex]
h_{\mu_{2l+1}}&0&\frac{9}{11}&\frac{16}{11}&\frac{21}{11}&\frac{24}{11}\\
&&&&&\\[-2.5ex]%\hline
&&&&&\\[-2.5ex]
h_{\sigma\mu_{2l+1}}& 2 & \frac{20}{11} & \frac{16}{11} &
\frac{10}{11} & \frac{13}{11}\\
&&&&&\\[-2.5ex]%\hline
&&&&&\\[-2.5ex]
h_{\sigma^2\mu_{2l+1}} & 2 &\frac{20}{11} & \frac{16}{11} &
\frac{10}{11} & \frac{13}{11}\\
  \end{array}
\end{displaymath}
Most of these lowest conformal energy eigenvalues are not
half-integers; this prohibits sharp  geometrical action of the 
inner-implementing representation 
$U^{\Loop{SU(9)}_2}$ (proposition \ref{prop:SpecUAR2pisharp}). 

Applying results of \Name{Longo} \cite{rL01}, we readily extract from (\ref{eq:su9e8br})  that $ \Loop{E(8)}_2$ is
completely rational, since $ \Loop{SU(9)}_2$ \cite{FG93,vL97,fX00a}
and $\Vir_{c(9)}$ \cite{KL02} are and the branching is finite. The
more interesting consequence of (\ref{eq:su9e8br}) is:

\begin{prop}\label{prop:43a}
  The extension $(A_{10},E_6)$ of $\Vir_{c(9)}$ is the maximal
  \Name{Coset} model associated with the chiral subnet
  $\Loop{SU(9)_2}\subset \Loop{E(8)}_2$. 

The decomposition of $\rho_{max}$, the localised representation of the
\Name{Coset} pair
$\lok{A}_{max}\coset\lok{C}_{max}\subset\Loop{E(8)}_2$ 
associated with $\Loop{SU(9)_2}\coset\Vir_{c(9)}\subset \Loop{E(8)}_2$,
is:
\begin{equation}\label{eq:branch43a}
  \rho_{max} \cong \bigoplus_{l=0}^4 \,\,
  \rho^{\lok{A}_{max}}_l\otimes\rho^{\lok{C}_{max}}_l\, , 
\end{equation}
where we have the following branchings:
\begin{eqnarray}
  \label{eq:branchA43a}
  \bar{\iota}^\lok{A}\rho^{\lok{A}_{max}}_l\iota^\lok{A}&\cong&
  \alpha_{L(\mu_{2l+1})} \oplus
  \alpha_{L(\sigma\mu_{2l+1})} \oplus
  \alpha_{L(\sigma^2\mu_{2l+1})} \zencom\\ 
  \label{eq:branchC43a}
  \bar{\iota}^\lok{C}\rho^{\lok{C}_{max}}_l\iota^\lok{C}&\cong&
  \phi^{(9)}_{2l+1,1} \oplus
   \phi^{(9)}_{2l+1,7} \zendot
\end{eqnarray}
The representations $\rho^{\lok{A}_{max}}_l$, $\rho^{\lok{C}_{max}}_l$
are all self-conjugate.
\end{prop}

\begin{pf}
Following proposition \ref{prop:lemma23} we read off (\ref{eq:su9e8br})
that the maximal \Name{Coset} model $\lok{C}_{max}$ associated with
$\Loop{SU(9)_2}\subset \Loop{E(8)}_2$ is a non-trivial local extension
of $\Vir_{c(9)}$. The decomposition of the vacuum
representation of $\lok{C}_{max}$ with respect to $\Vir_{c(9)}$ is
given by:
\begin{equation}
  \label{eq:rhoc9}
  \rho^\lok{C}\cong \phi_{1,1}^{(9)}\oplus \phi_{1,7}^{(9)}\,.
\end{equation}
 The extension is effected by a field of scaling dimension $8$, and the index
$[\lok{C}_{max}:\Vir_{c(9)}]$  coincides with the statistical
dimension of $\rho^\lok{C}$, which we calculate using
(\ref{eq:virstatdim}) as: $d(\rho^\lok{C})=3+\sqrt{3}$. There is only
one local extension of $\Vir_{c(9)}$ with this index, namely the
$(A_{10},E_6)$ extension \cite[table 3]{KL02}, which is hence
identified as a \Name{Coset} model.

$\lok{A}_{max}$ is a non-trivial local extension of
$\Loop{SU(9)_2}$, and it is completely rational, as $\Loop{SU(9)_2}$
is included with finite index \cite{rL01}. From arguments above 
we know that 
$\lok{A}_{max}\coset\lok{C}_{max}\subset \Loop{E(8)}_2$ is a normal
\Name{CTPS} (proposition \ref{prop:spatialitymax}). We 
want to derive the decomposition of 
$\rho_{max}$ from (\ref{eq:su9e8br}). It is clear that it has the  form of
(\ref{eq:coupmatrmax}), 
where each $\rho^{\lok{A}_{max}}_u$ ($\rho^{\lok{C}_{max}}_v$)  appears
only once, and for each $\rho^{\lok{A}_{max}}_u$
($\rho^{\lok{C}_{max}}_v$) the conjugate
$\rho^{\lok{A}_{max}}_{\bar{u}}$ ($\rho^{\lok{C}_{max}}_{\bar{v}}$)
has to appear as well \cite{khR00}. 
We know that all the representations occurring in (\ref{eq:su9e8br})
come from restrictions of the $\rho^{\lok{A}_{max}}_u$,
$\rho^{\lok{C}_{max}}_v$. 

Any sector of $\Vir_{c(9)}$ appearing in
(\ref{eq:su9e8br}) appears only once and they all are
self-conjugate. Looking at the restricted fusion rules of conjugate
sectors of $\lok{C}_{max}$, equation  (\ref{eq:restricfusion}), we
conclude: if a $\phi^{(9)}_{2l+1,1}$ appearing in  (\ref{eq:su9e8br}) is
the restriction of a $\rho^{\lok{C}_{max}}_v$, then its square has to
contain $\rho^\lok{C}$. But this is not the case, as one may verify
using the fusion rules (\ref{eq:vircmfusion}). This shows that the
restrictions of irreducible $\rho^{\lok{C}_{max}}_v$ contained in
$\rho_{max}$ are of the form
$\bar{\iota}^\lok{C}\rho^{\lok{C}_{max}}_v{\iota}^\lok{C}\cong
\phi^{(9)}_{2l+1,1} \oplus \phi^{(9)}_{2l+1,7}$, with $l$ depending on
$v$.

Now we may calculate the statistical dimensions of the irreducible
$\rho^{\lok{C}_{max}}_v$ contained in $\rho_{max}$ from (\ref{eq:su9e8br}),
(\ref{eq:virstatdim}) according 
to \cite[proposition 3.1]{sC02}:
\begin{equation}\label{eq:virc9dim}
      d(\bar{\iota}^\lok{C}\rho^{\lok{C}_{max}}_v\iota^\lok{C}) =
      d(\rho^{\lok{C}_{max}}_v) \,\,\, [\lok{C}_{max}: \Vir_{c(9)}] = 
      d(\rho^{\lok{C}_{max}}_v) \,\,\, d(\rho^\lok{C}) 
      \zendot
\end{equation}
The statistical dimensions of the representations of $\Vir_{c(9)}$ 
occurring in (\ref{eq:su9e8br}) are given by: 
\begin{equation}
  \label{eq:virstatV9}
  \begin{array}{c|ccc}
l&\phantom{\hspace{2em}}d^{(9)}_{2l+1,1}\phantom{\hspace{2em}}&\phantom{\hspace{2em}}d^{(9)}_{2l+1,7}\phantom{\hspace{2em}}&
(d^{(9)}_{2l+1,1}+d^{(9)}_{2l+1,7}) d(\rho^\lok{C})^{-1}\\ \hline 
0 & 1\phantom{.000000} & 3.732051 & 1\phantom{.000000}\\
1 & 2.682507 & 10.011252 & 2.682507 \\
2 & 3.513337 & 13.111953 & 3.513337 \\
3 & 3.228707 & 12.049700 & 3.228707 \\
4 & 1.918986 & 7.161753 & 1.918986 \\
  \end{array}
\end{equation}

The fusion rules of positive-energy representations $\Loop{SU(9)}_2$
are known \cite{aW98} and, therefore, the corresponding statistical
dimensions may be calculated as asymptotic dimensions
using formulae in, for example, \cite{KW88}. The dimensions for
representations occurring in (\ref{eq:su9e8br}) are: 
  \begin{equation}\label{eq:dimSU92}
    \begin{array}{lclclcl}
\multicolumn{5}{r}{d_{2\dot{\Lambda}_0} = d_{2\dot{\Lambda}_3} = d_{2\dot{\Lambda}_6}}&=&1\zencom\\
d_{\dot{\Lambda}_1+\dot{\Lambda}_8}&=&
d_{\dot{\Lambda}_4+\dot{\Lambda}_2}&=&
d_{\dot{\Lambda}_7+\dot{\Lambda}_5}&
\approx&2.682507\zencom\\
d_{\dot{\Lambda}_2+\dot{\Lambda}_7}&=&
d_{\dot{\Lambda}_5+\dot{\Lambda}_1}&=&
d_{\dot{\Lambda}_8+\dot{\Lambda}_4}&
\approx&3.513337\zencom\\
d_{\dot{\Lambda}_3+\dot{\Lambda}_6}&=&
d_{\dot{\Lambda}_6+\dot{\Lambda}_0}&=&
d_{\dot{\Lambda}_0+\dot{\Lambda}_3}&
\approx&3.228707\zencom\\
d_{\dot{\Lambda}_4+\dot{\Lambda}_5}&=&
d_{\dot{\Lambda}_7+\dot{\Lambda}_8}&=&
d_{\dot{\Lambda}_1+\dot{\Lambda}_2}&
\approx&1.918986\zendot\\
  \end{array}
\end{equation}
Below $\sqrt{6}$ the numbers which can occur as a statistical
dimension form a discrete 
set and the only value below $\sqrt{2}$ is $1$ \cite{vJ83,khR95}. This 
allows to write 
actual equalities in the first line, which implies:  
$[\lok{A}_{max}: \Loop{SU(9)}_2]= d(\rho^\lok{A}) = 3$.
Through the corresponding
version of equation (\ref{eq:virc9dim}) the statistical dimensions of
candidate $\rho^{\lok{A}_{max}}_u$ given by thinkable
decompositions of $\rho_{max}$ may be calculated. These have to
coincide with the dimension of the $\rho^{\lok{C}_{max}}_v$ they are
coupled to \cite[theorem 3.6]{khR00}. Comparing (\ref{eq:virstatV9})
and (\ref{eq:dimSU92}), this requirement admits the 
decomposition of $\rho_{max}$ only to be given by 
(\ref{eq:branch43a}), (\ref{eq:branchA43a}), (\ref{eq:branchC43a}).

The representations $\rho^{\lok{A}_{max}}_l$, $\rho^{\lok{C}_{max}}_l$
are all self-conjugate, because the $\phi^{(m)}_{r,s}$ all are and the
coupling matrix induces an isomorphism of fusion rules \cite{khR00}.
\end{pf}

Remark: According to  \cite[4.1.b]{KW88}, the $(A_{10},E_6)$ extension of
$\Vir_{c(9)}$ may as well be viewed as the maximal \Name{Coset} model
associated with the chiral subnet
$\Loop{E(8)}_3\subset\Loop{E(8)}_2\otimes\Loop{E(8)}_1$. 

Concerning the forth
exceptional local extension of a $\Vir_{c<1}$ model, namely 
the $(A_{28},E_8)$ extension of $\Vir_{c(27)}$, the classification of
current subalgebras with \Name{Coset} stress-energy tensor having
$c<1$ \cite{BG87} and the classification of conformal inclusions of
current algebras \cite{AGO87,BB87,SW86} suggests that this is probably
not a \Name{Coset} model associated with a current
subalgebra. Nevertheless, this chiral conformal theory is known to
exist by an abstract construction relying on \Name{DHR} data of
$\Vir_{c(27)}$ \cite{KL02} and thus appears to be a genuine
achievement of the conceptual approach of local quantum physics.

%%%%%%%%%%%%%%%%%%%%%%%%%%%%%%%%%%%%%%%%%%%%%%%%%%%%%%%%%%%%%%%%%%%%%%
%%%%%%%% A normal CTPS in LOOP E(8)_1 %%%%%%%%%%%%%%%%%%%%%%%%%%%%%%%%
%%%%%%%%%%%%%%%%%%%%%%%%%%%%%%%%%%%%%%%%%%%%%%%%%%%%%%%%%%%%%%%%%%%%%%

\subsection{A normal \Name{CTPS} in $ \Loop{E(8)}_1$}
\label{sec:e81ctps}

As an example for the structure discussed by \Name{M\"uger}
\cite{mM02} we discuss a \Name{Coset} pair in $ \Loop{E(8)}_1$. This
we will show first: Up to unitary equivalence the 
vacuum representation is the only locally normal representation of
$\Loop{E(8)}_1$. If we look only at the representations that stem from
integrating a unitarisable highest-weight
representation  of the corresponding current algebra, this is clear
because the \Name{Weyl}-alcove condition 
only admits one highest weight, namely the vacuum. But we need a
statement on the chiral model $\Loop{E(8)}_1$, which we obtain
applying well-known statements on completely rational models and our
current knowledge of the models $ \Loop{SU(n)}_k$. 

\begin{prop}\label{prop:E81trivsss}
  $\Loop{E(8)}_1$ is completely rational and its only sector is the
  vacuum sector.
\end{prop}

\begin{pf}
  Complete rationality of $\Loop{E(8)}_1$ holds, because
  $\Loop{SU(9)}_1\subset\Loop{E(8)}_1$ is a conformal
  inclusion. Indeed, the branching of this chiral subnet reads
  \cite{KS88}:
  \begin{equation}
    \label{eq:su91e81}
    L(\Lambda_0)\!\!\restriction \,\cong L(\dot{\Lambda}_0) +L(\dot{\Lambda}_3)
    +L(\dot{\Lambda}_6) \zendot
  \end{equation}
By the symbol $\restriction$ we indicate that the formula gives the
decomposition of  $L(\Lambda_0)$ as a
representation of $\Loop{SU(9)}_1$. 
  Again the notation is standard for representations of affine
  \Name{Kac-Moody} algebras (see \cite{KW88} for a summary):
  $L(\Lambda_0)$ is the vacuum representation of $\Loop{E(8)}_1$,
  $L(\dot{\Lambda}_0)$ is the vacuum representation  and
  $L(\dot{\Lambda}_{3,6})$ are inequivalent highest-weight representations of
  $\Loop{SU(9)}_1$.

  To prove that $\Loop{E(8)}_1$ has trivial superselection structure,
  we calculate its $\mu$-index which turns out to be $1$; this is a
  necessary and sufficient condition  for trivial superselection
  structure for completely rational chiral nets \cite[corollary
  32]{KLM01}. The identity $\mu_{\Loop{E(8)}_1}=1$ appears implicitly in 
  \cite{cS95} as well, but we prefer giving a self-contained proof. We
  know that the $\mu$-index of $ \Loop{SU(9)}_1$ is $9$ \cite[section
  4.4]{fX00a}  and since there are $9$
  locally normal  representations of $ \Loop{SU(9)}_1$ (cf remarks at
  the end of section  
  \ref{cha:currsub}.\ref{sec:hakacurr}) each of them
  has statistical dimension $1$. Hence, the index of the conformal
  inclusion is 
  determined to be: $[\Loop{E(8)}_1:\Loop{SU(9)}_1]=3$.

  The $\mu$-index of $\Loop{E(8)}_1$ may be calculated using the
  conformal inclusion above \cite[proposition 24]{KLM01}:
  \begin{equation}
    \label{eq:muE81}
    \mu(\Loop{SU(9)}_1) = [\Loop{E(8)}_1:\Loop{SU(9)}_1]^2
    \mu(\Loop{E(8)}_1) \zendot
  \end{equation}
  This yields $\mu(\Loop{E(8)}_1)=1$.
\end{pf}

\newpage
Now we present the example itself, the \Name{Coset} pair
$\Loop{SU(2)}_{16}\coset \Loop{SU(3)}_6 \subset \Loop{E(8)}_1$. It has
the branching \cite{KS88}:
\begin{eqnarray}
L(\Lambda_0)\restriction  &\cong& 
[L(16\dot{\Lambda}_0)\oplus L(16\dot{\Lambda}_1)]
\otimes 
[L(6\ddot{\Lambda}_0)\oplus L(6\ddot{\Lambda}_1)\oplus
 L(6\ddot{\Lambda}_2)]
\nonumber\\ 
&&
\oplus
[L(14\dot{\Lambda}_0+2\dot{\Lambda}_1)\oplus  
 L(2\dot{\Lambda}_0+14\dot{\Lambda}_1)\oplus
 2L(8\dot{\Lambda}_0+8\dot{\Lambda}_1)]\nonumber\\
&&\qquad
\otimes 
L(2\ddot{\Lambda}_0+2\ddot{\Lambda}_1+2\ddot{\Lambda}_2)
\nonumber\\ 
&&
\oplus[L(10\dot{\Lambda}_0+6\dot{\Lambda}_1)\oplus
 L(6\dot{\Lambda}_0+10\dot{\Lambda}_1)] \nonumber\\
&& \qquad \otimes 
[L(4\ddot{\Lambda}_0+\ddot{\Lambda}_1+\ddot{\Lambda}_2)\oplus 
 L(\ddot{\Lambda}_0+4\ddot{\Lambda}_1+\ddot{\Lambda}_2)\oplus
 L(\ddot{\Lambda}_0+\ddot{\Lambda}_1+4\ddot{\Lambda}_2)]
\nonumber\\
&&
\oplus 
[L(12\dot{\Lambda}_0+4\dot{\Lambda}_1)\oplus
 L(4\dot{\Lambda}_0+12\dot{\Lambda}_1)] \nonumber\\ 
&&\label{eq:KS88}
  \qquad 
\otimes 
[L(3\ddot{\Lambda}_0+3\ddot{\Lambda}_1)\oplus
 L(3\ddot{\Lambda}_0+3\ddot{\Lambda}_2)\oplus 
 L(3\ddot{\Lambda}_1+3\ddot{\Lambda}_2)] \zendot
\end{eqnarray}
By the symbol $\restriction$ we indicate that the formula gives the
decomposition of  $L(\Lambda_0)$ as a
representation of $\Loop{SU(2)}_{16}\otimes \Loop{SU(3)}_6$.

Again, the notation is the usual one for highest-weight
representations of affine \Name{Kac-Moody} algebras (summary eg in
\cite{KW88}): $\Lambda_0$ stands for the vacuum representation at
level $1$ of the  affine \Name{Kac-Moody} algebra with horizontal
subalgebra $E(8)$, 
$\dot{\Lambda}_i$ for the $i$th fundamental weight of the  affine
\Name{Kac-Moody} algebra with horizontal subalgebra isomorphic to the
\Name{Lie} algebra of $SU(2)$ and
$\ddot{\Lambda}_i$ for the $i$th fundamental weight of the  affine 
\Name{Kac-Moody} algebra with horizontal subalgebra isomorphic to the
\Name{Lie} algebra of $SU(3)$. If
one is interested in the highest weight of the modules of $SU(2)$ or
$SU(3)$ formed by the 
vectors of lowest energy, one simply ignores the multiple of
$\dot{\Lambda}_0$ and $\ddot{\Lambda}_0$ and takes $\dot{\Lambda}_i$,
$\ddot{\Lambda}_i$, $i\neq 0$, to stand for the corresponding
fundamental weights of $SU(2)$, $SU(3)$, respectively. 

The highest weight vectors of representations of $ \Loop{SU(2)}_{16}$
occurring in (\ref{eq:KS88}) have the following conformal energies:
%%%%%%%%%%%%%%%% 
\begin{displaymath}
  \begin{array}{lcclcclcc}
h(16\dot{\Lambda}_0)&=& 0\, , \,\,&h(14\dot{\Lambda}_0+2\dot{\Lambda}_1)&=& \frac{1}{9}\, ,
\,\,&h(2\dot{\Lambda}_0+14\dot{\Lambda}_1)&=&\frac{28}{9}
\zencom\\ 
&&&&&\\[-2.5ex]%\hline
&&&&&\\[-2.5ex]
h(16\dot{\Lambda}_1)&=& 4\, ,
\,\,&h(8\dot{\Lambda}_0+8\dot{\Lambda}_1)&=& \frac{10}{9}\, , \,\,\\ 
&&&&&\\[-2.5ex]%\hline
&&&&&\\[-2.5ex]
&&&h(10\dot{\Lambda}_0+6\dot{\Lambda}_1) &=&\frac{2}{3}\, ,
\,\,& h(6\dot{\Lambda}_0+10\dot{\Lambda}_1) &=& \frac{5}{3}\zencom
\\
&&&&&\\[-2.5ex]%\hline
&&&&&\\[-2.5ex]
&&&h(12\dot{\Lambda}_0+4\dot{\Lambda}_1)&=& \frac{1}{3}\, , \,\,& h(4\dot{\Lambda}_0+12\dot{\Lambda}_1)&=& \frac{7}{3} \zendot\\
  \end{array}
\end{displaymath}
Since most of these lowest energy eigenvalues are not half-integers,
the inner-implementing representation 
$U^{\Loop{SU(2)}_{16}}$ can not have sharp  geometrical action
(proposition \ref{prop:SpecUAR2pisharp}).  
%%%%%%%%%%%%%%%%%%%%%%%%%%%%%%%%%%%%%%%%%%%%5
\clearpage
We introduce some shorthand notation by defining, up to unitary
equivalence, representations of the respective chiral net localised in
some $I\Subset\Seins$:
\begin{eqnarray}
\alpha_{(0)}&:\cong&
L(16\dot{\Lambda}_0)\oplus L(16\dot{\Lambda}_1)
\zencom
\phantom{L\hspace{5em}L(4\ddot{\Lambda}_1+\ddot{\Lambda}_2)\oplus 
 L(\ddot{\Lambda}_0+4\ddot{\Lambda}_1+\ddot{\Lambda}_2)}
\nonumber\\
\alpha_{(1)}&:\cong&L(14\dot{\Lambda}_0+2\dot{\Lambda}_1)\oplus  
 L(2\dot{\Lambda}_0+14\dot{\Lambda}_1)\zencom\nonumber\\
\alpha_{(2)}&:\cong&L(8\dot{\Lambda}_0+8\dot{\Lambda}_1)\zencom\nonumber\\
\alpha_{(3)}&:\cong&L(8\dot{\Lambda}_0+8\dot{\Lambda}_1)\zencom \label{eq:defshortalphaE81}\\
\alpha_{(4)}&:\cong&L(10\dot{\Lambda}_0+6\dot{\Lambda}_1)\oplus
 L(6\dot{\Lambda}_0+10\dot{\Lambda}_1)\zencom\nonumber\\
\alpha_{(5)}&:\cong&L(12\dot{\Lambda}_0+4\dot{\Lambda}_1)\oplus
 L(4\dot{\Lambda}_0+12\dot{\Lambda}_1)\zendot\nonumber\\
&&\nonumber\\
  \zeta_{(0)}&:\cong&L(6\ddot{\Lambda}_0)\oplus L(6\ddot{\Lambda}_1)\oplus
 L(6\ddot{\Lambda}_2)\, ,\nonumber\\
  \zeta_{(1)}&:\cong&L(2\ddot{\Lambda}_0+2\ddot{\Lambda}_1+2\ddot{\Lambda}_2)\, ,\nonumber\\
  \zeta_{(2)}&:\cong&L(2\ddot{\Lambda}_0+2\ddot{\Lambda}_1+2\ddot{\Lambda}_2)\, ,\nonumber\\
  \zeta_{(3)}&:\cong&L(2\ddot{\Lambda}_0+2\ddot{\Lambda}_1+2\ddot{\Lambda}_2)\, ,\label{eq:defshortzetaE81}\\
  \zeta_{(4)}&:\cong&L(4\ddot{\Lambda}_0+\ddot{\Lambda}_1+\ddot{\Lambda}_2)\oplus 
 L(\ddot{\Lambda}_0+4\ddot{\Lambda}_1+\ddot{\Lambda}_2)\oplus
 L(\ddot{\Lambda}_0+\ddot{\Lambda}_1+4\ddot{\Lambda}_2)\, ,\nonumber\\
  \zeta_{(5)}&:\cong&L(3\ddot{\Lambda}_0+3\ddot{\Lambda}_1)\oplus
 L(3\ddot{\Lambda}_0+3\ddot{\Lambda}_2)\oplus 
 L(3\ddot{\Lambda}_1+3\ddot{\Lambda}_2)\, .\nonumber
\end{eqnarray}

All the models involved are completely rational and, therefore, we
know that the \Name{Coset} pair
$\lok{A}_{max}\coset\lok{C}_{max}\subset\Loop{E(8)}_1$ is a normal
\Name{CTPS}. It is straightforward to determine the 
decomposition of $\rho_{max}$, the localised, unitarily equivalent
representation:
\begin{prop}\label{prop:KS88}
  The branching for the \Name{Coset} pair
  $\lok{A}_{max}\coset\lok{C}_{max}\subset\Loop{E(8)}_1$ associated
  with $\Loop{SU(2)}_{16}\coset \Loop{SU(3)}_6 \subset \Loop{E(8)}_1$
  reads:
  \begin{equation}\label{eq:E81maxbranch}
    \rho_{max} \cong
    \bigoplus_{l=0}^5\,\,\rho^{\lok{A}_{max}}_l\otimes\rho^{\lok{C}_{max}}_l
    \zendot
  \end{equation}
  The branchings of the restricted representations read referring
  to (\ref{eq:defshortalphaE81}), (\ref{eq:defshortzetaE81}):
  \begin{displaymath}
    \bar{\iota}^\lok{A}\rho^{\lok{A}_{max}}_l\iota^\lok{A}\cong
  \alpha_{(l)}\,,\quad 
\bar{\iota}^\lok{C}\rho^{\lok{C}_{max}}_l\iota^\lok{C}\cong
  \zeta_{(l)} \zendot
  \end{displaymath}
\end{prop}

\begin{pf}
  The interpretation of the first line in
  (\ref{eq:KS88}) is clear from proposition \ref{prop:lemma23}. The
  argument from the proof of proposition
  \ref{prop:43a}, using the fusion rules
  (\ref{eq:vircmfusion}) only fixes the problem for $l=4,5$ and for
  $l=1,2,3$ we need to rely entirely on calculating and matching
  dimensions which turns out to be sufficient to determine the
  decomposition completely.

 We know that the statistical dimensions of irreducible
  representations of $\lok{A}_{max}$ and $\lok{C}_{max}$, which are
  coupled in $\rho_{max}$, have to coincide \cite[theorem
  3.6]{khR00}.  The statistical dimensions of candidate
  $\rho^{\lok{A}_{max}}_u$, 
  $\rho^{\lok{C}_{max}}_v$ contained in thinkable decompositions of
  $\rho_{max}$ may be calculated as follows: 
After having identified  $\rho^\lok{A}$, $\rho^\lok{C}$ in (\ref{eq:KS88})
(proposition  \ref{prop:lemma23}) we immediately get: 
  \begin{equation}\label{eq:indACE81}
    [\lok{A}_{max}: \Loop{SU(2)}_{16}]=d(\rho^\lok{A}) = 2 \,,\quad
    [\lok{C}_{max}: \Loop{SU(3)}_{6}]=d(\rho^\lok{C})=3 \zendot 
  \end{equation}
  For the statistical dimension of representations $\rho^{\lok{A}_{max}}_u$,
  $\rho^{\lok{C}_{max}}_v$ of $\lok{A}_{max}$, $\lok{C}_{max}$ we have the
  following (eg \cite[proposition 3.1]{sC02}):
  \begin{eqnarray}
    d(\bar{\iota}^\lok{A}\rho^{\lok{A}_{max}}_u\iota^\lok{A}) &=&
    d(\rho^{\lok{A}_{max}}_u) \,\, [\lok{A}_{max}:
    \Loop{SU(2)}_{16}] \zencom\nonumber\\
    \label{eq:AAmaxstatdim}
    d(\bar{\iota}^\lok{C}\rho^{\lok{C}_{max}}_v\iota^\lok{C}) &=&
    d(\rho^{\lok{C}_{max}}_v) 
    \,\,\, [\lok{C}_{max}: 
    \Loop{SU(3)}_{6}] \zendot
  \end{eqnarray}

Now the statistical dimensions of the irreducible representations of $
\Loop{SU(2)}_{16}$, $ \Loop{SU(3)}_6$ involved in 
  (\ref{eq:KS88}) can be calculated as 
  asymptotic dimensions (these coincide because the respective  fusion
  algebras are identical \cite{aW98}) using formulae in, eg,
  \cite{KW88}.  The results are\footnote{Below $\sqrt{6}$ numerical
    calculations of sufficient accuracy allow complete identification
    of the dimensions, since
  the allowed dimensions of localised representations below $\sqrt{6}$
  are known to form a discrete set \cite{vJ83,khR95}; the only
  value below $\sqrt{2}$ is $1$.}:
  \begin{equation}\label{eq:dimSU216}
    \begin{array}{llclcl}
\phantom{\hspace{8.5em}}&\multicolumn{3}{r}{d_{16\dot{\Lambda}_0} =
  d_{16\dot{\Lambda}_1}} &=& 1\zencom\\& 
   d_{14\dot{\Lambda}_0+2\dot{\Lambda}_1} &=&
   d_{2\dot{\Lambda}_0+14\dot{\Lambda}_1} &\approx& 2.879385\zencom\\&
   &&d_{8\dot{\Lambda}_0+8\dot{\Lambda}_1} &\approx& 5.758770\zencom\\&
      d_{\dot{10\Lambda}_0+6\dot{\Lambda}_1} &=&
   d_{\dot6{\Lambda}_0+10\dot{\Lambda}_1} &\approx& 5.411474\zencom\\&
   d_{12\dot{\Lambda}_0+4\dot{\Lambda}_1} &=&
   d_{4\dot{\Lambda}_0+12\dot{\Lambda}_1} &\approx& 4.411474\zendot
    \end{array}
  \end{equation}

  \begin{equation}\label{eq:dimSU36}
    \begin{array}{lclclcl}
\multicolumn{5}{r}{d_{6\ddot{\Lambda}_0} = d_{6\ddot{\Lambda}_1} = d_{6\ddot{\Lambda}_2}}&=&1\zencom\\
&&&&d_{2\ddot{\Lambda}_0+2\ddot{\Lambda}_1+2\ddot{\Lambda}_2}&\approx&8.638156\zencom\\
d_{4\ddot{\Lambda}_0+\ddot{\Lambda}_1+\ddot{\Lambda}_2}&=&d_{\ddot{\Lambda}_0+4\ddot{\Lambda}_1+\ddot{\Lambda}_2}&=&d_{\ddot{\Lambda}_0+\ddot{\Lambda}_1+4\ddot{\Lambda}_2}&\approx&5.411474\zencom\\
d_{3\ddot{\Lambda}_0+3\ddot{\Lambda}_1}&=&d_{3\ddot{\Lambda}_0+3\ddot{\Lambda}_2}&=&d_{3\ddot{\Lambda}_1+3\ddot{\Lambda}_2}&\approx&4.411474\zendot
  \end{array}
\end{equation}

If one compares the resulting statistical dimensions of
  candidate $\rho^{\lok{A}_{max}}_u$, $\rho^{\lok{C}_{max}}_v$
  coupled in $\rho_{max}$  according to the equations
  (\ref{eq:indACE81}), (\ref{eq:AAmaxstatdim}), (\ref{eq:dimSU216}), (\ref{eq:dimSU36})
  one finds that this only admits a decomposition of 
  $\rho_{max}$ of the form claimed above.
\end{pf}

Remarks: Simply by looking at the branching (\ref{eq:KS88}) one probably
conjectured a decomposition of $\rho_{max}$ into $4$ tensor products
of irreducibles rather than six.  \Name{M\"uger}'s
results \cite{mM02} show, that the connection between sectors coupled in the
branching (\ref{eq:E81maxbranch}) induces an
isomorphism of the \Name{DHR} subcategories generated by the sectors
of $\lok{A}_{max}$ and $\lok{C}_{max}$ which are contained in $\rho_{max}$.

%%%%%%%%%%%%%%%%%%%%%%%%%%%%%%%%%%%%%%%%%%%%%%%%%%%%%%%%%%
%% ENDE KAPITEL ENDLICHER INDEX %%%%%%%%%%%%%%%%%%%%%%%%%%
%%%%%%%%%%%%%%%%%%%%%%%%%%%%%%%%%%%%%%%%%%%%%%%%%%%%%%%%%%

%\input{NoSET.tex}
%%%%%%%%%%%%%%%%%%%%%%%%%%%%%%%%%%%%%%%%%%%%%%%%%%%%%%%%%%
%% KAPITEL KEIN EIT %%%%%%%%%%%%%%%%%%%%%%%%%%%%%%%%%%%%%%
%%%%%%%%%%%%%%%%%%%%%%%%%%%%%%%%%%%%%%%%%%%%%%%%%%%%%%%%%%
\chap{The conformally covariant derivatives of the $U(1)$ current}{The conformally covariant derivatives of the $U(1)$ current}{cha:noset}
%\chapter[Conformally covariant derivatives of $U(1)$-current]{Absence of stress-energy tensor in \Name{CFT}$_2$ models}
%\label{cha:noset}

%%%%%%%%%%%%%%%%%%%%%%%%%%%%%%%%%%%%%%%%%%%%%%%%%%%%%%%%%%%%%%%%%
%%%%%%%%%% INTRODUCTION %%%%%%%%%%%%%%%%%%%%%%%%%%%%%%%%%%%%%%%%%
%%%%%%%%%%%%%%%%%%%%%%%%%%%%%%%%%%%%%%%%%%%%%%%%%%%%%%%%%%%%%%%%%

%\sekt{Introduction}{Introduction}{sec:intro}
%\section{Introduction}
%\label{sec:intro}

Much of the present understanding of quantum field theories was
achieved by methods related to internal and spacetime
symmetries. There are reasons to be interested in objects connected
with symmetries which are of a local nature: the lemmas
\ref{lem:autUA}, \ref{lem:isoUA}, for example, were 
 proved  assuming the presence of a
stress-energy tensor. These two lemmas are the foundation of the
analysis on the isotony problem, they led to the concept of chiral
holography and should hold in quite general circumstances.  
In this chapter we establish limitations of the concept of a
stress-energy tensor and, hence, add reasons for being interested in
densities generating specific spacetime  symmetries in a more
general sense. 

Within the classical framework the relation between local objects and
continuous symmetries of a {Lagrange}an field theory is canonical
by \Name{Noether}'s theorem: to each such symmetry we have an explicitly
known conserved current, whose integrals over space, the charges,
generate the corresponding symmetry transformation. In quantum field
theory the situation is less satisfactory. If one quantises a classical
{Lagrange}an field theory, it may happen that some symmetries do
not survive at all because of renormalisation effects. Moreover,
there is no a priori knowledge of densities 
connected with continuous symmetries of a general
quantum field theory, although it is
possible to characterise such fields abstractly, of course. The nature
of conserved currents connected 
to symmetries at the quantum level (and of their charges in
particular) is hard to clarify in general. %We 
%just mention the sense in which integrals of these densities
%provide approximate generators for symmetries, cf eg \cite{mR76a}.

These problems are more accessible for the global conformal
spacetime symmetry in $\opo$ dimensions. Here we have an abundance of
models for which explicit constructions of a conserved \Name{Wightman}
quantum field are known, which serves as a density for the conformal
symmetry. When smeared out with suitable testfunctions,  this field
actually generates the conformal symmetry in the sense of
integrable \Name{Lie} algebra representations. Its interpretation as a
{\em stress-energy tensor} is in direct analogy with the classical
object. 
Depending on weak assumptions \Name{L\"uscher and Mack}
found that stress-energy tensors of conformally covariant quantum
field theories in $\opo$ dimensions always yield local formulations of the
\Name{Virasoro} algebra \cite{gM88,LM76,FST89}. In chapter
\ref{cha:currsub} we discussed current algebras as 
examples of this structure.

We prove: No such stress-energy tensor exists
in a class of completely 
well-behaved conformal theories in $\opo$ dimensions, the conformally
covariant derivatives of the $U(1)$ current. These
are constructed as fields on
\Name{Minkowski} space and possess  conformally 
covariant extensions on their own \Name{Fock} space, but
they do not transform covariantly with respect to the transformations
implementing global conformal symmetry of the $U(1)$ current. 

\Name{Yngvason}
\cite{jY94} studied  the conformally covariant derivatives as part of a  
broader class of derivatives of the $U(1)$ current and  established,
among other things, 
that they do not fulfill \Name{Haag}
duality on \Name{Minkowski} space\footnote{Essential duality, which
  is another name for \Name{Haag} duality on the conformal covering of
\Name{Minkowski} space, is a consequence of conformal symmetry
\cite{BGL93}.}.  
\Name{Guido, Longo and Wiesbrock} \cite{GLW98} studied locally normal
representations of these models and found representations of the
first derivative, which do not allow an 
implementation of global conformal symmetry. In a closing side-remark
they noted that this contradicts, by
unpublished results of \Name{D'Antoni and Fredenhagen}, presence of
diffeomorphism symmetry 
in these models.

This symmetry is not manifest in the  
commutation relations of local fields in this model, but as these only 
serve as ``field coordinates'' for the local quantum theory they 
generate, one has to have a closer look for a complete proof, see section
\ref{cha:app}.\ref{sec:locnorrep}. At this point we are 
interested in a direct argument that excludes presence of a
stress-energy tensor for the derivative models.

Most of  this chapter is also available as \cite{sK03b}.

%%%%%%%%%%%%%%%%%%%%%%%%%%%%%%%%%%%%%%%%%%%%%%%%%%%%%%%%%%%%%
%%%%%%%%%%%%% CHIRAL NETS OF PHIn %%%%%%%%%%%%%%%%%%%%%%%%%%%
%%%%%%%%%%%%%%%%%%%%%%%%%%%%%%%%%%%%%%%%%%%%%%%%%%%%%%%%%%%%%

\sekt{The chiral nets generated by the $\Phi^{(n)}$}{The chiral nets
  generated by the $\Phi^{(n)}$}{sec:phinchinets}
\label{sec:nucPn}

The $U(1)$ current in $\opo$ dimensions decomposes into two independent chiral
components, the chiral currents, and we shall discuss one of these
only (cf chapter \ref{cha:currsub}). The derivatives of the chiral
current $j$ are given as fields on the light-ray by
$\Phi^{(n)}(x) := \partial_x^n j(x)$, where we used
$\partial_x:= d/dx$. These fields are covariant with
scaling dimension $n+1$ when acted upon by the implementation of the
stabiliser group of $\infty$ for the 
$U(1)$-current theory. By looking
at their \Name{Wightman} functions one recognises that the
$\Phi^{(n)}$ possess 
conformally covariant extensions, if restricted to their own
\Name{Fock} space; the corresponding unitary representation of
$\PSL(2,\dopp{R})$ implementing global conformal symmetry leaves
invariant the vacuum and fulfills the spectrum condition. Each of
these extensions transforms covariantly 
with respect to a different representation  of the global conformal
group and lives on a different \Name{Fock} space.
From now on,  we look at the
fields as operators in their cyclic subrepresentation equipped with
their own representation of the global conformal group and we 
use the symbol $\Phi^{(n)}$ in this sense. 

By construction, the derivative fields $\Phi^{(n)}$ obey the following
commutation relation as fields on the light-ray:
$$\komm{\Phi^{(n)}(x)}{\Phi^{(n)}(y)} = \frac{i}{2\pi} (-)^n
\delta^{(2n+1)}(x-y)\Einsop\,\,.$$ 
We want to calculate the corresponding commutation relations for the
modes of the conformally extended fields on the compactified light-ray,
$\Seins$. These fields will be denoted $\widetilde{\Phi}^{(n)}$. The
calculation is done best in terms of smeared fields, for details see
section \ref{cha:currsub}.\ref{sec:hakacurr}. The testfunctions of
fields on $\Seins$ and their 
images living on the light-ray are connected by a transformation
 $f\mapsto \widehat{f}$ depending on the scaling dimension of
the respective field; its definition is induced by
$\widetilde{\Phi}^{(n)}(f)\equiv \Phi^{(n)}(\widehat{f})$. 
\begin{prop}\label{prop:Phinmodcomm}
  The modes $\Phi^{(n)}_m:=
\widetilde{\Phi}^{(n)}([z^{n+m}])$ have
the following commutation relations:
\begin{equation}\label{modcomrel}
  \komm{\Phi^{(n)}_m}{\Phi^{(n)}_{m'}} =
 % \komm{\widetilde{\Phi}^{(n)}(z^{n+m})}{\widetilde{\Phi}^{(n)}(z^{n+m'})} =
 \delta_{m,-m'}
   \Pi^{(n)}(m)\,\, ,
\end{equation}
if we set $\Pi^{(n)}(m):= \prod_{k=0}^{2n} (m-n+k)$.
\end{prop}
Remark: These relations imply that the modes $\Phi^{(n)}_m$,
$|m|\leq n$, are central, which in turn means that all
$L_0$-eigenspaces for eigenvalues $1,\ldots,n$ are null.

\begin{pf}
We use the
shorthand notations $\zeta:=(1+z)$, $d/dz=
\partial_\zeta$ and arrive at:
\begin{eqnarray}
  &&\komm{\widetilde{\Phi}^{(n)}(f)}{\widetilde{\Phi}^{(n)}(g)} \equiv
  \komm{\Phi^{(n)}(\widehat{f})}{\Phi^{(n)}(\widehat{g})}
\nonumber\\ 
&=&(-)^{2n+1} \oint \frac{dz}{2\pi i} f(z)
\zeta^{-2(n+1)}\klammer{\zeta^2\frac{d}{d\zeta}}^{2n+1}
\zeta^{-2n}g(z)\nonumber\\
&=& \oint \frac{dz}{2\pi i} g(z) \klammer{\frac{d}{dz}}^{2n+1} f(z)
  \zendot
\label{compcomrel}
\end{eqnarray}

The identity of the two integration kernels as distributions may be
proved inductively.  Applying the induction 
assumption we see that we have to prove:
%\begin{displaymath}
$  \zeta^{-2n} \partial_\zeta \zeta^2\partial_\zeta \zeta^{2n} \partial_\zeta^{2n-1} \zeta^{-2} = \partial_\zeta^{2n+1}$.
%\end{displaymath}
One may verify this identity for $n=1$ explicitly%; this one
%may be used to derive the \Name{Virasoro} algebra from the
%\Name{L\"uscher-Mack} algebra
. Then one proves by induction on
$n$: 
\begin{eqnarray*}
&&\zeta^{-2(n+1)} \partial_\zeta \zeta^2\partial_\zeta \zeta^{2(n+1)}
\partial_\zeta^{2n+1} \zeta^{-2}\\
&=&\zeta^{-2}\partial_\zeta^{2n+1}
%\\
%&&
\klammer{\zeta^2\partial_\zeta^2 -
  2(2n+1)\zeta^2\partial_\zeta \zeta^{-1}+(2n+1)(2n)} =
  \partial_\zeta^{2n+3} \zendot
\end{eqnarray*}  
\end{pf}

As we can see by looking at their canonical commutation relations,
the derivative fields may be treated as local quantum theories of
bounded operators in terms of \Name{Weyl} operators and their
relations (cf \cite{GLW98}). We take another approach
which was introduced  by \Name{Buchholz and
  Schulz-Mirbach} \cite{BS90} for the nets of the stress-energy tensor
and the $U(1)$ current: By  establishing
linear energy bounds referring to the conformal {Hamilton}ian
$L_0$  the \Name{Haag-Kastler}
axioms follow from \Name{Wightman}'s  set of axioms. 
In particular the
fields are essentially self-adjoint on the 
\Name{Wightman} domain, their bounded functions fulfill locality and
the local algebras generate  a dense subspace from
the vacuum. The local algebras are generated by unitaries $W(f):=
exp(i\closure{\widetilde{\Phi}^{(n)}(f)})$,
$\closure{\widetilde{\Phi}^{(n)}(f)}$ self-adjoint and 
$supp(f)\Subset\Seins$. The $W(f)$ are concrete representations
of the \Name{Weyl} operators. 
\begin{prop}
  The following defines the local algebras of the chiral net generated
  by the $\Phi^{(n)}$ fields:
  \begin{equation}
    \label{eq:deflokalgphin}
    \lok{A}_{\Phi^{(n)}} (I) := \menge{\closure{\widetilde{\Phi}^{(n)}(f)},
  \,\, supp(f) \subset I, f = \widetilde{\overline{\widehat{f}{\,}}}\,\,}'', \,\,
I\Subset \Seins \zendot
  \end{equation}
\end{prop}

\begin{pf}
The proof follows the lines indicated in \cite{BS90}.
If $\psi_N$ denotes an arbitrary eigenvector of $L_0$ with energy
$N$ and norm $1$, then $\Phi^{(n)}_m\psi_N$, $m>0$, is a multiple of a unit
vector of energy $N-m$, which we will call
$\psi_{N-m}$. Making use of a general estimate for positive, linear
functionals $\eta$ \cite{dB90}: $|\eta(Q)|^2\leq \eta(Q^*Q)$, we are led to
the following estimate: 
$$%\begin{eqnarray*}
  \|\Phi^{(n)}_m\psi_N\|^4 %&=&
  %\betrag{\skalar{\psi_N}{\Phi^{(n)}_{-m}\Phi^{(n)}_{m}\psi_N}}^2\\
\leqslant
%\skalar{\psi_N}{\Phi^{(n)}_{-m}\Phi^{(n)}_{m}\Phi^{(n)}_{-m}\Phi^{(n)}_{m}\psi_N}\\
%&=&
\|\Phi^{(n)}_m\psi_N\|^2 \|\Phi^{(n)}_m\psi_{N-m}\|^2+\Pi^{(n)}(m)
\|\Phi^{(n)}_m\psi_N\|^2 \zendot
$$%$\end{eqnarray*}
 For $m\neq0$ we set $\Pi'{}^{(n)}(m):= \frac{1}{m}\Pi^{(n)}(m)$ and 
we prove inductively using the spectrum condition:
%\begin{displaymath}
$
\|\Phi^{(n)}_m\Psi_N\|^2 \leq N \Pi'{}^{(n)}(m)$, $ m\geq 1
$.
%\end{displaymath}

For the generating modes we have:
%\begin{displaymath}
$$  \|\Phi^{(n)}_{-m}\Psi_N\|^2= \|\Phi^{(n)}_m\Psi_N\|^2 + \Pi^{(n)}(m) \leq
(N+m) \Pi'{}^{(n)}(m) \zendot
$$ 
%\end{displaymath}
The zeroth mode is central in the theory and is,
therefore, a multiple $q$ of the identity%; in fact it is a multiple of the
%vacuum expectation value of the field and this in turn is usually
%taken to be zero by convention, but we need not implement this
%here
. So we have: $\|\Phi^{(n)}_{0}\Psi_N\|^2 = q^2$, $q\in\dopp{R}$.   
For general $\Phi^{(n)}(f)$, $f\in C^\infty(\Seins)$, and a vector
$\Psi$ from the \Name{Wightman} domain we have the following estimate:
\begin{eqnarray}
  \norm{\Phi^{(n)}(f)\Psi} &\leqslant& \norm{(L_0+\Einsop)\Psi} \sum_{m\in\dopp{Z}}
  \betrag{f_m}\klammer{\betrag{m}+\Pi'{}^{(n)}(m)+\betrag{q}+1} \zendot
   \label{hbound}
\end{eqnarray}
This is the linear energy bound from which the \Name{Haag-Kastler} axioms
follow as discussed in the proof of theorem \ref{th:chcurrchlok}. 
\end{pf}

%%%%%%%%%%%%%%%%%%%%%%%%%%%%%%%%%%%%%%%%%%%%%%%%
%%%%%% Nuclearity and split property %%%%%%%%%%%
%%%%%%%%%%%%%%%%%%%%%%%%%%%%%%%%%%%%%%%%%%%%%%%%

\Name{Gabbiani and Fr\"ohlich} \cite{FG93} gave a formulation of the
nuclearity condition adequate for chiral nets, following
\cite{BDF87}. We start with a review on this 
formulation and relate  it to properties of $L_0$ in typical chiral
models. First, we define maps $\Theta_\beta$, $\beta>0$, from the local algebras of a chiral conformal model
$\lok{B}$ in its vacuum 
representation  into its vacuum \Name{Hilbert} space $\Hilb{H}$ by:
\begin{equation}
  \label{eq:defnucmap}
  \Theta_\beta: \lok{B}(I) \rightarrow \Hilb{H}\, , \,\, B\mapsto
  e^{-\beta L_0} B\Omega\, ,\,\, \beta>0 \zendot
\end{equation}

The functions $\Theta_\beta$ are required to be {\em nuclear}. This
means that for every $\Theta_\beta$  there exist a sequence of
vectors $\phi_\iota\in\Hilb{H}$ and a sequence of linear
functionals $\varphi_\iota\in\lok{B}(I)^*$ such that:
\begin{equation}
  \label{eq:defnuccond}
  \Theta_\beta (.) = \sum_\iota \varphi_\iota(.) \phi_\iota\, , \,\,
  \sum_\iota \norm{\varphi_\iota}\norm{\phi_\iota} \leqslant \infty
  \zendot
\end{equation}

Furthermore, defining the {\em trace-norm} of $\Theta_\beta$ by:
$\|\Theta_\beta\|_1:=\inf \sum_\iota \|\varphi_\iota\|\|\phi_\iota\|$, where
the infimum is to be taken over all sequences $(\phi_\iota)_\iota$,
$(\varphi_\iota)_\iota$ complying with the conditions above, there
shall be an asymptotic bound for $\beta\searrow 0$:
\begin{equation}
  \label{eq:nucbound}
  \|\Theta_\beta\|_1 \leqslant e^{(\frac{\beta_0}{\beta})^n}\, , \,\,
  \beta_0, n\,>\, 0 \zendot
\end{equation}
$\beta_0, n$ are constants, depending on $I\Subset\Seins$, possibly.

In typical chiral models, $L_0$ has a total set of eigenvectors,
$e^{-\beta L_0}$ is trace-class for  $\beta>0$, and its trace has good
asymptotic properties for $\beta\searrow 0$ (eg
\cite[(13.13.14)]{vKbook} and below). With these structures it is
straightforward to establish the nuclearity condition
(\ref{eq:defnuccond}), (\ref{eq:nucbound}). If $\{\phi_{N,i}\}_{N,i}$ is
an orthonormal basis of eigenvectors of $L_0$-eigenvalue $N$ labelled
by an additional multiplicity index $i$, then the expansion in
equation (\ref{eq:defnuccond}) reads: 
\begin{equation}
  \label{eq:nuceigexp}
  \Theta_\beta (.) = \sum_{N,i}  \skalar{e^{-\beta
      L_0}\phi_{N,i}}{.\Omega}\ket{\phi_{N,i}} \equiv \sum_{N,i}
  \varphi_{N,i}(.) \ket{\phi_{N,i}} \zendot
\end{equation}
and the following estimate holds:
%\begin{equation}
%  \label{eq:nucest}
$  \sum_{N,i} \norm{\varphi_{N,i}} \norm{\phi_{N,i}} \leq Tr (e^{-2\beta
      L_0}) < \infty $. %\zendot
%\end{equation}
This inequality holds for $\|\Theta_\beta\|_1$ as well and the
asymptotic properties of $Tr(e^{-2\beta L_0})$ yield a bound of the
form (\ref{eq:nucbound}).

Now we turn to the nuclearity condition for the conformally covariant
derivatives of the $U(1)$ current. Each of their \Name{Fock} spaces is
the quotient of a {\em {Verma} module} modulo the space of null
vectors. The 
\Name{Verma} module has at each energy level $N$ a basis given by
vectors $\Phi^{(n)}_{-m_1}\ldots\Phi^{(n)}_{-m_k}\Omega$, $\sum_i m_i
=N$, $0< m_1\leq \ldots \leq m_k$. 
Since null vectors reduce the multiplicity of $L_0$
eigenvalues, the trace of $e^{-\beta L_0}$ in the
vacuum representation of the derivative models is
dominated by the corresponding $L_0$-character for the $U(1)$ current, whose
\Name{Verma} module does not contain any null vectors. 
The following discussion
applies for the same reason to all theories defined by a
stress-energy tensor, and to the $U(1)$-current algebra itself, of course. 

The trace of
$e^{-\beta L_0}$ in the vacuum representation of the $U(1)$ current
coincides with the {\em combinatorial 
  partition function} $p(e^{-\beta})$, which is directly
connected to {\em {Dedekind}'s $\eta$-function}:
\begin{displaymath}
  p(e^{-\beta})^{-1} = \prod_{m\geq 1} (1-e^{-\beta m}) =
  e^{-\frac{\beta}{24}}\eta(i\beta/(2\pi)) \zendot
\end{displaymath}
For the nuclearity condition we have to check the asymptotic behaviour
for $\beta \searrow 0$. This behaviour is determined by the
transformation law of $\eta$ for $\tau \mapsto 1/\tau$. This reads
\cite[III.\S3]{bS74}%, \cite{KW88} (1.7.5)
:
%\begin{displaymath}
$  \sqrt{\beta/2\pi} \,\eta(i\beta/(2\pi)) = \eta(i2\pi/\beta)$.
%\end{displaymath}
We have with $\beta_0>-1+\pi^2/6$ and $n=1$:
\begin{equation}
  \label{eq:nuclcond}
   \lim_{\beta \searrow 0} p(e^{-\beta})
   e^{-\klammer{\frac{\beta_0}{\beta}}^n} = 0 \zendot
\end{equation}
This estimate is a special form of a nuclearity condition and ensures
the split property for 
all models under consideration by arguments as given in 
\cite[lemma 2.12]{FG93}.

%%%%%%%%%%%%%%%%%%%%%%%%%%%%%%%%%%%%%%%%%%%%%%%%%%%%%%%%%%%%%%%%%%%%%%
%%%%%% NO SET IN PHIn %%%%%%%%%%%%%%%%%%%%%%%%%%%%%%%%%%%%%%%%%%%%%%%%
%%%%%%%%%%%%%%%%%%%%%%%%%%%%%%%%%%%%%%%%%%%%%%%%%%%%%%%%%%%%%%%%%%%%%%

\sekt{No stress-energy tensor in $\Phi^{(n)}$-models}{No stress-energy tensor in $\Phi^{(n)}$-models}{sec:prep}
%\section{No stress-energy tensor in $\Phi^{(n)}$-models}
%\label{sec:prep}

We seek for a stress-energy tensor in the theories defined by
conformally covariant  derivatives of degree $n$ of the $U(1)$ current
in $\opo$ 
dimensions. We assume the stress-energy tensor to deserve its name and
therefore it should be a
local, covariant, conserved, symmetric, traceless quantum field
$\Theta$ of scaling dimension 
$2$, which is relatively local to the $\Phi^{(n)}$ under consideration
and a density for its infinitesimal conformal
transformations. Because all models involved factorise into chiral
components, we shall discuss the situation on the
compactified light-ray, ie the fields  live on
$\Seins$. 

According to the analysis of \Name{L\"uscher and Mack} the commutation
relations of $\widetilde{\Theta}$ have a very 
specific form \cite{LM76,gM88} \cite[theorem 3.1]{FST89}. $\widetilde{\Theta}$ is a 
\Name{Lie} field with an extension proportional to $c$, the {\em
  central charge} of $\widetilde{\Theta}$:
$$\frac{c}{12}
\oint \frac{dz}{2\pi i} f'''(z) g(z) = \komm{\widetilde{\Theta}(f)}{\widetilde{\Theta}(g)} -
\widetilde{\Theta}(f'g-fg')\,\,.$$
% %For thesis:
% In the proof of the \Name{L\"uscher Mack} theorem \cite{gM88,FST89},
% $c$ is a translational invariant local quantum field. $\Omega$
% is the only translational invariant vector (up to scalar
% multiplication) and through clustering (eg \cite{rZ84}, theorem
% 2.2.20) one readily sees, that
% $\norm{(c-\skalar{\Omega}{c\Omega})\Omega} = 0$. Now,
% $\Omega$ is separating for local fields, which yields
% $c=\skalar{\Omega}{c\Omega}\Einsop$. A straightforward
% calculation shows:
% %End of addition for thesis
$c/2$ is the normalisation constant of the two
point function of $\widetilde{\Theta}$, hence we have $c\in\dopp{R}_+$,
and, by the
\Name{Reeh-Schlieder} theorem, 
%\footnote{See for example
%  \cite{rJ65} (lemma 2, sect. V.3.B or for a direct argument
%  \cite{fG86a}.}
 $\widetilde{\Theta}=0$ if and only if  $c=0$. 

%%%%%%%%%
%%%%%%%%%
\begin{prop}\label{prop:noset}
  The conformally covariant derivatives of the $U(1)$ current in $\opo$
  dimensions do not contain a stress-energy tensor.
\end{prop}

\begin{pf}
Looking at the commutation
relations of the modes of $\Phi^{(n)}$ (equation \ref{modcomrel}), we
learn that the eigenspaces 
of the conformal {Hamilton}ian $L_0$ associated with energy
$1,\ldots,n$ are all null.  If 
$n\geq 2$ this yields for  $L_{-2}=\widetilde{\Theta}([z^{-1}])$: 
$c/2 = \|L_{-2}\Omega\|^2 = 0$, and hence $\widetilde{\Theta}=0$. 

In the case $n=1$ all vectors of energy $2$ are multiples of
$\Phi^{(1)}_{-2}\Omega$. If there is a stress-energy tensor $\widetilde{\Theta}$
then we have: $ \gamma L_{-2} \Omega = \Phi^{(1)}_{-2}\Omega$,
$c|\gamma|^2 = 12$. Obviously, 
$\widetilde{\Phi}^{(1)}- \gamma \, \widetilde{\Theta}$ is a
quasi-primary field and 
its two-point function is determined by conformal covariance up to a
constant, $C\geq 0$:
\begin{displaymath}
  \langle \Omega,\, 
(\left.\widetilde{\Phi}^{(1)}\right.^\dagger(z)- \bar{\gamma}
\widetilde{\Theta}^\dagger(z))  
(\widetilde{\Phi}^{(1)}(w)- \gamma \widetilde{\Theta}(w)) \, \Omega \rangle
= 
C (z_> -w)^{-4} \zendot
\end{displaymath}
In particular, we have:
$
%\begin{displaymath}
  C = \|(\Phi^{(1)}_{-2}-\gamma
    L_{-2})\Omega\|^2 = 0
%\end{displaymath}
$.
By the \Name{Reeh-Schlieder} theorem, the field
$\widetilde{\Phi}^{(1)}- \gamma \widetilde{\Theta}$ is zero. Since
$\gamma^{-1}\widetilde{\Phi}^{(1)}$ is not a 
stress-energy tensor, the claim holds for $n=1$ as well.  
\end{pf}

%{\bf Remark:} corollary \ref{prelprop} contains an alternative proof
%for the case $n=1$.

%%%%%%%%%%%%%%%%%%%%%%%%%%%%%%%%%%%%%%%%%%%%%%%
%%%%%%%%%%% DISCUSS %%%%%%%%%%%%%%%%%%%%%%%%%%%
%%%%%%%%%%%%%%%%%%%%%%%%%%%%%%%%%%%%%%%%%%%%%%%

%\sekt{Discussion}{Discussion}{sec:nosdis}
%\section{Discussion}
%\label{sec:dis}

We have shown that the quantum field theory of the conformally covariant
derivatives of the 
$U(1)$ current in $\opo$ dimensions does not contain a stress-energy
tensor. This adds another detail to their character as archetypes of
conformal theories in $\opo$ dimensions: In spite of being completely
well-behaved, they do not exhibit special properties of 
other comparatively simple models such as strong additivity or
presence of a stress-energy tensor.

If there
is a local density associated in some sense with the conformal
symmetry of these models, it has to be of a different
nature. The proofs of the lemmas \ref{lem:autUA} and \ref{lem:isoUA},
which are the foundation of chapter \ref{cha:netend}, indicate why
such densities are desirable. We discuss possible generalisations
in the following chapter.

\chap{Discussion}{Discussion}{cha:disc}
%\chapter[Discussion]{Discussion}
%\label{cha:disc}
%\fancyhead[LO]{VII}
%\fancyhead[RE]{VII}
%\fancyhead[CO]{Discussion}
%\fancyhead[CE]{Discussion}

% Conformal incl problem in section \ref{subsec:bosappl}

\sekt{Summary}{Summary}{sec:sum}

We studied chiral conformal subnets $\lok{A}\subset\lok{B}$, their
\Name{Coset} models $\lok{C}\subset\lok{B}$, ie subnets of $\lok{B}$
which commute with $\lok{A}$, and found objects and structures
intrinsically associated with the subnet $\lok{A}\subset\lok{B}$. Most
of our results hold on general grounds and are not directly
related to a specific class of models, 
although our work is inspired by current subalgebras and their
\Name{Coset} models, compare chapter \ref{cha:currsub}. We abstracted
from these examples when we formulated our general assumptions, see
section \ref{cha:cospa}.\ref{sec:genass}.

In section \ref{cha:cospa}.\ref{sec:bosug}, we found that there is a
unique inner-implementing representation $U^\lok{A}$ with the
following properties: $U^\lok{A}$ is affiliated with $\lok{A}$
in a global sense, its non-trivial unitaries are genuine global
observables of the subnet $\lok{A}$. The adjoint action of $U^\lok{A}$
implements global conformal symmetry on the subnet
$\lok{A}\subset\lok{B}$. $U^\lok{A}$ is a unitary, strongly continuous
representation of $\PSL(2,\dopp{R})^\sim$ of positive energy which is
complemented by another representation $U^{\lok{A}'}$ by unitaries in
the commutant of the global algebra $\lok{A}=\bigvee_{I\Subset\Seins}
\lok{A}(I)$. $U^{\lok{A}'}$ is a positive-energy representation of
$\PSL(2,\dopp{R})^\sim$, too, and yields a factorisation of $U$, the
implementation of global conformal symmetry in the vacuum
representation of $\lok{B}$, as $U\circ\symb{p}= U^\lok{A}
U^{\lok{A}'}$, $\symb{p}: \PSL(2,\dopp{R})^\sim \rightarrow
\PSL(2,\dopp{R})$ the covering projection. Both $U^\lok{A}$ and
$U^{\lok{A}'}$ leave the vacuum invariant.

$U^\lok{A}$ is intrinsic for the subnet $\lok{A}\subset\lok{B}$ and it
is directly related to its \Name{Coset} structure: the local
observables of $\lok{B}$ which commute with all unitaries of
$U^\lok{A}$ form the maximal \Name{Coset} model $\lok{C}_{max}$; the
local algebras of $\lok{C}_{max}$ contain the local observables of all
\Name{Coset} models $\lok{C}\subset\lok{B}$ associated with the subnet
$\lok{A}\subset\lok{B}$ (lemma \ref{lem:cosmax}).

We raised the question whether \Name{Coset} models are of a {\em local
nature}, ie if there is an alternative characterisation not by global
data like the unitaries of $U^\lok{A}$ but by local data, more
specifically if the local algebras $\lok{C}_{max}(I)$  coincide with
the local relative commutants $\lok{C}_I =
\lok{A}(I)'\cap\lok{B}(I)$. The task was to show that the $\lok{C}_I$
increase with the localisation region $I$ and we referred to this as
the {\em isotony problem}. This problem exists for embeddings of
theories in any spacetime and we argued that this problem must be
taken seriously. We took a new perspective by studying it through the
action of the inner-implementing representation $U^\lok{A}$ on the
local observables of the ambient theory $\lok{B}$.

The construction of $U^\lok{A}$ relies
on the fact that $\PSL(2,\dopp{R})^\sim$ is, as $\PSL(2,\dopp{R})$ and
all other global conformal groups \cite{BGL93},
generated by translation subgroups which fulfill the spectrum
condition in every positive-energy representation of the global
conformal group. A result of \Name{Borchers} 
\cite{hjB66} ensures the existence of globally inner implementations
of automorphism
groups of a \Name{v.Neumann} algebra $\lok{A}$ induced by the adjoint
action of a positive-energy representation of a translation group. We
derived from these implementations of 
translation automorphisms the globally inner implementation of all
automorphisms associated with conformal symmetry.
This construction generalises within its limits the
\Name{Sugawara} construction of stress-energy tensors for current
algebras and we introduced the name {\em Borchers-Sugawara
construction} for it since its key building block is the result of
\Name{Borchers}. 

The \Name{Borchers-Sugawara} construction of $U^\lok{A}$ is completely
general but it does 
not relate to the local structure of the chiral subnet. We found that a
single {\em Additional Assumption} on the way $U^\lok{A}$ is generated
by local observables of $\lok{A}$ (see page \pageref{ass}) is sufficient
in order to determine the geometrical character of the adjoint action
of $U^\lok{A}$ on local observables of $\lok{B}$ and to tackle the
isotony problem: If $U^\lok{A}$ is generated by integrals of a
stress-energy tensor contained in $\lok{A}$ both problems can be
solved.

So, our analysis applies directly to a large class of examples, the
current subalgebras and their \Name{Coset} models which contain the
\Name{Coset} stress-energy tensor, because the \Name{Sugawara}
construction yields stress-energy tensors as explicit quadratic
functions of the currents of the respective current algebra (theorem
\ref{th:SegSugCon}). The isotony problem is of
interest for this class of examples because the \Name{Coset} models
are not known to be strongly additive in general.

Even in presence of a stress-energy tensor we needed to pin down the
problems of determining the geometric impact of $U^\lok{A}$ on local
observables of $\lok{B}$ and of establishing isotony for the local
relative commutants by general arguments. The \Name{Additional
Assumption} was only needed to establish two crucial, but natural
lemmas (lemma \ref{lem:autUA} and lemma \ref{lem:isoUA}). Moreover, we
did not use  specific structures of current subalgebras, as the
\Name{L\"uscher-Mack} theorem shows that the structures we need hold
for any subtheory $\lok{A}$ which contains a stress-energy tensor provided the
action of the stress-energy tensor on the vacuum \Name{Hilbert} space
of $\lok{B}$ has some weak technical properties (cf section
\ref{cha:currsub}.\ref{sec:sugcurralg} and proof of theorem
\ref{th:SegSugCon}).  

Chapter \ref{cha:netend} contains the main results of this work: We
established on grounds of the \Name{Additional Assumption} that the
adjoint action of dilatations $U^\lok{A}(\tilde{D}(t))$,
$t\in\dopp{R}$, induces automorphisms of $\lok{B}(\Seins_+)$ (lemma
\ref{lem:autUA}) and that this feature, which we call {\em
net-endomorphism property}, allows a satisfactory analysis of the
geometric impact of $U^\lok{A}$ on local observables of the ambient
theory $\lok{B}$ (proposition \ref{prop:netend}). In particular, this
ensures that for given $I\Subset\Seins$ the algebras 
$Ad_{U^\lok{A}(\tilde{g})} \lok{B}(I)$  consist of local operators as
long as $\tilde{g}$ is sufficiently close to the identity, in spite of 
the $U^\lok{A}(\tilde{g})$ being genuine global observables for
$\tilde{g}\neq id$. This was achieved through a converse of
\Name{Borchers}' theorem on half-sided translations (theorem
\ref{bowiesatz}), ie by an interplay of positivity of energy and
modular theory.

The net-endomorphic action of $U^\lok{A}$ on $\lok{B}$ found its
interpretation by {\em chiral holography}:  $U^\lok{A}$
and $U^{\lok{A}'}$ may be taken to be implementations of chiral
conformal transformations on either light-cone coordinate in $\opo$
dimensions which generate from the chiral conformal theory $\lok{B}$ 
a $\opo$-dimensional holographic extension $\lok{B}^\opo$. The
original net $\lok{B}$ is contained as the
net of observables associated with time-zero double cones, and $\lok{B}^\opo$
fulfills all properties of a conformal theory in $\opo$ dimensions
with one exception: it's 
not the translations in futurelike directions but in right spacelike
directions which have positive generators (theorem
\ref{th:chihol}). For this reason we called $\lok{B}^\opo$ a {\em
quasi-theory}.

$\lok{B}^\opo$ allows to take a geometrical point of view with respect
to chiral subnets and their \Name{Coset} models: the maximal
\Name{Coset} model $\lok{C}_{max}$ and the chiral subnet
$\lok{A}_{max}$, consisting of all local observables of $\lok{B}$ on
which $U^\lok{A}$ implements conformal covariance, are contained in
$\lok{B}^\opo$ as subtheories of all left, respectively right chiral
observables. This establishes a direct connection between chiral
subnets and their \Name{Coset} models and inclusions of chiral
observables in $\opo$-dimensional conformal theories as studied before
by \Name{Rehren} \cite{khR00}. 

Chiral holography shows that the symmetries which yield chiral subnets
and their \Name{Coset} models as fixed-point subtheories is a
 spacetime symmetry given by the chiral conformal
transformations on the respective other light-cone coordinate. There
is no need to broaden the symmetry concept in order to understand
these inclusions, one only has to go into the chirally holographic
picture.  

The interpretation of $\lok{A}_{max}$, $\lok{C}_{max}$ as chiral
observables in $\lok{B}^\opo$ directly leads to a spatial
identification of these subnets: the \Name{Hilbert} subspaces they
generate from the vacuum coincide with the spaces of $U^{\lok{A}'}$-
and $U^\lok{A}$-invariant vectors, respectively \cite[lemma
2.3]{khR00}. We gave an alternative proof using the net-endomorphism
property of $U^\lok{A}$ directly and showed that the subnet generated
by a chiral subnet $\lok{A}\subset\lok{B}$ and any  \Name{Coset}
model, $\lok{C}$, is the tensor product $\lok{A}\otimes\lok{C}$ of both
theories (proposition \ref{prop:lemma23}). This has interesting
applications (see below).

In section \ref{cha:netend}.\ref{sec:isoprob} we solved the isotony
problem in our setting. The \Name{Additional Assumption} was used to
establish that the dilatations $U^\lok{A}(\tilde{D}(t))$,
$t\in\dopp{R}$, are contained in the algebra
$\lok{A}(\Seins_+)\vee\lok{A}(\Seins_-)$ (lemma
\ref{lem:isoUA}). After being complemented with the net-endomorphism
property and some general arguments this proved sufficient to
establish isotony for the local relative commutants; we summarised
this in our main result, theorem \ref{th:main}. 

On the way to our solution of the isotony problem we
used structures of the chiral conformal group,
$\PSL(2,\dopp{R})$, and of the group of orientation preserving
diffeomorphisms of the circle, 
$\Diff_+(\Seins)$, or rather their universal covering groups and of their
(exponentiated) positive-energy representations. The
latter group comes into play by the theorem of \Name{L\"uscher and
Mack} \cite{FST89,gM88,LM76} on the commutation relations of
stress-energy tensors in chiral and $\opo$-dimensional conformal
quantum field theory. A technical result of \Name{Toledano-Laredo}
\cite{vL99a} on the exponentiation of these commutation relations was
vital for proving the \Name{Additional Assumption} in presence of
stress-energy tensors (compare proof of theorem \ref{th:SegSugCon}).  

Our approach is independent of the properties of strong additivity and
finiteness of index for $\lok{A}\coset\lok{C}_{max}$
($\lok{A}\subset\lok{B}$ {\em cofinite}), which are central for other
studies on chiral subnets and their \Name{Coset} models and which may be
established for current subalgebras by taking advantage of their
specific structure, compare in
particular \cite{fX00,fX99}, \cite{rL01}, \cite{KL02},
\cite{tL94}. Both approaches are complementary, and our analysis of
the action of the inner-implementation $U^\lok{A}$ on the ambient
theory $\lok{B}$ yields new insights for cofinite inclusions of
current algebras as well:

The  spatial identification of $\lok{A}_{max}$ and $\lok{C}_{max}$
(proposition \ref{prop:lemma23}) allowed us interpret {\em branchings}
for particular current subalgebras $\lok{A}\subset\lok{B}$ and
concrete \Name{Coset} models $\lok{C}$ which are 
available from group-theoretic analysis \cite{GKO86,KW88,KS88} as
restricted branchings of normal canonical 
tensor product subfactors (\Name{CTPS}) \cite{khR00} given by
$\lok{A}_{max}\coset\lok{C}_{max}\subset\lok{B}$. We gave a
\Name{Coset} construction of the third exceptional local extension of
$\Vir_{c<1}$ models as classified in \cite{KL02} (section
\ref{cha:finind}.\ref{sec:exepext}) and found a normal \Name{CTPS} in
an ambient theory with trivial superselection structure, $
\Loop{E(8)}_1$, a setting discussed by \Name{M\"uger} \cite{mM02} (see
section \ref{cha:finind}.\ref{sec:e81ctps}).   

We obtained some results for general cofinite subnets
$\lok{A}\subset\lok{B}$ as well: in this setting the
net-endomorphism property alone is sufficient to solve the isotony
problem (proposition \ref{prop:UACconf}). Moreover, for \Name{Coset}
pairs $\lok{A}\coset\lok{C}\subset\lok{B}$ which are {\em spatial}
(definition \ref{def:spatialcospa}) the chiral rigid rotations by
$2\pi$, $U^\lok{A}(\tilde{R}(2\pi))$, induce automorphisms of each
local algebra $\lok{B}(I)$ (proposition \ref{prop:AutchirRot}), which
improves the general knowledge on the net-endomorphic action of
$U^\lok{A}$ on $\lok{B}$ (cf proposition \ref{prop:netend}).

We found that for spatial \Name{Coset} pairs
$\lok{A}\coset\lok{C}\subset\lok{B}$ of finite index the  quasi-theory
$\lok{B}^\opo$ can fulfill timelike 
commutativity and hence can be reinterpreted directly  as a genuinely
physical model only if the
spectrum of $U^\lok{A}(\tilde{R}(2\pi))$ is contained in $\{\pm
1\}$ (proposition \ref{prop:SpecUAR2pisharp}). This necessary
condition excludes {\em sharp geometrical 
action} of $U^\lok{A}$ on the chiral theory $\lok{B}$, which is
equivalent with timelike commutativity of 
$\lok{B}^\opo$ (proposition \ref{prop:quasisharp}), for all
(known) current subalgebras except the cases one can make up trivially.

In section \ref{cha:currsub}.\ref{sec:cocosub} we applied  our
new perspective of analysing the action of the inner-implementation of a
subtheory on the observables of the ambient theory when we gave a
direct characterisation of conformal inclusions of current
algebras. Moreover, one may recognise from this discussion that it
is of limited use to know explicitly the commutation relations of the
stress-energy tensor of a subtheory on the fields of the ambient
theory, which we derived for current subalgebras (equation 
\ref{geneitcomm}). 

Chapter \ref{cha:noset}, finally, was to show that there are
well-behaved conformal models which do not possess a stress-energy
tensor, namely the conformally covariant derivatives of the
$U(1)$ current. Since these models live in $\opo$ dimensions but
decompose into independent chiral components, they provide examples of
this structure in both settings. If there is a local density
associated with the conformal symmetry of these models, it has to be of
a different nature, and this asks for possible generalisations of our
discussion. 

\sekt{Outlook}{Outlook}{sec:outlook}

A very general quantum version of \Name{Noether}'s theorem exists
\cite{BDL86} on grounds of the split property, which was  
established easily for the   conformally
covariant derivatives in section \ref{cha:noset}.\ref{sec:phinchinets}. Here,
symmetries are implemented 
on local algebras by operators which are localised in a somewhat
enlarged region. \Name{Carpi} \cite{sC99a} took point-like limits of
 local implementers 
applying methods of \Name{J\"or\ss{} and Fredenhagen} \cite{FJ96} and reconstructed the stress-energy tensor
of some models in this way. Hence he gave  an explicit account of the
relation of local 
implementers to densities as needed in chapter \ref{cha:netend}.
 We
comment briefly of the possible application of local implementers in
the context of chiral subnets and the
isotony problem.

Let us begin with a short summary of the quantum \Name{Noether} theorem
of \Name{Buchholz, Doplicher and Longo} \cite{BDL86}. 
The split property of the subnet $\lok{A}$, which definitely holds if
$\lok{B}$ is split (proposition \ref{prop:subsplit}), proves that for
given $I\Subset\Seins$ and each $J\Subset\Seins$ which contains the
closure of $I$ there is a faithful, normal product state
$\eta_{I\subset J}$ for 
$\lok{A}(I)e_\lok{A}\vee\lok{A}(J')e_\lok{A}$ \cite{dB74}. This state
is then taken to define the {\em universal localisation map}
$\Psi_{I\subset J}$ which maps each bounded operator
 on $e_\lok{A}\Hilb{H}$ 
to a local observable in $\lok{A}(J)e_\lok{A}$ and acts trivially on
$\lok{A}(I)e_\lok{A}$. The universal localisation map is a
$*$-homomorphism of norm $1$ and the operators 
$\Psi_{I\subset J}(U(g)e_\lok{A})$,
$g\in\PSL(2,\dopp{R})$,  form a 
representation of $\PSL(2,\dopp{R})$ with  the same spectral
properties as $Ue_\lok{A}$. For $g\in\PSL(2,\dopp{R})$ and
$I_0\Subset\Seins$ which satisfy $gI_0\subset I$, the adjoint action
of $ \Psi_{I\subset J}(U(g)e_\lok{A})$ implements  the
automorphism $\alpha_g$ on $\lok{A}(I_0)e_\lok{A}$, and for this
reason the operators $ \Psi_{I\subset J}(U(g)e_\lok{A})$
are called {\em local implementers}.

By local normality  of the embedding $\lok{A}\subset\lok{B}$ we can
lift the local implementers from the vacuum subrepresentation to the
subnet, and we write the corresponding operator in $\lok{A}(J)$ by
$\Psi^{\lok{A}\subset\lok{B}}_{I\subset J}(U(g)e_\lok{A})$.
These operators are local implementers for the subtheory in the
obvious sense. Taking advantage of the group structure of
$\PSL(2,\dopp{R})$ one can use the locality of $\lok{B}$ to show that
every local observable of $\lok{B}(I_0)$ which commutes with the local
implementers   
$\Psi^{\lok{A}\subset\lok{B}}_{I\subset J}(U(g)e_\lok{A})$
 for some neighbourhood of the identity commutes with all of
$\lok{A}(I_0)$. This allows a characterisation of the maximal
\Name{Coset} model analogous to the one given in lemma
\ref{lem:cosmax}. 

If we want to deal with the isotony problem from this angle, however,
we have to look at the limit $J\rightarrow I_0$, and it seems
unlikely that it might prove possible to establish non-trivial weak
limit points for the local implementers themselves for $g\neq id$. But
looking at the proof of lemma \ref{lem:isoUA}, equation
(\ref{eq:weakscalconv}) in particular, one might hope that such limit points
exist for bilocal products of local implementers as
\begin{equation}\label{eq:bilocimp}
\Psi^{\lok{A}\subset\lok{B}}_{I_-\subset J_-}(U(g_-)e_\lok{A})
\Psi^{\lok{A}\subset\lok{B}}_{I_+\subset J_+}(U(g_+)e_\lok{A}) \,,
\,\, I_\pm, J_\pm \subset\Seins_\pm \,, 
\,\, g_\pm \in\PSL(2,\dopp{R}) \zendot
\end{equation}
in the limit $I_\pm\rightarrow\Seins_\pm$, $g_\pm\rightarrow
D(t)\neq id$ (for small $t$). To establish such limit 
points one has to relate the product states to the subnet
structure. In case there are weak limit points which agree with
$U^\lok{A}(\tilde{D}(t)$ up to a phase (cf equation
(\ref{eq:weakscalconv})), one directly gets a solution for the isotony
problem by the arguments in the proof of lemma \ref{lem:isoUA}.

It was proved in \cite{ADF87} that there are choices for the enlarged
regions (intervals $J$ above) such that  
the universal localisation maps converge point wise
strongly to the identity map in the limit of the localisation regions
exhausting spacetime (in our case $I$ tending to $\dopp{R}$), provided
the union of all local algebras acts 
irreducibly on \Name{Hilbert} space \cite{ADF87}; but there is no
method available to establish weak limit points for bilocal operators
of a subtheory.

We end our speculations on the possible role of local implementers in
the context of chiral subnets with a simple observation: Provided one
can establish  weak limit points for bilocal operators as in
(\ref{eq:bilocimp}) which agree with $U^\lok{A}(\tilde{D}(t))$ for
$t\neq 0$ up to a phase and weak limit points for corresponding local
implementers of the ambient theory $\lok{B}$ which agree with the
dilatations $U(D(t))$ up to a phase, it is clear that
$U^{\lok{A}'}(\tilde{D}(t)) = U(D(t)) U^\lok{A}(\tilde{D}(t))^*$
is affiliated with the maximal \Name{Coset} model. 

If one uses the
representatives of localised diffeomorphisms instead of local
implementers, such a connection is established by (a corresponding
version of)  equation
(\ref{eq:weakscalconv}), if both $\lok{A}$ and $\lok{B}$ contain a
stress-energy tensor.   
In fact, if $U^{\lok{A}'}(\tilde{D}(t))$ is affiliated with
$\lok{C}_{max}$, covariance  directly leads to the identity
$U^{\lok{A}'} = U^{\lok{C}_{max}}$. Hence, the isotony problem appears to be
intimately related to the conformal inclusion problem which we
discussed in section \ref{cha:cospa}.\ref{subsec:bosappl} (compare proposition
\ref{prop:UACconf} for a similar connection for cofinite subnets). 

It is the author's opinion that one should start investigating the
relation of local implementers to the problems outlined above by investigating
possible weak limit points of bilocal products of local 
implementers in the vacuum representation of the conformally covariant
derivatives of the $U(1)$ current. These are particularly simple
models as they satisfy canonical commutation relations and the  split
property was established very easily for 
these models (section \ref{cha:noset}.\ref{sec:nucPn}). 

Finally, we comment briefly on possible generalisations of our
analysis to subtheories in other spacetimes: Our analysis directly
generalises to $\opo$ dimensions, because in this context any
stress-energy tensor decomposes into two independent chiral parts
(by the \Name{L\"uscher-Mack} theorem), the global conformal group is
a factor group of $\PSL(2,\dopp{R})^\sim\times\PSL(2,\dopp{R})^\sim$, and
double cones and their causal complements in $1+1$-dimensional
covering space are Cartesian products of proper
 intervals on the (simply connected covering of the) circle.

Adaptations of the \Name{Borchers-Sugawara} construction for conformal
groups of higher-dimensional spacetimes are possible 
 (see discussion in  section \ref{cha:cospa}.\ref{sec:bosugsum}) and
provided one can establish the corresponding versions of the two
crucial lemmas (lemma \ref{lem:autUA}, lemma \ref{lem:isoUA}) our
general strategy will work because it only involves general properties
of conformal groups and the causal structure of conformal coverings of
spacetime (cf \cite{BGL93}). Since there is no equivalent of the
\Name{L\"uscher-Mack} theorem in higher dimensions, one probably has
to use local implementers in order to establish substitutes of the
lemmas \ref{lem:autUA} and \ref{lem:isoUA}.

For subnets with lesser spacetime symmetry like \Name{Poincar\'e}
invariant theories in $\opd$ dimensions one has to resort to
different methods. In this context, \Name{Carpi and Conti}
solved the isotony problem by methods less direct
than ours, but very general ones \cite{CC01,CC03}. Thus, the quest
for the heart of this problem still awaits further investigation.  

%%%%%%%%%%%%%%%%%%%%%%%%%%%%%%%%%%%%%%%%%%%%%%%%%%%%%%%%%%
%% ENDE KAPITEL DISKUSSION %%%%%%%%%%%%%%%%%%%%%%%%%%%%%%%
%%%%%%%%%%%%%%%%%%%%%%%%%%%%%%%%%%%%%%%%%%%%%%%%%%%%%%%%%%

%>>>>>>>>>>>>>>>>>>>>>>>>>>>>>>>>>>>>>>>>>>>>>>>>>>>>>>>>>>>
%<<<<<<<<<<<<<<<<<<<<<<<<<<<<<<<<<<<<<<<<<<<<<<<<<<<<<<<<<<<

\begin{appendix}

%>>>>>>>>>>>>>>>>>>>>>>>>>>>>>>>>>>>>>>>>>>>>>>>>>>>>>>>>>>>>>
%<<<<<<<<<<<<<<<<<<<<<<<<<<<<<<<<<<<<<<<<<<<<<<<<<<<<<<<<<<<<<

%\input{App.tex}  
\chap{Appendix}{}{cha:app}
\fancyhead[LO]{\thechapter.\thesection}
\fancyhead[RE]{\thechapter.\thesection}
%\chapter[Appendix]{Appendix}
%\label{cha:app}

\sekt{Lemmas on $\PSL(2,\dopp{R})$ and $\PSL(2,\dopp{R})^\sim$}{Lemmas on $\PSL(2,\dopp{R})$ and $\PSL(2,\dopp{R})^\sim$}{pslapp}

This appendix contains some statements on the group of global
chiral conformal transformations,
$\PSL(2,\dopp{R})$, and its universal covering group,
$\PSL(2,\dopp{R})^\sim$. For a short summary on the properties of
the group  $\PSL(2,\dopp{R})$ and its action on $\Seins$ see
\cite{FG93}; a very short introduction is contained in section
\ref{cha:cospa}.\ref{sec:introcongeo}.  

%%%%%%%%%%%%%%%%%%%%%%%%%%%%%%%%%%%%%%%%%%%%%%%%%%%
%%%% LEMMA ON INVARIANT VECTORS W.R.T. PSL(2,R) %%%
%%%%%%%%%%%%%%%%%%%%%%%%%%%%%%%%%%%%%%%%%%%%%%%%%%%

The following lemma is crucial for the solution of the isotony problem
(theorem \ref{th:main}) and it is involved in the proof of proposition
\ref{prop:lemma23} as well. Its statement is a variation of lemma B.2 in
\cite{GL96} and 
its proof is a detailed version of the argument indicated by
\Name{Guido and Longo}.

\begin{lemma}\label{lem:invvec}
Let $\Hilb{H}$ be a separable \Name{Hilbert} space and $V$ a unitary, strongly
continuous representation  of $\PSL(2,\dopp{R})^\sim$  on
$\Hilb{H}$. If $H\subset \PSL(2,\dopp{R})^\sim$ is a subgroup having
closed, non-compact image in  $\PSL(2,\dopp{R})$    under the action of the
covering projection $\symb{p}$, then each $V|_H$-invariant vector is
in fact $V$-invariant. If $V$ is a representation of positive energy,
then each vector which is invariant with respect to $V(\tilde{R}(.))$
is $V$-invariant as well.  
\end{lemma}
\begin{pf}
As a first step we reduce  $V$ by its trivial subrepresentation,
$\Einsop_{\Hilb{H}_0}$, and $\Hilb{H}$ by the \Name{Hilbert} subspace
of $V$-invariant vectors,
$\Hilb{H}_0$. For any vector in $\Hilb{H}_0$ the statements of the
lemma are true, obviously, and we only have to discuss the
complementary representation $V^\perp$  on the \Name{Hilbert} subspace
orthogonal to $\Hilb{H}_0$, which we denote by $\Hilb{H}^\perp$. We
look at an  arbitrary  $V|_H$-invariant vector $\psi\in
\Hilb{H}^\perp$  and prove that any such vector has to vanish. A
technical complication arises as $\PSL(2,\dopp{R})^\sim$ has an
infinite centre. This forces us to apply a trick included in the
proof of \cite[corollary  B.2.]{GL96}. 

 We decompose $V^\perp$  into a direct integral of irreducible representations
 $V_x$ (see eg \cite{BR77}). The corresponding components $\psi_x$ of $\psi$  
are almost all $V_x|_H$-invariant  and almost all $V_x$ are different from the 
trivial representation. Next, we take the tensor product of a non-trivial 
$V_x$ and its conjugate representation, $\overline{V_x}$.  We prove: 
$V_x\otimes\overline{V_x}$ does not contain the trivial representation.  
To this end we assume that there is a partial isometry 
$W: \Hilb{H}_x\otimes\overline{\Hilb{H}_x}\rightarrow \dopp{C}$ intertwining 
$V_x\otimes\overline{V_x}$ and the trivial representation on $\dopp{C}$.
 Using \cite{KR83} (theorem 2.4.1) we define an operator 
$T: \Hilb{H}_x \rightarrow \Hilb{H}_x$ with $\|T\|\leq 1$ by 
$\langle \eta_2, T\eta_1\rangle_x := W \eta_1\otimes \overline{\eta_2}$, 
$\eta_{1,2}\in\Hilb{H}_x$. One readily checks that $T$ commutes with $V_x$ 
and hence is of the form $e^{i\varphi}\|T\|\Einsop$. Choosing an orthonormal 
basis $\{e_n\}_{n\in\dopp{N}}$ this yields for 
$\chi_N:=\sum_{n=1}^N \frac{1}{n} e_n\otimes \overline{e_n}$ the following: 
$W \chi_N = e^{i\varphi} \|T\| \sum_{n=1}^N \frac{1}{n}$. Since $\Hilb{H}_x$ 
has infinite dimension \cite{dG93} this contradicts the assumption: 
 $\lim_{N\rightarrow\infty} \chi_N\in \Hilb{H}_x\otimes
 \overline{\Hilb{H}_x}$ but $W \chi_N$  is unbounded. 

We know that $V_x\otimes\overline{V_x}$
 does not admit a  non-trivial
invariant vector and it forms a representation of
$\PSL(2,\dopp{R})$ as the kernel of $V_x\otimes\overline{V_x}$ contains
 the centre of $\PSL(2,\dopp{R})^\sim$. Now we are in the position to
 apply  \cite[theorem 2.2.20]{rZ84} and thus we have for any
 $\xi_x\in\Hilb{H}_x$:           
\begin{displaymath}
  \lim_{\symb{p}(\tilde{g})\rightarrow \infty}
  \betrag{\skalar{V_x(\tilde{g})\xi_x}{\xi_x}}^2 =
  \lim_{\symb{p}(\tilde{g})\rightarrow \infty}
  \skalar{(V_x\otimes\overline{V_x})(\tilde{g})\,\, 
  \xi_x\otimes\overline{\xi_x}}{\xi_x\otimes\overline{\xi_x}}  
  = 0 \zendot 
\end{displaymath}
If we apply this to a $V_x|_H$-invariant vector $\psi_x$, we readily
see: $\psi_x=0$. Integrating over $x$ yields the first statement of
the lemma. 

The result on rigid conformal rotations may be deduced in the same
manner: The irreducible representations $V_x$ are almost all of
positive energy and the only irreducible representation of
$\PSL(2,\dopp{R})^\sim$ having positive energy and containing a
non-trivial $\tilde{R}(.)$-invariant vector is the trivial
representation \cite{dG93}.                 
\end{pf}

%%%%%%%%%%%%%%%%%%%%%%%%%%%%%%%%%%%%%%%%%%%%%%%%%%
%%%% DIL INV FINDIM SPACES %%%%%%%%%%%%%%%%%%%%%%%
%%%%%%%%%%%%%%%%%%%%%%%%%%%%%%%%%%%%%%%%%%%%%%%%%%

The following lemma is a 
variant of \cite[lemma B.3]{GL96}:
\begin{lemma}\label{lem:diltransspace}
  $V$ a unitary, strongly continuous representation of the
  translation-dilatation group on a \Name{Hilbert} space
  $\Hilb{H}$. Then every finite-dimensional subspace which is left
  invariant globally by $V(D(\tau))$ for some $\tau\neq 0$ consists of
  translation invariant vectors.
\end{lemma}

\begin{pf}
Let $\Hilb{K}$ denote the finite-dimensional subspace of $\Hilb{H}$
left invariant by  $V(D(\tau))$. Then $V(D(\tau))$ may be diagonalised
on $\Hilb{K}$ and we take some eigenvector $\psi$ satisfying
$V(D(\tau))\psi= a\psi$. We treat the case $\tau>0$, else one has to
look at $V(D(\tau))^*\psi= a^{-1}\psi$:
\begin{displaymath}
  \skalar{\psi}{V(T(t))\psi} =
  \skalar{a^{-1}V(D(\tau))\psi}{V(T(t))a^{-1}V(D(\tau))\psi} =
  \skalar{\psi}{V(T(e^{-\tau}t))\psi} \zendot
\end{displaymath}
$n$-fold iteration 
yields the identity
$\skalar{\psi}{V(T(t))\psi}=\skalar{\psi}{\psi}$ in the limit
$n\rightarrow \infty$. It is elementary to 
check that this implies $\norm{\psi-V(T(t))\psi}^2=0$, ie $\psi= V(T(t))\psi$.
\end{pf}

The identity components $\SO(s,1)^\uparrow_+$ of the \Name{Lorentz}
groups in $s+1$ 
dimensions, $\SO(s,1)$, are known to be
{\em exponential}, ie the exponential map is surjective for these
groups \cite{mM94,mM97}. This applies to $\PSL(2,\dopp{R})\cong
\SO(2,1)_+^\uparrow$, but we consider it worthwhile to give a straightforward
elementary proof. 
\begin{prop}\label{prop:PSLexp}
  Every element in $\PSL(2,\dopp{R})$ lies on at least one
  one-parameter subgroup.
\end{prop}
\begin{pf}
  We solve the problem using the formulation of
  $\PSU(1,1)\cong\PSL(2,\dopp{R})$ as automorphisms of the open unit
  disc $D_1\subset\dopp{C}$ (see eg \cite{FL80}). Such automorphisms, $f$, are
  parametrised in terms of their zero, $z_0$, and a complex phase,
  $e^{i\phi}$:
  \begin{displaymath}
    f(z) = e^{i\phi} \frac{z-z_0}{1-\overline{z_0}z}\, , \,\, z\in D_1
    \zendot
  \end{displaymath}
  It easy to see, that the action of $\PSU(1,1)$ on $\Seins$, equation  
  (\ref{eq:psuaction}), is  exactly of this form; the inverse $f^{-1}$ of an
  automorphism $f$ has inverse phase, $e^{-i\phi}$. Hence, we may
  restrict attention to $0\leq \phi \leq \pi$.

  We will prove that it is possible to take the square root of every 
  automorphism of $D_1$ and that iterating
  this process eventually yields an $n$th root arbitrarily close to the
  identity ($z_0=0$, $e^{i\phi}=1$). In a neighbourhood of the
  identity the exponential map is known to be a (local) diffeomorphism
  from the \Name{Lie} algebra into the group (eg
  \cite[Korollar III.2.10]{HN91}).

  Given $f$, we seek a $g$, parametrised by $w_0$, $e^{i\psi}$, which
  solves $f=g^2$ and which is closer to the identity. $g$ has to solve
  the following equations:
\begin{eqnarray*}
  \frac{w_0(e^{i\psi}+1)}{e^{i\psi}+|w_0|^2} &\stackrel{!}{=}&
  z_0\zencom\\
\frac{e^{i\psi}(e^{i\psi}+|w_0|^2)}{
  1+|w_0|^2e^{i\psi}-z\overline{w_0}(1+e^{i\psi})}  &\stackrel{!}{=}&
\frac{e^{i\phi}}{1-\overline{z_0}z}\,,\quad z\in D_1 \zendot
\end{eqnarray*}

The case $e^{i\phi}=1$ leads to elementary quadratic equations whose
solutions have again $e^{i\psi}=0$. The iterated square roots can
readily be seen to tend towards the identity. Solutions satisfying
$w_0=z_0$ can only occur for $z_0=0$; 
in this case $f$ is a pure rotation and lies on a one-parameter
group. 

In the other cases, one can derive the following, equivalent pair of
equations for $w_0$, $e^{i\psi}$: 
\begin{eqnarray*}
   w_0&=&e^{-i\frac{\psi}{2}}e^{i\frac{\phi}{2}}
  z_0 \frac{\sin \frac{\psi}{2}}{\sin \frac{\phi}{2}}\zencom\\
\sin(\frac{\phi}{2}-\psi)\,\, |1-e^{i\phi}| &\stackrel{!}{=}& (1-\cos
  \psi) |z_0|^2 \zendot
\end{eqnarray*}
Looking at the possible graphs of the functions equated in the second
formula one readily sees that there always is a solution with
$\psi<\frac{\phi}{2}$. Taking this solution, one immediately gets
$w_0$ with $|w_0|^2<\frac{1}{2}|z_0|^2$. Again, this is just what we need.
\end{pf}

The group $\SL(2,\dopp{R})$ can easily be seen not to be exponential:
there is no $g\in\SL(2,\dopp{R})$ solving the equation
$diag(-e^{\tau},-e^{-\tau})=g^2$.    

%%%%%%%%%%%%%%%%%%%%%%%%%%%%%%%%%%%%%%%%%%%%%%%%%%%%%%%%%%%%%%%
%%%%%% DIFF_+S^1 APPENDIX %%%%%%%%%%%%%%%%%%%%%%%%%%%%%%%%%%%%%
%%%%%%%%%%%%%%%%%%%%%%%%%%%%%%%%%%%%%%%%%%%%%%%%%%%%%%%%%%%%%%%

\sekt{Lemmas on $\Diff_+(\Seins)$ and $\Diff_+(\Seins)^\sim$}{Lemmas on $\Diff_+(\Seins)$ and $\Diff_+(\Seins)^\sim$}{diffapp}

This appendix contains some technical lemmas on the group of
orientation preserving diffeomorphisms of the circle,
$\Diff_+(\Seins)$, and its universal covering group,
$\Diff_+(\Seins)^\sim$. For short summaries on its general properties
and more  statements related to local quantum physics see
\cite{jM83,PS86,tL94}. 

The first lemma is a simple statement on the position of scale
transformations in $\Diff_+(\Seins)^\sim$; it is needed in the proofs of
lemma \ref{lem:autUA} and lemma \ref{lem:isoUA}.

\begin{lemma}\label{lem:diffD}
For a fixed scale transformation $D(t)\neq id$, $t$ small, there exist
diffeomorphisms  
$g_\delta,\,\, g_\varepsilon\in \Diff_+(\Seins)$ which are localised
in arbitrarily small 
neighbourhoods of $+1$ and $-1$, respectively, and which agree with
$D(t)$ close to $+1$ and $-1$, respectively, such
that, by defining 
\begin{displaymath}
  g^{\tau_1}_\delta 
:= D(\tau_1)g_\delta D(\tau_1)^{-1}
\, , \,\,
  g^{\tau_2}_\varepsilon 
:= D(\tau_2)^{-1}g_\varepsilon D(\tau_2)\zencom
\end{displaymath}
 we have for all $\tau_{1,2}\in\dopp{R}_+$:
 \begin{equation}\label{eq:scalposdiff}
   D(t) = g_+^{\tau_1,\tau_2} 
g_-^{\tau_1,\tau_2}   
g^{\tau_1}_\delta     g^{\tau_2}_\varepsilon
\zendot
 \end{equation}
Here, the diffeomorphisms $g_+^{\tau_1,\tau_2}$, $ g_-^{\tau_1,\tau_2}$
are uniquely specified by their being localised in the upper and lower
half circle, respectively. After a local identification of
$\Diff_+(\Seins)$ with a sheet of $\Diff_+(\Seins)^\sim$ containing
the identity, equation (\ref{eq:scalposdiff}) still holds
for the respective images in $\Diff_+(\Seins)^\sim$.
\end{lemma}

\begin{pf}
If $I_1$ and $I_2$ are
neighbouring intervals, the ``completed union'' which consists of
$I_1\cup I_2$ and the common boundary point will be denoted
$I_1\overline{\cup} I_2$.

Choose a set $\{I^0_\iota, \iota= +,-,\delta,\varepsilon\}$ 
of proper, disjoint 
intervals such that ${I^0_\pm}\subset \Seins_\pm$, $+1\in
I^0_\delta$, 
$-1\in I^0_\varepsilon$, the $I^0_\iota$ are separated by proper
intervals $I_{a}$,.., $I_d$  and a covering of $\Seins$ by
proper intervals $I^1_\iota$ is defined through:
%%%%%%%%%%%%%%%%%%%%%%%%
\begin{displaymath}
  I^1_+ := I_a\overline{\cup}I^0_+\overline{\cup}I_b\,,\,\,
  I^1_- := I_c\overline{\cup}I^0_-\overline{\cup}I_d\,,\,\, 
I^1_\delta:=I_a\overline{\cup}I^0_\delta\overline{\cup}I_d\,,\,\,
I^1_\varepsilon:=I_c\overline{\cup}I^0_\varepsilon\overline{\cup}I_b\zendot
\end{displaymath}

 For fixed $t$, one can choose these intervals such that  $D(t)$
 satisfies $D(t) I^0_\iota \Subset I^1_\iota$. Since $D(t) \Seins_\pm \subset
\Seins_\pm$, we may choose $g_\delta$,
$g_\varepsilon$ close to $id$ such that $g_\delta$ agrees with $D(t)$
 on $I^0_\delta$ 
and with $id$ on $I^1_\delta{}'$ and $g_\varepsilon$ agrees with $D(t)$
on $I^0_\varepsilon$ 
and with $id$ on $I^1_\varepsilon{}'$. Referring to this choice we
set:
\begin{displaymath}
  g_\pm\restriction{}{I^1_\pm}:= D(t) g_\delta^{-1} g_\varepsilon^{-1}\restriction{}{\Seins_\pm}
  \,\,,\quad
g_\pm\restriction{}{\overline{\Seins_\mp}}  
:= id\restriction{}{\overline{\Seins_\mp}}\,.   
\end{displaymath}
Then we have $D(t)=g_+g_-g_\delta g_\varepsilon$. We may now apply the
definitions in the lemma to this choice 
and recognise the results to satisfy equally well the assumptions of
the construction 
just given. 

For a neighbourhood of the identity the covering projection $\symb{p}:
\Diff_+(\Seins)^\sim \rightarrow \Diff_+(\Seins)$ is a
homeomorphism. If we apply $\symb{p}^{-1}$ to $D(t)$,
$g_\delta$, $g_\varepsilon$, $g_+$, $g_-$, we have
$\symb{p}^{-1}(D(t)) = \symb{p}^{-1}(g_+) \symb{p}^{-1}(g_-)
\symb{p}^{-1}(g_\delta)  \symb{p}^{-1}(g_\varepsilon)$. For small
$\tau_1$, $\tau_2$ the equality (\ref{eq:scalposdiff}) holds with the
corresponding replacements, and the same is true for all
$\tau_{1,2}\in\dopp{R}_+$ by continuity: 
denoting the covering
projection from $\dopp{R}$ onto $\Seins$ by $\symb{p}$, all the group elements
involved belong to the identity component of the subgroup
of $\Diff_+(\Seins)^\sim$ which  stabilises $\symb{p}^{-1}(+1)$ and
$\symb{p}^{-1}(-1)$, ie we never leave the first sheet of the covering.   
\end{pf}

 Basically, the following lemma says: For all
elements $\phi$ in a neighbourhood of the identity in
$\Diff_+(\Seins)$ there is a presentation as a
finite product of localised diffeomorphisms $\Xi_i(\phi)$, which are
continuous and unital functions of the
group element $\phi$. Its general ideas are due to \Name{D'Antoni and
  Fredenhagen}\footnote{The author is indebted to  Prof. Fredenhagen
  (Hamburg) for a discussion on related ideas of an unpublished joint work
  he had undertaken together with Prof. D'Antoni \cite{DF97}.}.
\label{apdiff}
\begin{lemma}
  \label{locmap}Let $\{I_i\}_{i\in\dopp{Z}_m}$ be a finite covering  of
  the circle by three or more proper intervals having the following
  additional properties:
$I_i\cap I_{i+1}=:I_{i,i+1}\Subset\Seins$, $I_i\cap I_j = \varnothing$
if $j\not \in\{i\pm 1,i\}$.

We choose a neighbourhood $\Geb{U}_\varepsilon\subset \Diff_+(\Seins) $
containing the identity and depending on $\varepsilon>0$,
$1>\delta>0$, such that for all $\phi\in \Geb{U}_\varepsilon$ the following
conditions are fulfilled:
\begin{displaymath}
  \begin{array}{r@{:\,\,}rclcl}
(i)&\qquad \min_{z\in\Seins}\{[arg \phi(z)]'\}&>&\delta\zencom\\
(ii)&\qquad \betrag{arg\phi (z)-arg z}&<& \varepsilon \delta \zendot
  \end{array}
\end{displaymath}

Then there are $\varepsilon$ such that there exist continuous
localisation mappings $\Xi_i: \Geb{U}_\varepsilon \longrightarrow
\Diff_{\widetilde{I}_i}(\Seins)$, $\widetilde{I}_i \Subset\Seins$,
$i=1,..,m$, 
with the following features: 
$$%\begin{displaymath}
\phi = \prod_{i=1}^m \Xi_i (\phi)\, , \quad \Xi_i (id) = id\,.
$$%\end{displaymath}
\end{lemma}

\begin{pf}
Equivalently we look at periodic diffeomorphisms of the real axis:
$\varphi\in C^\infty(\dopp{R})$, $\varphi'(x)>0$,
$\varphi(x+2\pi)=\varphi(x) + 2\pi$.  We denote the analogue of
$\varphi$ in $\Diff_+(\Seins)$ by $\check{\varphi}$. The preimage of an
interval 
$I\Subset\Seins$ 
under the covering projection $\symb{p}$ will be called $\hat{I}$. We choose a
smooth partition $\mu$ of unity on 
$\Seins$ satisfying $1\geq\mu_i\geq 0$,
$supp(\mu_i)\subset I_i$. On the covering space we define
$\lambda_i(x) := \mu_i(\symb{p}(x))$. %According to the 
%choice of intervals $I_i$ for any given $x\in \dopp{R}$ at most two
%$\lambda_i(x)$ can differ from zero. 

We set:
%\begin{displaymath}
$  \Psi_k[\varphi](x) := x + \sum_{i=1}^k \lambda_i(x)(\varphi(x)-x)$,
$ k=0,\ldots,m$. 
%\end{displaymath}
%Obviously $\Psi_k[\varphi]$ is smooth and periodic. 
 $\Psi_k[\varphi]$ coincides with $\varphi$
on $\hat{M}_k:= \{\bigcup_{j=k+1}^mI_j\}'$ and with $id$ on $\hat{N}_k :=
\{\bigcup_{j=1}^kI_j\}'$, since the sum $\sum_{i=1}^k \lambda_i(.)$
takes the values $1$ and $0$, respectively. 
On $\hat{I}_{m,1}$ we have (corresponding bound
for  $\hat{I}_{k,k+1}$): 
\begin{eqnarray}
  \Psi_k[\varphi]'(x)% &=& 1 - \lambda_1(x)+
%  \lambda_1'(x)(\varphi(x)-x) + \lambda_1(x) \varphi'(x)\\
&=&\lambda_m(x)+\lambda_1(x)\varphi'(x) +
\lambda_1'(x)(\varphi(x)-x)\nonumber\\
%&\geq& min_{1,\varphi'(x)} + \lambda_1'(x)(\varphi(x)-x)\\
&\geqslant& \min\{1,\varphi'(x)\} - \max_{\xi\in
  \hat{I}_{m,1}}\betrag{\lambda_1'(\xi)} |\varphi(x)-x| \label{eq:locbound}
\end{eqnarray}
With
$\varepsilon^{-1}:=\max_{\xi\in\dopp{R}}\sum_{k=1}^m|\lambda_k'(\xi)|$,
(\ref{eq:locbound}), ($i$),
($ii$) imply 
$ \Psi_k[\varphi]'> 0$, which means that $\Psi_k[\varphi]$ is a
periodic diffeomorphism..

A closer look at the action of the $\Psi_k[\varphi]$, $k<m$, reveals
that $\check{\Psi}_k[\varphi]\circ\check{\Psi}_{k-1}[\varphi]^{-1}$ is
localised in $N_k'\Subset\Seins$ and that
$\check{\Psi}_m[\varphi]\circ\check{\Psi}_{m-1}[\varphi]^{-1}$ is
localised in $\check{\varphi}(M_{m-1})'$.
We set:
$\widetilde{I}_{k}' := N_{m+1-k}$ for $k=2,\ldots, m$, and
$\widetilde{I}_1{}':= \check{\varphi}(M_{m-1})$. 

We define the {\em localising maps} by
%\begin{displaymath}
$  \Xi_k(\varphi) :=
\Psi_{m+1-k}[\varphi]\circ\Psi_{m-k}[\varphi]^{-1}$, $k=1,\ldots,m$.
%\end{displaymath}
Continuous dependence of  $\Xi_k(\varphi)$ on $\varphi$ is obvious,
$\Psi_k[id]=id$ yields
$\Xi_k(id)=id$. Finally, with $\Psi_{m}[\varphi]= \varphi$ and
$\Psi_{0}[\varphi]= id$:
\begin{displaymath}
  \prod_{k=1}^m \Xi_k(\varphi) =
  \Psi_{m}[\varphi]\circ\Psi_{m-1}[\varphi]^{-1}\circ
  \Psi_{m-1}[\varphi]\circ\ldots\circ \Psi_{0}[\varphi]^{-1}
= \varphi\circ id
\end{displaymath}  
\end{pf}

%%%%%%%%%%%%%%%%%%%%%%%%%%%%%%%%%%%%%%%%%%%%%%%%%%%%%%%%%%%%%%%%%%%%%%%
%%%%%%%% SET-PROOF OF LEMMA AUTUA %%%%%%%%%%%%%%%%%%%%%%%%%%%%%%%%%%%%
%%%%%%%%%%%%%%%%%%%%%%%%%%%%%%%%%%%%%%%%%%%%%%%%%%%%%%%%%%%%%%%%%%%%%%%

\sekt{Alternative argument for lemma \ref{lem:autUA}}{Alternative argument for lemma \ref{lem:autUA}}{sec:altset}

The the net-endomorphism property of $U^\lok{A}$ can be shown to hold
using only the stress-energy tensor $\Theta^\lok{A}$ and 
without making any (explicit) reference to the exponentiated representation of
$\Diff_+(\Seins)^\sim$.
% , if $\Theta^\lok{A}$ is a \Name{Wightman} field
% with the following properties: when smeared out with real test
% functions $f$, it yields operators 
% $\Theta^\lok{A}(f)$ which are 
% essentially self-adjoint on the \Name{Wightman} domain and whose closures
% $\closure{\Theta^\lok{A}(f)}$ are local in the sense of self-adjoint
% operators.  
We indicate the 
alternative argument proving lemma \ref{lem:autUA} briefly.

If $\Theta^\lok{A}$ is smeared with real test functions $f$, the 
 respective closures 
$\closure{\Theta^\lok{A}(f)}$ are essentially
self-adjoint 
on the \Name{Wightman} domain,  linear in the test function and fulfill
locality in the sense of self-adjoint operators 
\cite{BS90}. Let ${\Theta^\lok{A}(f)}^{-}$, $f(x)=x$, denote the generator of
dilatations
in the representations $\Upsilon^\lok{A}$, $U^\lok{A}$ (cf theorem
\ref{th:SegSugCon}).  

We  
cover the circle $\Seins$ by four proper intervals $I_\pm$,
$I_\varepsilon$, $I_\delta$, satisfying $\overline{I_\pm}\subset \Seins_\pm$,
$-1 \in I_\varepsilon\Subset \Seins$, $+1 \in I_\delta\Subset \Seins$,
where $I_\varepsilon$ and $I_\delta$ are taken to be arbitrarily
small. We take a partition of unity subordinate to this covering by
real test functions and thus gain the corresponding decomposition of
$\widetilde{f}$ into test functions with compact support
in the respective 
proper intervals: $\widetilde{f}=f_+ + f_- +f_\delta +
f_\varepsilon$. If one smears $\widetilde{\Theta}^\lok{A}$ with
$f_+$, $f_-$, $f_\delta$, $f_\varepsilon$ or a real linear
combination,  one gets an essentially 
self-adjoint operator on the \Name{Wightman} domain.

According to \Name{Trotter}'s formula \cite[theorem VIII.31]{RS72}
we have in the strong operator topology:
\begin{displaymath}
  \lim_{n\rightarrow \infty}
  \klammer{e^{i\frac{t}{n}{\Theta^\lok{A}(f)}^{-}}
e^{-i\frac{t}{n}\closure{\widetilde{\Theta}^\lok{A}(f_++f_-)}}}^n =
e^{-it\closure{\widetilde{\Theta}^\lok{A}(f_\delta)}}
  e^{-it\closure{\widetilde{\Theta}^\lok{A}(f_\varepsilon)}} \zendot
\end{displaymath}

We introduce a new notation: $g_\tau(z):= [d(D(-\tau)z)/dz]^{-1}
g(D(-\tau) z)$. Covariance of $\Theta^\lok{A}$ with respect to
dilatations then reads as: 
$$Ad_{U^\lok{A}(D(\tau))}  (e^{is\closure{\widetilde{\Theta}^\lok{A}(g)}}) =
e^{is\closure{\widetilde{\Theta}^\lok{A}(g_\tau)}}\,\,.$$  
One can verify easily by looking at the $n$th step of the sequence
in \Name{Trotter}'s formula that the following holds:
\begin{equation}\label{eq:SETnetend}
  e^{it\closure{\Theta^\lok{A}(f)}} = 
e^{-it \closure{\widetilde{\Theta}^\lok{A}(f_\delta)}} 
e^{-it \closure{\widetilde{\Theta}^\lok{A}(f_\varepsilon)}}
e^{it \closure{\widetilde{\Theta}^\lok{A}(f_++f_\delta+f_\varepsilon)}}
e^{it \closure{\widetilde{\Theta}^\lok{A}((f_-+f_\delta+f_\varepsilon)_{-t})}}
\zendot
\end{equation}
It is elementary to see that the strong convergence of the
\Name{Trotter} formula  entails weak (and hence strong) convergence of
the operators involved in equation (\ref{eq:SETnetend}).
This formula yields for $\overline{I}\subset\Seins_+$, $I\Subset\Seins$:
  \begin{equation}\label{eq:altautUA}
%$
    U^\lok{A}(\tilde{D}(t)) \lok{B}(I) U^\lok{A}(\tilde{D}(t))^* \subset
    \bigcap_{J\supset\overline{\Seins_+}} \lok{B}(J) = \lok{B}(\Seins_+) 
\zendot
%$.
  \end{equation}
The latter equality is due to continuity from the outside. The
remainder follows as in the proof of lemma \ref{lem:autUA}.

%%%%%%%%%%%%%%%%%%%%%%%%%%%%%%%%%%%%%%%%%%%%
%%%%%%% BMT THEOREM %%%%%%%%%%%%%%%%%%%%%%%%
%%%%%%%%%%%%%%%%%%%%%%%%%%%%%%%%%%%%%%%%%%%%

\sekt{Detailed proof of the \Name{BMT} theorem}{Detailed proof of the
  {BMT} theorem}{sec:BMT}

 The original proof of \cite[theorem 1]{BMT88} (\Name{BMT} theorem) is very
brief and we consider it worthwhile to make 
available a detailed proof. We give a formulation  for the chiral
situation, the generalisation to other conformal scenarios is
straightforward. The author thanks Prof. D. Buchholz 
(G\"ottingen) for mentioning the relation of this theorem to the
inner-implementing representation of translations. 

\begin{prop}\label{prop:BMT}
  Assume $\pi$ to be a covariant representation of a chiral conformal
  theory $\lok{B}$ and that there exists a vector $\phi$ in the
  representation \Name{Hilbert} space $\Hilb{H}_\pi$ which is cyclic
  for $\pi(\lok{B})=\bigvee_{I\Subset\Seins}\pi_I(\lok{B}(I))''$. The unitary, 
  strongly continuous representation 
  $U_\pi$ of $\PSL(2,\dopp{R})^\sim$, which implements conformal
  covariance of $\pi$, shall be of positive energy.

Then, $\pi$ is locally normal and unitarily equivalent to a
representation localised in an arbitrary $I_0\Subset\Seins$.
\end{prop}

\begin{pf}
$U_\pi$ defines automorphisms of $\pi(\lok{B})$ and satisfies the
spectrum condition. Thus, there is a representation $V_{\pi(\lok{B})}$ of the
translations which is inner in $\pi(\lok{B})$, satisfies the spectrum
condition and implements the translations on $\pi$ \cite{hjB66,wA74}
(cf \cite[theorem 3.2.46]{BR87}). The positive generator of $V_{\pi(\lok{B})}$
shall be called $H$ and for some $t>0$ we set:
$\psi:=e^{-tH}\phi$. $\psi$ is analytic for $V_{\pi(\lok{B})}$ in a
full neighbourhood of the identity.

We use the spectral decomposition of $H$ to define $M_n := \int_0^n
dE_\lambda e^{\lambda t}$. These operators are contained in
$\pi(\lok{B})$. Obviously, we have $X\phi= \lim_{n\rightarrow\infty}
XM_n \psi$ for every $X\in\pi(\lok{B})$. This proves that $\psi$ too is
cyclic for $\pi(\lok{B})$. By the usual \Name{Reeh-Schlieder} argument
\cite{hjB68} the vector $\psi$ is cyclic for some
$\pi_{I_\infty}(\lok{B}(I_\infty))''$ where $I_\infty$ is supposed to
contain the point $\infty$. This implies, because of locality, that
$\psi$ is separating for each $\pi_{I_0}(\lok{B}(I_0))''$, if $I_0$
and $I_\infty$ are disjoint. Covariance of $\pi$ implies now that each
$\pi_{I}(\lok{B}(I))''$ has a separating vector.

Every \Name{v.Neumann} algebra which possesses a separating vector is
$\sigma$-finite \cite[I.1.4.,Prop. 6]{jD81} and every
$*$-homomorphism from a $\sigma$-finite, properly infinite 
\Name{v.Neumann} algebra 
into a $\sigma$-finite \Name{v.Neumann} algebra is automatically
normal \cite[V.5, Theorem 5.1]{mT79} (the local algebras $\lok{B}(I)$,
which are type $\III_1$ factors, are $\sigma$-finite and properly
infinite). The argument of 
\cite[theorem  2.4.24]{BR87} shows that the representations $\pi_I$
are faithful, too.  Normality implies the identity $\pi_I(\lok{B}(I)) =
\pi_I(\lok{B}(I))''$  and adding faithfulness we get: 
$\pi_I(\lok{B}(I))\cong \lok{B}(I)$. Since $\lok{B}(I)$ is a type
$\III$ factor this isomorphy is in fact spatial, ie: it may be
implemented by a unitary from $\Hilb{H}_\pi$ onto $\Hilb{H}$ (eg
\cite[II.4.6., Theorem]{jS67}). If we pick a $\lok{B}(I_0)$ and denote
the  unitary implementer of the isomorphism between
$\pi_I(\lok{B}(I_0))$ and $\lok{B}(I_0)$ by $W$, then $\rho_\pi(.) := Ad_W
(\pi(.))$ defines a unitarily equivalent, localised representation on
$\Hilb{H}$. 
\end{pf}

%\input{Apextra.tex}

%%%%%%%%%%%%%%%%%%%%%%%%%%%%%%%%%%%%%%%%%%%%%%%%%%%%%%%%%%%%%%%%%%%%%%%%%%%%%
%%%% Covariance of locally normal representations of diffeomorphism %%%%%%%%%
%%%% invariant local chiral conformal quantum theories              %%%%%%%%%
%%%%%%%%%%%%%%%%%%%%%%%%%%%%%%%%%%%%%%%%%%%%%%%%%%%%%%%%%%%%%%%%%%%%%%%%%%%%%
\sekt{$\Diff_+(\Seins)$-Symmetry and $\PSL(2,\dopp{R})$-Covariance}{$\Diff_+(\Seins)$-Symmetry and $\PSL(2,\dopp{R})$-Covariance}{sec:locnorrep}
%\section{$\Diff_+(\Seins)$-Symmetry and $\PSL(2,\dopp{R})$-Covariance} 
%\label{sec:locnorrep}

We deduce
covariance with respect to global conformal transformations for all
locally normal representations of  diffeomorphism covariant chiral components
of a factorising conformal theory in $\opo$ dimensions. As an
application we prove the statement of \cite{GLW98} saying that the
theory of the first conformally covariant derivative of the $U(1)$
current does not contain a 
stress-energy tensor.
 
Diffeomorphism covariance of a chiral net $\lok{B}$
means that there is a strongly continuous map $\Upsilon_0$ from
the orientation preserving diffeomorphisms of the circle,
$\Diff_+(\Seins)$, into the unitaries on $\Hilb{H}$, the
representation space of the vacuum representation of $\lok{B}$,
implementing a geometric automorphic action
$\alpha$ of $\Diff_+(\Seins)$: $$\Upsilon_0(\phi) \lok{B}(I)
\Upsilon_0(\phi)^* \equiv \alpha_\phi(\lok{B}(I)) = \lok{B}(\phi(I))\, , \,\, 
I\Subset\Seins\, , \,\, \phi\in \Diff_+(\Seins)\,\,.$$  If $\phi$ acts
trivially on an interval 
$I'$, then 
$Ad_{\Upsilon_0(\phi)}$ is to implement the trivial automorphism of
$\lok{B}(I')$; such $\phi$ is said to be localised in $I$ and gives
rise to a local operator $\Upsilon_0(\phi)\in \lok{B}(I)$, by \Name{Haag} duality of $\lok{B}$. $\Upsilon_0$ defines a
ray representation, as the cocycles
$\Upsilon_0(\phi_1)\Upsilon_0(\phi_2)\Upsilon_0(\phi_1\phi_2)^*$ commute with
$\lok{B}$ and $\lok{B}$ is irreducible. We require
$\Upsilon_0(id)=\Einsop$ and 
$\alpha\!\restriction\!{\PSL(2,\dopp{R})}$ to be identical to the global conformal
covariance of $\lok{B}$.
 Since $\Diff_+(\Seins)$ is a simple group (theorem of
\Name{Epstein, Herman, Thurston}, cf \cite{jM83}) the whole representation
$\Upsilon_0$ is contained in $\lok{B}_{uni}$, the universal $C^*$-algebra
generated by the local algebras $\lok{B}(I)$, $I\Subset\Seins$.

In models having a stress-energy tensor, the restricted
representation $\Upsilon_0\!\restriction\!{\PSL(2,\dopp{R})}$  is in fact a
representation of $\PSL(2,\dopp{R})$ (cf theorem \ref{th:SegSugCon}). The
further analysis does not require the answer to the cohomological question
whether this may be achieved always by a proper choice of phases for
$\Upsilon_0$ and we shall, therefore, not concern ourselves with this problem.

We deal with a locally normal representation $\pi$ of
$\lok{B}$, 
ie a family of normal 
representations $\pi_I$ of the local algebras $\lok{B}(I)$ by bounded
operators on a \Name{Hilbert} space $\Hilb{H}_\pi$, which is
required to be consistent with isotony: $I\subset J \Rightarrow
\pi_J\!\restriction\!{\lok{B}(I)}= \pi_I$. This family lifts
uniquely to a representation $\pi$ of $\lok{B}_{uni}$
                                %(cf \cite{GL96});
and the $\pi_I$ are given in terms of
the embeddings $\iota_I: \lok{B}(I)\hookrightarrow \lok{B}_{uni}$
by $\pi_I = \pi\circ \iota_I$ \cite{kF90, GL92}. 

It is easy to see, that $\pi\circ
\Upsilon_0$  implements the
automorphic action $\alpha$ and 
represents the group laws of $\Diff_+(\Seins)$ up to multiplication
with operators in the 
centre of $\pi(\lok{B}_{uni})$. The restrictions of $\pi\circ \Upsilon_0$ to  
subgroups of localised diffeomorphisms are weakly and thus strongly
continuous by local normality and the 
local cocycles are phases, since local algebras are factors. We note
that $\pi$ is unital because of local normality. 

We will now restrict our attention to the subgroup of global conformal
transformations, $\PSL(2,\dopp{R})$,
and construct a unitary, strongly continuous representation of its
universal covering group 
${\PSL(2,\dopp{R})}^\sim$ from $\pi\circ \Upsilon_0\!\restriction\!{\PSL(2,\dopp{R})}$. This
representation will implement the automorphic 
action $\alpha$ of $\PSL(2,\dopp{R})$ on $\lok{B}$ in the
representation $\pi$ and will be inner in the global sense, ie it
will be contained in the \Name{v.Neumann} algebra of global
  observables,
$\pi(\lok{B}):=\bigvee_{I\Subset\Seins}\pi_I(\lok{B}(I))$. The line of
argument will be very similar to the one leading to theorem \ref{hauptsatz}.

We can write every $g\in \PSL(2,\dopp{R})$ in the form 
$g = {T}(p_g) {D}(\tau_g) {R}(t_g)$, where each term 
depends continuously on $g$ (\Name{Iwasawa} decomposition,
\cite[appendix I]{FG93}). In fact, any $g\in \PSL(2,\dopp{R})$ may
be written as a 
product of four translations and four special conformal
transformations, each single of them depending continuously on
$g$, if one uses the identities:
\begin{eqnarray}
  {D}(\tau) &=& {S}(-(e^{\frac{\tau}{2}}-1)e^{-\frac{\tau}{2}})\,
  {T}(1)\, {S}(e^{\frac{\tau}{2}}-1) \,
  {T}(-e^{-\frac{\tau}{2}}) \zencom\label{eq:Dtsdiff}\\
{R}(2t) &=& {S}((-1+\cos t)(\sin t)^{-1}) \,
{T}(\sin t) \, {S}((-1+\cos t)(\sin t)^{-1}) \zendot
\label{eq:Rtsdiff}
\end{eqnarray}

According to lemma \ref{locmap}, there are continuous, identity preserving localisation maps $\Xi_j$,
$j=1,..,m$, which map a neighbourhood of the identity,
$\Geb{U}_\varepsilon\subset\Diff_+(\Seins)$, into groups of localised
diffeomorphisms such that we have $\prod_{j=1}^m \Xi_j(\phi) = \phi$,
$\phi\in \Geb{U}_\varepsilon$.
If we specialise to translations, this means that
there is an open interval $I_\varepsilon$ containing  $0$ for which
the mapping $t\mapsto \prod_j \pi\circ \Upsilon_0 (\Xi_j(T(t)))$ is unital
and strongly continuous. We extend this map to all
of $\dopp{R}$ through a choice of a $\tau\in I_\varepsilon$, $\tau>0$,
 defining $n_t\in\dopp{Z}$ by its properties $t= n_t \tau+ (t-n_t
\tau)$, $t-n_t 
\tau \in [0,\tau[$, and setting $$T^{\pi(\lok{B})}(t):=
\klammer{\prod_j \pi\circ 
  \Upsilon_0 (\Xi_j(T(\tau)))}^{n_t} \prod_j \pi\circ \Upsilon_0
(\Xi_j(T(t-n_t\tau)))\,\,.$$
One can easily check that this is indeed a weakly and thus strongly
continuous map into the unitaries on $\Hilb{H}_\pi$ by recognising
that the mappings involved are continuous and unital ($\pi(\Einsop)=
\Einsop$, $\Xi_i(id)=id$). 

%%%%%%%%%%%%%%%%%%%%%%%%%%%%%%%%%

This procedure applies to the special conformal transformations as
 well, and we may use the result, the \Name{Iwasawa} decomposition
 and (\ref{eq:Dtsdiff}), (\ref{eq:Rtsdiff}) to define for each
${g}\in\PSL(2,\dopp{R})$: 
\begin{equation}
  \label{eq:pib}
  \pi^\lok{B}({g}) := \prod_{i=1}^{4}
  T^{\pi(\lok{B})}(t^{(i)}_{{g}}) S^{\pi(\lok{B})}(n^{(i)}_{{g}})\,
  ,\quad {g}\in\PSL(2,\dopp{R}) \zendot
\end{equation}
We have $\pi^\lok{B}(id)=\Einsop$.
The following lemma asserts that the
$\pi^\lok{B}({g})$ define an  inner-implementing
representation up to a cocycle in the centre of $\pi(\lok{B})$. To this end
we define operators sensitive to 
the violation of the group multiplication law:
$ z^\lok{B}({{g}},{{h}}):= \pi^\lok{B}({{g}}) \pi^\lok{B}({{h}}) \pi^\lok{B}({{g}}{{h}})^*,\,
{{g}}, {{h}}\in\PSL(2,\dopp{R})$. 

\begin{lemma}\label{lem:wordsphin}
  $\pi^\lok{B}: g\mapsto \pi^\lok{B}(g)$ defines a strongly continuous
  mapping with unitary values in $\pi(\lok{B})$. The adjoint action of
  $\pi^\lok{B}(g)$, $g\in \PSL(2,\dopp{R})$, on $\pi(\lok{B})$ implements the
  automorphism $\alpha_g$. $z^\lok{B}: (g,h)\mapsto z^\lok{B}(g,h)$
  defines a strongly continuous 2-cocycle with unitary values in
  $\pi(\lok{B})'\cap\pi(\lok{B})$.
\end{lemma}

\begin{pf} 
Unitarity is obvious. Strong continuity follows since we multiply
continuous functions. The
implementing property of the $\pi^\lok{B}(g)$ follows immediately by the
decomposition  $g=\prod_{i=1}^4 T(t^{(i)}_g) S(s^{(i)}_g)$ , the subsequent
decomposition 
of these into products of localised diffeomorphisms, the definition of
$\pi^\lok{B}(g)$ and the implementation property of the (generalised)
ray representation $\pi\circ \Upsilon_0$ of $\Diff_+(\Seins)$. At
this point all
but the cocycle properties of $z^\lok{B}$ follow immediately from
its definition. If
we look at $\pi^\lok{B}(f)\pi^\lok{B}(g)\pi^\lok{B}(h)$, insert some
identities appropriately, we find:
$  z^\lok{B}(f,gh)z^\lok{B}(g,h)=z^\lok{B}(f,g)z^\lok{B}(fg,h)$. 
Even more immediate are the equalities 
$z^\lok{B}(id,g)=z^\lok{B}(g, id) = \Einsop$.
\end{pf}

We write the
abelian \Name{v.Neumann} algebra generated by the cocycle operators
$z^\lok{B}(g,h)$ as follows:
$
  \lok{Z}^\lok{B} \equiv \{z^\lok{B}(g,h),
      z^\lok{B}(g,h)^* | g,h \in \PSL(2,\dopp{R})\}''
$. Obviously $\lok{Z}^\lok{B}$ is contained in the centre of $\pi(\lok{B})$.
Now we are prepared to realise the construction itself:
\begin{lemma}\label{lem:constrphin}
  For every $\tilde{g}\in {\PSL(2,\dopp{R})}^\sim$ there exists a unitary operator
  $z^\lok{B}(\tilde{g})\in\lok{Z}^\lok{B}$ such that
  \begin{equation}
    \label{eq:defUAphin}
    U_\pi(\tilde{g}):= z^\lok{B}(\tilde{g})\pi^\lok{B}(\symb{p}(\tilde{g}))
  \end{equation}
defines a unitary, strongly continuous
representation, whose adjoint action
implements the automorphic action
$\alpha\circ \symb{p}$ on $\pi(\lok{B})$.
\end{lemma}

\begin{pf} The proof of lemma \ref{lem:constr} applies word for word.
\end{pf}

The outcome of the construction presented above proves the main result
of this section. It was
known already, perhaps not in the present formulation, to
\Name{D'Antoni and Fredenhagen} \cite{DF97}; its
uniqueness statement is a simple consequence
of the fact that 
${\PSL(2,\dopp{R})}^\sim$ is a perfect group (proposition \ref{prop:inneruni}):
\begin{theo}\label{covdiffrep}
  Let $\lok{B}$ be a chiral conformal, diffeomorphism covariant
  theory. Then any locally normal representation 
  $\pi$ of $\lok{B}$ is covariant with respect to the automorphic
  action of $\PSL(2,\dopp{R})$. The implementing representation may be
  chosen to be the unique globally $\pi(\lok{B})$-inner, implementing
  representation $U_\pi$ of ${\PSL(2,\dopp{R})}^\sim$.
\end{theo}

The construction given here for diffeomorphism covariant theories is
more general than the \Name{Borchers-Sugawara} construction
(section \ref{cha:cospa}.\ref{sec:bosug}), if these possess locally
normal representations which 
violate positivity of 
energy. For representations with finite 
  statistical dimension the spectrum condition is always fulfilled
  because of the theorem we have just
  derived and results of \cite{BCL98}. For infinite index
  representations there exists a criterion 
  for strongly additive theories; it was given in \cite{BCL98}, too.
 In presence
of the spectrum condition the construction given here and the
\Name{Borchers-Sugawara} construction agree by uniqueness.

% The general line of argument for our construction was known already to
% \Name{D'Antoni and Fredenhagen} and the author is grateful to
% K. Fredenhagen for him discussing these ideas with him. 

As an application we have the following corollary, which provides an
explicit proof for the closing remark of \cite{GLW98} and is an
alternative for the proof of proposition \ref{prop:noset} in the case
$n=1$; we make the same assumptions on stress-energy tensors as in
section \ref{cha:noset}.\ref{sec:prep}:
\begin{cor}\label{prelprop}
There is no stress-energy tensor in the $\Phi^{(1)}$-model.
\end{cor}

\begin{pf} Lets assume that there was a stress-energy tensor in this
  model. Then the \Name{Fock} space of
$\Phi^{(1)}$ decomposes completely, as representation space of the
\Name{Virasoro} algebra, into irreducible highest-weight
representations \cite[proposition 11.12.c]{vKbook}. Since for fixed
$\symb{c}<1$ 
there are only finitely many allowed ground states and the  energy
eigenspaces are finite-dimensional, this decomposition would be
finite. For $\symb{c}<1$, the theory generated by the stress-energy
tensor, $\lok{B}_\Theta$, is {\em completely rational} \cite{KL02}, a
property it 
would pass on to the theory generated by $\Phi^{(1)}$, denoted
$\lok{B}_{\Phi^{(1)}}$ \cite{rL01}. 
In particular, $\lok{B}_{\Phi^{(1)}}$ would be strongly additive, which
it is not \cite{jY94}. 

%Include in extended version:
%
% Alternatively we can exclude $\symb{c}<1$ by a simple observation. The
% set of vectors $\Phi^{(1)}_{-m_1}\ldots \Phi^{(1)}_{-m_k}\Omega$,
% $m_1\geq\ldots\geq m_k> 1$,  provide a dense set of mutually
% orthogonal vectors according to $\Phi^{(1)}$'s commutation relations
%   (\ref{modcomrel}).   On the contrary the
%   vectors $L_{-m_1}\ldots
% L_{-m_k}\Omega$, $m_1\geq\ldots\geq m_k> 1$, given in terms of
% generating modes of a stress-energy tensor with an admissible central
% charge less than $1$, ie $c=1-6/(m+2)(m+3)$,
% $m\in \dopp{N}$, fail to generate all vectors at level $\sum_i m_i
% =(m+1)(m+2)$, because at this level there is a null vector
% \cite{DMS96}(sect. 8.1.1). Thus there had to be a highest-weight
% vector for the \Name{Virasoro} algebra of the same energy.  By the
% following estimate on the admitted highest weights this is impossible
% ($1\leq q\leq p\leq m+1$):    
% \begin{displaymath}
%   h_{p,q}(m) =  \frac{[(m+3)p-(m+2)q]^2-1}{4(m+2)(m+3)}\leqslant
%   \frac{1}{4} (m+1)(m+2)\,.
% \end{displaymath}

The particular shape of \Name{Fock} space teaches us the following%as
                                                                  %well
:
From the representation theory of the 
\Name{Virasoro} algebra for $\symb{c}=1$ \cite{aR85} and for
$\symb{c}>1$ \cite[lemma 2]{rL88} we learn that the set of vectors 
$L_{-m_1}\ldots
L_{-m_k}\Omega$, $m_1\geq\ldots\geq m_k> 1$, at a particular level
$\sum_i m_i$ is linearly independent, ie such a stress-energy tensor
would generate a dense
set of vectors from the vacuum as does $\Phi^{(1)}$. The same holds for
the local quantum theories generated by both fields, $\lok{B}_\Theta$ and
$\lok{B}_{\Phi^{(1)}}$, respectively. 

By conformal
covariance and the \Name{Bisognano-Wichmann} property for 
chiral conformal theories the local algebras
$\lok{B}_\Theta(I)$ are modular covariant subalgebras of the 
local algebras $\lok{B}_{\Phi^{(1)}}(I)$. By results of
\Name{Takesaki} \cite{mT72}, \Name{Jones} \cite{vJ83} and the
\Name{Reeh-Schlieder} theorem we know
that the projection $e_\Theta$ onto $\overline{\lok{B}_\Theta\Omega}$
completely characterises  
$\lok{B}_\Theta$ through $\lok{B}_\Theta(I) = \{e_\Theta\}'\cap
\lok{B}_{\Phi^{(1)}}(I)$. We have just deduced $e_\Theta=
\Einsop$, and thus the two local quantum theories coincide. Both fields
have to be regarded as different coordinates of the same theory.

The representation of the
\Name{Virasoro} algebra defined by the commutation relations of the
stress-energy tensor integrates to a projective representation of
$\Diff_+(\Seins)$ \cite{GW85}. %, \cite{vL99a}
  A generating set of the local
algebras $\lok{B}_\Theta(I)$ is given by all one-parameter groups
$exp(it \Theta(f))$, $supp(f)\subset I$, $\Theta(f)$
symmetric, % (see 
           % eg \cite{KR91}, 5.7.53), 
  which represent one-parameter subgroups of
$\Diff_+(\Seins)$ \cite{BS90,tL94}.
 This shows diffeomorphism covariance of
$\lok{B}_\Theta$ and, by assumption, of $\lok{B}_{\Phi^{(1)}}$.  

By theorem \ref{covdiffrep} any locally normal representation of
$\lok{B}_{\Phi^{(1)}}$ would be covariant, but \cite{GLW98}
  have given \Name{DHR}-automorphisms for this model, which are not
  covariant. This contradicts the assumption.
\end{pf}

%>>>>>>>>>>>>>>>>>>>>>>>>>>>>>>>>>>>>>>>>>>>>>>>>>>>>>>>><
\end{appendix}
%<<<<<<<<<<<<<<<<<<<<<<<<<<<<<<<<<<<<<<<<<<<<<<<<<<<<<<<<<

\clearpage
\fancyhead[LO]{}
\fancyhead[RE]{}
\fancyhead[CO]{Bibliography}
\fancyhead[CE]{Bibliography}
%\input{strcos.bbl}

%\bibliography{../../biblio}

\clearpage
\chapter*{Frequently used symbols}
  \label{chap-symbols}
\thispagestyle{empty}
\addcontentsline{toc}{chapter}{{Frequently used symbols}}

Most of the notation in this dissertation is in wide-spread
use. Therefore, we only give a list of frequently occurring 
symbols of special importance. 

\vspace{2ex}

\begin{minipage}[t]{14cm}
  \begin{tabular}[]{lll}
\multicolumn{1}{c}{Symbol} & \multicolumn{1}{c}{Description} &
\multicolumn{1}{c}{Reference}\\  
\\
$T$, $D$, $S$, $R$&one-parameter groups in $\PSL(2,\dopp{R})$&section
\ref{cha:cospa}.\ref{sec:introcongeo}\\ 
$\tilde{T}$, $\tilde{D}$, $\tilde{S}$, $\tilde{R}$&one-parameter
groups in $\PSL(2,\dopp{R})^\sim$&section
\ref{cha:cospa}.\ref{sec:introcongeo}\\ 
$P,K$&generators of one-parameter groups& section
\ref{cha:cospa}.\ref{sec:introcongeo}\\ 
$L_0$&conformal Hamiltonian& section
\ref{cha:cospa}.\ref{sec:introcongeo}\\
\\ 
$I\Subset\Seins$, $I'$&proper interval in $\Seins$, its causal
complement&section 
\ref{cha:cospa}.\ref{sec:introcongeo}\\
$\Seins_+$, $\Seins_-$&upper and lower half-circle&section
\ref{cha:cospa}.\ref{sec:introcongeo}\\
$\dopp{C}_1$&phases in $\dopp{C}$&section
\ref{cha:cospa}.\ref{sec:introcongeo}\\
\\
$\lok{A}\subset\lok{B}$&chiral subnet&page \pageref{def:chsubnet}\\
$\lok{A}$&net of subalgebras \& its global algebra&pages \pageref{def:chsubnet}, \pageref{ind:globalg}\\
$e_\lok{A}$&cyclic projection of subnet $\lok{A}\subset\lok{B}$&page \pageref{ind:eA}\\
$U^\lok{A}$, $U^{\lok{A}'}$&inner implementations for $\lok{A}$,
$\lok{A}'$&section  \ref{cha:cospa}.\ref{sec:bosug}\\
\\
$\lok{A}\coset\lok{C}\subset\lok{B}$&\Name{Coset} pair&page
\pageref{def:cospa}\\ 
$\lok{A}\otimes\lok{C}$&vacuum representation of a \Name{Coset} pair&
page \pageref{ind:ACtimes}\\
$\lok{C}_I$& local relative commutant& page
\pageref{def:cospa}\\ 
$\lok{C}_{max}$&maximal \Name{Coset} model&page \pageref{lem:cosmax}\\
$\lok{A}_{max}$&maximal covariance extension of
$\lok{A}\subset\lok{B}$&page \pageref{eq:Amax}\\
\\
$ \Loop{G}_k$&model of loop group $ \Loop{G}$ at level
$k$&page \pageref{ind:loop}\\
$\Vir_c$&model of a stress-energy tensor& page \pageref{ind:virc}\\
$\Diff_I(\Seins)$&subgroup of localised diffeomorphisms& page
\pageref{ind:locdiff}\\ 
\\
$\closure{\Phi(f)}$
&closure of smeared quantum field ${\Phi(f)}$&\\ 
$\widetilde{f}$, $\widehat{g}$&transformations on test functions&page
\pageref{eq:licocompchii}\\ 
% &&\pageref{}\\
 \end{tabular}
\end{minipage}

\clearpage

\chapter*{Danksagungen}
  \label{chap-acknowledgements}
\thispagestyle{empty}
\addcontentsline{toc}{chapter}{{Danksagungen}}

Ich m\"ochte Herrn Professor K.-H.\ Rehren f\"ur die Gelegenheit danken,
 eine interessante Fragestellung der
mathematischen Physik zu untersuchen, f\"ur viele n\"utzliche
Diskussionen, die wir  miteinander hatten, und f\"ur seine
best\"andige und inspirierende Unterst\"utzung. Ich hoffe, dass diese
Arbeit seiner 
Betreuung Ehre macht. 

 Herrn Professor D.\ Buchholz m\"ochte ich Dank sagen f\"ur stete
 Ermunterungen und f\"ur seine Bereitschaft,  als Referent zur
Verf\"ugung zu stehen. 

Weiterhin danke ich dem Evangelischen Studienwerk Villigst f\"ur seine
Unterst\"utzung durch ein Promotionsstipendium im Rahmen des
F\"orderschwerpunktes \glqq{}Wechselwirkung\grqq{} und der
Berliner-Ungewitter-Stiftung f\"ur die Gew\"ahrung von Reisemitteln. 

Meine Arbeit und ich verdanken allen Angeh\"origen des Institutes
f\"ur Theoretische Physik der Universit\"at G\"ottingen und dem guten
Arbeitsklima gerade in der Arbeitsgruppe \glqq{}Quantenfeldtheorie und
Statistische Mechanik\grqq{} viele Anregungen und unverzichtbare
Unterst\"utzung.    

Schlie{\ss}lich w\"aren all diejenigen zu nennen, die mich in den
Stand gesetzt und im Stande erhalten haben, diese Arbeit zu
verfertigen. Ich kann sie hier gar nicht alle nennen und ihnen an dieser
Stelle angemessen danken, vor allem  denjenigen nicht, denen ich
besonders viel zu verdanken habe -- sie werden es wissen.   

\clearpage
\thispagestyle{empty}

\vspace*{\fill}
\vfill

\mbox{$\quad$}

\vspace*{\fill}
\vfill

\clearpage

\chapter*{Lebenslauf}
  \label{chap-currvit}
\thispagestyle{empty}
%\addcontentsline{toc}{chapter}{{Acknowledgements}}

{\bf S\"oren K\"oster}

\noindent wurde geboren am 2. November 1968 in Aachen \newline 
und ist seitdem deutscher Staatsangeh\"origer. 

\vspace{5ex}

\hspace{-2em}\begin{minipage}[t]{14cm}
  \begin{tabular}[t]{ll}
Juni 1988 & Abitur an der Johannes-Brahms-Schule in Pinneberg\\
\\
Oktober 1990 -- &Studium der Physik an der Universit\"at
Hamburg\\
\phantom{O} September 1993 & \\
\\
August 1992 & Vordiplom Physik\\
\\
Oktober 1993 -- & ``Advanced course M.Sc. in elementary particle
theory''\\
\phantom{O}September 1994 & an der University of Durham (UK)\\
\\
Dezember 1994 & Verleihung M.Sc.\\ 
\\
Oktober 1994 -- & Studium der Physik an der Universit\"at
Hamburg\\
\phantom{O}September 1997\\ 
\\
April 1998 -- & Aufbaustudium Physik an der\\
&Georg-August-Universit\"at G\"otingen,\\
\phantom{O}M\"arz 2003 &Bearbeitung des Promotionsthemas\\
&\glqq{}Struktur von
\Name{Coset}-Theorien\grqq{}\\
& unter der Betreuung von Prof. K.-H.~ Rehren. 
 \end{tabular}
\end{minipage}

\clearpage

\end{document}